\documentclass[aps,prc,onecolumn,amsmath,floatfix,nofootinbib]{revtex4-2}
\usepackage{graphicx}
\usepackage{physics}
\usepackage[urlcolor=green]{hyperref}
\usepackage{amsfonts}

\newcommand{\srap}{y}
\newcommand{\D}{\mathrm{D}}
\renewcommand{\d}{\textrm{d}}
\newcommand{\sNN}{s_{_{N\hspace{-.1em}N}}}

\begin{document}
\title{Statistical analysis of initial state and final state response in heavy-ion collisions}

\author{Nicolas Borghini} \email{borghini@physik.uni-bielefeld.de}
\author{Marc Borrell} \email{marcborrell@physik.uni-bielefeld.de}
\author{Nina Feld} \email{nkersting@uni-bielefeld.de}
\author{Hendrik Roch} \email{hroch@physik.uni-bielefeld.de}
\author{S\"oren Schlichting} \email{sschlichting@physik.uni-bielefeld.de}
\author{Clemens Werthmann} \email{cwerthmann@physik.uni-bielefeld.de}

\affiliation{Fakult\"at f\"ur Physik, Universit\"at Bielefeld, D-33615 Bielefeld, Germany}

\begin{abstract}
We develop a general decomposition of an ensemble of initial density profiles in terms of an average state and a basis of modes that represent the event-by-event fluctuations of the initial state. 
The basis is determined such that the probability distributions of the amplitudes of different modes are uncorrelated.
Based on this decomposition, we quantify the different types and probabilities of event-by-event fluctuations in Glauber and Saturation models and investigate how the various modes affect different characteristics of the initial state. 
We perform simulations of the dynamical evolution with K\o MP\o ST and MUSIC to investigate the impact of the modes on final-state observables and their correlations. 
\end{abstract}

\maketitle

\tableofcontents

\section{Introduction}
\label{sec:intro}

Experiments with relativistic heavy-ion collisions at facilities like the Relativistic Heavy Ion Collider and the Large Hadron Collider provide the opportunity to study deconfined QCD matter  which is dynamically evolving in an out-of-equilibrium state.
Over the last years it has become emergent that the bulk dynamics of the evolving QCD matter can be described by relativistic dissipative fluid dynamics~\cite{Teaney:2009qa,Busza:2018rrf,Romatschke:2017ejr}.

Evidently, the description of the Quark Gluon Plasma (QGP) as a relativistic viscous fluid requires the knowledge of an initial condition for fluid-dynamical fields, as well as the QCD equation of state and transport coefficients to close the system of equations. While the QCD equation of state can be obtained from lattice QCD simulations~\cite{Borsanyi:2010cj,Borsanyi:2013bia,HotQCD:2014kol,Bazavov:2017dsy}, a first-principles calculation of QCD transport coefficients represents an outstanding theoretical challenge, as does the calculation of the initial energy deposition in high-energy heavy-ion collisions. 

Generally, this so-called initial state in heavy-ion collisions is obtained from theoretical models~\cite{Miller:2007ri,Loizides:2014vua,Broniowski:2007nz,Drescher:2006ca,Schenke:2012wb,Paatelainen:2012at,Moreland:2014oya,Giacalone:2019kgg}, which ultimately provide profiles of the hydrodynamic fields that fluctuate on an event-by-event basis, with a varying degree of sophistication and rooting in the underlying theory of QCD. During the early pre-equilibrium stage, where the system approaches local thermal equilibrium to assure the subsequent applicability of fluid dynamics, these initial-state fluctuations get modified, which despite the short duration of this period of $\lesssim 1\;\textrm{fm}/c$ can have an impact on the observables in the final state \cite{Schlichting:2019abc,Gale:2021emg,Giacalone:2019ldn}. Subsequently, over the course of the nonlinear hydrodynamic evolution, the fluctuations in the initial state can again either be washed out or intensified~\cite{Heinz:2013th,Luzum:2013yya}, before eventually leaving an experimentally observable imprint on the final-state observables. 

Since neither the QCD transport properties nor the initial state can be directly inferred from experimental observations, it has thus become customary to simultaneously extract properties of the initial state and QCD transport properties from statistical model/data comparisons~\cite{Novak:2013bqa,Sangaline:2015isa,Bernhard:2016tnd,Bernhard:2019bmu,Devetak:2019lsk,Nijs:2020ors,Nijs:2020roc}. However, it is intuitively clear and empirically proven, that in such global analyses, the different aspects are strongly correlated~\cite{Bernhard:2016tnd,Sangaline:2015isa} and different assumptions about the properties of the initial state can lead to different extractions of QCD transport properties~(see, e.g., Ref.~\cite{Nijs:2022rme} for a recent example).

By now there exist a sizable number of different initial-state models, which are based on different underlying degrees of freedom, ranging from models derived within effective theories of high-energy QCD, such as IP-Glasma~\cite{Schenke:2012wb,Schenke:2012hg} or EKRT~\cite{Paatelainen:2013eea,Niemi:2015qia}, to purely parametric models, such as the Monte Carlo Glauber model~\cite{Miller:2007ri} or T$_\textrm{R}$ENTo~\cite{Moreland:2014oya}. However, despite this plethora of choice, up to now there exists no systematic framework for the general characterization of these models. 
While for some specific observables, such as, e.g., the flow harmonics $v_n$~\cite{Voloshin:1994mz,Bhalerao:2020ulk}, characterizations of the initial state or initial-state estimators of these quantities, such as the eccentricities $\varepsilon_{n}$ have been empirically derived~\cite{Ollitrault:1992bk,Alver:2010gr,Teaney:2010vd,Gardim:2011xv,Teaney:2012ke,Niemi:2012aj,Plumari:2015cfa,Noronha-Hostler:2015dbi}, it is often times not clear what aspect of a particular model is favored or disfavored by certain experimental observations~\cite{*[{See, e.g., the study of the sensitivity of various observables to the granularity of the initial state in }] [{}] Gardim:2017ruc}.

Beyond such event-by-event hydrodynamic simulations of final-state observables from fluctuating initial states, some effort has been invested to describe the initial state in terms of fluctuations around an average profile~\cite{Floerchinger:2013rya,Floerchinger:2013vua,Floerchinger:2013hza,Floerchinger:2013tya,Floerchinger:2014fta,Floerchinger:2018pje,Floerchinger:2020tjp}, which allows to scrutinize the effects of different kinds of initial-state fluctuations on final-state observables~\cite{Mazeliauskas:2015vea}. While this mode-by-mode approach can also reduce the computational cost of the analysis, in practice it is typically limited to the linear response, and has not been systematically tested against event-by-event simulations.

The central objective of this paper is to develop an optimal decomposition of the initial state into fluctuations around an average background, which can serve as a systematic framework for the characterization of the initial-state and final-state response in high-energy heavy-ion collisions. By performing a diagonalization of the covariance matrix of different fluctuating modes, this framework serves to quantify the structure and statistical importance of different types of fluctuations. Since the construction of this basis is in a certain sense optimized for the mode-by-mode approach, it also allows us to calculate the linear and quadratic response of initial- and final-state observables. We will exemplify the use of this framework with the example of two different initial-state models --- a Glauber model and a Saturation model --- which further gives us the possibility to study their differences within our framework. By performing state-of-the-art dynamical simulations in K{\o}MP{\o}ST~\cite{Kurkela:2018vqr} and MUSIC~\cite{Schenke:2010nt,Schenke:2010rr,Paquet:2015lta}, we further investigate the impact of the mode-by-mode linear and quadratic response on final-state observables, and further compare the results of the mode-by-mode approach to the more commonly used event-by-event simulations.

This paper is organized as follows.
In Sec.~\ref{sec:stat_caracterization_IS} we introduce the theoretical framework of the density-matrix formalism, as well as the two models used in the following.
Then we present the results of the statistical characterization of the background and the fluctuating modes and possible characterization schemes thereof.
In Sec.~\ref{sec:response} we introduce the (non-)linear response theory for the fluctuation modes and our simulation setup for the dynamical system evolution. 
We present our results for the linear and quadratic mode-by-mode response, in particular the responses of the flow coefficients to the initial-state eccentricities. 
In Sec.~\ref{sec:fluctuations_observables} we turn to the variances and correlations of different characteristics.
We introduce a prediction of joint probability distributions for observables using a Gaussian statistics ansatz within the mode-by-mode approach and compare it to the event-by-event results.
In Sec.~\ref{sec:conclusions} we draw our conclusions.


\section{Statistical characterization of the initial state}
\label{sec:stat_caracterization_IS}

Below we introduce our decomposition of initial states in terms of an average state and fluctuation modes, and we apply the decomposition to initial states from two different models. 
The construction of a basis of uncorrelated fluctuation modes about an average event, starting from a random sample of events, is explained in Sec.~\ref{subsec:theory_basis}. 
We then describe in Sec.~\ref{subsec:models} the two initial-state models that will be used as illustration in this paper, namely a Monte-Carlo Glauber model and a Saturation model.
For both models and collisions at fixed parameter, namely, either $b=0$ or $b=9$~fm,  we present in Sec.~\ref{subsec:results_stat_charact_IS} the respective average states and fluctuation modes.
A few global characteristics of the latter are then introduced and discussed (Sec.~\ref{subsec:classification_modes}), first in terms of eccentricities and angular-integrated radial profiles, then with a Bessel--Fourier decomposition.

\subsection{Mode decomposition of the initial state}
\label{subsec:theory_basis}

The starting point of our analysis is a set of $i=1,\dots,N_\mathrm{ev}$ configurations $\{\Phi^{(i)}\}$ from a given initial-state model, that includes fluctuations on an event-by-event basis.
The configurations $\Phi^{(i)}$ may for instance be energy or entropy density profiles, i.e., a function of the position $\textbf{x}$, corresponding either to events at a fixed impact parameter $b$ or within a certain centrality class. 
We show that one can introduce an average configuration $\bar{\Psi}$ and appropriate set of modes $\{\Psi_l\}$ such that every configuration $\Phi^{(i)}$ can be written as the sum of $\bar{\Psi}$ and a linear combination of the modes:
\begin{equation}
\label{eq:evt_vs_modes}
\Phi^{(i)}(\textbf{x}) = \bar{\Psi}(\textbf{x}) + \sum_l c_l^{(i)} \Psi_l(\textbf{x}),
\end{equation}
with expansion coefficients $\{c_l^{(i)}\}$ that are realizations of centered, uncorrelated random variables with unit variance, i.e.,
\begin{gather}
\expval{c_l} = 0, \label{eq:<c_l>=0} \\
\expval{c_l c_{l'}} = \delta_{ll'}, \label{eq:<c_lc_l'>}
\end{gather}
where $\expval{\cdots}$ denotes a statistical average over events. 
For brevity we shall omit the position variable $\textbf{x}$ in the remainder of this section. 
We assume for simplicity that the $\{\Phi^{(i)}\}$ are real-valued, as holds for thermodynamical densities. 

The ``background'' or ``average event'' is most naturally defined by averaging the individual configurations $\{\Phi^{(i)}\}$ over the $N_\mathrm{ev}$ events, i.e.,
\begin{equation}
\bar{\Psi} \equiv \frac{1}{N_\mathrm{ev}} \sum_{i=1}^{N_\mathrm{ev}} \Phi^{(i)}.
\label{eq:avg_state}
\end{equation} 
The  ``events'' $\{\Phi^{(i)}\}$ can be viewed as elements of a Hilbert space of functions of $\textbf{x}$,\footnote{The inner product is the usual one for square-integrable functions, namely, their overlap integral over the whole space.} on which we consider an arbitrary orthonormal basis $\{\tilde{\chi}_l\}$. 
In principle, the Hilbert space is infinite-dimensional, but in practice we shall discretize the energy density on a finite grid, in which case the Hilbert space is finite-dimensional. 
In that case, we consider the trivial basis of grid points, that is $\tilde{\chi}_l(\textbf{x}) = \delta_{l,\textbf{x}}$, with the standard inner product $\sum_\textbf{x}f(\textbf{x})g(\textbf{x})$.
Each event, or more precisely its deviation from the average event, can be decomposed over that basis
\begin{equation}
\label{eq:evt_vs_chi_l}
\Phi^{(i)} = \bar{\Psi} + \sum_{l} \tilde{c}_{l}^{(i)} \tilde{\chi}_l.
\end{equation}
Clearly, the expansion coefficients obey
\begin{equation}
\frac{1}{N_\mathrm{ev}}\sum_{i=1}^{N_\mathrm{ev}}\tilde{c}_{l}^{(i)}=0.
\label{eq:average_cl}
\end{equation}

If we now view the events $\{\Phi^{(i)}\}$ as equiprobable states, then for every $l$ the coefficients $\{\tilde{c}_{l}^{(i)}\}$ can be interpreted as realizations of a random variable $\tilde{c}_l$, with $\expval{\tilde{c}_l} = 0$ thanks to Eq.~\eqref{eq:average_cl}.
To those equiprobable states one can associate a density matrix
\begin{equation}
\frac{1}{N_\mathrm{ev}}\sum_i  \Phi^{(i)} \Phi^{(i)\mathsf{T}},
\end{equation}
which according to Eq.~\eqref{eq:evt_vs_chi_l} and \eqref{eq:average_cl} equals
\begin{equation}
\frac{1}{N_\mathrm{ev}}\sum_i \Phi^{(i)} \Phi^{(i)\mathsf{T}} = 
\bar{\Psi}\bar{\Psi}^\mathsf{T} + \sum_{l,l^\prime} \langle \tilde{c}_{l} \tilde{c}_{l^\prime} \rangle \tilde{\chi}_l\tilde{\chi}_{l^\prime}^\mathsf{T}.
\end{equation}
To obtain a density matrix $\rho$ reflecting solely the fluctuations about the average event, we consider
\begin{equation}
\rho \equiv \frac{1}{N_\mathrm{ev}}\sum_i  \Phi^{(i)} \Phi^{(i)\mathsf{T}} - \bar{\Psi}\bar{\Psi}^\mathsf{T} = 
\sum_{l,l^\prime} \langle \tilde{c}_{l} \tilde{c}_{l^\prime} \rangle \tilde{\chi}_l\tilde{\chi}_{l^\prime}^\mathsf{T}.
\label{eq:rho}
\end{equation}

Next, we diagonalize the density matrix
\begin{equation}
\label{eq:diagonalization}
\rho \tilde{\Psi}_l = \lambda_{l} \tilde{\Psi}_l
\end{equation}
and sort the (orthonormal) basis vectors $\{\tilde{\Psi}_l\}$ according to the magnitude of their eigenvalues $\lambda_{l}$.
The latter quantifies the relative weight of the contribution of $\tilde{\Psi}_l$ to the $N_\textrm{ev}$ random events. 
The spectral decomposition of the density matrix thus reads
\begin{equation}
\rho = \sum_l \lambda_l  \tilde{\Psi}_l  \tilde{\Psi}_l^\mathsf{T}.
\label{eq:diagonalized}    
\end{equation}
Comparing Eqs.~\eqref{eq:rho} and \eqref{eq:diagonalized}, we see that choosing the basis of eigenvectors $\{\tilde{\Psi}_l\}$ of $\rho$ as arbitrary basis $\{\tilde{\chi}_l\}$ in decomposition~\eqref{eq:evt_vs_chi_l}, the random variables defined by the expansion coefficients obey 
\begin{equation}
\label{eq:<tilde(c)_l.tilde(c)_l'>}
\expval{\tilde{c}_{l} \tilde{c}_{l'} } = \lambda_l \delta_{ll'}
\end{equation}
for all $l$ and $l'$, in addition to the property $\expval{\tilde{c}_l} = 0$ mentioned above. 

As final step, we rescale each eigenvector of $\rho$ by a factor determined by the associated eigenvalue:
\begin{equation}
\Psi_l\equiv \sqrt{\lambda_l}\tilde{\Psi}_l,
\label{eq:renormeigenvectors}
\end{equation}
such that the ``modes'' $\{\Psi_l\}$ form an orthogonal basis but are not uniformly normalized to unity. 
Physically they represent the fluctuation modes contributing to initial-state configurations, and the normalization $\sqrt{\lambda_l}$ measures by how much $\tilde{\Psi}_l$ typically contributes to a configuration.
Redefining in parallel the expansion coefficients as
\begin{equation}
c_l \equiv \frac{ \tilde{c}_l}{\sqrt{\lambda_l}},
\end{equation}
so that the decomposition~\eqref{eq:evt_vs_chi_l} of the $i$th event takes the form~\eqref{eq:evt_vs_modes}, the property $\expval{\tilde{c}_l} = 0$ becomes at once Eq.~\eqref{eq:<c_l>=0}, while relation~\eqref{eq:<tilde(c)_l.tilde(c)_l'>} yields Eq.~\eqref{eq:<c_lc_l'>}.
Thus we have attained our goal. 

We note that in practice, one can directly process a set of ``events'' --- in the following energy density profiles --- to compute the average state~\eqref{eq:avg_state} and the density matrix via Eqs.~\eqref{eq:avg_state} and \eqref{eq:rho}. Subsequently, the density matrix is diagonalized, from where one obtains the modes $\{\Psi_l\}$ and the eigenvalues $\lambda_l$. 
We further note that the average event is not part of the basis of modes, which means that $\bar{\Psi}$ can be decomposed over this basis if needed.

In the present paper we only investigate energy density profiles in the transverse plane to reduce the numerical complexity.
It is a straightforward procedure to extend the decomposition method to a third spatial dimension just by introducing an additional index in the numerical computation of the average state and the density matrix for the longitudinal direction.
However, this increases substantially the amount of computing power required.
The current derivation of our decomposition method includes a single scalar field density for the initial state.
From a purely hydrodynamic point of view, one could also consider initial conditions involving the entire energy-momentum tensor in the same manner as a third dimension by introducing more indices in the numerical computation for the different tensor entries.
Any nontrivial rotational properties of the latter quantities do not require modifications of the decomposition, because they emerge naturally in the procedure.

\subsection{Models}
\label{subsec:models}

While the mode decomposition in Eqs.~\eqref{eq:evt_vs_modes}--\eqref{eq:<c_lc_l'>} is completely general, we now apply it to the study of profiles of the initial energy density in the transverse plane of a heavy-ion collision. We illustrate this at the example of two different models for the initial state of Pb--Pb collisions at $\sqrt{\sNN}=5.02$ TeV, namely a Glauber model and a Saturation model, which we now briefly introduce.  

Both models rely on a Monte Carlo (MC) sampling of the Pb nuclei from a spherically symmetric Woods--Saxon distribution~\cite{Woods:1954zz} with half-density radius $R=6.62$~fm and diffusivity $a=0.546$~fm.
The two nuclei generated for one Pb--Pb event are then shifted by the impact parameter of the collisions, which is in all our simulations oriented along the $x$-axis and has a fixed value, mostly either $b=0$ or $b=9$~fm.
The center of each nucleus is defined by the center of mass of the resulting nucleon configuration.
In this exploratory study we consider fixed impact parameters to study the effects of a rotationally symmetric average state and one with broken rotational symmetry. It is also straightforward to apply the same procedure on centrality-selected events.

With these MC models we produce two-dimensional energy-density profiles $e(\textbf{x})$, that are suitable as initialization for longitudinally boost invariant dynamics as will be presented in Sec.~\ref{sec:response}. 
For the present exploratory study we did not try to optimize the choice of parameters of the models such that they yield the same value for a given quantity under identical conditions, e.g., the same average total energy or multiplicity in collisions at $b=0$.

\subsubsection{Glauber model}
\label{subsubsec:Glauber_model}

We use a MC Glauber model in which we sample the positions of nucleons. 
To mimic their Fermi repulsion, we implemented a minimum nucleon separation of 0.4~fm. 

To generate the energy density profiles in the overlap region of the two sampled Pb nuclei, we compute the local number $N_\mathrm{part}(\textbf{x})$ of participants and the local number $N_\mathrm{coll}(\textbf{x})$ of binary collisions.
For the collision between two nucleons we use a geometric criterion, namely the distance between their transverse positions has to be less than $\sqrt{\sigma_\mathrm{inel}^{_{N\hspace{-.1em}N}}/\pi}$. 
Since in this paper we only consider nuclear collisions at $\sqrt{\sNN}=5.02$~TeV, we take $\sigma_\mathrm{inel}^{_{N\hspace{-.1em}N}}= 67.6$~mb for the inelastic nucleon-nucleon cross section.

The values of the numbers of participants and binary collisions are stored on a two-dimensional grid with a spacing of 0.1~fm. 
For $N_\mathrm{part}(\textbf{x})$ we use the grid points closest to the respective positions of the nucleon centers, while for $N_\mathrm{coll}(\textbf{x})$ the grid point closest to the halfway point between the two nucleons is used.
In a second step, the initial energy density profile is assumed to be given by a linear combination of contributions from soft processes, represented by $N_\mathrm{part}(\textbf{x})$ and hard processes, represented by $N_\mathrm{coll}(\textbf{x})$~\cite{dEnterria:2020dwq}:
\begin{equation}
e_\textrm{d}(\textbf{x}) \propto (1-\alpha)\frac{N_\mathrm{part}(\textbf{x})}{2} + \alpha N_\mathrm{coll}(\textbf{x}),
\label{eq:Glauber1}
\end{equation}
with a fraction from binary scatterings $\alpha = 0.2$.
To obtain a smooth profile, we redistribute the energy density $e_{\d}$ at each grid point $(x_i,y_j)$ in its vicinity using a Gaussian smearing $\propto \textrm{e}^{[(x-x_i)^2+(y-y_j)^2]/2\sigma^2}$ with width $\sigma = 0.4$~fm~\cite{Holopainen:2012id}.
The overall normalization factor ($1246\;\mathrm{GeV}/\mathrm{fm}^2$) of the energy density in Eq.~\eqref{eq:Glauber1} was roughly matched to obtain the charged hadron multiplicity at midrapidity for central events.
This gives us our profile $e(\textbf{x})$, which we identify with the energy density at midrapidity
\begin{equation}
e(\textbf{x}) \equiv 
\left.\frac{\d E}{\tau_0\,\d^2\textbf{x}\,\d\srap}\right\vert_{\srap=0} 
\label{eq:Glauber2}
\end{equation}
where $\tau_0$ is the initialization time of the system, which later on in Sec.~\ref{sec:response} will be the starting time of the K\o MP\o ST evolution, and $\srap$ denotes the spatial rapidity.

\subsubsection{Saturation model}
\label{subsubsec:Saturation_model}

The second model that we consider for the initial state is based on the color glass condensate (CGC) effective field theory for QCD at high energies~\cite{Gelis:2010nm,Iancu:2003xm}.
To compute the initial energy deposition in the collision between nuclei $A$ and $B$, we start from the $k_\mathrm{T}$-factorization formula~\cite{Lappi:2017skr,Blaizot:2010kh}
\begin{equation}
\frac{\d N_g}{\d^2\textbf{x}\,\d^2\textbf{P}\,\d Y\,\d\srap} = 
\frac{g^2N_c}{4\pi^5\textbf{P}^2(N_c^2-1)} \delta(Y-\srap) 
\int\!\frac{\d^2\textbf{k}}{(2\pi)^2}\,\Phi_A\!\left(\textbf{x}+\frac{\textbf{b}}{2},\textbf{k}\right) \Phi_B\!\left(\textbf{x}-\frac{\textbf{b}}{2},\textbf{P}-\textbf{k}\right)
\quad 
\label{eq:gluon_spectrum}
\end{equation}
for the initial transverse momentum ($\bf P$) spectrum of gluons produced per unit  rapidity ($Y$) at transverse position $\bf x$.
Here $g$ is the Yang--Mills coupling, $N_c=3$ the number of colors, $\textbf{b}$ the  impact parameter of the nucleus-nucleus collision, and $\Phi_{A/B}(\textbf{x},\textbf{k})$ the unintegrated gluon distribution of nucleus $A$ or $B$.
Within the Golec-Biernat and W\"usthoff (GBW) model~\cite{Golec-Biernat:1998zce} the latter is parameterized as
\begin{equation}
\Phi_{A/B}(\textbf{x},\textbf{k}) = 4\pi^2\frac{N_c^2-1}{g^2N_c} 
\frac{\textbf{k}^2}{Q^2_{s,A/B}}\,\textrm{e}^{-\textbf{k}^2/Q^2_{s,A/B}},
\label{eq:unintegrated_gluon_dist}
\end{equation}
where $Q^2_{s,A/B} = Q^2_{s,A/B}(x,\textbf{x})$ is the (adjoint) saturation scale~\cite{Albacete:2014fwa} for the nucleus, which depends on the transverse position $\textbf{x}$ and the longitudinal momentum fraction $x\equiv |\textbf{P}|\,\textrm{e}^{\pm Y}/\sqrt{\sNN}$.
From Eqs.~\eqref{eq:gluon_spectrum} and~\eqref{eq:unintegrated_gluon_dist} we can analytically derive the gluon spectrum within the GBW model and compute the initial transverse energy density per unit rapidity in a heavy-ion collision via
\begin{equation}
[e(\textbf{x})_{}\tau]_0 = 
\int\!\d Y\!\int\!\d^2\textbf{P}\,\abs{\textbf{P}} \frac{\d N_g}{\d^2\textbf{x}\,\d^2\textbf{P}\,\d Y\,\d\srap}.
\end{equation}
By assuming that the energy density is dominated by $|\textbf{P}|\simeq Q_{s,A/B}$, i.e.
\begin{equation}
x = \frac{Q_{s,A/B}(x,\textbf{x})\,\textrm{e}^{\pm Y}}{\sqrt{\sNN}},
\label{eq:x_QsAB_rel}
\end{equation}
the integrals in the GBW model can be evaluated analytically and yield
\begin{equation}
[e(\textbf{x})_{}\tau]_0 =  \frac{N_c^2-1}{4g^2N_c\sqrt{\pi}}
\frac{Q^2_{s,A}Q^2_{s,B}}{(Q^2_{s,A}+Q^2_{s,B})^{5/2}}\left[2Q^4_{s,A}+7Q^2_{s,A}Q^2_{s,B}+2Q^4_{s,B}\right]\!.
\label{eq:edens_saturation}
\end{equation}
We can then compute the energy density at each point in the transverse plane from the saturation scales of the nuclei at the given position.

We parameterize the latter as
\begin{equation}
Q^2_{s,A/B}(x,\textbf{x}) = Q^2_{s,p}(x)\,\sigma_0\,T_{A/B}(\textbf{x}),
\label{eq:sat_scale_nucleus}
\end{equation}
where $\sigma_0\,T_{A/B}(\textbf{x})$ effectively counts the number of nucleons at transverse position $\bf x$, with $T_{A/B}(\textbf{x})$ the nuclear thickness function and $\sigma_0=2\pi B_G$ where the nucleon size $B_G=4\;\mathrm{GeV}^{-2}$ is determined from fits~\cite{Kowalski:2003hm,Rezaeian:2012ji} to HERA data.
For the thickness function we use a MC Glauber sampling of nucleon positions $\textbf{x}_i$ inside each nucleus and summing over all nucleons we compute
\begin{equation}
T_{A/B}(\textbf{x}) = \sum_{i\in A/B}\! T_p(\textbf{x}-\textbf{x}_i),
\end{equation}
with a Gaussian proton thickness function $T_p(\textbf{x})$~\cite{McLerran:2015qxa}
\begin{equation}
T_p(\textbf{x}) = \frac{1}{2\pi B_G}\textrm{e}^{-\textbf{x}^2/2 B_G}.
\end{equation}

For the average saturation scale $Q^2_{s,p}(x)$ of the proton in Eq.~\eqref{eq:sat_scale_nucleus}, we use
\begin{equation}
Q_{s,p}^2(x) = Q_{s,0}^2\, x^{-\lambda}\,(1-x)^\delta
\label{eq:sat_scale_proton}
\end{equation}
with $Q_{s,0}^2 = 0.63\,\text{GeV}^2$, $\lambda=0.36$ and $\delta=1$. 
We solve Eqs.~\eqref{eq:x_QsAB_rel} and~\eqref{eq:sat_scale_nucleus} self-consistently in the limit $x\ll 1$, yielding
\begin{equation}
x = \left(\frac{Q^2_{s,0}\,\sigma_0T_{A/B}(\textbf{x})\,\textrm{e}^{\pm 2Y}}{\sNN}\right)^{\!\frac{1}{2+\lambda}},
\end{equation}
which can then be inserted in Eq.~\eqref{eq:sat_scale_nucleus} to obtain the saturation scale of the respective nucleus, and thereby the initial-state energy density in Eq.~\eqref{eq:edens_saturation}.
Since the energy density in Eq.~\eqref{eq:edens_saturation} is computed at leading order, we allow for an additional rescaling of the energy density of order one to reproduce the charged hadron multiplicity at midrapidity in central events.

\subsection{Mode decomposition for the Glauber and Saturation models}
\label{subsec:results_stat_charact_IS}

To apply the mode decomposition introduced in Sec.~\ref{subsec:theory_basis} to energy density profiles obtained within the models described in Sec.~\ref{subsec:models}, we generated for each model and for each impact-parameter value $N_\mathrm{ev} = 2^{21}$ random profiles. 
This large number of ``events'' allows us to reduce statistical uncertainties and also to better assess the possible degeneracy between modes with closely lying eigenvalues.

From each set of random events we computed the corresponding average state $\bar{\Psi}$ [Eq.~\eqref{eq:avg_state}] and density matrix $\rho$ [Eq.~\eqref{eq:rho}]. 
To decompose the latter on a finite basis, we introduce a new orthogonal spatial grid with $N_\textrm{s}^2 = 128\times 128$ sites with a spacing of 0.19 resp.\ 0.21~fm for the Glauber resp.\ Saturation model.\footnote{Due to the high computational demand for the generation of the profiles, especially for diagonalizing the density matrix, we decided to coarse grain the resolution of the Glauber model using a bilinear interpolation, and we set the resolution of the Saturation model accordingly.}
The density matrix $\rho$ is then constructed on the trivial orthonormal basis whose $N_\textrm{s}^2$ elements have a unit weight localized at a single grid site and vanish elsewhere, that is $\tilde{\chi}_l = \delta_{l,\textbf{x}}$.
Diagonalizing this $(N_\textrm{s}^2\times N_\textrm{s}^2)$-dimensional representation of $\rho$ is the most time-intensive step in the calculation of the modes $\{\Psi_l\}$, which are thereby determined together with their respective eigenvalues $\{\lambda_l\}$.

\begin{figure*}[!t]
	\includegraphics[width=\linewidth]{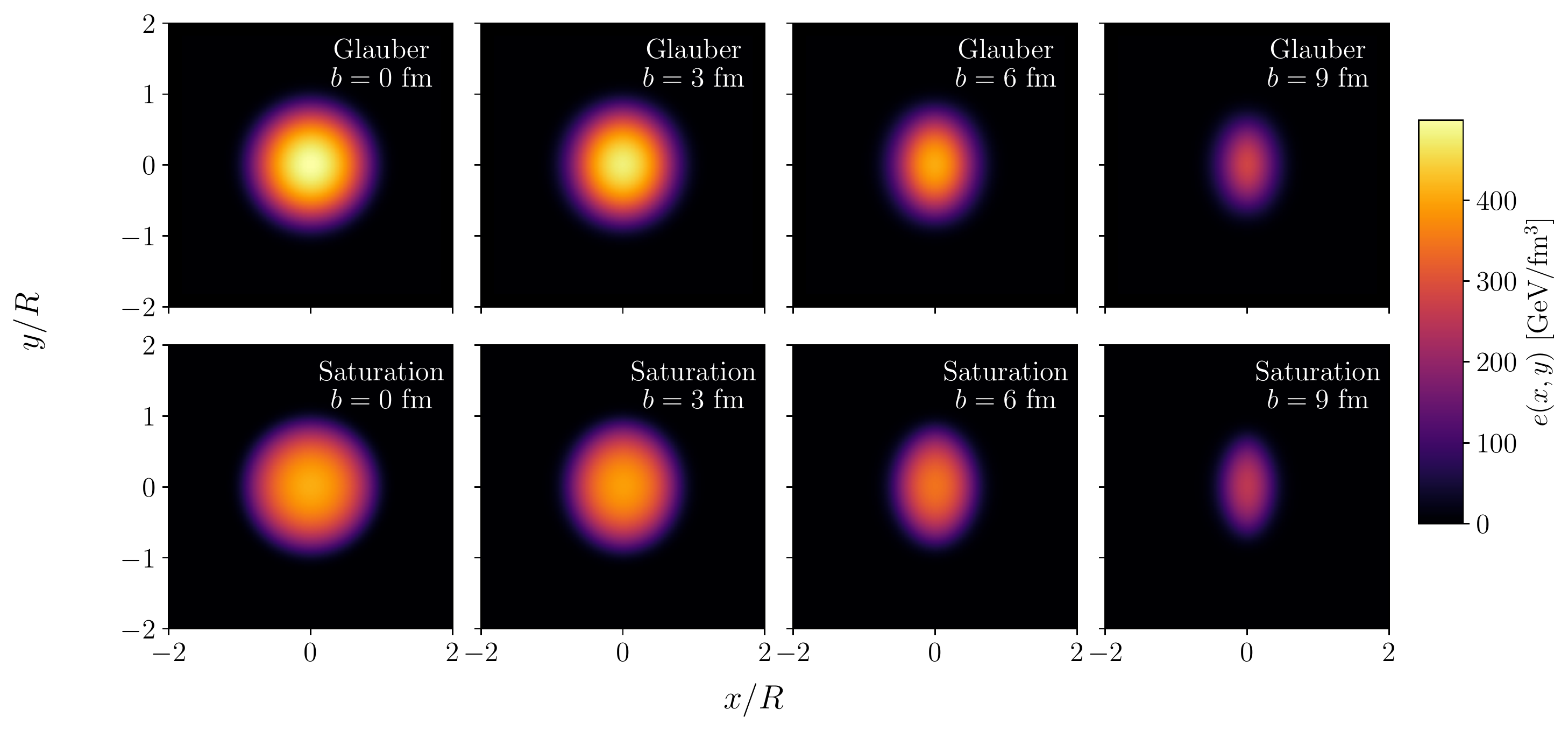}
	\vspace{-7mm}
	\caption{Energy density profiles of the average event $\bar{\Psi}$ for different impact parameters (from left to right: $b=0$, 3, 6, 9~fm) in the Glauber (top) and Saturation (bottom) models.}
	\label{fig:avg_sates_density}
\end{figure*}

In Fig.~\ref{fig:avg_sates_density} we show energy density profiles of the average event $\bar{\Psi}$, Eq.~\eqref{eq:avg_state}, computed for both models at four different values of the impact parameter.
We emphasize that in our study the impact parameter 
is always oriented along the $x$-direction. Alternatively, one could let the orientation vary randomly on an event-by-event basis, such that the average state would always have an azimuthal rotation symmetry. By fixing the direction of the impact parameter, the leading contributions to (anisotropic) flow observables are captured by the average state, such that the fluctuations on top of this state are small in size and perturbation theory for observables can be applied  without the need to go to high orders in the expansion. 
Conversely, in an expansion around an azimuthally symmetric average event for collisions at nonzero impact parameter, the large elliptic deformation of the initial profiles has to be entirely captured by the fluctuation modes, whose relative contribution to the total energy density of each individual event is thus much more important.
Accordingly, a decomposition around a rotationally symmetric average state would complicate the perturbation theory of the observables introduced later on.
Here and in later figures, the coordinates are given in units of the radius of the Pb nucleus used for nucleon sampling, i.e., $R=6.62$~fm.
As was to be expected, these profiles are azimuthally symmetric at $b=0$ and become more and more elliptic with increasing $b$. 
In fact, the average event seems to be more elliptic in the Saturation model than in the Glauber model, which will be confirmed hereafter.
The energy density at the center takes larger values in the Glauber model, yet this is not really significant since the parameters of the two models were not calibrated so as to yield equal results for a global quantity like the total energy of $\bar{\Psi}$ --- accordingly, when we compare the radial profiles in Sec.~\ref{subsec:classification_modes} below, we rescale the average events by their respective total energies.

\begin{figure*}[!t]
	\includegraphics[width=0.495\linewidth]{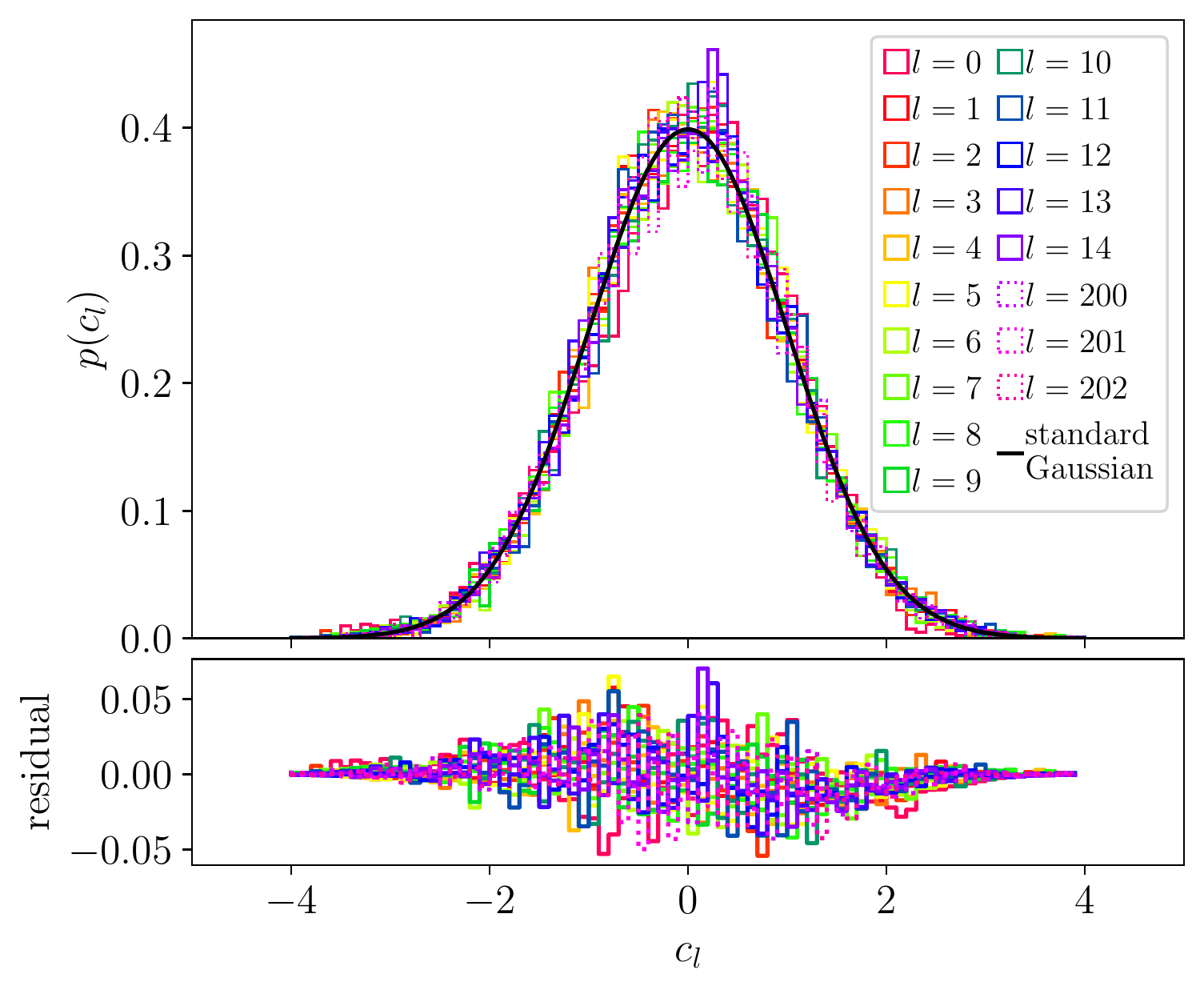}
	\includegraphics[width=0.495\linewidth]{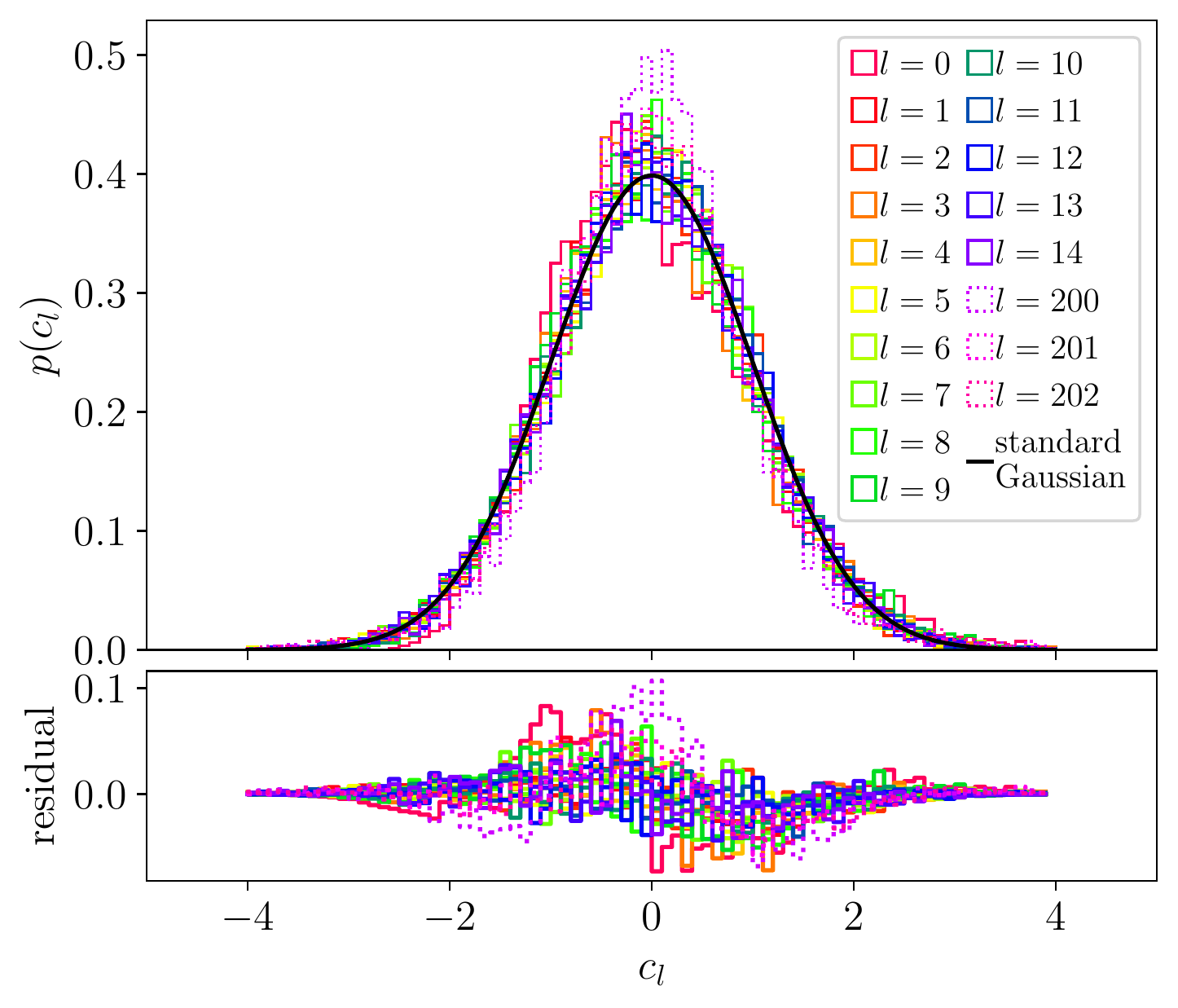}
	\vspace{-7mm}
	\caption{Relative frequency of the expansion coefficients $c_l$ (histograms) computed from 8192 events in the Glauber model at $b=0$ (left) and $b=9$~fm (right), compared with a standard Gaussian distribution (full black line).}
	\label{fig:cl_hist_Glauber}
\end{figure*}

Interestingly, the average event $\bar{\Psi}$ in the MC Glauber model at a given impact parameter is almost identical to the energy density profile given by an optical Glauber model with the same scaling with $N_\textrm{part}$ and $N_\textrm{coll}$ and the same parameters. 
A qualitative check done by plotting the two densities on top of each other reveals only small differences in the outer regions of the profile.
This was confirmed more quantitatively by a Bessel--Fourier expansion of the densities, which will be described in Sec.~\ref{subsubsection:Bessel_Fourier} in further detail.

Before we present the fluctuation modes and their eigenvalues, let us discuss the expansion coefficients $\{c_l\}$. 
We picked 8192 random events among the $N_\textrm{ev}$ used to determine the average event and the modes, and we decomposed them according to Eq.~\eqref{eq:evt_vs_modes}. 
For each value of $l\geq 0$, which labels the modes in order of decreasing eigenvalue $\lambda_l$, we thus obtained 8192 values of the expansion coefficients $c_l$. 
In Fig.~\ref{fig:cl_hist_Glauber} we show relative frequency histograms for the coefficients $c_l$ of the 15 modes ($0\leq l\leq 14$) with the largest eigenvalues and of a few higher modes ($l\in\{200,201,202\}$), computed for the Glauber model at impact parameters $b=0$ (left panel) and $b=9$~fm (right).\footnote{Results for the Saturation model are very similar and not shown.}
We also display a Gaussian distribution with unit variance for comparison. 

\begin{figure*}[!htb]
\includegraphics[width=\linewidth]{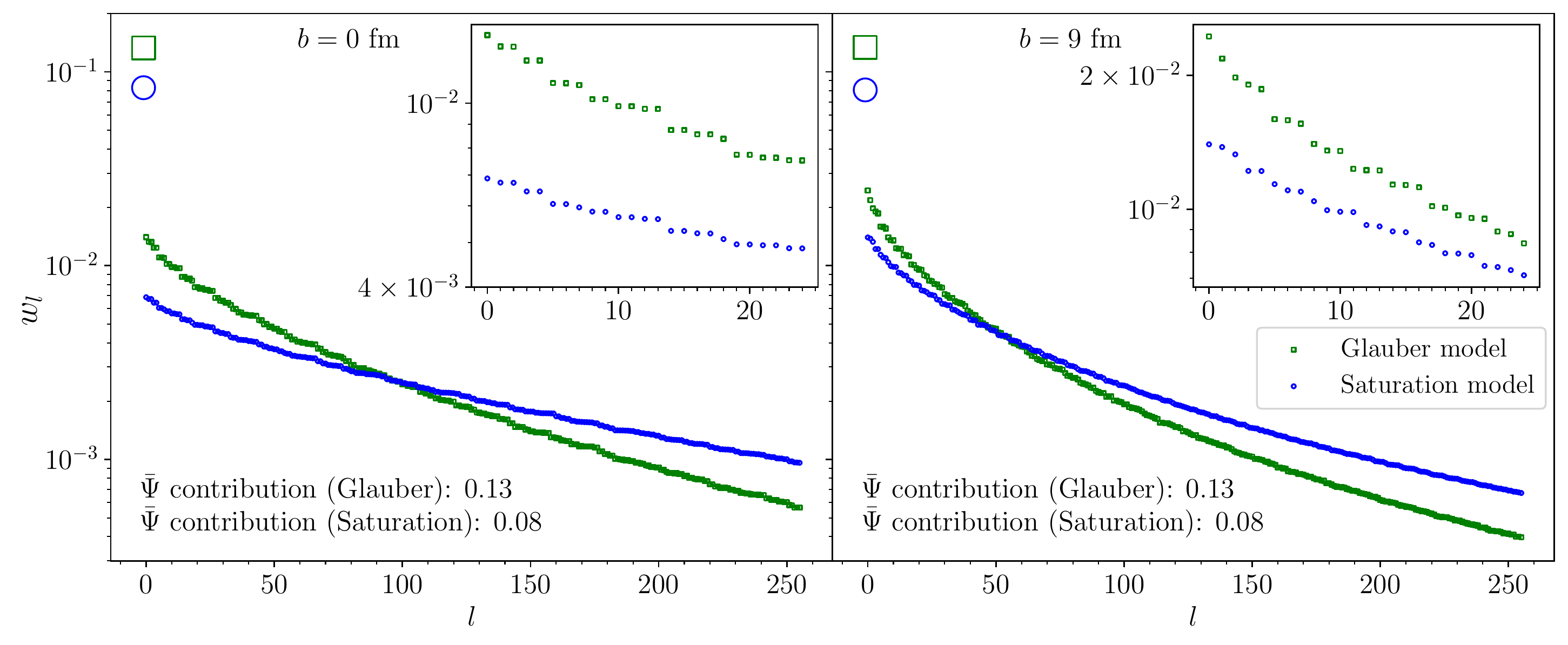}
\vspace{-7mm}
\caption{Relative weights~\eqref{eq:w_l} of fluctuation modes ($w_l$ for $0\leq l\leq 255$) and of the average event ($\bar{w}$, larger symbols at $l=-1$) at impact parameter $b=0$ (left) and $b=9$~fm (right) in the Glauber (squares) and the Saturation (circles) models.}
\label{fig:eigenvalues}
\end{figure*}

At vanishing impact parameter, the probability distributions $p(c_l)$ of all modes, irrespective of $l$, are very close to being Gaussian. 
Indeed the residual difference between the relative frequencies and the standard normal distribution is at most about 0.05 in absolute value. 
At $b=9$ fm, the  expansion coefficients $c_l$ are still almost Gaussian-distributed, although the deviations are larger than at $b=0$, with residual differences ranging up to 0.1 for the shown modes. 
Some distributions, as e.g., $p(c_0)$, seem to be skewed even for small $l$ values.
In turn, the tails of the modes around $l=200$ are thinner than that of the Gaussian distribution, hinting at a positive excess kurtosis.

For a more quantitative comparison to the standard normal distribution, we computed the first four moments of the $p(c_l)$ distributions obtained from the 8192 events. 
These moments are presented in Appendix~\ref{appendix:moments_cl_coeff} for both Glauber and Saturation models at $b=0$ and $b=9$~fm.

A natural measure of the relative importance of the fluctuation modes $\{\Psi_l\}$ is via the eigenvalues $\{\lambda_l\}$, which quantify their contributions to the density matrix $\rho$, Eq.~\eqref{eq:diagonalized}: more precisely, $\norm{\Psi_l} = \sqrt{\lambda_l}$ by construction.
To include the average event $\bar{\Psi}$, which is not an eigenvector of $\rho$, in the comparison, we define
\begin{equation}
w_l \equiv \frac{\sqrt{\lambda_l}}{\sum_l \sqrt{\lambda_l} + \norm{\bar{\Psi}}}
\quad\text{and}\quad
\bar{w} \equiv \frac{\norm{\bar{\Psi}}}{\sum_l \sqrt{\lambda_l} + \norm{\bar{\Psi}}},
\label{eq:w_l}
\end{equation}
where the denominator is the sum of the norms of all modes --- in our calculation we sum over all 16384 eigenvalues --- and of the average event. By construction, the weights $\{w_l\}$ and $\bar{w}$ sum up to unity, such that each one can be regarded as a measure of the relative importance of a given mode.

In Fig.~\ref{fig:eigenvalues} we show the first 256 (i.e., the 256 largest) relative weights $w_l$ and that of the average event for both initial-state models at $b=0$ and $b=9$~fm. 
At both impact-parameter values the contribution of $\bar{\Psi}$ is $\bar{w}=13\%$ in the Glauber model, $\bar{w}=8\%$ in the Saturation model.
At $b=0$, the fluctuation modes have a relative weight of less than about 1\%, which decreases quickly with increasing mode number $l$.
At $b=9$ fm the relative weights $w_l$ of the first modes are slightly larger than at $b=0$, ranging up to around 2\%--3\%. 
Events at large impact parameters have larger density fluctuations compared with the average event which is precisely what we find here.

Comparing both models, in the Glauber model the relative weights $w_l$ fall off with a steeper slope while the average event has in general a larger contribution than in the Saturation model. 
This reflects the fact that the energy density has a more detailed, finer structure in the Saturation model, which makes higher-order perturbations more probable.
How much this result is affected by the smearing radius in the Glauber model, or if one applies the Glauber picture at the valence-quark level, is not further investigated in this paper.

Eventually, one sees that at $b=0$ there often come pairs of degenerate modes with the same eigenvalue, for instance $(1,2)$, $(3,4)$, $(5,6)$, and so on. Since this degeneracy can be attributed to the rotational symmetry of the system at zero impact parameter, it is partially lifted at finite impact parameter $b=9$~fm, as will be discussed in more detail in Sec.~\ref{subsec:classification_modes}.

\begin{figure*}[!t]
\includegraphics[width=0.8\linewidth]{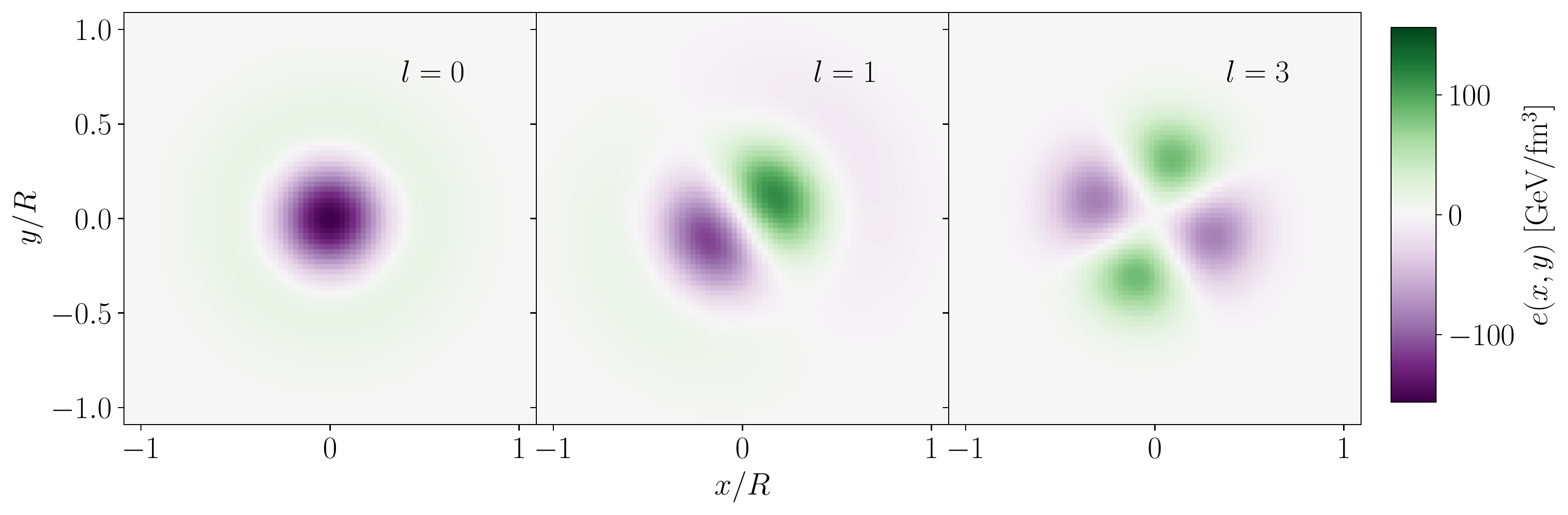}
\includegraphics[width=0.8\linewidth]{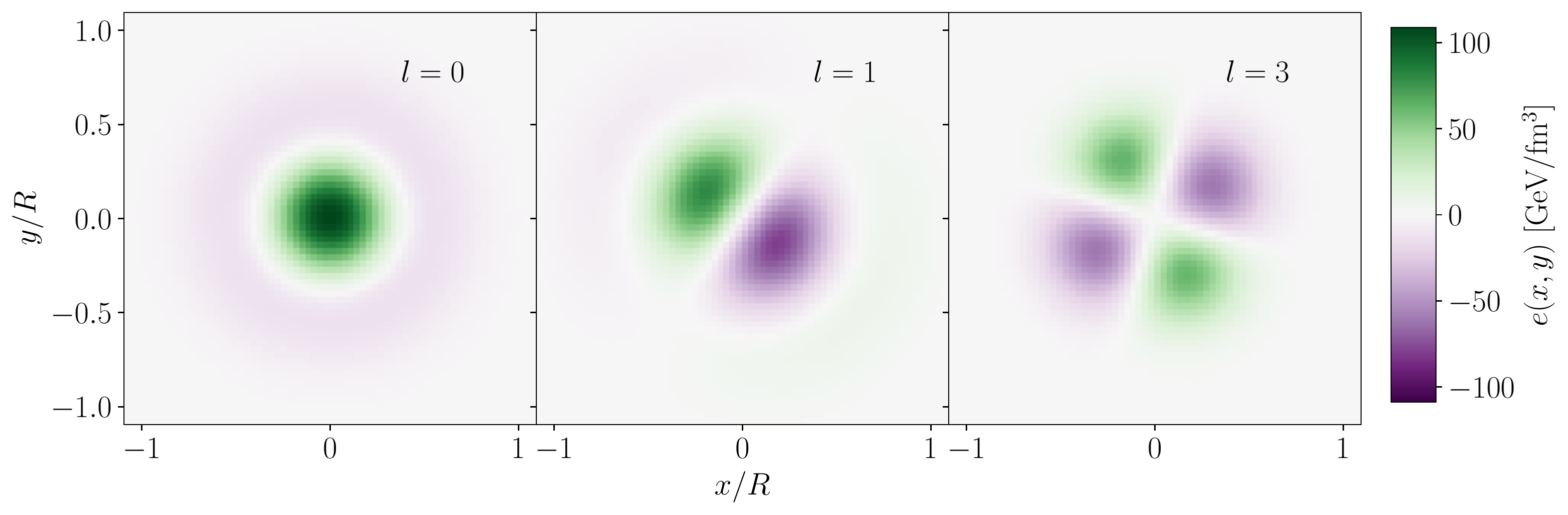}
\vspace{-3mm}
\caption{Density plots of modes $l=0$, 1, 3 at impact parameter $b=0$ in the Glauber (top) and Saturation (bottom) models. Note that the overall sign of the mode is arbitrary.}
\label{fig:mode_examples}
\end{figure*}

By construction, the modes $\{\Psi_l\}$ represent the event-by-event fluctuations of the initial state about the average event $\bar{\Psi}$.
Thus they indicate which random fluctuations of the energy density profile are more or less likely to occur. 
In Fig.~\ref{fig:mode_examples} we show three examples ($l=0,1,3$) of modes at $b=0$ in the Glauber (top row) and Saturation (bottom row) models. 
The first mode ($l=0$) is radially symmetric, the second one ($l=1$) has a dipole structure and the mode $l=3$ has a quadrupole structure.

We note that although the energy density of each mode can take both positive and negative values, this is in itself unproblematic. They contribute ``in addition'' to the average event, which is significantly larger in absolute value, so that the sum of $\bar{\Psi}$ and $c_l$ times $\Psi_l$ is non-negative everywhere as long as $c_l$ is not much larger than~1. 
This explains why the modes take smaller values in the Saturation model than in the Glauber model, since the same ordering holds for the respective average events of the two models. 

Since the negative $-\Psi_l$ is an eigenvector of the density matrix $\rho$ with the same eigenvalue (and the same norm) as $\Psi_l$, the overall sign of a mode has no physical meaning. 
This holds in particular for the apparent opposite signs of the modes $l=0$ in the two models in Fig.~\ref{fig:mode_examples}.

\subsection{Characterization of the average event and the modes for the Glauber and Saturation models}
\label{subsec:classification_modes}

Now that we have established the basic features of the decomposition, we will move on to discuss the structure of the modes, and in particular characterize their geometric shapes. In Appendix~\ref{appendix:mode_plots} we show the first 60 eigenvectors of the density matrix for the Glauber (Figs.~\ref{fig:60_modes_b0_Glauber}, \ref{fig:60_modes_b9_Glauber}) and Saturation (Figs.~\ref{fig:60_modes_b0_Saturation}, \ref{fig:60_modes_b9_Saturation}) models at $b=0$ and $b=9$~fm.
To allow a better comparison between the eigenvectors, they all have the same norm, i.e., they correspond to the $\{\tilde{\Psi}_l\}$ of Sec.~\ref{subsec:theory_basis}.  

At $b=0$ one clearly observes eigenvectors with rotational invariance, like $l=0$, 7, 18, or 33 (in the Glauber model). 
For the same eigenvectors, one also sees that the number of zero crossings with increasing distance from the center differs, growing with $l$. 
In turn, there are pairs of eigenvectors that can be deduced from each other by a rotation by a integer fraction of $180^\textrm{o}$, e.g., for $l=1$ and 2 (rotation by $\pi/2$), $l=3,4$ (rotation by $\pi/4$), $l=5,6$ (rotation by $\pi/6$), and so on. 
Each of these pairs consists of degenerate orthogonal eigenvectors with the same eigenvalue, see Fig.~\ref{fig:eigenvalues}, which span a two-dimensional space of eigenvectors with arbitrary orientation.\footnote{Some of the eigenvectors have a more complicated profile, like $l=40$ in the Glauber model or $l=33$ in the Saturation model. Generally, this happens when more than two eigenvectors are degenerate (within statistical uncertainty), such as, for instance, the eigenvector $l=33$ of the Saturation model is a radial one (like in the Glauber model) with a small admixture of eigenvectors invariant under rotations by $\pi/4$ like those with $l=34, 35$.}

Conversely, at finite impact parameter, rotational symmetry is broken, and the eigenvectors are no longer radially symmetric. 
Instead, most of the eigenvectors now admit the $x$- and $y$-directions as reflection-symmetry or antisymmetry axes. 
In parallel, the degeneracy of the eigenvalues at $b=0$ is partially lifted. 
Overall, the profiles at $b=9$~fm look significantly more complicated than at $b=0$, which is why we now introduce a number of quantities to characterize their profiles as well as those of the average event.

\subsubsection{Azimuthal and radial dependence}

\begin{figure*}[!t]
\includegraphics*[width=\linewidth]{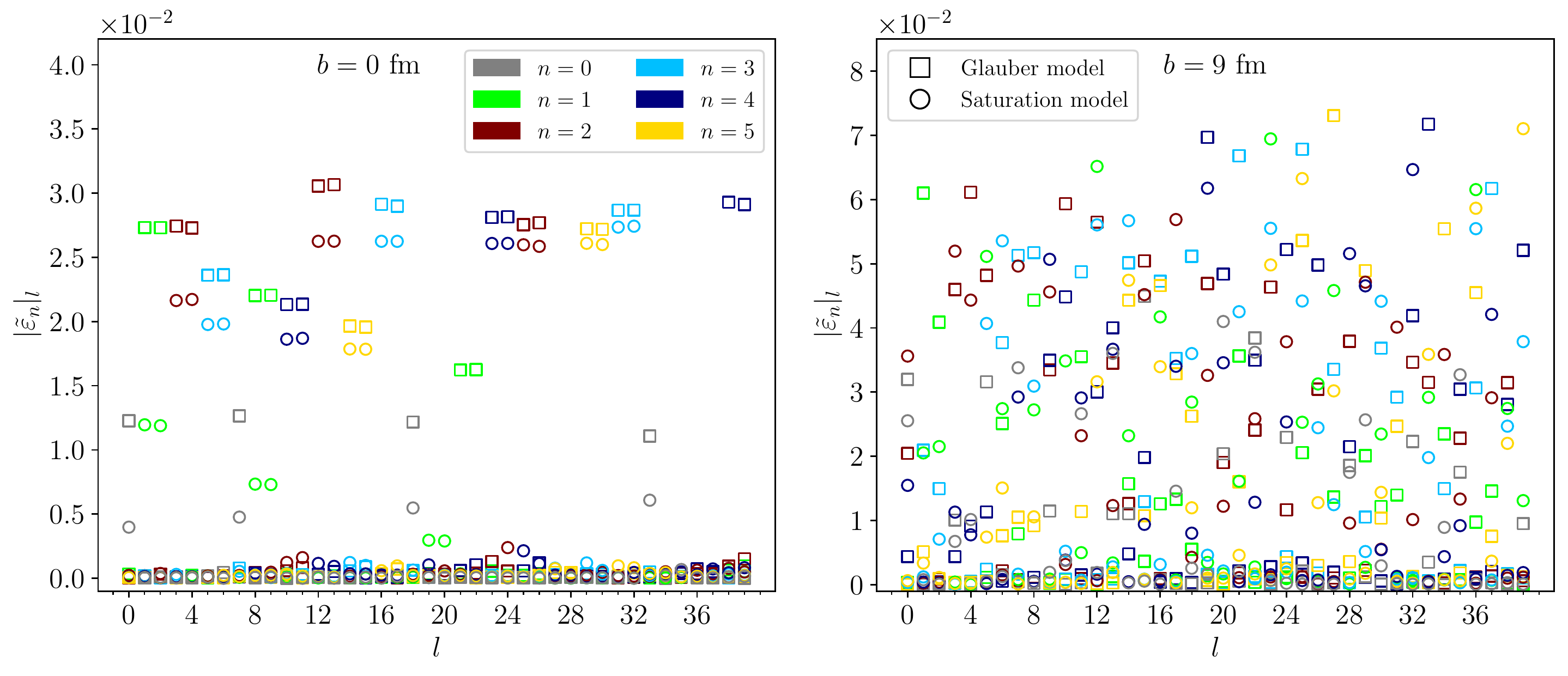}
\vspace{-7mm}
\caption{Mode eccentricities~\eqref{eq:mode_eccentricities1} and \eqref{eq:mode_eccentricities2} and relative energy content for both models at $b=0$ (left) and $b=9$ fm (right).}
\label{fig:mode_eccentricities}
\end{figure*}

To characterize the modes more precisely, we introduce several quantities.
The first one is the total energy of a mode $\Psi_l(r,\theta)$ given by
\begin{equation}
\mathcal{E}_l \equiv \tau_0\!\int\! \Psi_{l}(r,\theta)\;r\,\d r\, \d\theta.
\label{eq:mode_energy}
\end{equation}
Since the modes represent fluctuations about the average state, they should have a small amount of energy in comparison with the latter.
Similarly, to describe the overall azimuthal dependence of the modes we introduce complex ``eccentricities''
\begin{equation}
\tilde{\varepsilon}_1(\Psi_l) \equiv 
-\frac{\displaystyle\int\!r^3 \textrm{e}^{\textrm{i}\theta} \Psi_l(r,\theta)\;r\,\d r\,\d\theta}%
{\displaystyle\int\!r^3 \bar{\Psi}(r,\theta)\;r\,\d r\,\d\theta}
\quad\text{for }n=1 \label{eq:mode_eccentricities1}
\end{equation}
and
\begin{equation}
\tilde{\varepsilon}_n(\Psi_l) \equiv 
-\frac{\displaystyle\int\! r^n \textrm{e}^{\textrm{i}n\theta}\Psi_l(r,\theta)\;r\,\d r\,\d\theta}%
{\displaystyle\int\! r^n\bar{\Psi}(r,\theta)\;r\,\d r\,\d\theta}
\quad\text{for }n\geq 2.
\label{eq:mode_eccentricities2}
\end{equation}
We denote by $\abs{\tilde{\varepsilon}_n}_l$ the modulus of these eccentricities for mode $\Psi_l$.
Note that the definitions differ from the usual ones~\eqref{eq:eccentricities1} and \eqref{eq:eccentricities2} for initial-state eccentricities, in that we use the average state $\bar{\Psi}$ instead of $\Psi_l$ in the denominator. 
This ensures that the latter is always nonzero.
Definition~\eqref{eq:mode_eccentricities2} also makes sense with $n=0$, yielding the ratio $\abs{\tilde{\varepsilon}_0}_l = \mathcal{E}_l/\bar{\mathcal{E}}$ of the energy of mode $\Psi_l$ to that of the average state.

Evaluating $\tilde{\varepsilon}_n(\bar{\Psi})$ gives the traditional eccentricities of the average event, whose values we give in Table~\ref{tab:average_state_eccentricities} for our two models at $b=0$ and 9~fm.
At $b=0$ rotational symmetry should result in vanishing spatial anisotropies.
However, due to limited numerical precision and the finite number of events, the values are not exactly zero, but of the order $10^{-4}$ or smaller.
\begin{table}[!t]
\caption{\label{tab:average_state_eccentricities} Eccentricities of the average states in both models at $b=0$ and $b=9$ fm.}
\begin{ruledtabular}
\begin{tabular}{ccccc}
$|\varepsilon_1|$ & $|\varepsilon_2|$ & $|\varepsilon_3|$ & $|\varepsilon_4|$ & $|\varepsilon_5|$ \\ \hline
Glauber & $b=0$ & & & \\
$3.0\times 10^{-5}$ & $3.9\times 10^{-5}$ & $1.1\times 10^{-4}$ & $3.3\times 10^{-5}$ & $4.5\times 10^{-5}$ \\
\hline
Saturation & $b=0$ & & & \\
$6.0\times 10^{-5}$ & $9.3\times 10^{-5}$ & $5.5\times 10^{-5}$ & $4.7\times 10^{-5}$ & $3.9\times 10^{-5}$ \\
\hline
Glauber & $b=9$ fm & & & \\
$5.7\times 10^{-5}$ & $0.29$ & $2.6\times 10^{-4}$ & $9.4\times 10^{-2}$ & $9.0\times 10^{-5}$ \\
\hline
Saturation & $b=9$ fm & & & \\
$1.7\times 10^{-4}$ & $0.40$ & $7.4\times 10^{-5}$ & $0.20$ & $1.7\times 10^{-4}$ \\
\end{tabular}
\end{ruledtabular}
\end{table}

In contrast, at $b=9$~fm the average state has clearly nonzero eccentricities with even $n$, namely $\varepsilon_2$ and $\varepsilon_4$ of order $10^{-1}$, in both models.
The eccentricities with odd $n$ are of the same magnitude as in the case $b=0$, i.e., due to numerical fluctuations.
Comparing the models, $\varepsilon_2$ in the Saturation model is about 40\% larger and $\varepsilon_4$ is more than twice as large as in the Glauber model.
Larger average eccentricities in the Saturation model based on $k_{\rm T}$-factorization compared with the Glauber model have been observed before and reflect the sharper edges of the density distribution in the former model~\cite{Lappi:2006xc,Hirano:2009ah}.

Turning to the modes, we display in Fig.~\ref{fig:mode_eccentricities} the absolute values $\abs{\tilde{\varepsilon}_n}_l$ of their five first eccentricities~\eqref{eq:mode_eccentricities1}--\eqref{eq:mode_eccentricities2} and the ratio $\abs{\tilde{\varepsilon}_0}_l$ of their energy content compared with that of the average state.

At zero impact parameter, the modes with a clear nonvanishing energy, of the order of 1\% of that of the average state, are those with rotational invariance (see Fig.~\ref{fig:mode_examples}, $l=0$), simultaneously characterized by very small eccentricities $\abs{\tilde{\varepsilon}_n}_l\lesssim 10^{-3}$ for $n\geq 1$.
Apart from these ``radial modes,'' the other ones contain roughly hundred times less energy, $\abs{\tilde{\varepsilon}_0}_l < 10^{-4}$. 
This corresponds most probably to a vanishing energy content, the finite value being due to numerical precision like grid artifacts that make it impossible to resolve the modes exactly.

At $b=0$ the eccentricities with a given $n\geq 1$ occur in mode pairs: 
for example, modes $l=1,2$ have a sizable $\tilde{\varepsilon}_1$ followed by $l=3,4$ with a finite $\tilde{\varepsilon}_2$.
This reflects the existence of (quasi-)degenerate modes with a nearly identical profile up to a rotation, and is actually required to respect the absence of a preferred direction at $b=0$. 
The eccentricities of the two members of a mode pair sometimes slightly differ, which can again be ascribed to numerical (in)accuracy. 
Another point one can notice is that some modes, e.g., modes 27 and 28 in either the Glauber or the Saturation model, at first seem to have neither an eccentricity nor to contain energy.
However, looking at the eigenvectors themselves in Fig.~\ref{fig:60_modes_b0_Glauber} or Fig.~\ref{fig:60_modes_b0_Saturation}, one sees that in fact these modes have a rotational symmetry of order 7, i.e., a nonzero $\tilde{\varepsilon}_7$, which is not shown in Fig.~\ref{fig:mode_eccentricities}.
Eventually, let us also note that the eccentricities in the Glauber model are in general larger compared with those in the Saturation model but the structure of the modes is the same in both models.

For collisions at $b=9$ fm in the right panel of Fig.~\ref{fig:mode_eccentricities}, the breaking of rotational symmetry has several consequences. 
First of all, the modes no longer come in pairs with regard to the eccentricities $\abs{\tilde{\varepsilon}_n}_l$ with $n\geq 1$, which reflects the lifting of the degeneracy of their eigenstates. 
Second, the lack of rotational symmetry leads to the absence of purely radial modes.
This is also visible in the density plots shown in Fig.~\ref{fig:60_modes_b9_Glauber} or Fig.~\ref{fig:60_modes_b9_Saturation} for $b=9$ fm: the modes are elongated along the $x$- or the $y$-direction, so that there are more modes which have a nonzero $\varepsilon_2$.
In fact, the modes seem to have not a single nonzero eccentricity as is generally the case at $b=0$, but rather either all odd or all even eccentricities. 
Namely, a single mode has sizable $\varepsilon_1$, $\varepsilon_3$, $\varepsilon_5$ and zero $\varepsilon_2$ and $\varepsilon_4$, or the other way around.%
\footnote{Once again there are some exceptions, such as, e.g., $l=15$ for the Glauber model, for which both odd and even eccentricities can be clearly seen. By going back to the right panel of Fig.~\ref{fig:eigenvalues}, one sees that the corresponding eigenvalue is (within statistical uncertainty) degenerate with a neighboring mode, which results in some mixing between the eigenvectors.}
This property reflects the overall invariance of the system under parity, such that each individual mode must have a definite parity, if there is no degeneracy.
The typical eccentricity values are almost twice as large as at $b=0$.
Simultaneously, many modes now contain up to a few percent of the average-state energy.
Eventually, comparing the two models we observe that the order of the modes is no longer the same as it was in the rotational symmetric case.

To assess the transverse profiles of the average state and the modes, we rotate each of them such that the argument of the complex eccentricity $\tilde{\varepsilon}_n$ [Eqs.~\eqref{eq:mode_eccentricities1} and \eqref{eq:mode_eccentricities2}] with the largest modulus lies along the $x$-direction, thereby maximizing the real part of $\tilde{\varepsilon}_n$ while the imaginary part of $\tilde{\varepsilon}_n$ vanishes (up to numerical fluctuations). 
With the rotated mode --- and with the average state (denoted with a subscript $l=\bar{\Psi}$) ---, we define
\begin{align}
\mathcal{C}_l(r)&\equiv -\frac{\tau_0}{\bar{\mathcal{E}}}\! 
\int\! \Psi_l(r,\theta)\cos(n\theta)\,\d\theta, \label{eq:C_r}\\
\mathcal{S}_l(r)&\equiv -\frac{\tau_0}{\bar{\mathcal{E}}}\! 
\int\! \Psi_l(r,\theta)\sin(n\theta)\,\d\theta,
\end{align}
where the angle $\theta$ is measured from the $x$-axis while $\bar{\mathcal{E}}$ is the energy of the average state. We note that these definitions can be regarded as differential version of the spatial eccentricities $\tilde{\varepsilon}_n$ in Eq.~\eqref{eq:mode_eccentricities2}, which can be obtained (up to a factor) from the quantities $\mathcal{C}_l(r)$ and $\mathcal{S}_l(r) $ by radial integration with a weight $r^n$.

\begin{figure}[!t]
	\includegraphics[width=0.5\linewidth]{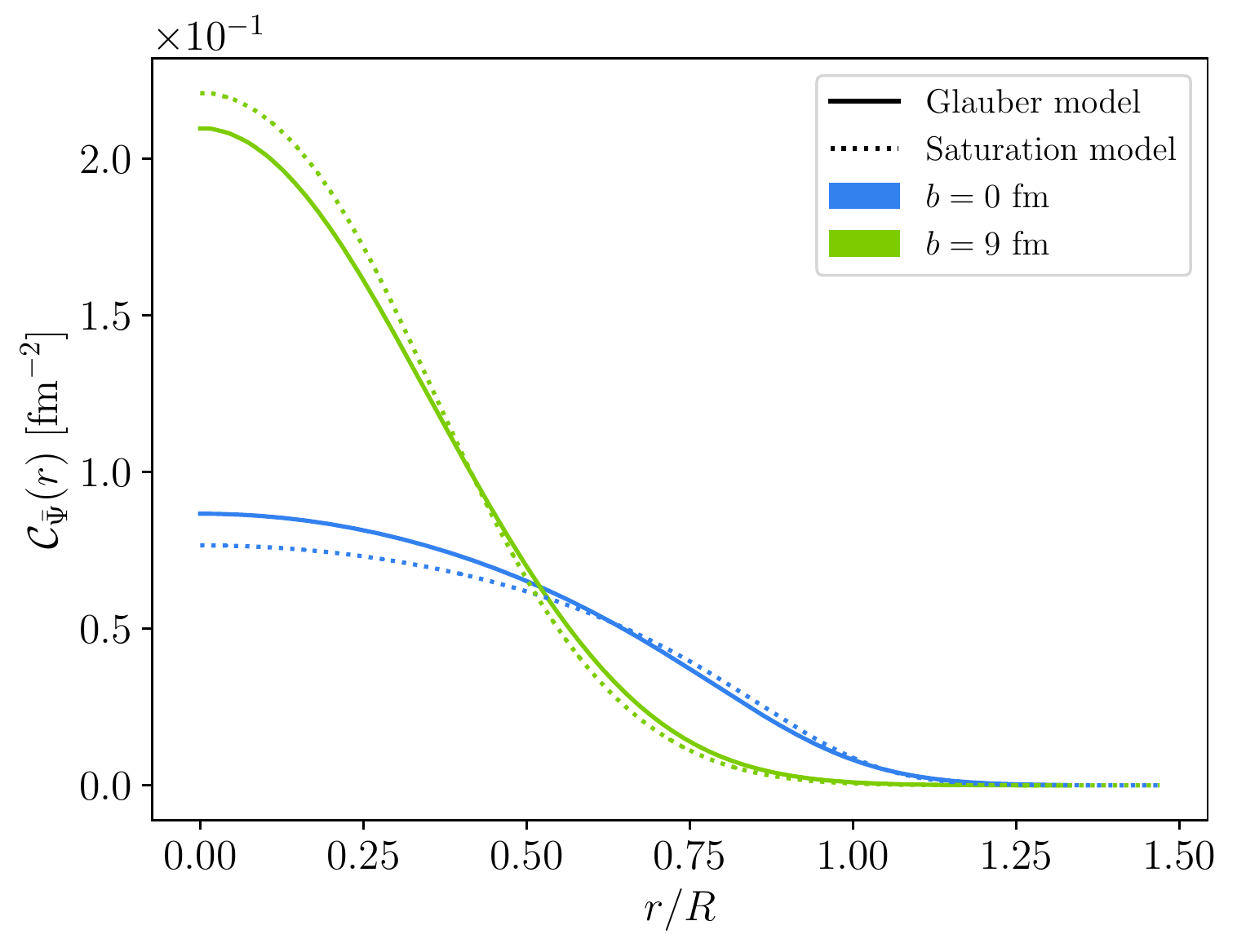}
	\vspace{-3mm}
	\caption{Radial profile $\mathcal{C}_{\bar{\Psi}}(r)$, Eq.~\eqref{eq:C_r}, of the average states in the Glauber (full lines) and the Saturation model (dotted lines) at $b=0$ and 9~fm.}
	\label{fig:avg_state_theta_int}
\end{figure}

\begin{figure}[!t]
\includegraphics[width=0.5\linewidth]{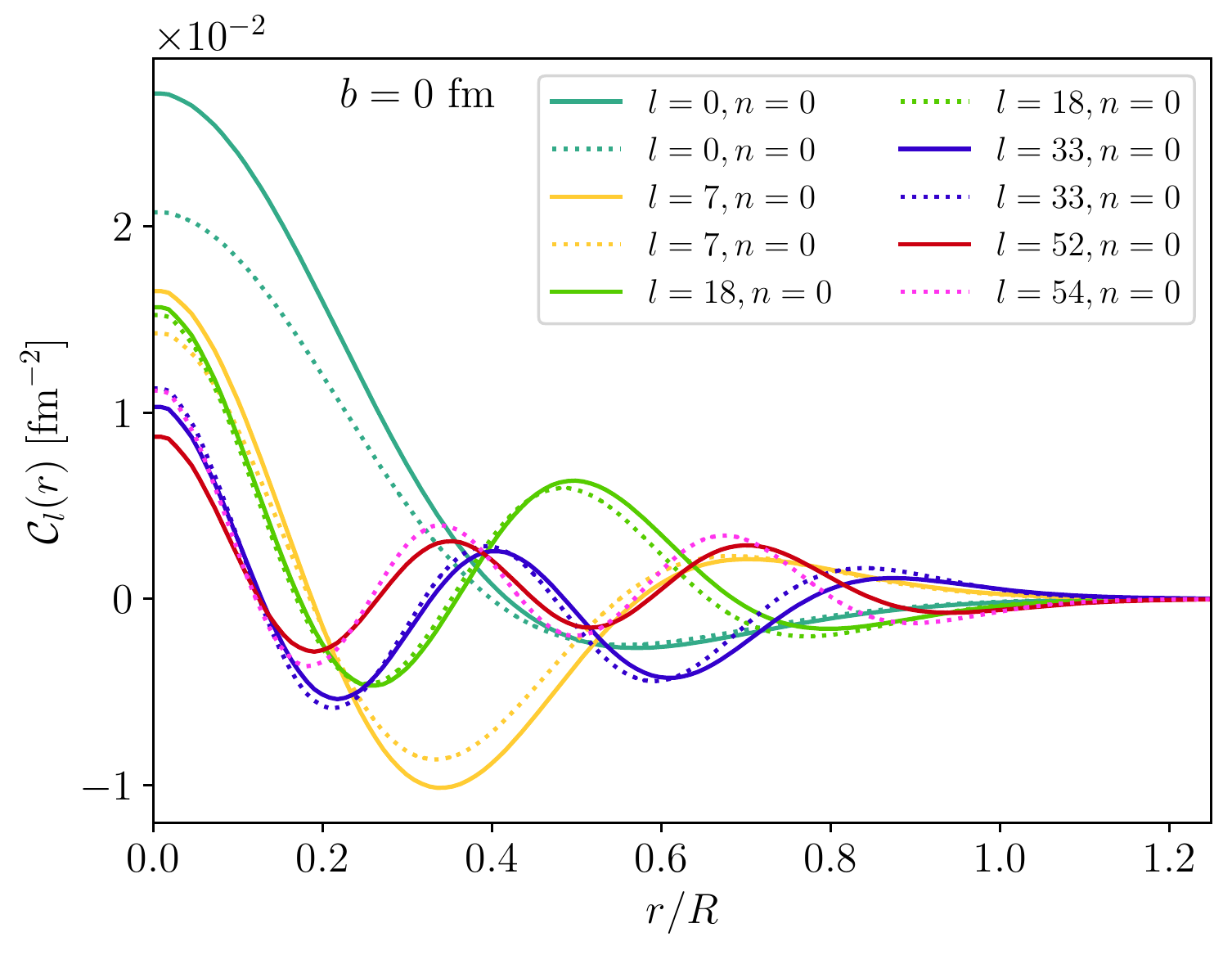}
\vspace{-3mm}
\caption{Radial profile $\mathcal{C}_l(r)$ of the rotationally invariant modes in the Glauber (full lines) and Saturation (dashed lines) models at $b=0$ fm.}
\label{fig:modes_rotated_radial_A}
\end{figure}

\begin{figure*}[!hbt]
\includegraphics[width=\linewidth]{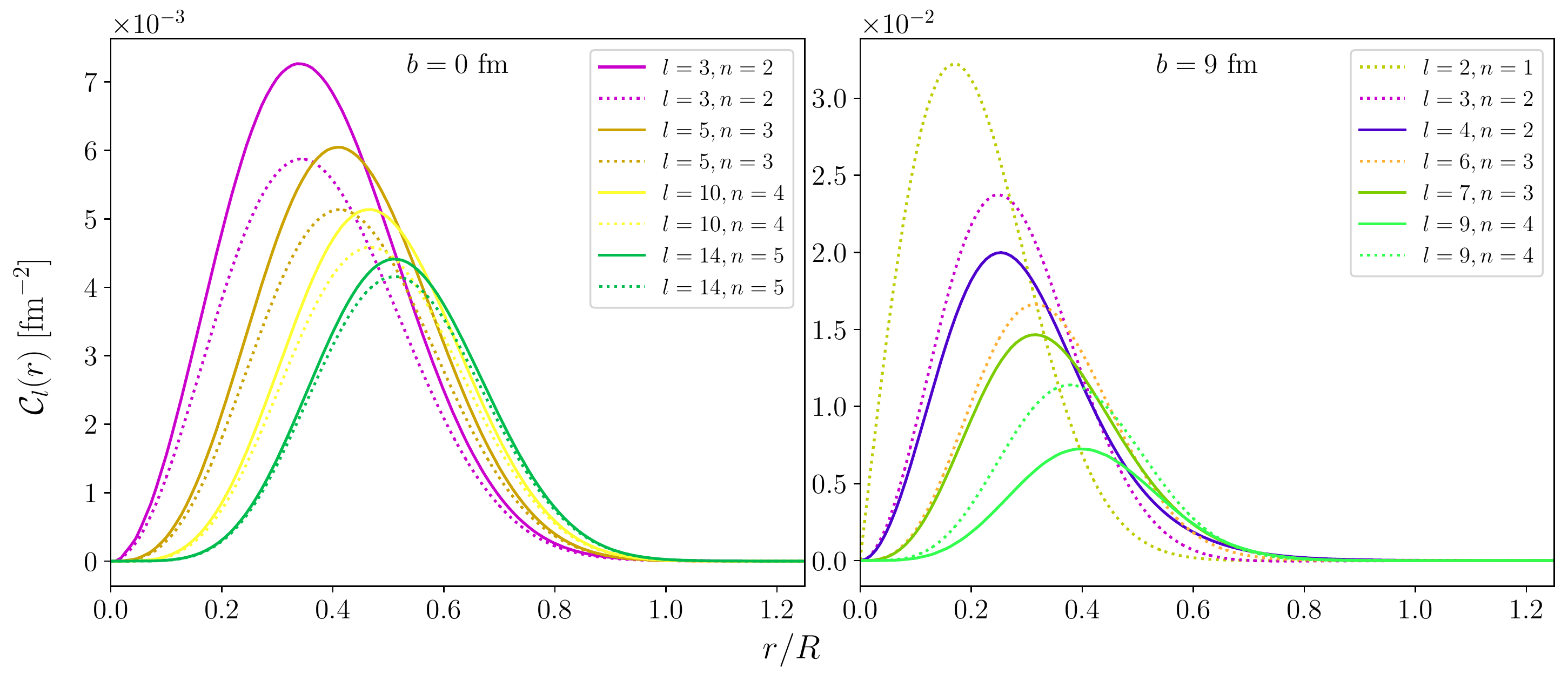}
\vspace{-7mm}
\caption{Radial profile $\mathcal{C}_l(r)$ of the ``first excitation modes'' for each harmonic $n$, corresponding to the largest $\tilde{\varepsilon}_n$, at $b=0$ (left) and $b=9$ fm (right), for the Glauber (full lines)  and Saturation (dotted lines) models.}
\label{fig:modes_rotated_1st_A}
\end{figure*}

\begin{figure*}[!htb]
\includegraphics[width=\linewidth]{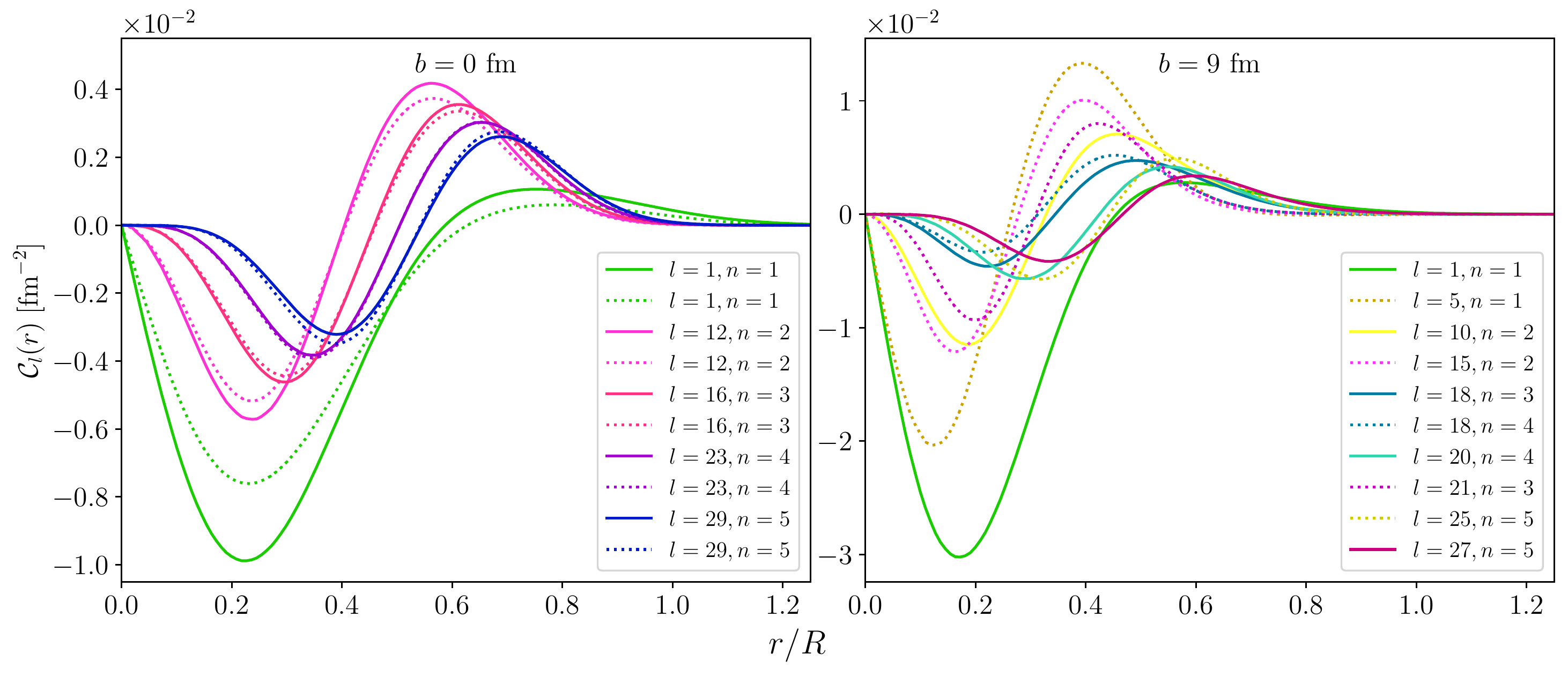}
\vspace{-7mm}
\caption{Radial profile $\mathcal{C}_l(r)$ of the ``second excitation modes'' for each harmonic $n$, corresponding to the largest $\tilde{\varepsilon}_n$, at $b=0$ (left) and $b=9$ fm (right), for the Glauber (full lines)  and Saturation (dotted lines) models.}
\label{fig:modes_rotated_2nd_A}
\end{figure*}

Figure~\ref{fig:avg_state_theta_int} shows $\mathcal{C}_{\bar{\Psi}}(r)$ for the average states in the two models.
These radial profiles are positive everywhere, as they should be.
The average states have a similar extent in the radial direction in both models.
The profiles slightly differ: at $b=0$ the Glauber model has more energy at the center and drops faster for larger $r$, while the reverse behavior is found at finite impact parameter.

Turning to the radial profiles $\mathcal{C}_l(r)$ of the modes, we find that they may now change sign with $r$ --- which they indeed should, if two modes with the same rotational symmetry are to be orthogonal. 
We shall call ``$k$th excitation'' (for a given rotational symmetry) the modes for which $\mathcal{C}_l(r)$ changes sign $k-1$ times.

Figure~\ref{fig:modes_rotated_radial_A} shows the radial profiles of the first five modes with rotational symmetry found at $b=0$ in both initial-state models. 
For better readability, we multiplied by $-1$ the modes that have a negative value at $r=0$ in Figs.~\ref{fig:60_modes_b0_Glauber} and \ref{fig:60_modes_b0_Saturation} --- for instance the mode $\Psi_0$ of the Saturation model.
All radial modes shown have a finite value at $r=0$ and change sign at least once: 
the profiles $\mathcal{C}_{l=0}$ have a single zero, those with $l=7$ two zeros, three zeros for $\mathcal{C}_{l=18}$, and so on. 
Within a given model, the values $\mathcal{C}_l(r=0)$ decrease with growing $l$, which is partly due to the decreasing norm $\norm{\Psi_l}=\sqrt{\lambda_l}$ quantifying the importance of the modes. 

In Fig.~\ref{fig:modes_rotated_1st_A} we show the transverse profiles of first excitation modes, with constant sign, whose largest $\tilde{\varepsilon}_n$ is that of order $n\in\{1,\dots,5\}$.
At $b=0$, these modes (for $n$ ranging from 2 to 5) happen to have the same label $l$ in the two models; no mode with $n=1$ and no sign change along the radial direction was found in either model. 
At $b=9$~fm there is a mode with constant-sign radial profile and $n=1$ in the Saturation model, but not in the Glauber model. 
In addition, no mode with $n=5$ and constant sign $\mathcal{C}_l$ was found.
At both impact-parameter values, the maximum of $\mathcal{C}_l$ moves to increasingly larger radius $r$ and its value decreases with increasing $n$ (and $l$) within a given model.

The latter behavior also generally holds for the positions of the local extrema and the hierarchy of the corresponding absolute values of $\mathcal{C}_l$ for the second excitation modes shown in Fig.~\ref{fig:modes_rotated_2nd_A}, with the exception of the modes whose largest eccentricity is $\tilde{\varepsilon}_1$.
The latter also differ from the modes with $n\in\{2,\dots,5\}$ in that their second extremum --- the maximum, for the convention on the sign of $\Psi_l$ used in the figure --- is much smaller in amplitude than the first extremum. 
In contrast, for the modes with $n\in\{2,\dots,5\}$ the values of $\mathcal{C}_l$ at the minimum and the maximum are similar in magnitude.

Eventually, the values of $\mathcal{S}_l(r)$ we found are at least one order of magnitude smaller than those of  $\mathcal{C}_l(r)$, and thus much more affected by numerical precision which is the reason why they are not presented here.

\subsubsection{Comparison with Bessel--Fourier decomposition}
\label{subsubsection:Bessel_Fourier}

As an alternative to the characteristics introduced in the previous subsection, one can also decompose the average event $\bar{\Psi}$ and the modes $\{\Psi_l\}$ found in a given model on a basis chosen \textit{a priori\/}, i.e., not ``optimized'' as is that consisting of the modes. 

For the two-dimensional energy-density profiles with finite support given by the two models of Sec.~\ref{subsec:models}, a convenient basis is that underlying the Bessel--Fourier decomposition, which was already used in the context of initial-state characterization in the past~\cite{Coleman-Smith:2012kbb,Floerchinger:2013rya,Floerchinger:2013vua,Floerchinger:2014fta}.
An advantage of such a decomposition is data reduction: anticipating the following, we shall see that at least the first modes can be to a very good approximation characterized by ${\cal O}(10)$ expansion coefficients each, which is significantly less expensive to store than the ${\cal O}(10^4)$ values per mode on the ``trivial'' basis attached to the computational grid. 
Additionally, the Bessel-Fourier decomposition is also well suited for (semi-)analytical calculations in a mode-by-mode approach~\cite{Floerchinger:2013rya,Floerchinger:2013vua,Floerchinger:2013hza}.

The basis underlying the Bessel--Fourier decomposition consists of the functions
\begin{equation}
\chi_{n,k}(r,\theta) = \frac{1}{J_{|n|+1}( j_{n,k})} J_n\bigg(\frac{r}{r_0} j_{n,k} \bigg)_{}\textrm{e}^{\textrm{i}_{}n\theta}
\end{equation}
with $n\in\mathbb{Z}$ that characterizes the angular dependence and $k$ a positive integer that determines the granularity of the radial profile. 
$J_n$ denotes the $n$-th Bessel function of first kind and $j_{n,k}$ its $k$-th zero.
Eventually, $r_0$ is the radius of the domain to which the expansion is restricted, such that the function to be decomposed vanishes at every point with $r=r_0$. 
In our calculations we take $r_0 = 12$~fm, equal to half of the grid width. 
Every function $f(r,\theta)$ on the transverse plane can then be decomposed in the form
\begin{equation}
f(r,\theta)= \sum_{n,k} A_{n,k} \chi_{n,k}(r,\theta)
\end{equation} 
with the complex expansion coefficients
\begin{equation}
A_{n,k}=\frac{1}{\pi r_0^2} \int\! f(r,\theta) \chi^*_{n,k}(r,\theta)\, r\,\d r\,\d\theta.
\end{equation}
If $f$ is real-valued, as is the case of the energy-density distributions we consider, then $A_{n,k} = A^*_{-n,k}$.
Hereafter we only present coefficients $\abs{A_{n,k}}$ for the average event and modes computed within the Glauber model as an illustration. 
Similar results regarding the Bessel--Fourier decomposition were found within the Saturation model.

\begin{figure*}[!htb]
	\includegraphics[width=0.4\linewidth]{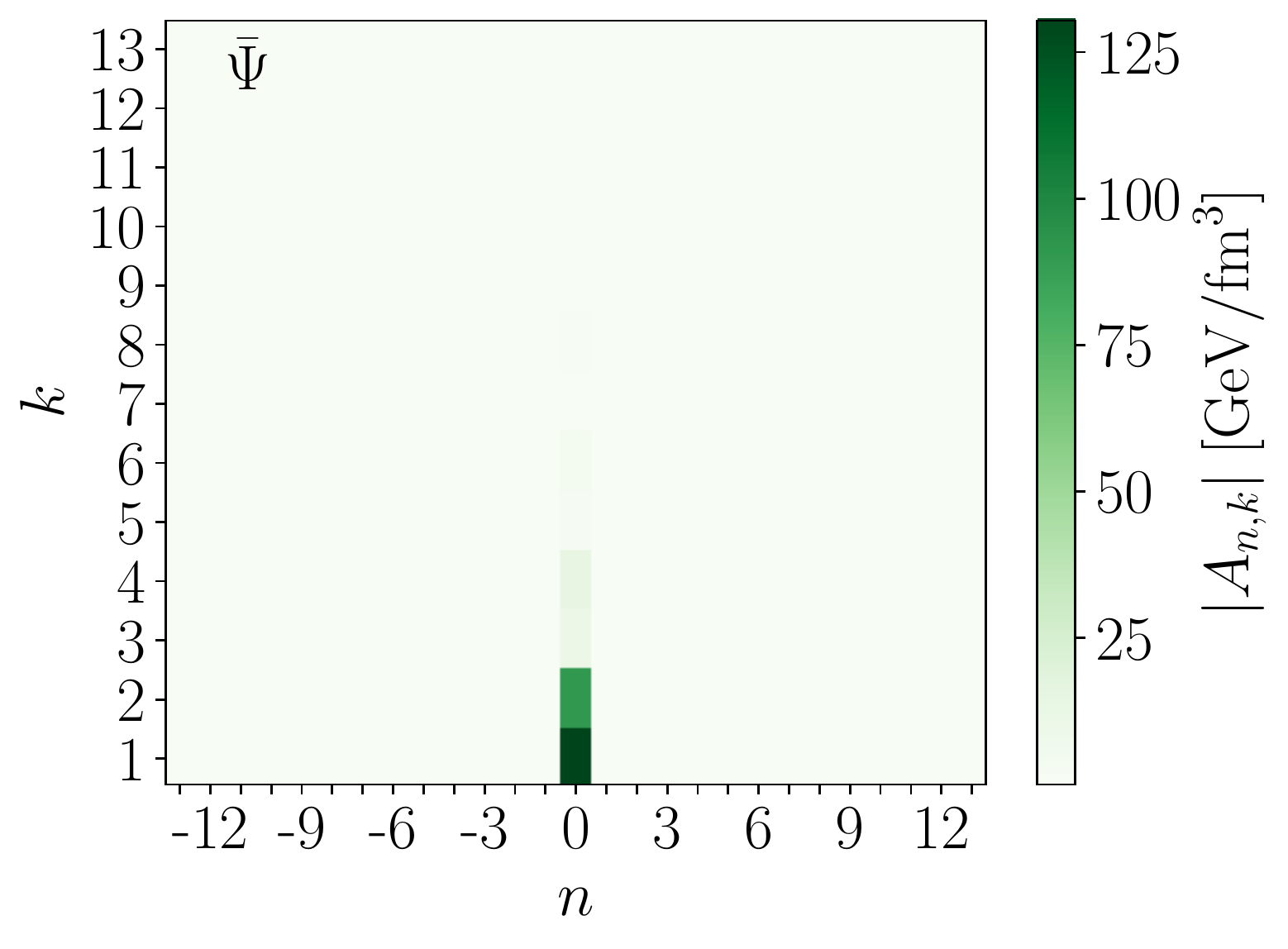}
	\includegraphics[width=0.4\linewidth]{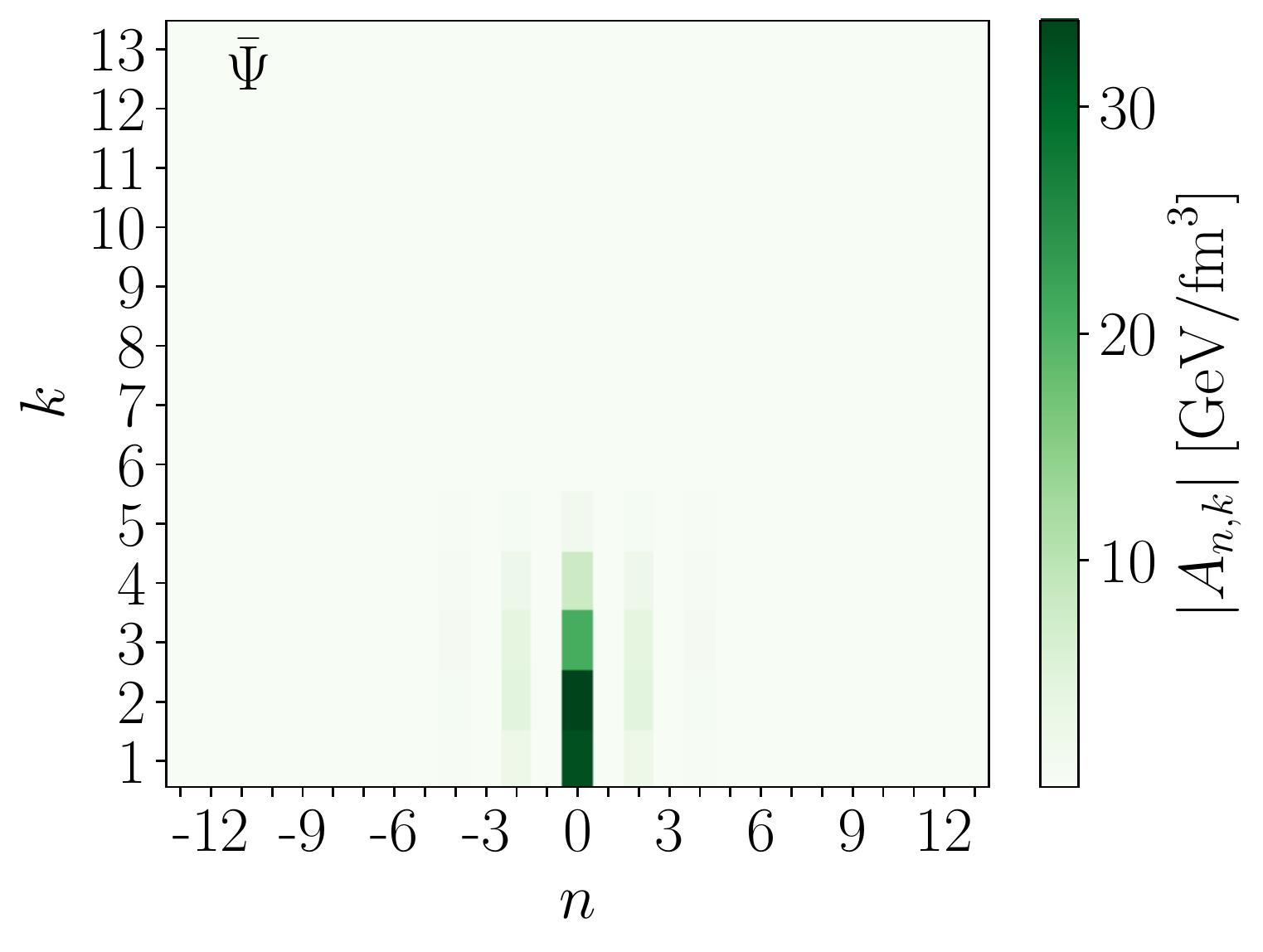}
	\vspace{-3mm}
	\caption{Absolute values $|A_{n,k}|$ of the Bessel--Fourier expansion coefficients for the average states at $b=0$ (left) and $b=9$ fm (right) in the Glauber model.}
	\label{fig:BesselFourier_Glauber_AvSt}
\end{figure*}

\begin{figure*}[!htb]
\includegraphics[width=0.325\linewidth]{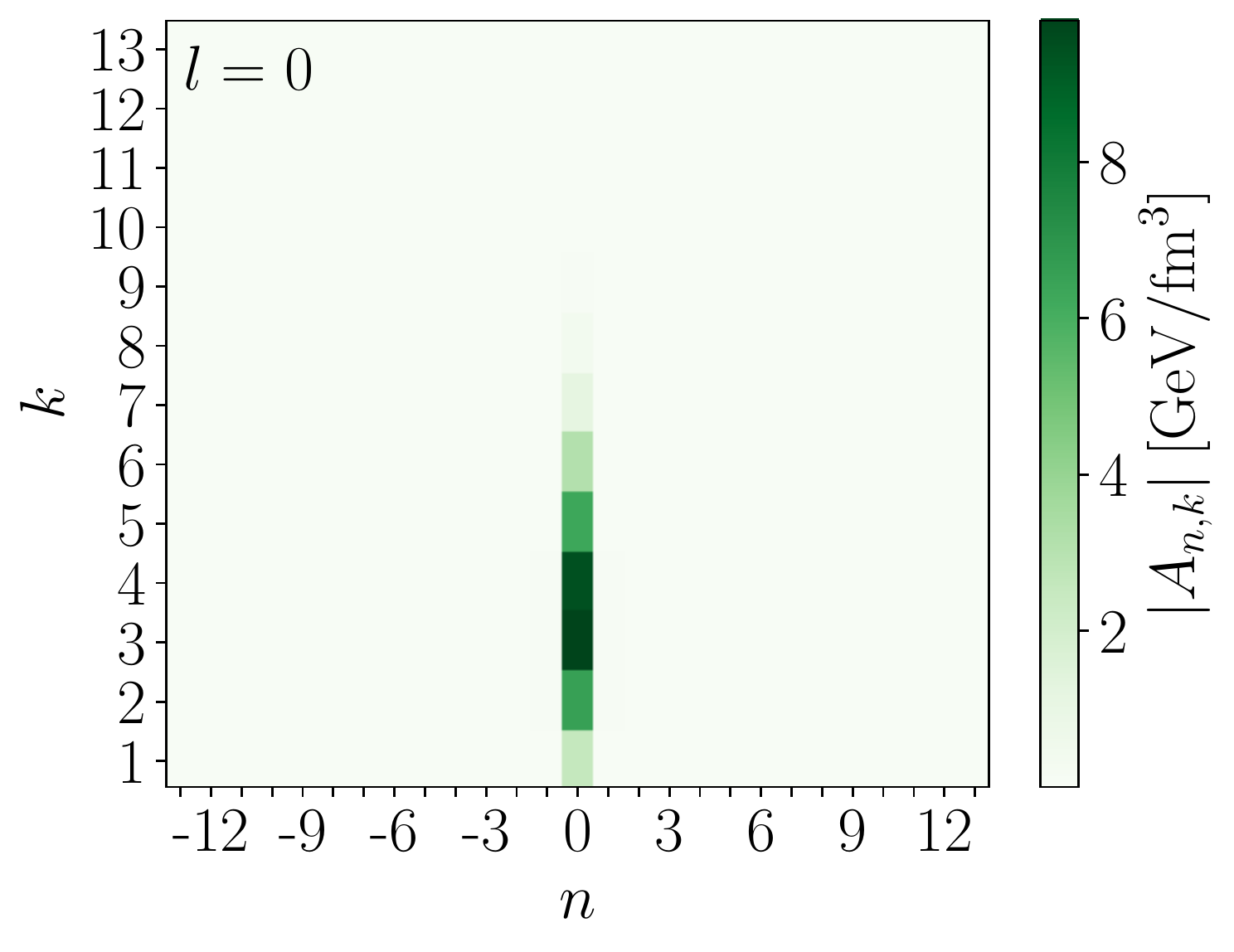}
\includegraphics[width=0.325\linewidth]{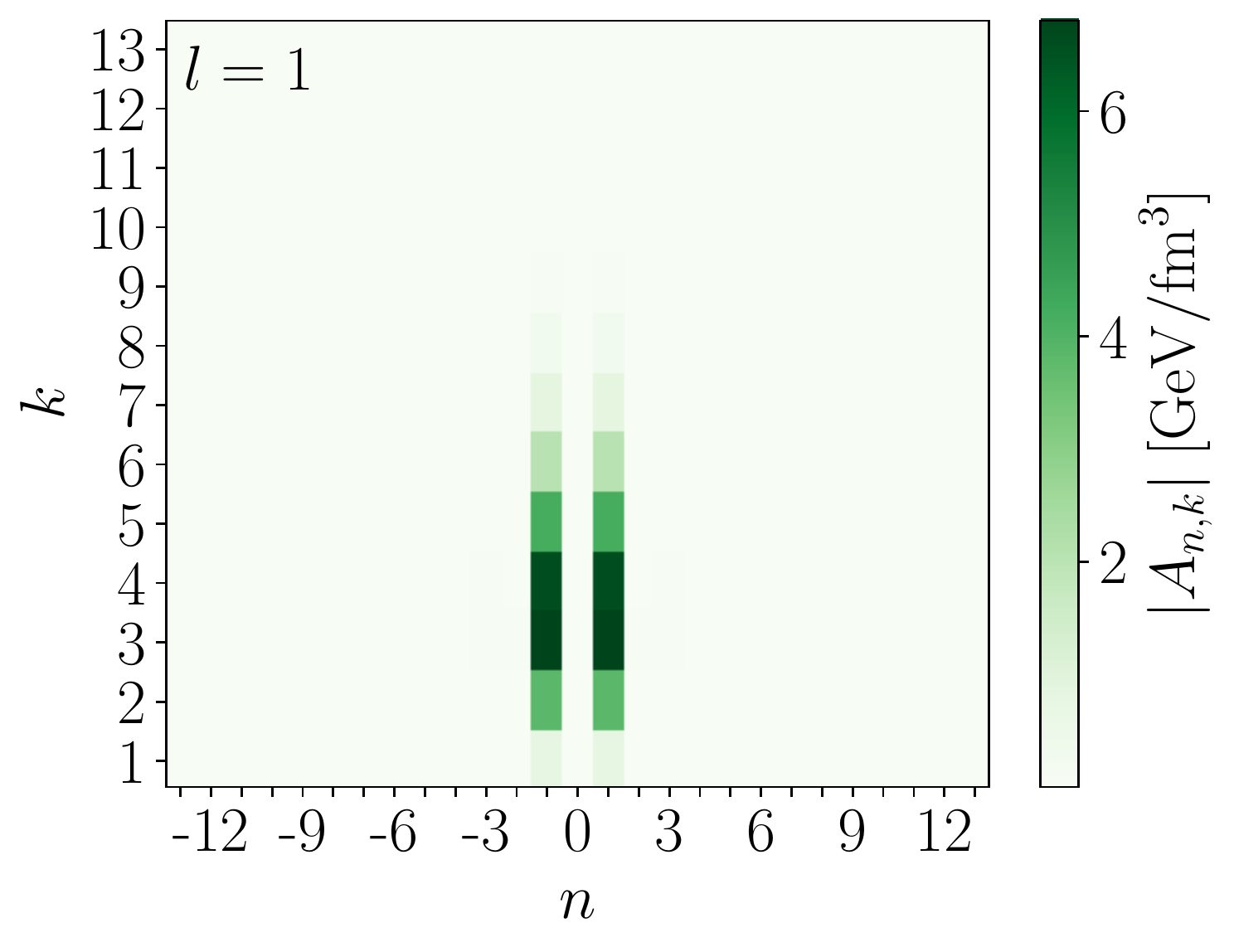}
\includegraphics[width=0.325\linewidth]{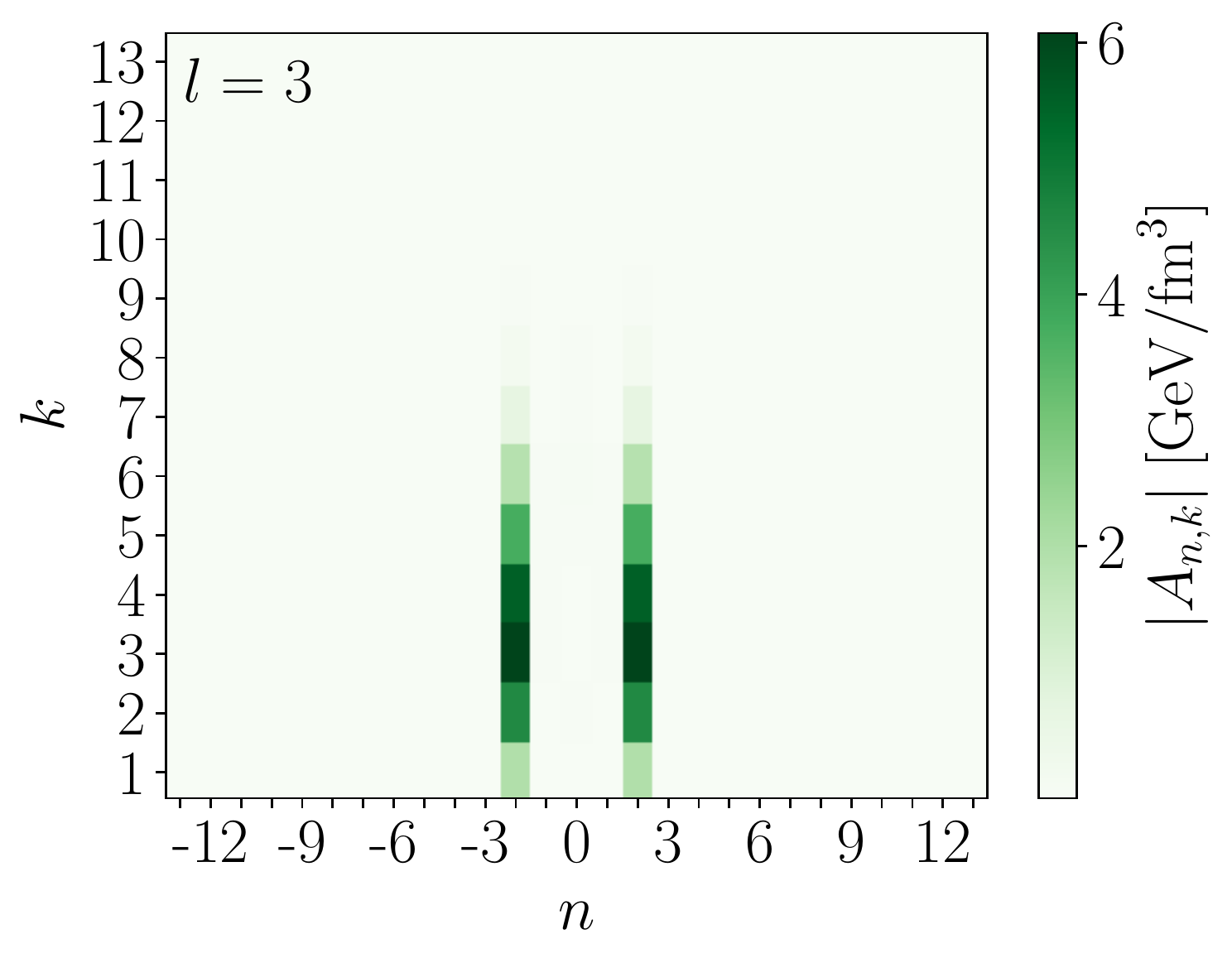}
\includegraphics[width=0.325\linewidth]{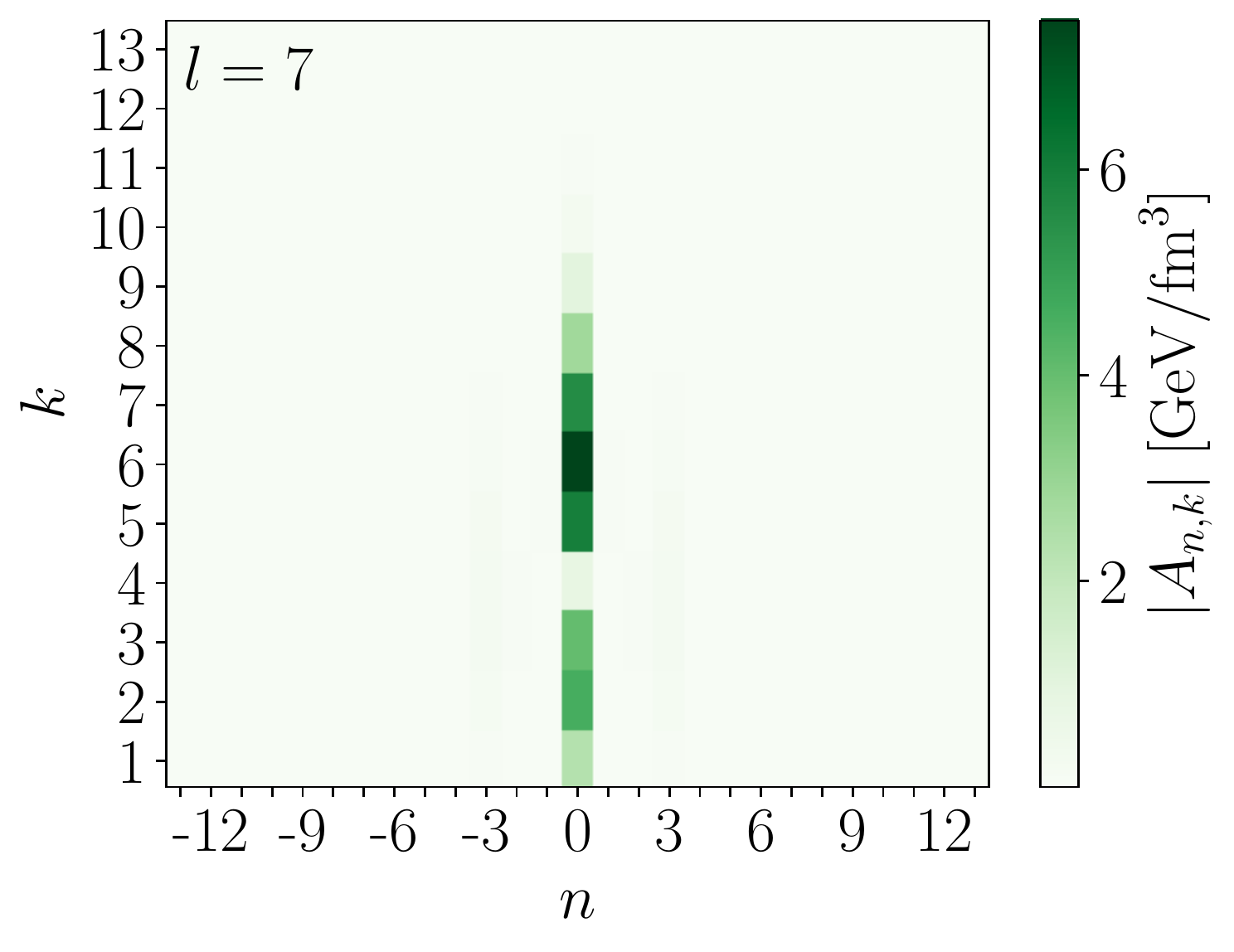}
\includegraphics[width=0.325\linewidth]{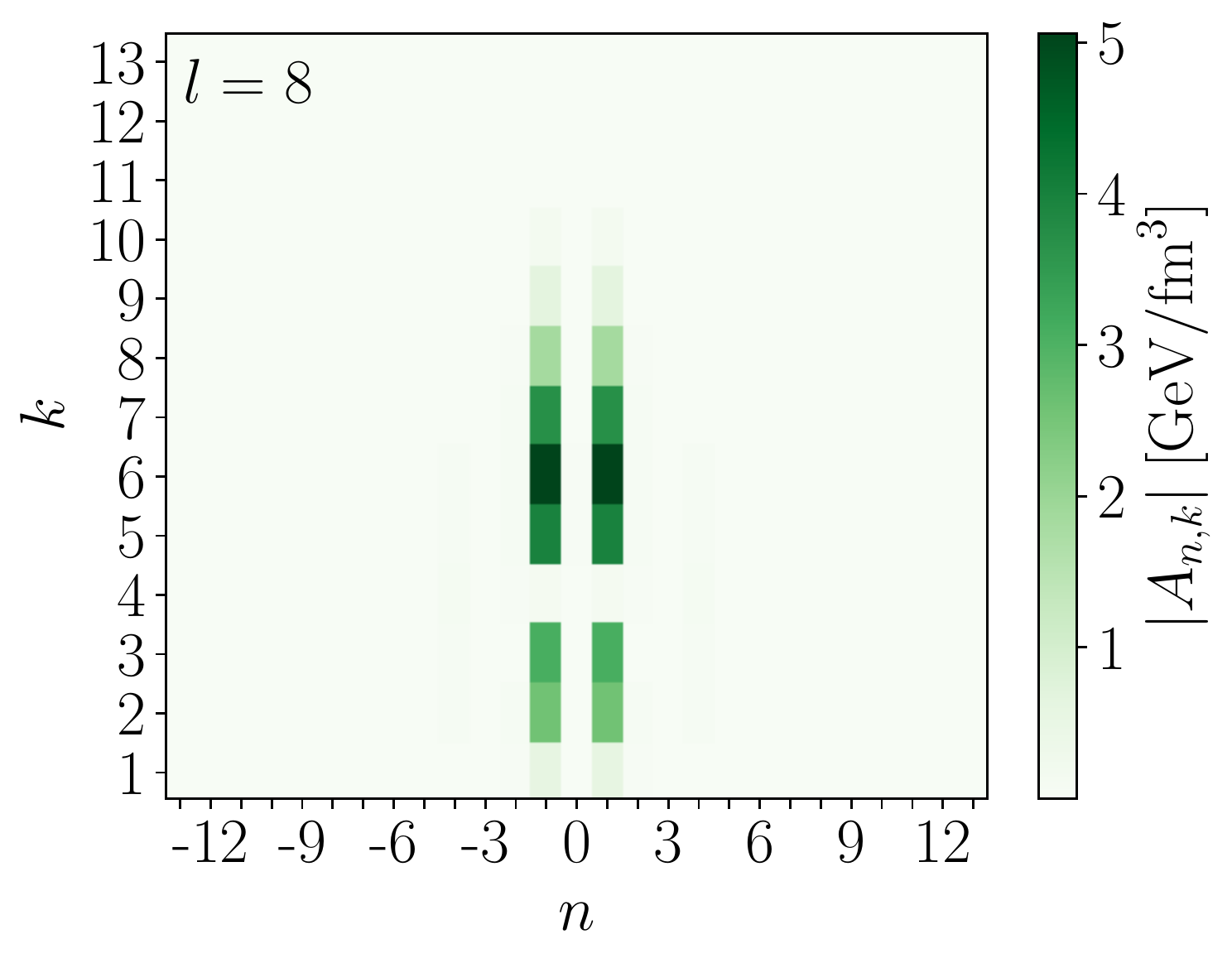}
\includegraphics[width=0.325\linewidth]{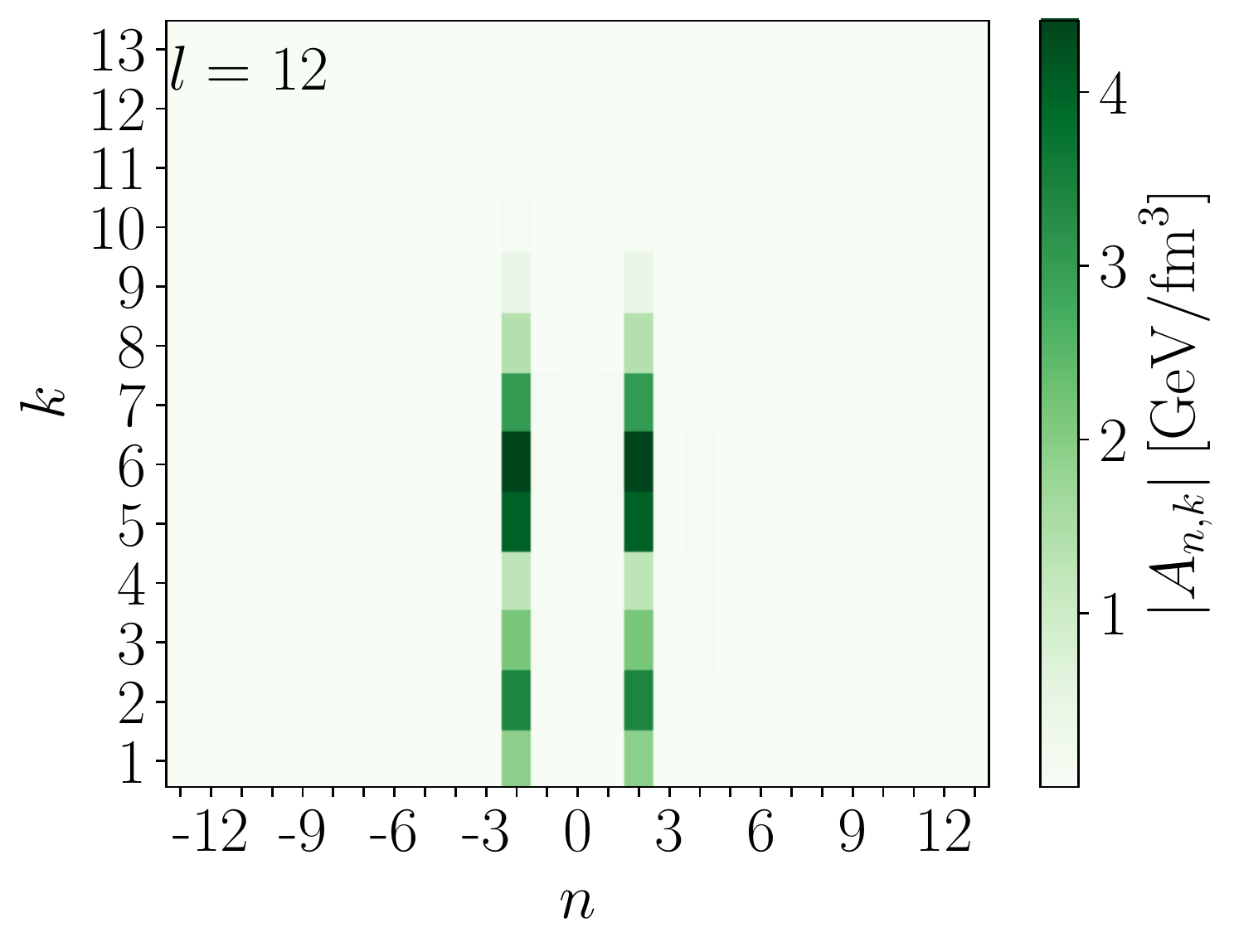}
\vspace{-3mm}
\caption{Absolute values $|A_{n,k}|$ of the Bessel--Fourier expansion coefficients for radial modes (left), $\varepsilon_1$ modes (center) and $\varepsilon_2$ modes (right) at $b=0$ in the Glauber model. The upper panels show the lowest-excitation modes, the lower panels the next excited modes in the corresponding harmonic.}
\label{fig:BesselFourier_Glauber_b0}
\end{figure*}
\begin{figure*}[!htb]
\includegraphics[width=0.325\linewidth]{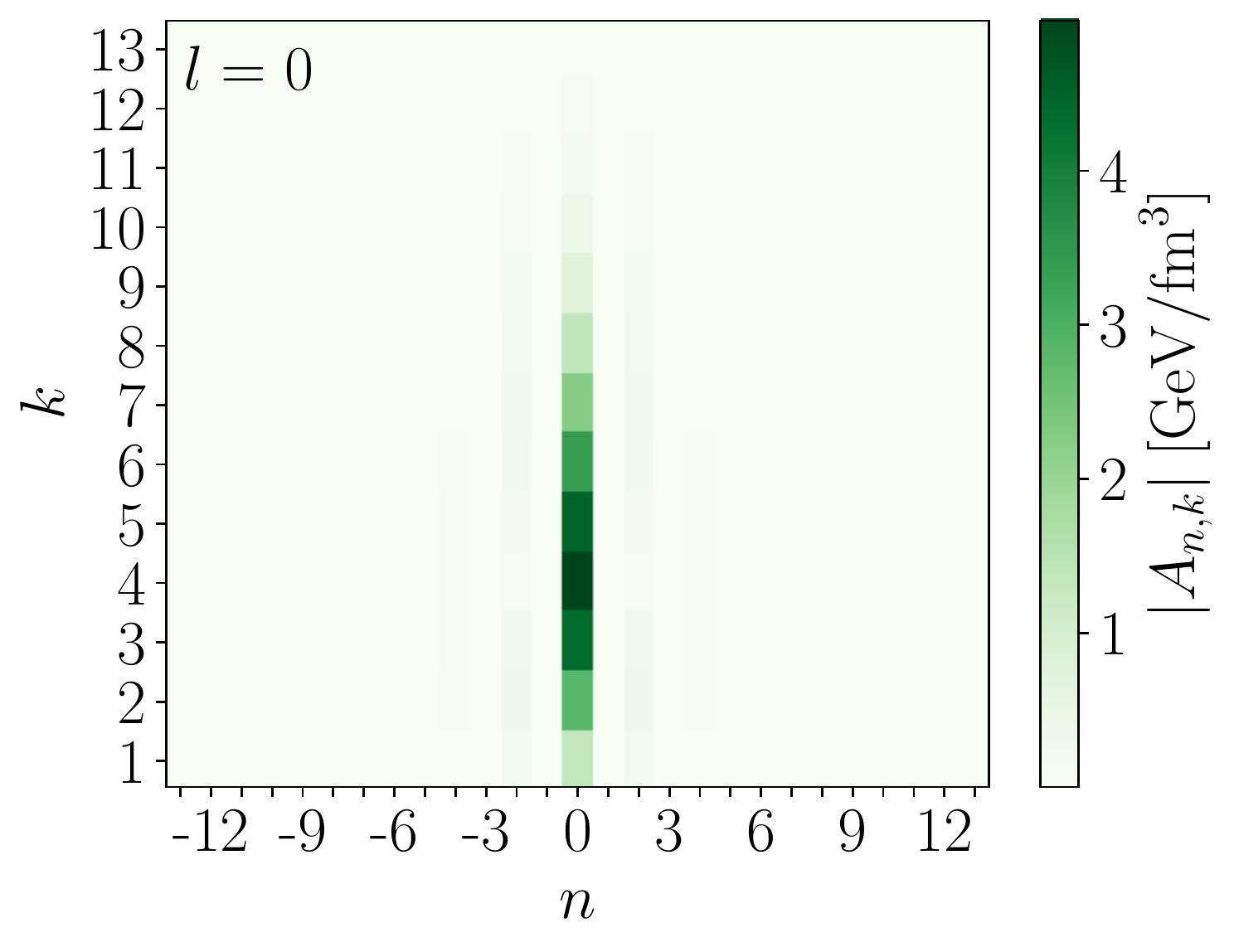}
\includegraphics[width=0.325\linewidth]{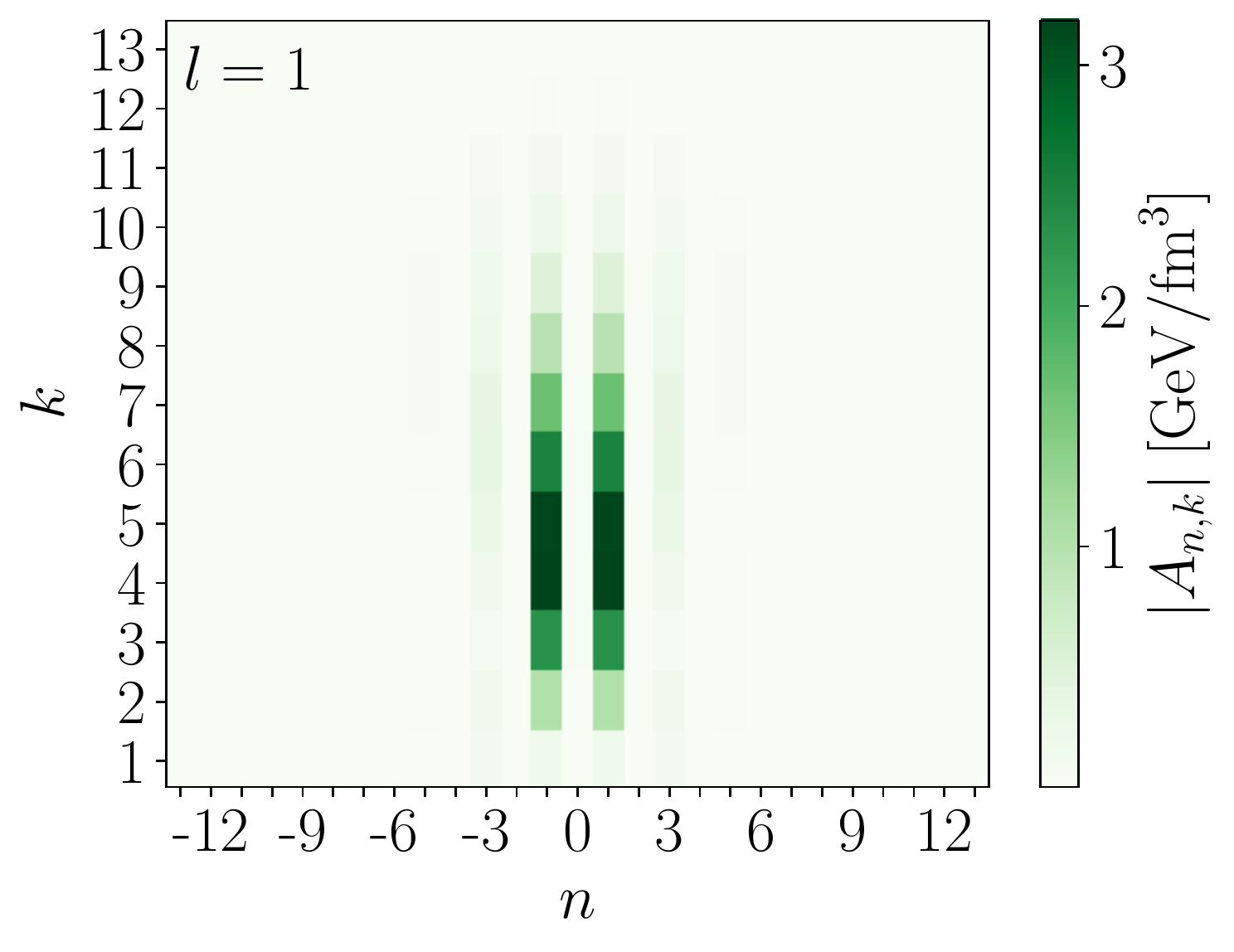}
\includegraphics[width=0.325\linewidth]{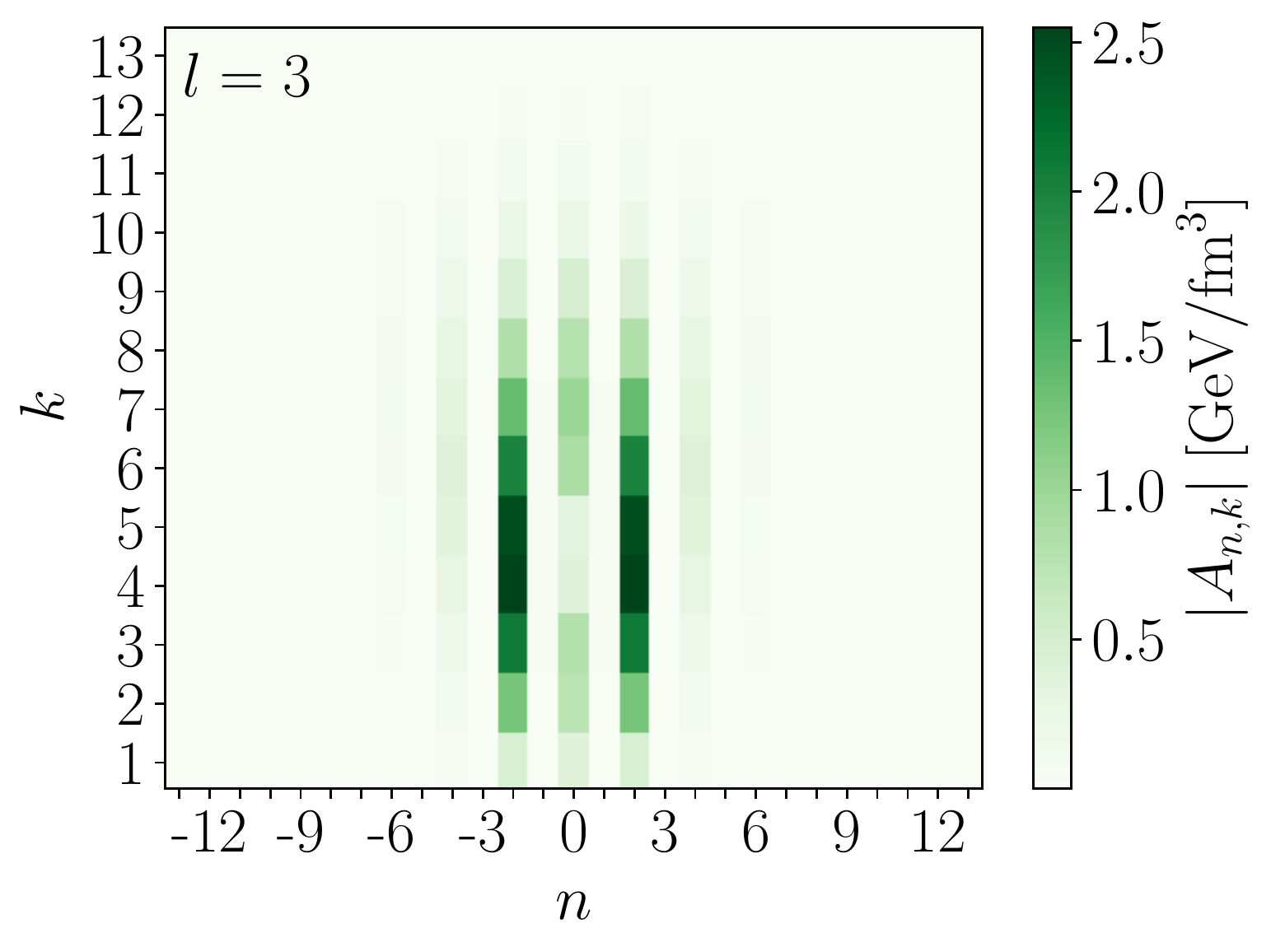}
\includegraphics[width=0.325\linewidth]{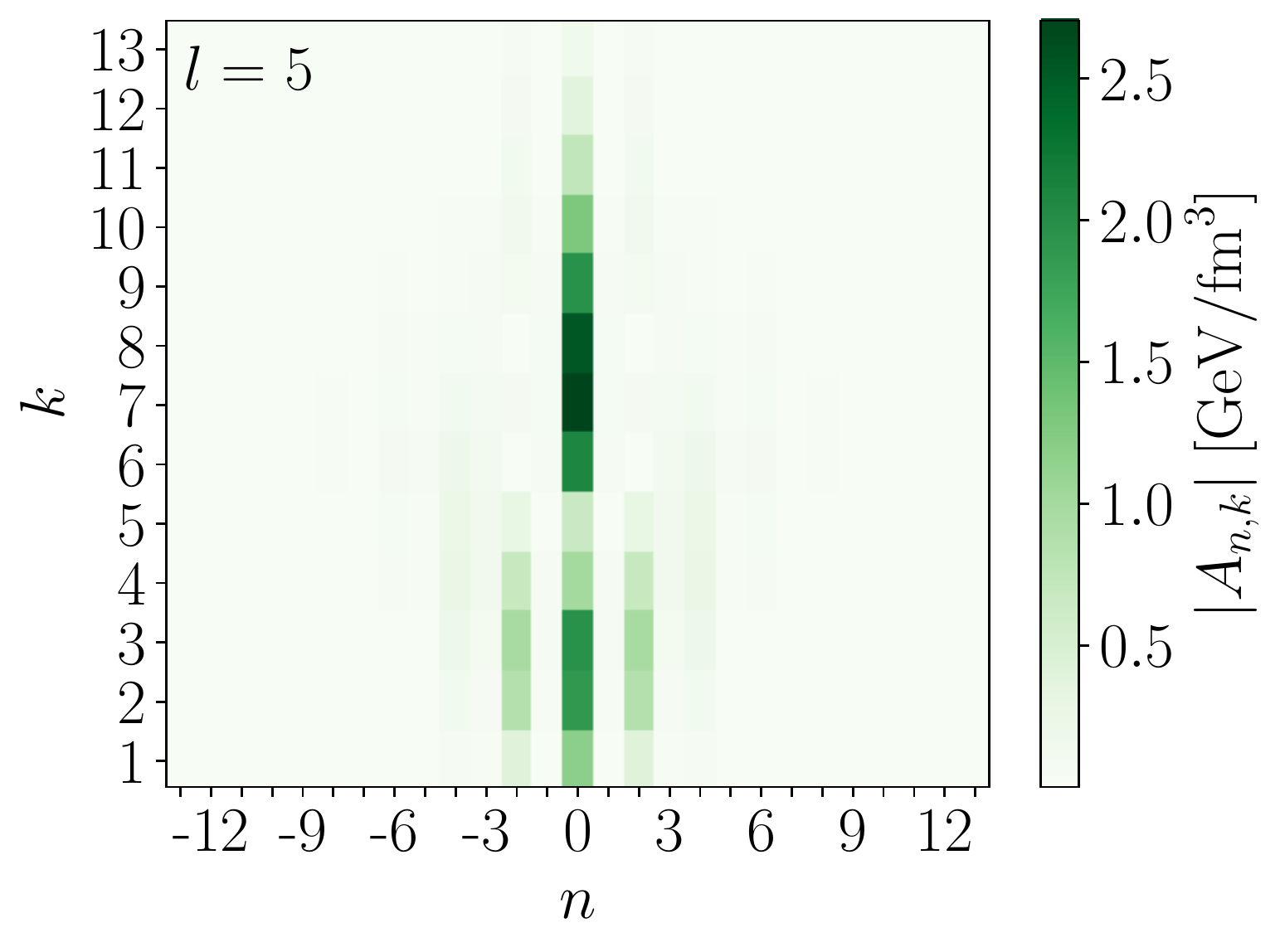}
\includegraphics[width=0.325\linewidth]{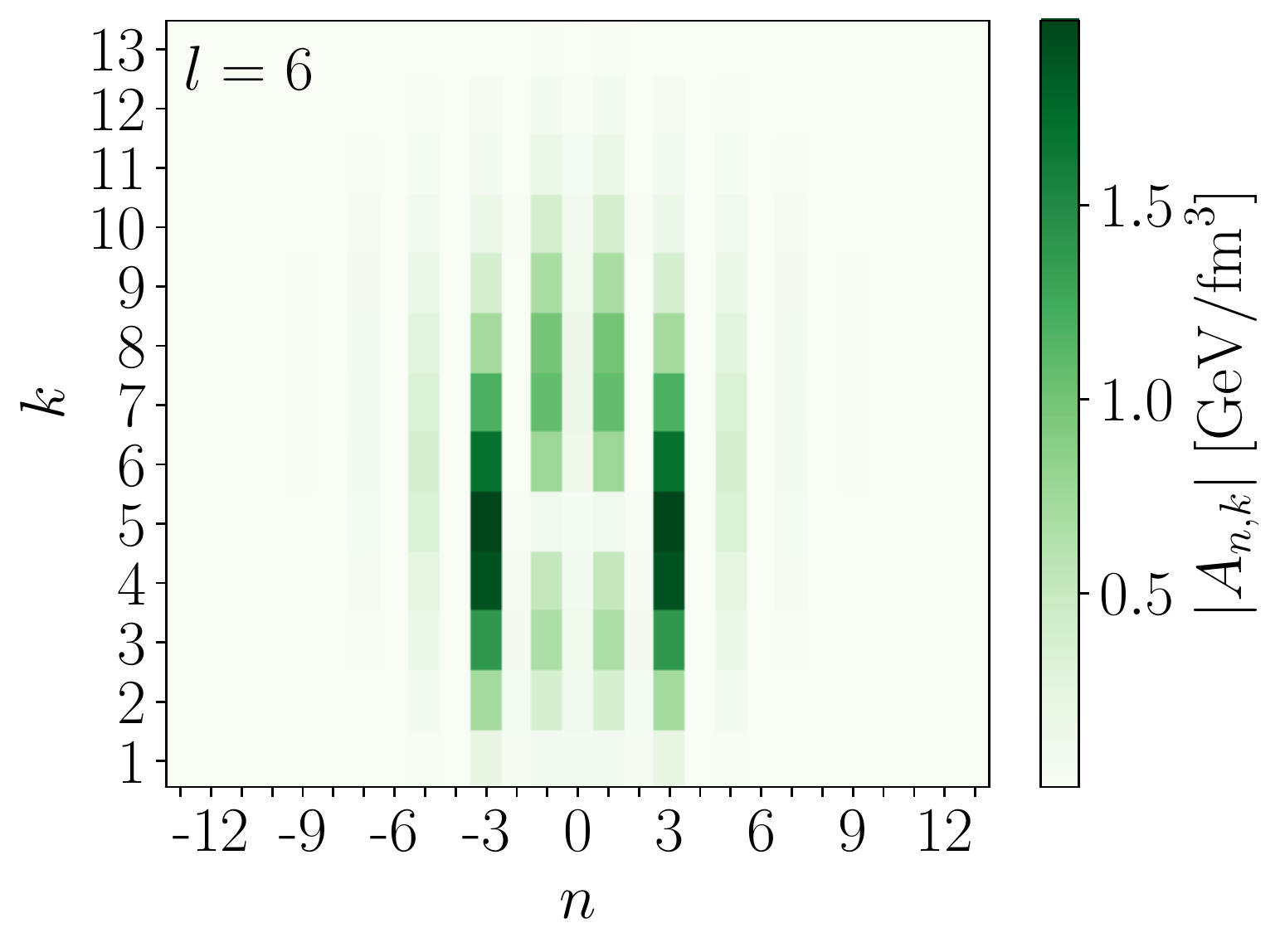}
\includegraphics[width=0.325\linewidth]{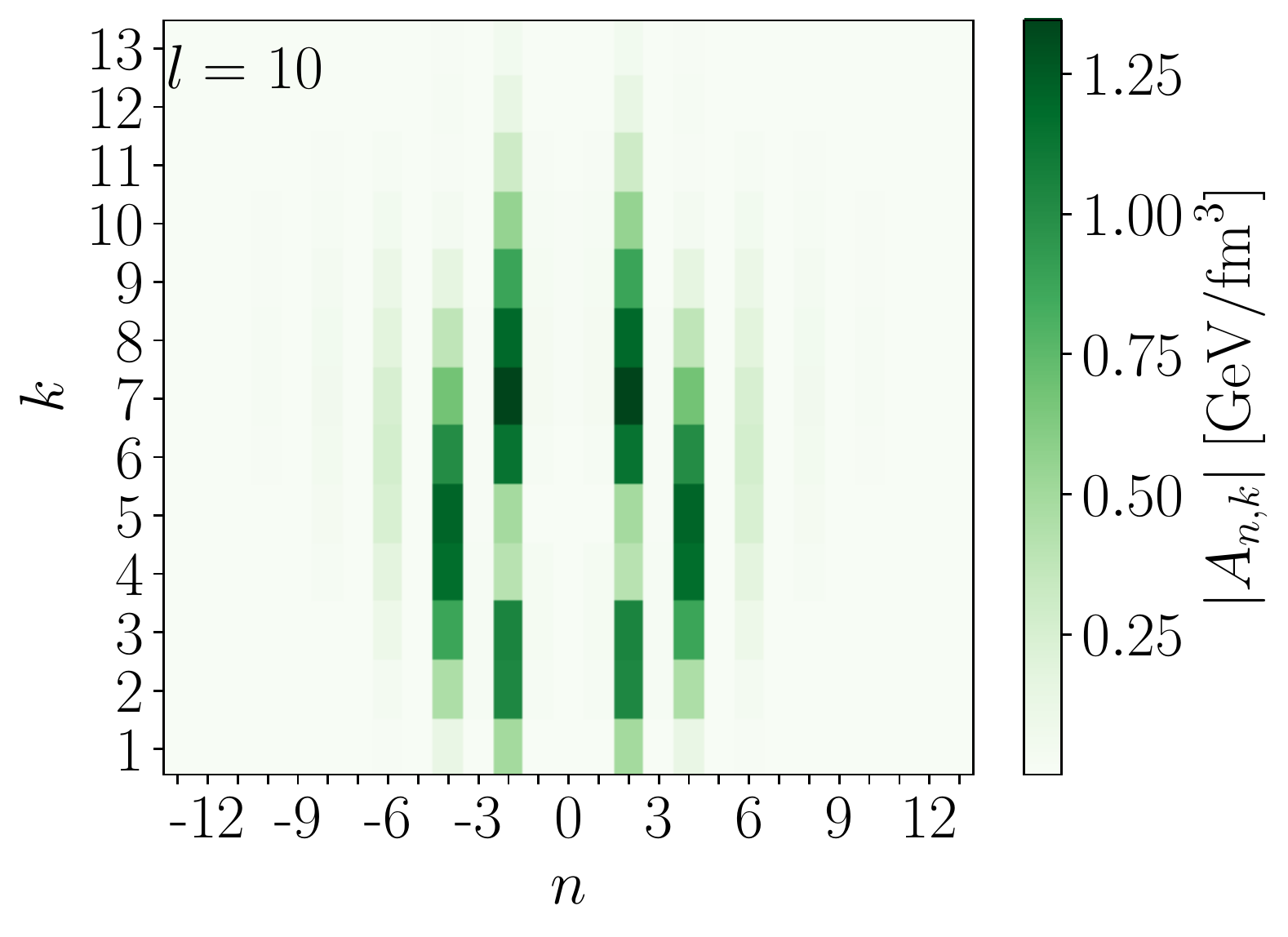}
\vspace{-3mm}
\caption{Absolute values $|A_{n,k}|$ of the Bessel--Fourier expansion coefficients for a few modes at $b=9$~fm in the Glauber model.}
\label{fig:BesselFourier_Glauber_b9}
\end{figure*}

We first show in Fig.~\ref{fig:BesselFourier_Glauber_AvSt} the absolute values of the expansion coefficients for the average states at $b=0$ and $b=9$~fm of Fig.~\ref{fig:avg_sates_density}.
As was to be anticipated, at vanishing impact parameter the average state consists of radially symmetric components only, with coefficients $A_{0,k}$ that decrease rapidly with $k$.
Going to $b=9$~fm, the description of the average state necessitates more sizable coefficients.
These include first a large contribution with $n=0$, similar to that at $b=0$, with the interesting difference that the maximum coefficient is no longer that with $k=1$ but rather $A_{0,2}$.
Additionally, there are smaller but still sizable coefficients with $n=2$ and 4, corresponding to the nonzero $\varepsilon_2$ and $\varepsilon_4$ of the average state at $b=9$~fm reported in Table~\ref{tab:average_state_eccentricities}.

In Figs.~\ref{fig:BesselFourier_Glauber_b0} and \ref{fig:BesselFourier_Glauber_b9} we show the absolute values $|A_{n,k}|$ of the Bessel--Fourier coefficients for a few chosen modes at $b=0$ and $b=9$~fm respectively. 
At vanishing impact parameter we saw previously that there are radially symmetric modes, as well as modes with a single nonvanishing eccentricity $\tilde{\varepsilon}_n$. 
We start with radially symmetric modes in the left panels of Fig.~\ref{fig:BesselFourier_Glauber_b0}, namely the coefficients for the mode $l=0$ resp.\ $l=7$ in the top resp.\ bottom panel. 
Only coefficients with $n=0$ are sizable. 
In addition, the largest coefficient (in absolute value) lies at higher $k$ than for the average state: the maximum is at larger $k$ for mode $l=7$, which changes sign twice along the radial direction, than for mode $l=0$, which changes sign only once. 
The latter property means that higher-excited modes have more structure on smaller length scales, reflecting their sign changes, and thus more weight $|A_{n,k}|$  at higher $k$-values. 
Accordingly, it also holds when one looks at modes with a dipole structure, i.e., a sizable $\tilde{\varepsilon}_1$ (middle panels of Fig.~\ref{fig:BesselFourier_Glauber_b0}) or modes with an $\tilde{\varepsilon}_2$ (right panels).
Quite normally, these modes only have sizable coefficients $A_{\pm 1,k}$ or $A_{\pm 2,k}$, respectively.
Eventually, one can note that for the modes in the lower panels the $|A_{n,k}|$ have two successive maxima as $k$ increases.

At $b=9$~fm we have seen that the modes typically have several nonzero eccentricities $\tilde{\varepsilon}_n$. 
This reflects itself in that the Bessel--Fourier coefficients $A_{n,k}$ are now sizable for several values of $n$, as illustrated in Fig.~\ref{fig:BesselFourier_Glauber_b9}. 
Calling for brevity ``mostly $\varepsilon_n$'' a mode whose largest eccentricity is that in the $n$-th harmonic, we show a mostly $\varepsilon_1$ mode (middle upper panel), two mostly $\varepsilon_2$ modes (right panels), and a mostly $\varepsilon_3$ mode (middle lower panel).
We also show the $\{|A_{n,k}|\}$ for ``quasi-radial'' modes, for which the maximum coefficients are with $n=0$ (left). 
Although all these modes are low-lying ones, with $l\leq 10$, their description necessitates significantly more expansion coefficients than at $b=0$. 
This holds not only for the azimuthal dependence ($n$), but also radially ($k$), which shows that there is more structure at smaller length scales.

\section{Mode-by-mode response}
\label{sec:response}

The fluctuation modes introduced above generally influence the systems characteristics, be it in the initial state, where the modes are defined, or in the final state following some dynamical evolution. 
To describe this influence, we introduce in Sec.~\ref{subsec:response_theory} coefficients quantifying the linear and quadratic effects of a given mode on an arbitrary observable.
After specifying in Sec.~\ref{subsec:observables} the observables we shall investigate in the following, we briefly describe in  Sec.~\ref{subsec:time_evol} our setup for the system evolution using K\o MP\o ST and MUSIC.
We then present in Sec.~\ref{subsec:lin_quad_response} the mode-by-mode response of the system characteristics, both at linear and quadratic order, for initial conditions from the MC Glauber model. 
Eventually, we focus on the linear and nonlinear dynamic response of anisotropic-flow coefficients to the asymmetry in the initial-state geometry in Sec.~\ref{subsec:response_coeff}.

\subsection{Linear and quadratic response of observables}
\label{subsec:response_theory}

Consider a generic set of observables $\{O_\alpha\}$, where the index $\alpha$ labels different observables, such as, e.g., $v_n,dN_\textrm{ch}/d\eta,\dots$. Energy density profiles $\Phi$ of every event, can be decomposed according to Eq.~\eqref{eq:evt_vs_modes} into an average state $\bar{\Psi}$ and fluctuation modes $\{\Psi_l\}$ 
\begin{equation}
\Phi = \bar{\Psi} + \sum_l c_l \Psi_l,
\end{equation}
where the expansion coefficients $c_l$ are typically of order unity. By performing a Taylor expansion of an observable $O_\alpha$ around the average state ($\bar{\Psi}$), we can then express the value of an observable $O_\alpha(\Phi)$ in a given event as
\begin{equation}
O_\alpha(\Phi) = 
O_\alpha(\bar{\Psi}) + 
\sum_l\frac{\partial O_\alpha}{\partial c_l} \bigg|_{\bar{\Psi}} c_l + \frac{1}{2} \sum_{l,l'}\frac{\partial^2 O_\alpha}{\partial c_l\, \partial c_{l'}} \bigg|_{\bar{\Psi}}  c_l c_{l^\prime} + \mathcal{O}(c_l^3) \equiv  \bar{O}_\alpha + \sum_l L_{\alpha,l} c_l +  
\frac{1}{2}\!\sum_{l,l'}Q_{\alpha,l l'}c_l c_{l'} + \mathcal{O}(c_l^3),
\label{Oseries}
\end{equation}
where $\bar{O}_\alpha=O_\alpha(\bar{\Psi})$ denotes the value of the observable $O_\alpha$ in the average state $\bar{\Psi}$, while $L_{\alpha,l}=\frac{\partial O_\alpha}{\partial c_l} \big|_{\bar{\Psi}}$ resp.\ $Q_{\alpha,ll^\prime}=\frac{\partial^2 O_\alpha}{\partial c_l\, \partial c_{l'}}\big|_{\bar{\Psi}}$ is the linear- resp.\ quadratic-response coefficient of $O_\alpha$ to the mode $l$ resp.\ the modes $l$ and $l'$.

By truncating the Taylor expansion at second order, the statistical average of an observable $O_\alpha$ over events then reads
\begin{equation}
\expval{O_\alpha} \simeq \expval{ \bar{O}_\alpha + \sum_l L_{\alpha,l} c_l +
\frac{1}{2} \sum_{l,l'} Q_{\alpha,ll'} c_l c_{l'}} = \bar{O}_\alpha + \frac{1}{2} \sum_l Q_{\alpha,ll},
\label{eq:<O_alpha>}
\end{equation}
where we used $\expval{\bar{O}_\alpha}=\bar{O}_\alpha$ and the properties~\eqref{eq:<c_l>=0} and \eqref{eq:<c_lc_l'>} of the statistics of the coefficients $\{c_l\}$.
Interestingly, this average value does not involve the linear-response coefficients $L_{\alpha,l}$, nor the quadratic coefficients with $l\neq l'$. 
In turn, the covariance of two observables $O_\alpha$ and $O_\beta$ at quadratic order in $\{c_l\}$ also follows naturally from Eq.~\eqref{Oseries}. 
Invoking Eq.~\eqref{eq:<O_alpha>} one finds
\begin{equation}
\expval{(O_\alpha-\expval{O_\alpha})(O_\beta-\expval{O_\beta})} \simeq 
\sum_l L_{_\alpha,l}L_{\beta,l}.
\label{eq:covariance}     
\end{equation}
In particular setting $\beta=\alpha$ yields the variance of $O_\alpha$.
That is, the covariances are entirely determined by the linear-response coefficients $L_{_\alpha,l}$ (up to corrections of order $c_l^3$).

To compute the linear- and quadratic-response coefficients, which are first and second derivatives respectively, we introduce for every mode $\Psi_l$ the states
\begin{equation}
\Psi^+_l \equiv \bar{\Psi} + \delta \Psi_l,
\quad
\Psi^-_l \equiv \bar{\Psi}-\delta \Psi_l,
\label{eq:psi+/-}
\end{equation}
where $\delta$ is a small parameter.
Correspondingly, we compute the observables for these states: $O^\pm_{\alpha,l}\equiv O_\alpha(\Psi_l^\pm)$, and then estimate
\begin{gather}
L_{\alpha,l}=\frac{O^+_{\alpha,l}-O^-_{\alpha,l}}{2\delta},
\label{eq:Lalpha} \\
Q_{\alpha, l l}=\frac{O^+_{\alpha,l} + O^-_{\alpha,l} -2\bar{O}_\alpha}{\delta^2},
\label{eq:Qalpha}
\end{gather}
by finite difference formulas for the first-order and second-order centered derivatives.

As a final remark, Eqs.~\eqref{Oseries}--\eqref{eq:covariance} can naturally be extended to higher orders in the coefficients $\{c_l\}$, which could make sense since in practice these coefficients are by construction of order unity.
However, when going to higher order in $c_l$ one encounters 3-point averages (or higher) of the $\{c_l\}$, which in contrast to Eqs.~\eqref{eq:<c_l>=0} and \eqref{eq:<c_lc_l'>} are not directly fixed by our construction and have to be extracted from the sample of events. We shall see hereafter that restricting oneself to order $c_l^2$ is already a sufficient approximation for a number of observables, for which the quadratic response is already subleading with respect to the linear response.  
The reason is that the $\{c_l\}$ are of order unity because we have absorbed the magnitude and thereby the importance of the fluctuations in the normalization of the modes, by scaling with $\sqrt{\lambda_l}$.
Physically, the actual expansion can be understood to proceed in terms of $\tilde{c}_l = \sqrt{\lambda_l}c_l$, so that higher modes typically contribute less (see Fig.~\ref{fig:eigenvalues}).
Similarly, higher orders in the $\{\tilde{c}_l\}$ generally correspond to small contributions compared with that of the average state.

\subsection{System characteristics}
\label{subsec:observables}

In our investigations of mode-by-mode response we consider multiple observables, which we now list. 

We first compute a number of characteristics of the initial state, from the centered energy density $e(r,\theta)$. 
For the latter, we consider either the average state $\bar{\Psi}$ or the states $\Psi^+_l$, $\Psi^-_l$ defined in Eq.~\eqref{eq:psi+/-} for the first 256 modes and for a few values of $\delta$ that are specified in Sec.~\ref{subsec:linearitycheck}, or ``full'' initial states as given by the Glauber or Saturation model.  
From the calculations with $\bar{\Psi}$ and $\Psi^\pm_l$ we obtain the response coefficients~\eqref{eq:Lalpha} and \eqref{eq:Qalpha}.

Integrating $e(r,\theta)$ over the whole transverse plane yields the total energy per unit rapidity $\d E/\d y$
\begin{equation}
\label{eq:dE/dy}
\frac{\d E}{\d y} \equiv \tau_0\!\int\!e(r,\theta)\,r\,\d r\,\d\theta,
\end{equation}
where the notation anticipates the fact that we consider a longitudinally boost-invariant system. 
Note that from this definition it is obvious that the response of $\d E/\d y$ to the addition of a fluctuation mode is purely linear, so that we can already anticipate that the corresponding quadratic-response coefficients $Q_{\alpha,ll}$ will vanish.

Next is the average square radius $\{r^2\}$, where the curly brackets $\{\dots\}$ denote an average over the centered energy density:
\begin{equation}
\{r^2\} \equiv \frac{\displaystyle\int\!r^2 e(r,\theta)\,r\,\d r\,\d\theta}%
{\displaystyle\int\! e(r,\theta)\,r\,\d r\,\d\theta}.
\label{eq:mean_rsquared}
\end{equation}
To characterize the asymmetry of the energy density profiles, we compute the spatial eccentricities~\cite{Teaney:2010vd,Gardim:2011xv} 
\begin{align}
\varepsilon_1 \textrm{e}^{\textrm{i}\Phi_1}&\equiv 
-\frac{\displaystyle\int\! r^3 \textrm{e}^{\textrm{i}\theta}e(r,\theta)\,r\,\d r\,\d\theta}%
{\displaystyle\int\! r^3 e(r,\theta)\,r\,\d r\,\d\theta}\quad\text{for } n=1, \label{eq:eccentricities1}\\
\varepsilon_n \textrm{e}^{\textrm{i}n\Phi_n}&\equiv 
-\frac{\displaystyle\int\! r^n \textrm{e}^{\textrm{i}n\theta}e(r,\theta)\,r\,\d r\,\d\theta}%
{\displaystyle\int\! r^{n}e(r,\theta)\,r\,\d r\,\d\theta}\quad\text{for }\ n\geq 2,
\label{eq:eccentricities2}
\end{align}
which we will quantify in terms of the cosine and sine parts $\varepsilon_{n,\mathrm{c}}$, $\varepsilon_{n,\mathrm{s}}$ given by
\begin{align}
\varepsilon_n \textrm{e}^{\textrm{i}n\Phi_n} = \varepsilon_{n,\mathrm{c}}+\textrm{i}_{}\varepsilon_{n,\mathrm{s}}
\end{align}
for $n\in\{1,\dots,5\}$. We note that, in contrast with the ``mode eccentricities'' $\tilde{\varepsilon}_n$ introduced in Sec.~\ref{subsec:classification_modes}, the energy density in Eqs.~\eqref{eq:mean_rsquared}--\eqref{eq:eccentricities2} is the same in the numerator and denominator. Since the full energy density enters the denominators of Eqs.~\eqref{eq:mean_rsquared}--\eqref{eq:eccentricities2}, $\{r^2\}$, the eccentricities do not necessarily respond linearly to the addition of a fluctuation mode and nonvanishing quadratic-response coefficients $Q_{\alpha,ll}$ become possible. 

After letting the system evolve from the initial state as we describe in Sec.~\ref{subsec:time_evol}, we obtain a final state consisting of hadrons. 
All final-state observables are computed at midrapidity, namely over the pseudorapidity range $\abs{\eta}\leq 0.5$.
From the charged hadron distribution $\d N_\textrm{ch}/p_\textrm{T}\,\d p_\textrm{T}\,\d\varphi_\textrm{p}\,\d\eta$ of each event, we compute first the charged multiplicity per unit pseudorapidity 
\begin{equation}
\frac{\d N_\textrm{ch}}{\d\eta} = 
\int\!\frac{\d N_\textrm{ch}}{p_\textrm{T}\,\d p_\textrm{T}\,\d\varphi_\textrm{p}\,\d\eta} \, p_\textrm{T}\,\d p_\textrm{T}\,\d\varphi_\textrm{p},
\label{eq:dNchdeta}
\end{equation}
with $\varphi_\textrm{p}$ being the particle momentum azimuth.
Here and below the integral over transverse momentum $p_\textrm{T}$ is performed in the range 0.01--3~GeV/$c$.

Next we compute the event-by-event average transverse momentum $[p_\textrm{T}]$ of particles
\begin{equation}
[p_\textrm{T}] \equiv 
\frac{\displaystyle\int\! p_\textrm{T}\frac{\d N_\textrm{ch}}{p_\textrm{T}\,\d p_\textrm{T}\,\d\varphi_\textrm{p}\,\d\eta}\, p_\textrm{T}\,\d p_\textrm{T}\,\d\varphi_\textrm{p}}%
{\displaystyle\int\! \frac{\textrm{d}N_\textrm{ch}}{p_\textrm{T}\,\d p_\textrm{T}\,\d\varphi_\textrm{p}\,\d\eta}\, p_\textrm{T}\,\d p_\textrm{T}\,\d\varphi_\textrm{p}}.
\label{eq:av_pt}
\end{equation}
The last type of observables we consider are the integrated anisotropic flow coefficients~\cite{Voloshin:1994mz}
\begin{equation}
v_n \textrm{e}^{\textrm{i}n\Psi_n} \equiv 
\frac{\displaystyle\int\! \textrm{e}^{\textrm{i}n\varphi_\textrm{p}} \frac{\textrm{d}N_\textrm{ch}}{p_\textrm{T}\,\d p_\textrm{T}\,\d\varphi_\textrm{p}\,\d\eta}\, p_\textrm{T}\,\d p_\textrm{T}\,\d\varphi_\textrm{p}}%
{\displaystyle\int\! \frac{\textrm{d}N_\textrm{ch}}{p_\textrm{T}\,\d p_\textrm{T}\,\d\varphi_\textrm{p}\,\d\eta}\, p_\textrm{T}\,\d p_\textrm{T}\,\d\varphi_\textrm{p}}.
\label{eq:flow_harmonics}
\end{equation}
Consistent with the initial-state observables, we also consider the cosine and sine parts $v_{n,\mathrm{c}}$, $v_{n,\mathrm{s}}$ according to $v_n \textrm{e}^{\textrm{i}n\Psi_n}=v_{n,\mathrm{c}}+\textrm{i}_{}v_{n,\mathrm{s}}$.
In fact, considering the sine and cosine parts separately also has the advantage that they behave more smoothly around $\delta=0$, since they can take both positive or negative values. 
In contrast the absolute values $\varepsilon_n=\sqrt{(\varepsilon_{n,\mathrm{c}})^2+(\varepsilon_{n,\mathrm{s}})^2}$ or $v_n=\sqrt{(v_{n,\mathrm{c}})^2+(v_{n,\mathrm{s}})^2}$ are by definition always non-negative, such that their first derivative with respect to some of the $c_l$ may be undefined at $\delta=0$.

\subsection{Time evolution of the system}
\label{subsec:time_evol}

We let the system evolve from the initial energy density profile using two numerical successive frameworks.
For the pre-equilibrium evolution we use the effective kinetic description K\o MP\o ST~\cite{Kurkela:2018vqr} and for the subsequent evolution we employ the relativistic dissipative hydrodynamics code MUSIC~\cite{Schenke:2010nt,Schenke:2010rr,Paquet:2015lta}.

With the energy density profiles obtained from the Glauber or Saturation models we generate an initial energy-momentum tensor
\[
\textstyle T^{\mu\nu}(\tau_0,x,y) = \mathrm{diag}(e(x,y),\frac{1}{2}e(x,y),\frac{1}{2}e(x,y),0)
\]
as in Ref.~\cite{Kurkela:2018vqr} and assume longitudinal boost invariance. 
The points with coordinates $(x,y)$ are now the nodes of a grid with spacing 0.1~fm: 
to obtain the corresponding values of the energy density, we used a bilinear interpolation scheme to reduce the coarser grids on which we computed the average events and the modes. 
We start the pre-equilibrium stage at $\tau_0=0.2$~fm/$c$ and let the system evolve with an effective shear viscosity to entropy density ratio $\eta/s=0.16$ until $\tau_\mathrm{hydro}=1.1$~fm/$c$.

The output energy momentum tensor of K\o MP\o ST is then used as initial condition for the fluid-dynamical evolution.
MUSIC is run in its boost-invariant mode, so effectively it is a $2+1$ dimensional evolution.
The equation of state is taken from lattice QCD results by the hotQCD collaboration~\cite{HotQCD:2014kol}.
For the first-order transport coefficients we use a constant value of $\eta/s=0.16$ and vanishing bulk viscosity.
Particlization is performed at the level of the distribution function with the Cooper--Frye prescription at a fixed temperature $T_\mathrm{fo} = 155$~MeV, using the Cornelius algorithm~\cite{Huovinen:2012is} to find the hypersurface and including $\delta f$ corrections in the MUSIC code. 
For this step we include 320 particle species and compute their momentum distributions.
The particles are then allowed to further decay, but further hadronic interactions are not included.
At the end of the resonance decays, all observables are computed from the charged-hadron single-particle distribution using the equations presented in Sec.~\ref{subsec:observables}.

\subsection{Response of observables in the Glauber model}
\label{subsec:lin_quad_response}

\subsubsection{Linearity check}
\label{subsec:linearitycheck}

\begin{figure*}[!t]
	\includegraphics[width=0.495\linewidth]{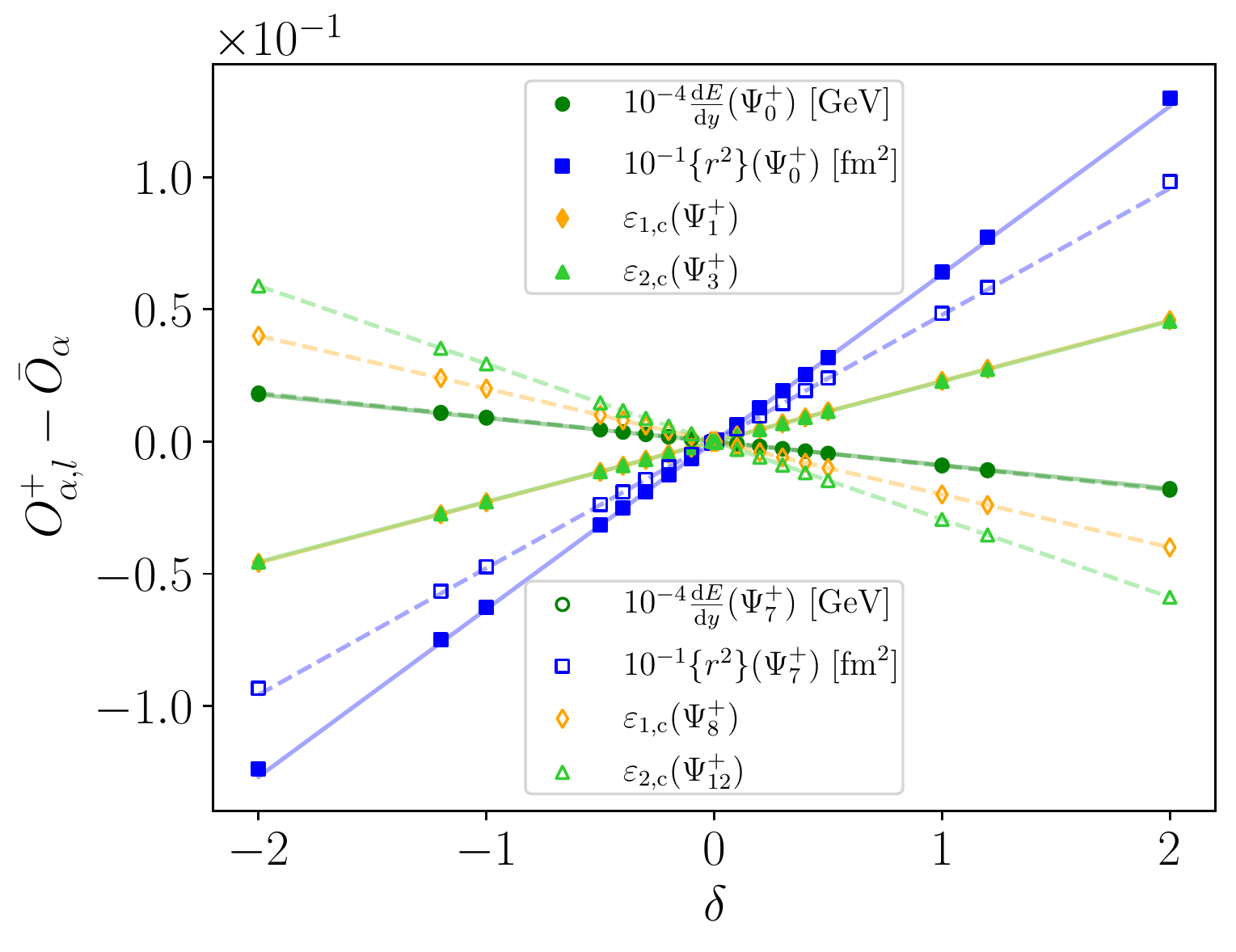}
	\includegraphics[width=0.495\linewidth]{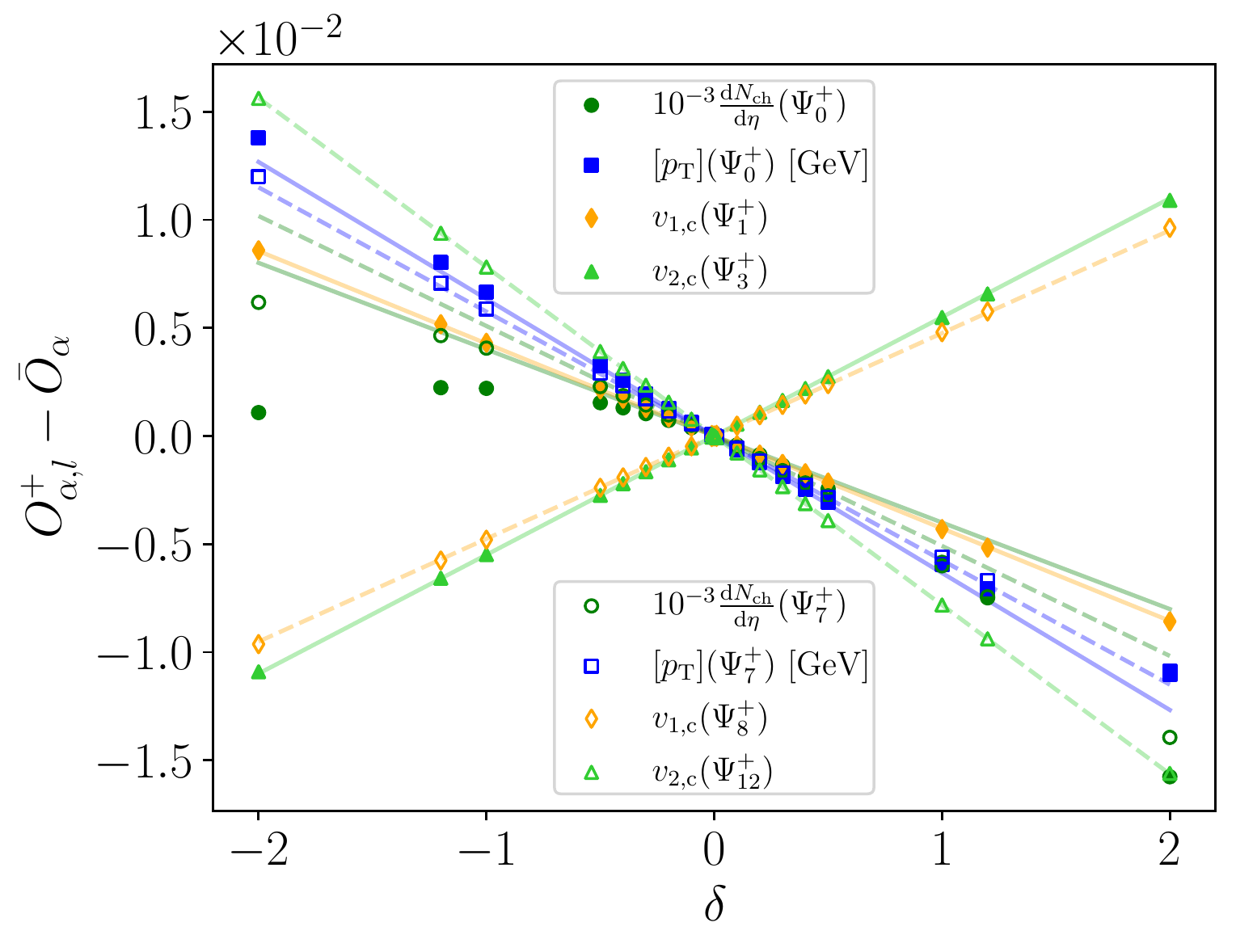}
	\includegraphics[width=0.495\linewidth]{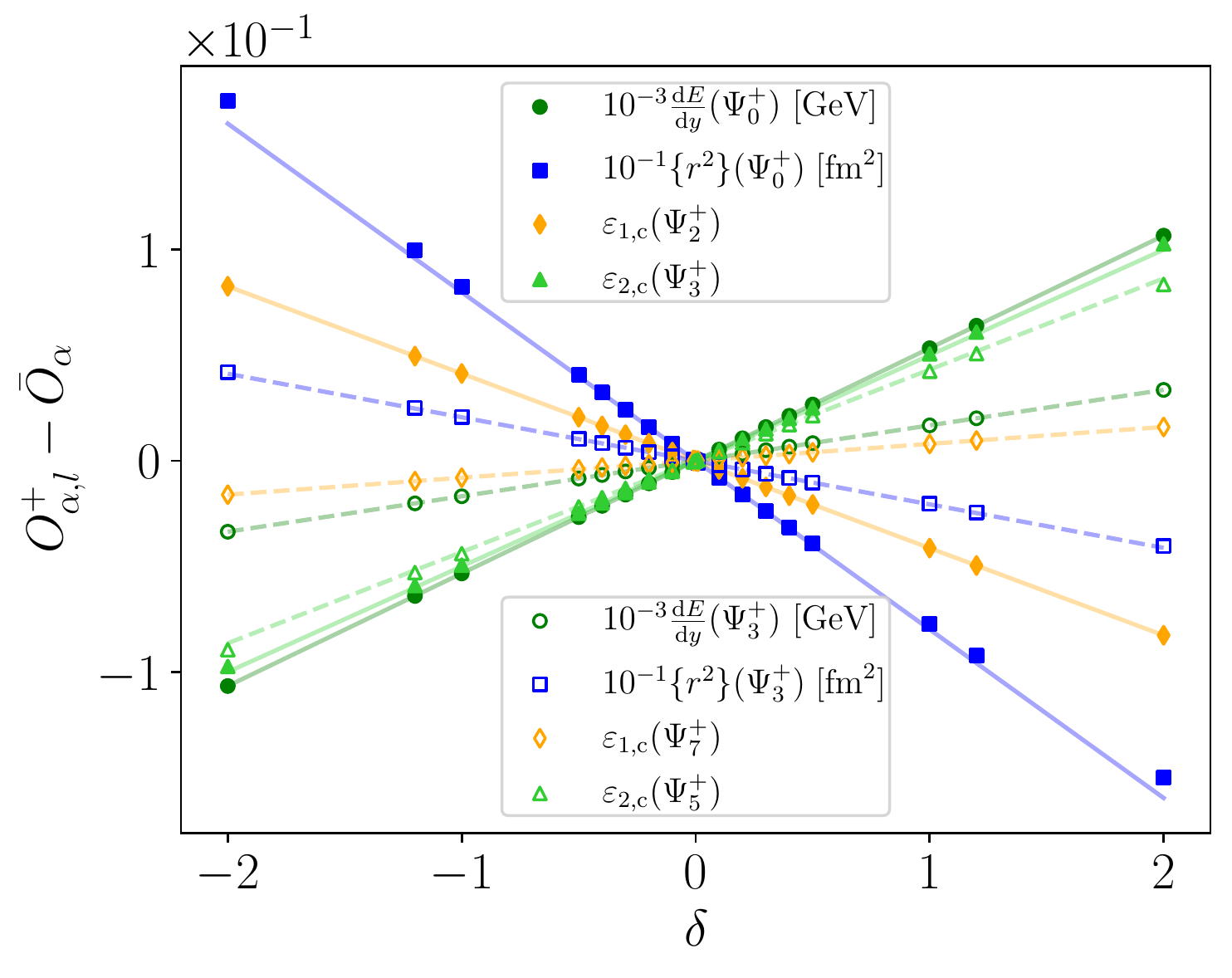}
	\includegraphics[width=0.495\linewidth]{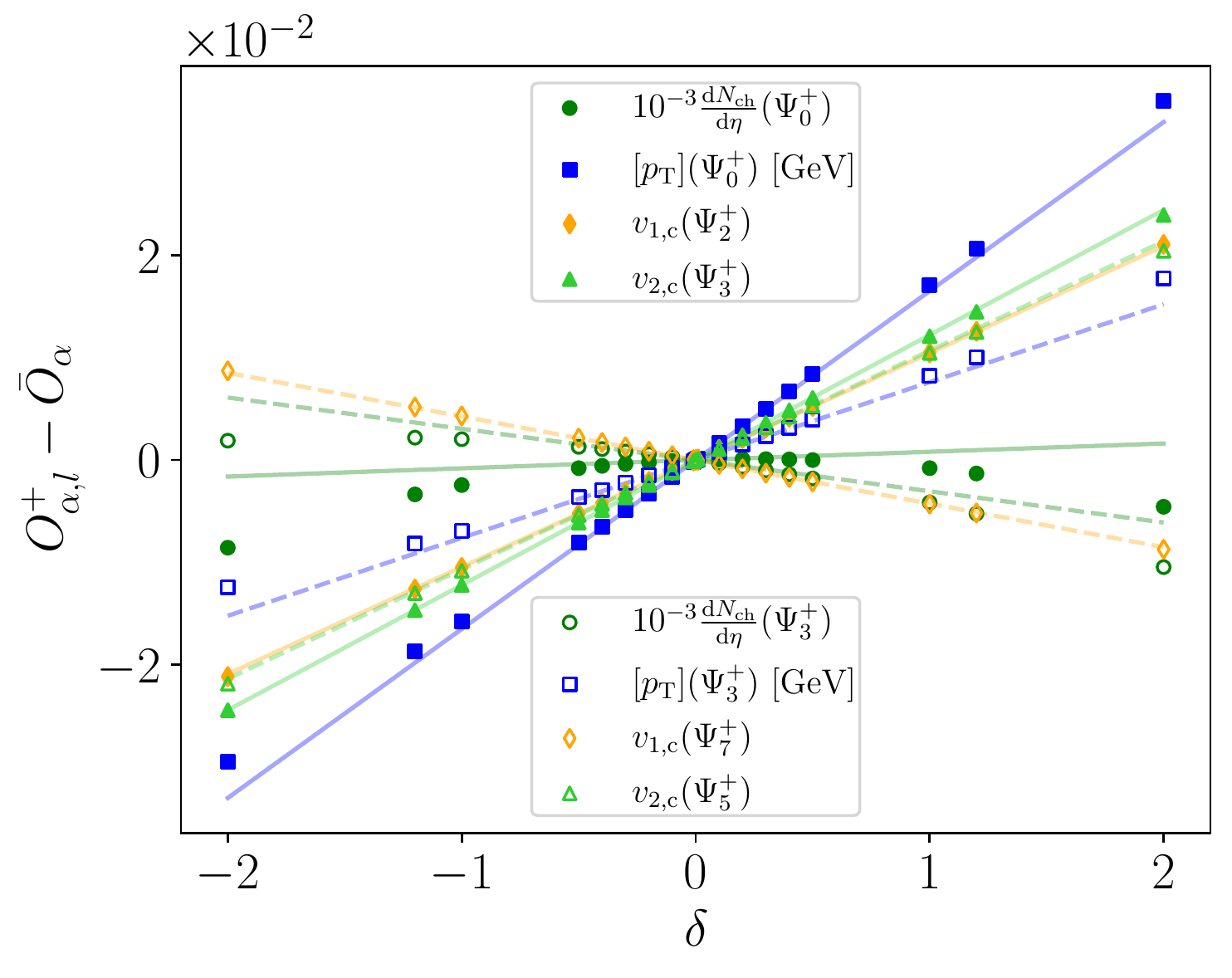}
	\vspace{-7mm}
	\caption{Variation with $\delta$ of $O^+_{\alpha,l}-\bar{O}_\alpha$ for initial-state (left) and final-state (right) observables, together with linear fits to the points with $|\delta|\leq 0.01$, using the Glauber model at $b=0$ (top) and $b=9$ fm (bottom)
		Closed symbols and full lines correspond to the first modes contributing to the respective observable, while open symbols and dashed lines are for the second modes.}
	\label{fig:Linearity_test_Glauber}
\end{figure*}

As a first assessment of how the observables $O_\alpha$ vary when fluctuation modes $\Psi_l$ are added to the average state, we compute $O^+_{\alpha,l}-\bar{O}_\alpha \equiv O_\alpha(\bar{\Psi}+\delta\Psi_l)-O_\alpha(\bar{\Psi})$ for various values of $\delta$ ranging between $-2$ and 2 and a number of modes.
In Fig.~\ref{fig:Linearity_test_Glauber} we show the dependence on $\delta$ of $O^+_{\alpha,l}-\bar{O}_\alpha$ for initial-state (left) and final-state (right) characteristics in the Glauber model,\footnote{The corresponding investigation for observables with initial states from the Saturation model are presented in Appendix~\ref{appendix:linearity_check}, with similar results.} for collisions at $b=0$ (top) and $b=9$~fm (bottom).
As examples, we chose the first (full symbols) and second (open symbols) modes with a nonzero contribution to $\d E/\d y$, $\{r^2\}$, $\varepsilon_{1,\mathrm{c}}$ and $\varepsilon_{2,\mathrm{c}}$ in the initial state. 
These result in a nonvanishing response for $\d N_\textrm{ch}/\d\eta$, $[p_\textrm{T}]$, $v_{1,\mathrm{c}}$ and $v_{2,\mathrm{c}}$ in the final state.
For each observable and mode, we also display straight lines obtained by fitting the points with $\delta\in\lbrace 0,\pm 0.001,\pm 0.01\rbrace$.

At $b=0$ the initial-state characteristics mostly depend linearly on $\delta$.
The only exception is the average squared radius $\{r^2\}$, which slightly deviates from the linear fit at the largest values of $|\delta|$.
This departure from linearity reflects the fact that the denominator in Eq.~\eqref{eq:mean_rsquared} is itself dependent of $\delta$, making the ratio nonlinear.
In the final state, the flow coefficients also depend linearly on $\delta$. 
In contrast the charged hadron multiplicity and the average transverse momentum show a marked departure from the linear behavior, to which we shall come back in the next Section.

Very similar results hold at finite impact parameter $b=9$~fm. 
In the initial state, the nonlinearity in $\{r^2\}$ is larger than at $b=0$ and there is also a small nonlinearity in $\varepsilon_{2,\mathrm{c}}$.
Looking at the final-state observables, the nonlinearity of $\varepsilon_{2,\mathrm{c}}$ translates into a slightly nonlinear $v_{2,\mathrm{c}}$, less striking than for $\textrm{d}N_\textrm{ch}/\textrm{d}\eta$ or $[p_\textrm{T}]$.

\subsubsection{Linear- and quadratic-response coefficients}
\label{subsec:L_Q_Glauber}

\begin{figure*}[!htb]
\includegraphics[width=0.495\linewidth]{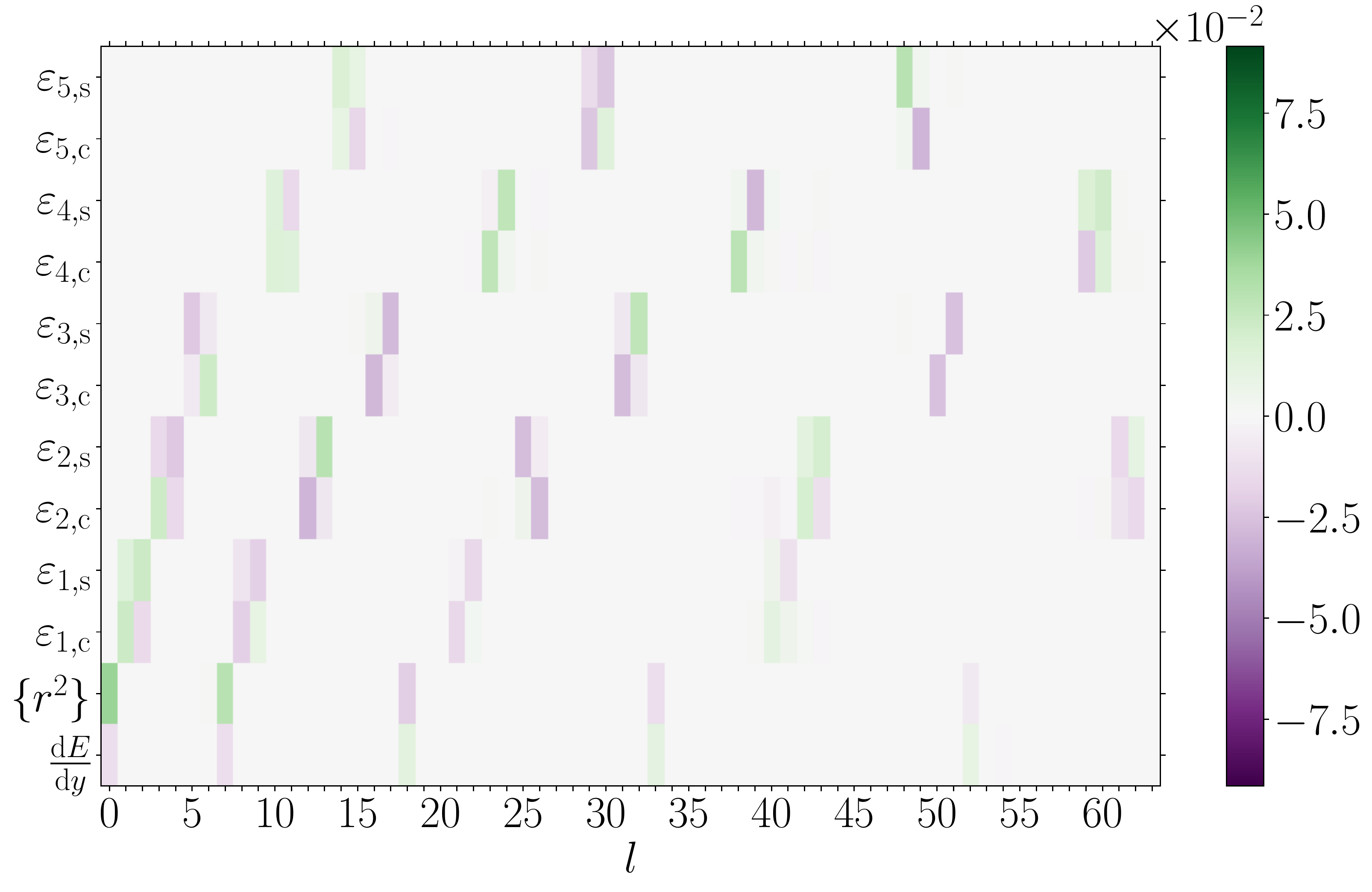}
\includegraphics[width=0.495\linewidth]{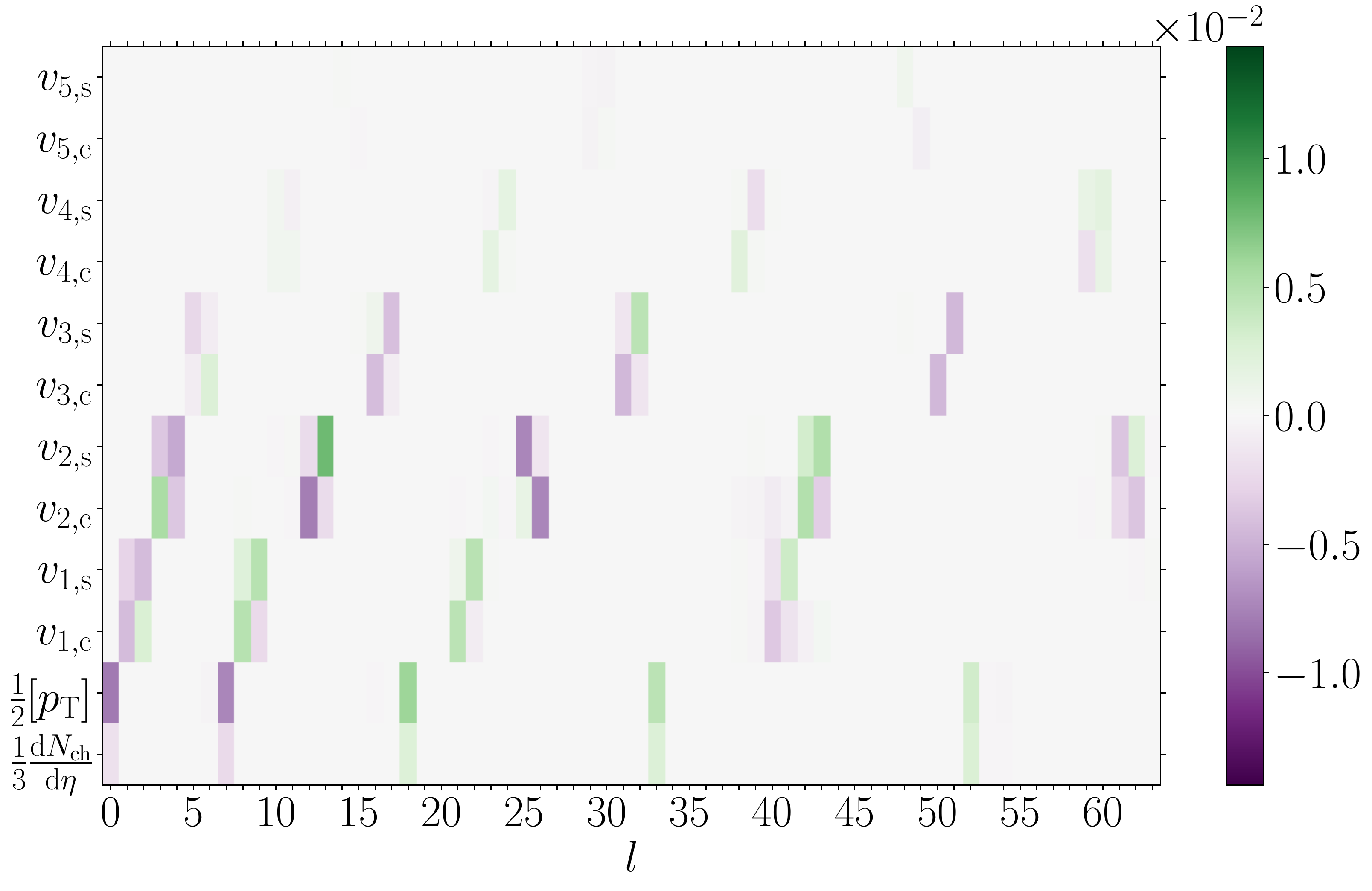}
\includegraphics[width=0.495\linewidth]{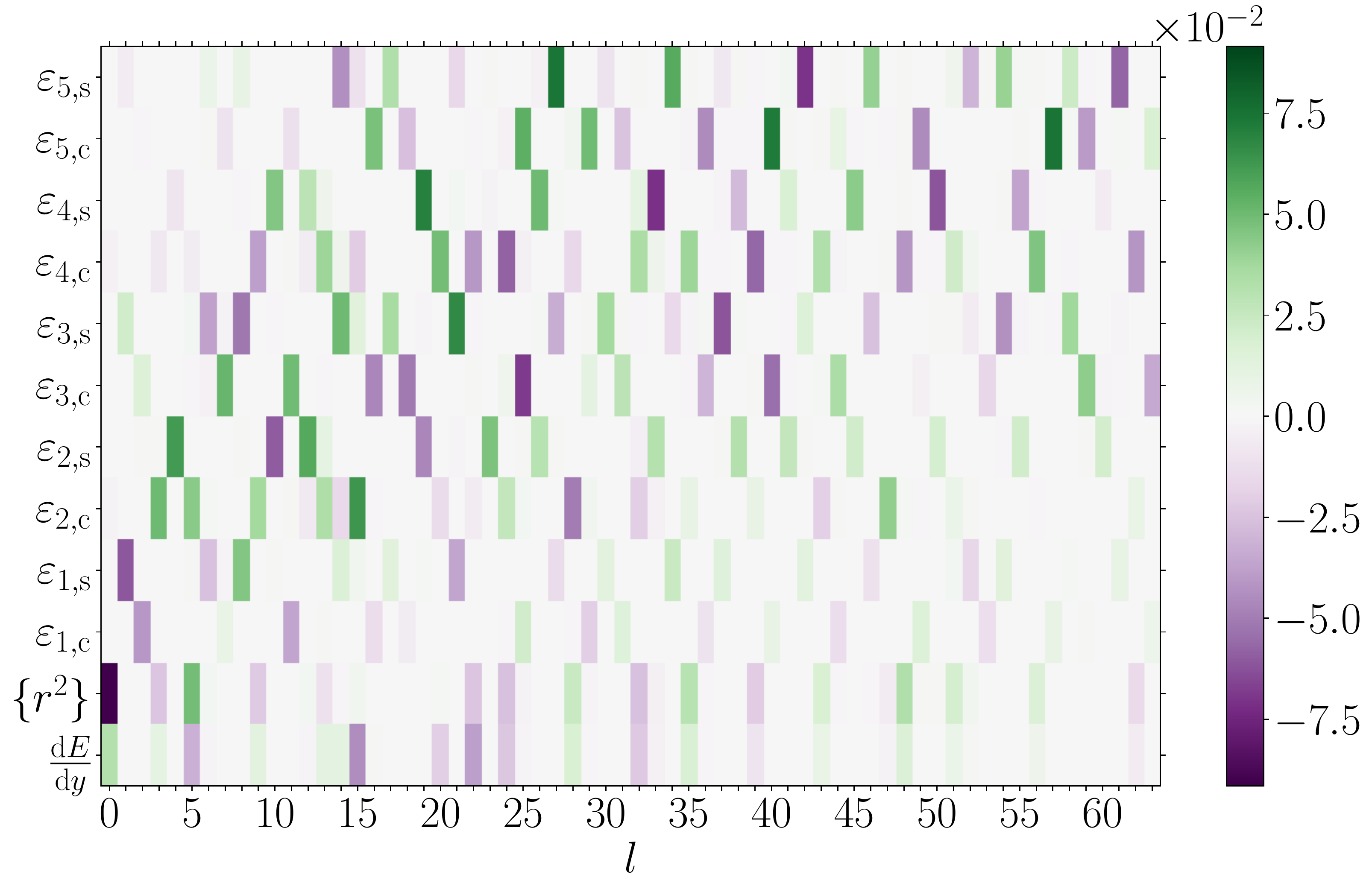}
\includegraphics[width=0.495\linewidth]{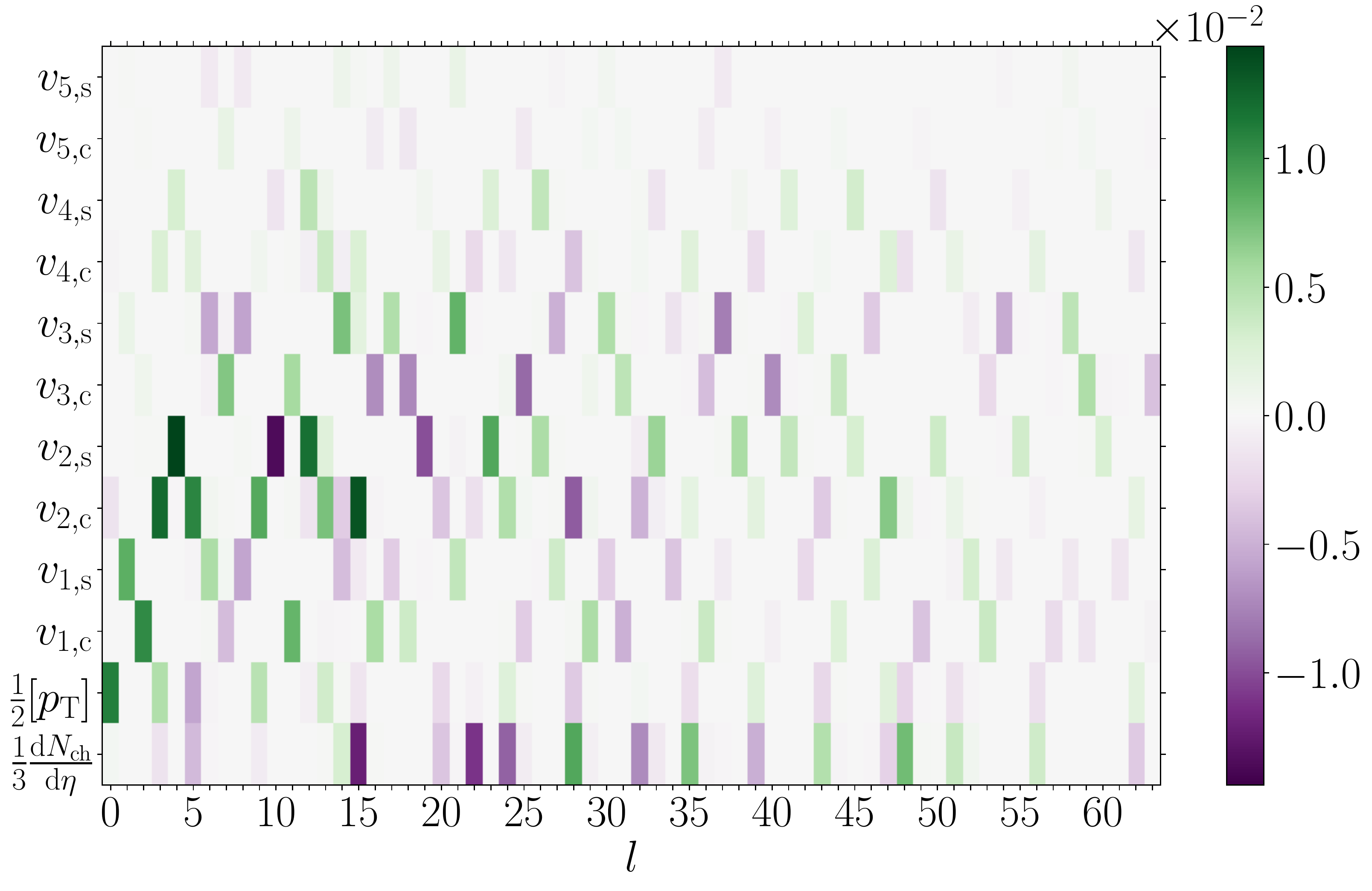}
\vspace{-5mm}
\caption{Linear-response coefficients $L_{\alpha,l}$ for initial-state characteristics (left) and final-state observables (right) in the Glauber model at $b=0$ (top) and $b=9$~fm (bottom). The coefficients for dimensionful observables and multiplicity have been divided by $\bar{O}_\alpha$.}
\label{fig:L_IS_FS_Glauber}
\end{figure*}

\begin{figure*}[!htb]
\centering
\includegraphics[width=0.495\linewidth]{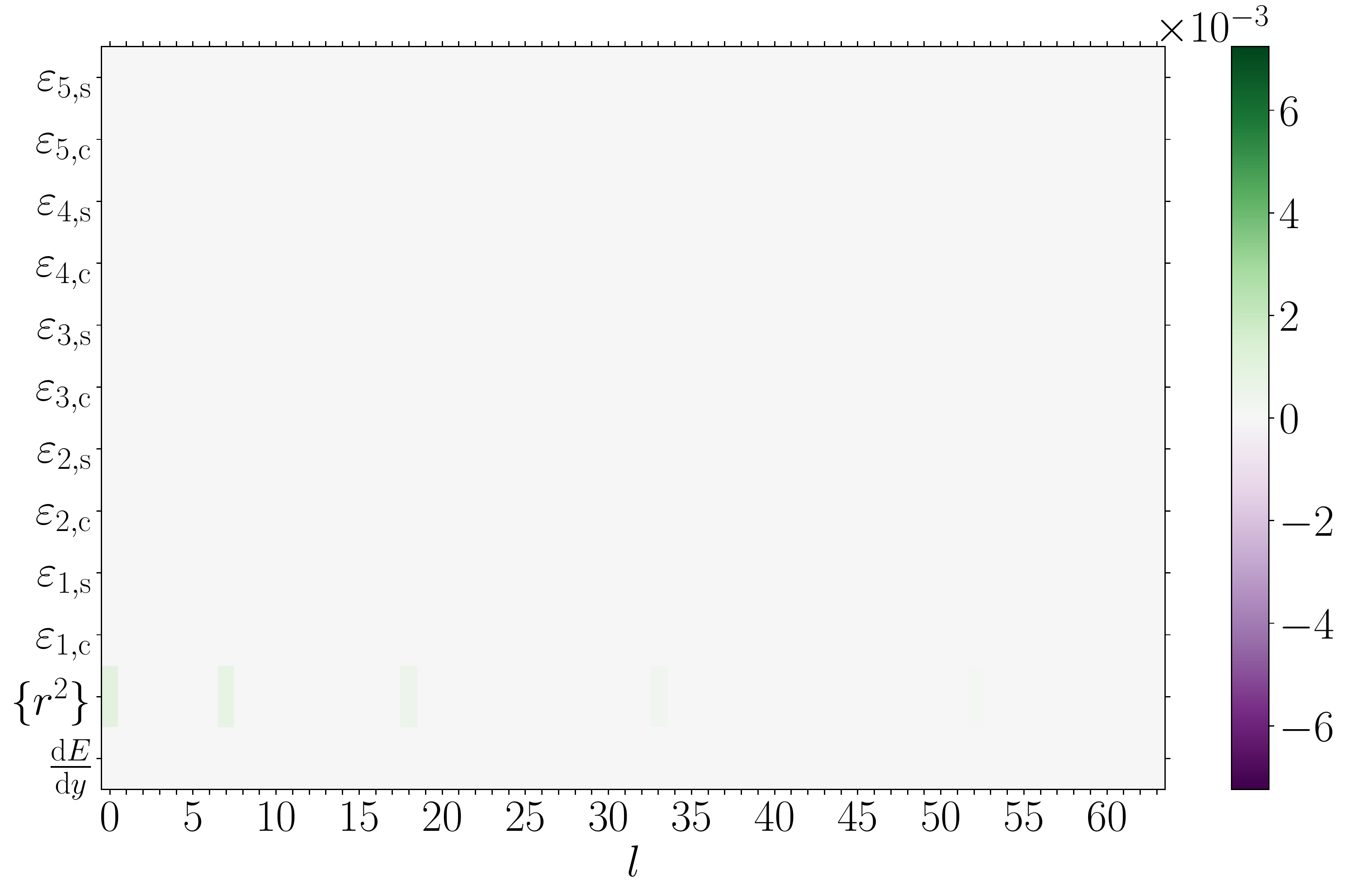}
\includegraphics[width=0.495\linewidth]{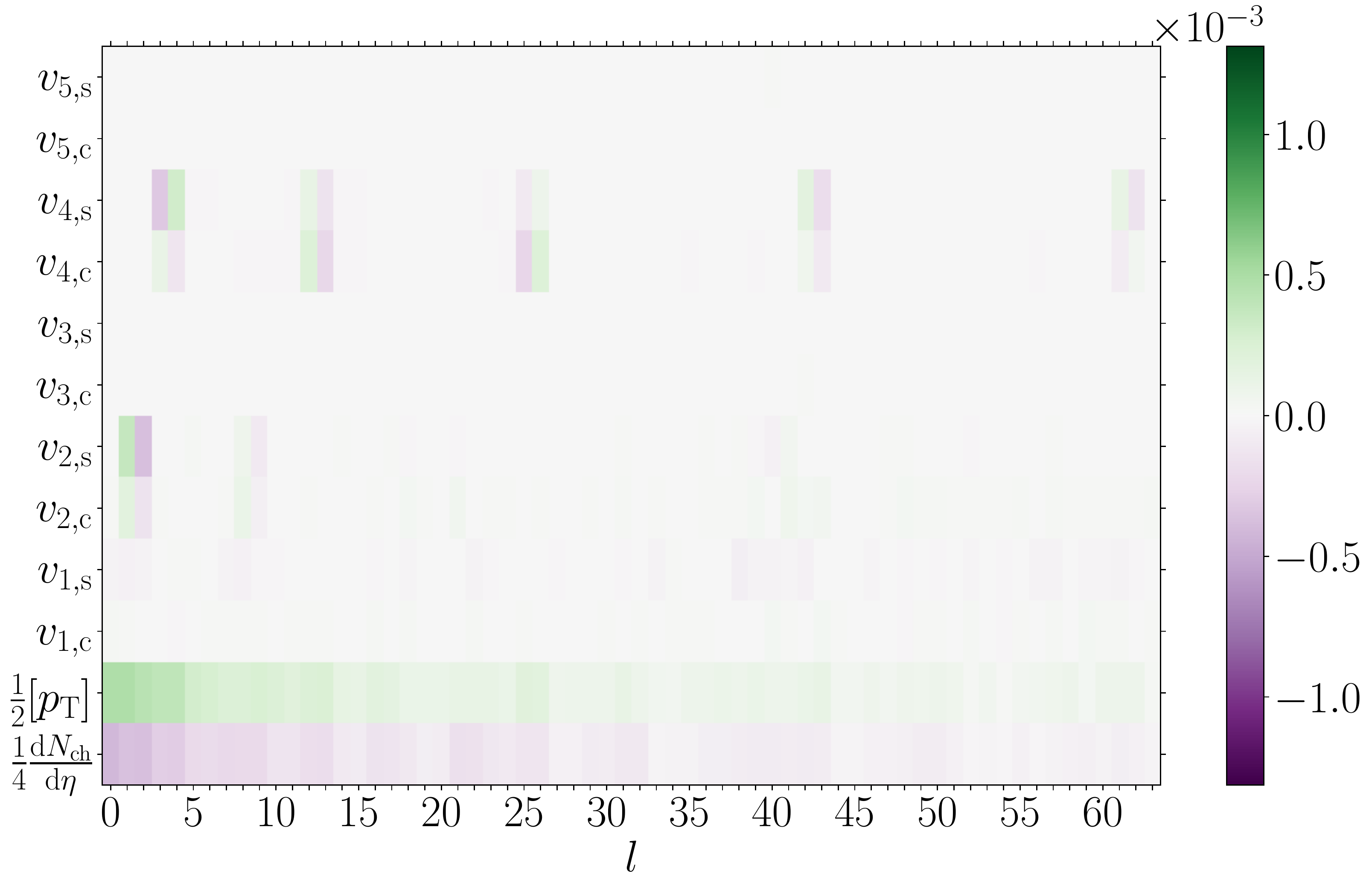}
\includegraphics[width=0.495\linewidth]{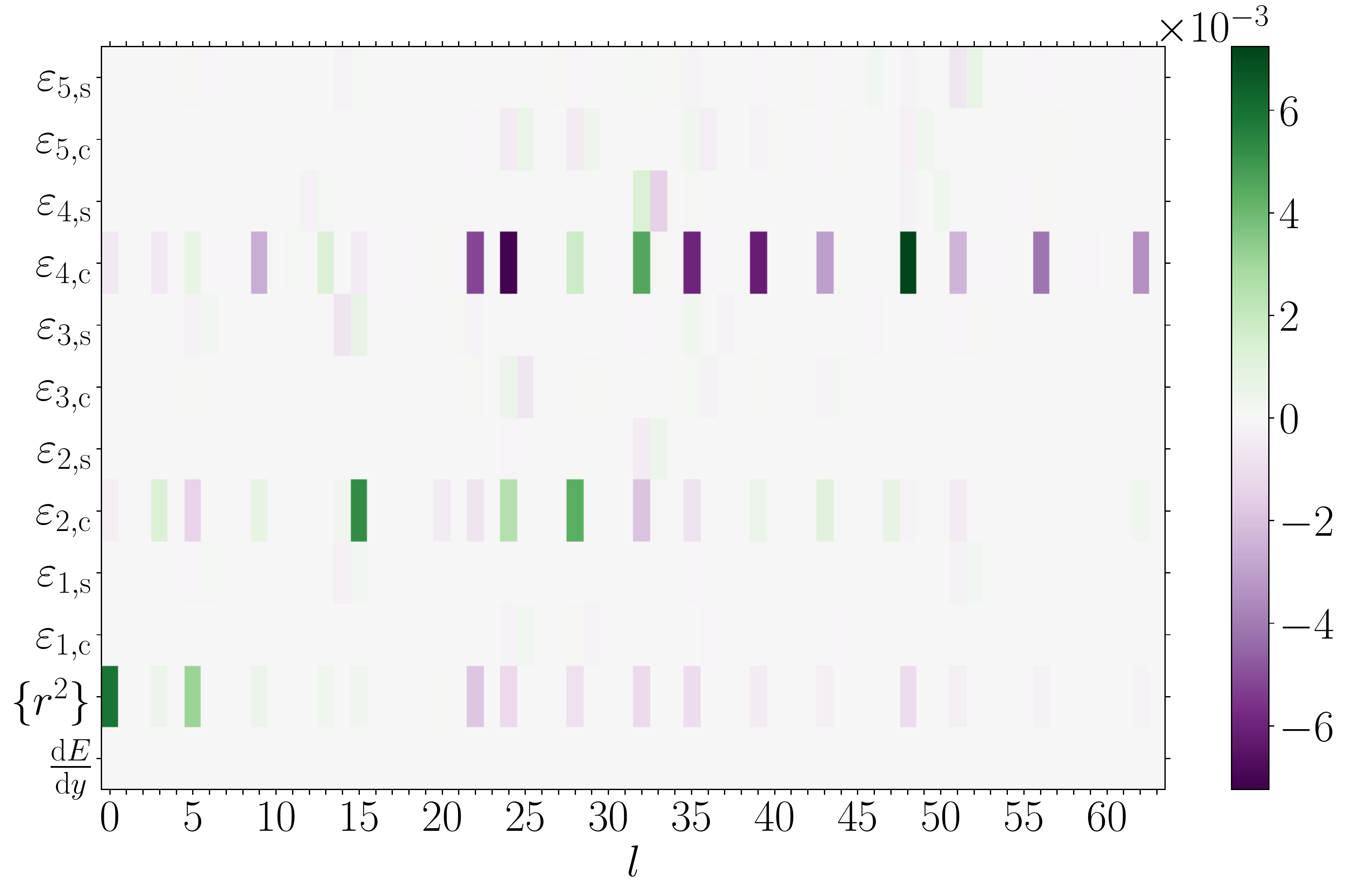}
\includegraphics[width=0.495\linewidth]{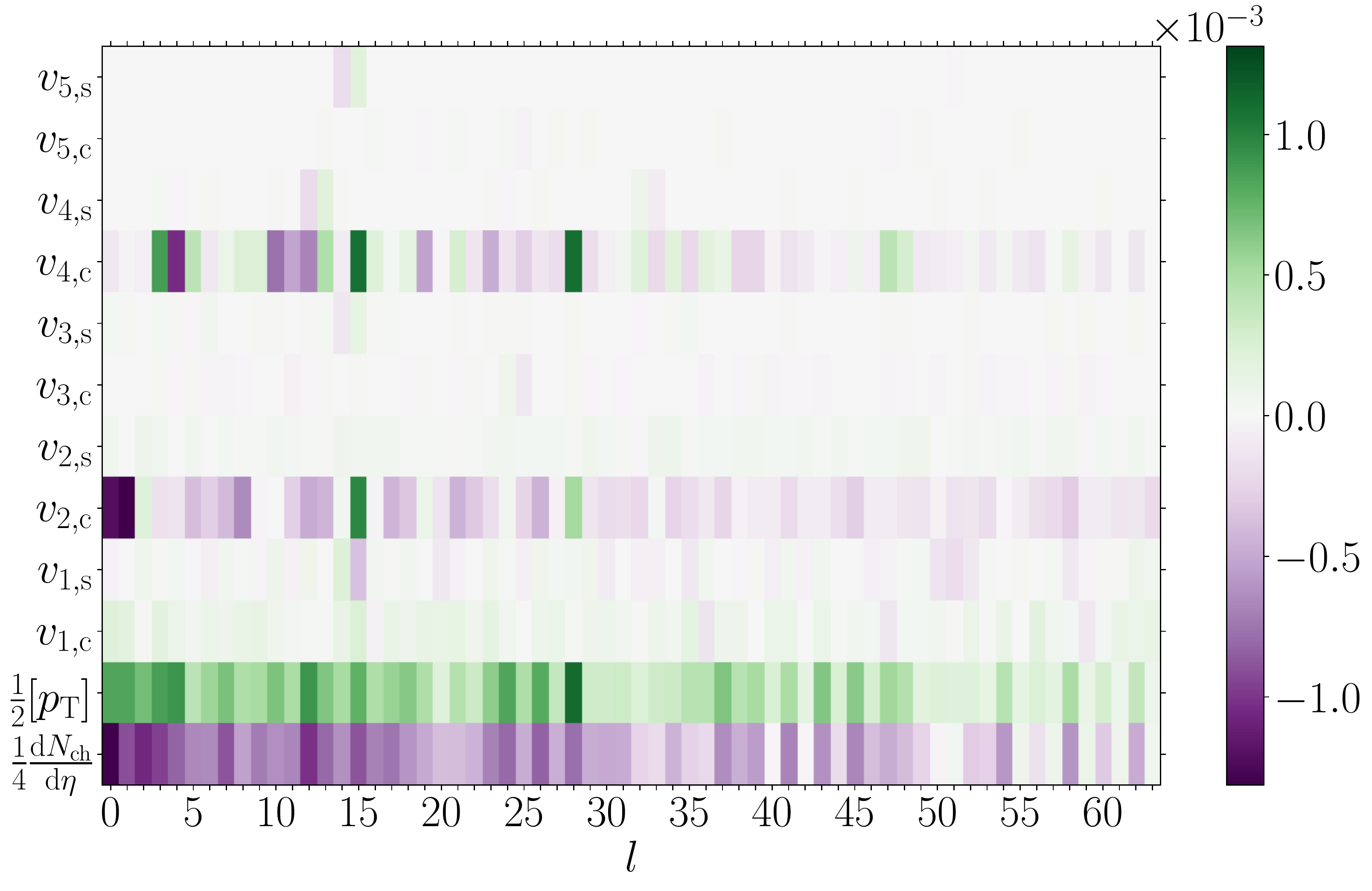}
\vspace{-5mm}
\caption{Diagonal quadratic-response coefficients $Q_{\alpha,ll}$ for initial-state (left) and final state observables (right) in the Glauber model at $b=0$ (top) and $b=9$~fm (bottom). The coefficients for dimensionful observables and multiplicity have been divided by $\bar{O}_\alpha$.}
\label{fig:Q_IS_FS_Glauber}
\end{figure*}

For a more quantitative description of the linear and nonlinear contributions of fluctuation modes to system characteristics, we now discuss the linear- and (diagonal) quadratic-response coefficients $\{L_{\alpha,l}$\}, $\{Q_{\alpha,ll}\}$ introduced in Eq.~\eqref{Oseries} and computed via Eqs.~\eqref{eq:Lalpha} and \eqref{eq:Qalpha} with $\delta=0.1$.
In this Section we show results for collisions with initial states from the Glauber model, while those obtained with the Saturation model --- which are qualitatively similar --- are presented in Appendix~\ref{appendix:L_Q_Saturation}.

The linear- resp.\ quadratic-response coefficients are shown in Fig.~\ref{fig:L_IS_FS_Glauber} resp.\ Fig.~\ref{fig:Q_IS_FS_Glauber}. 
In both figures, the top panels are for collisions at $b=0$ and the bottom panels for $b=9$~fm. 
In turn, the panels on the left display response coefficients for initial-state characteristics, while those on the right correspond to final-state observables. 
We have divided the coefficients $\{L_{\alpha,l}\}$, $\{Q_{\alpha,ll}\}$ for the dimensionful observables ($\d E/\d y$, $\{r^2\}$, $[p_\textrm{T}]$) and the multiplicity by the value $\bar{O}_\alpha$ of the observable in the average state to obtain dimensionless ``reduced'' coefficients of the same magnitude as those for the eccentricities or flow coefficients.
When showing either linear or quadratic response for a given set of observables, we deliberately use the same scale for the coefficients at both impact parameter values, to allow for a direct comparison.
This explains why, for example, the upper left plot of Fig.~\ref{fig:Q_IS_FS_Glauber} (or equivalently Fig.~\ref{fig:Q_IS_FS_Saturation} in Appendix~\ref{appendix:L_Q_Saturation} for the Saturation model) looks almost empty: 
the respective response coefficients at $b=0$ are significantly smaller than those at $b=9$~fm displayed in the lower left panel. 
Eventually, remember that the response coefficients will multiply expansion coefficients $c_l$ (or their square) of order~1 when contributing to an observable, see Eq.~\eqref{Oseries}.
Since the linear coefficients (Fig.~\ref{fig:L_IS_FS_Glauber}) are typically one order of magnitude larger than the quadratic ones (Fig.~\ref{fig:Q_IS_FS_Glauber}), the contribution from quadratic response is generally smaller than that from linear response. 

Let us begin with the initial-state characteristics at $b=0$ (upper left). 
Each fluctuation mode generally contributes either to the eccentricities $\varepsilon_{n,\rm c/s}$ in a single harmonic,\footnote{An exception is mode $l=40$, which contributes to both $\varepsilon_1$ and $\varepsilon_2$: as was already noted at the beginning of Sec.~\ref{subsec:classification_modes}, this mostly-$\varepsilon_1$ mode is quasi-degenerate with several others, including some ($l=42$ and 43) with a quadrupolar structure, i.e., an $\varepsilon_2$.} in which case the response is purely linear, or to the system energy and mean square transverse radius, where the latter shows a small quadratic response.
For instance, adding mode $l=1$ to the average state contributes both $\varepsilon_{1,\rm c}$ and $\varepsilon_{1,\rm s}$, while mode $l=0$ affects $\d E/\d y$ and $\{r^2\}$. 
Note that no response coefficient is visible for a few modes, like e.g., $l=19$ and 20, which in fact turn out to have only an $\varepsilon_6$ or higher-order eccentricity. 

Since the average state at $b=0$ is radially symmetric, the coefficients $\{L_{\alpha,l}\}$ for a given eccentricity directly yield (up to multiplication by $c_l$ of order~1) the eccentricity of the initial state $\bar{\Psi}+c_l\Psi_l$, which is thus of the order of a few percent. 
In turn, the coefficients for energy density and $\{r^2\}$ yield the relative change of the corresponding quantity, which is also of a few percent, with respect to its value in the average state. 

The linear-response coefficients for eccentricities show a marked feature: pairs of neighboring fluctuation modes show both $\varepsilon_{n,\rm c}$ and $\varepsilon_{n,\rm s}$ for a given $n$, three of which have the same sign while the fourth has the opposite sign. 
This $2\times2$-structure reflects the arbitrary orientation of the symmetry-plane angle $\Phi_n$, which need not be along the $x$- or $y$-axis due to the rotational symmetry at $b=0$, and is rotated by $\pi/2n$ for degenerate modes with otherwise the same profile. 
It would in principle be possible to rotate simultaneously both modes such that one only contributes to $\varepsilon_{n,\rm c}$ and the other to $\varepsilon_{n,\rm s}$ with equal response coefficients (in absolute value, since the overall sign of a mode is arbitrary). 

Eventually, the absence of sizable quadratic-response coefficients for the eccentricities tells us that the contribution of the nonradially-symmetric modes to the denominators of Eqs.~\eqref{eq:eccentricities1}--\eqref{eq:eccentricities2} must be very small. 
This is consistent with the fact that the azimuthal modulation of the corresponding eigenvectors in Fig.~\ref{fig:60_modes_b0_Glauber} seem to be oscillating about zero, so that the integral of $r^n\Psi_l(r,\theta)$ over $\theta$ at fixed $r$ already vanishes for any $n$.

Turning to the response coefficients for final-state observables at $b=0$ (top right panel), they now include the dynamical response of the system to the initial-state characteristics. 
The linear coefficients follow a similar pattern as those for the initial characteristics, in that the modes contribute either to the multiplicity and the average momentum, or the anisotropic-flow coefficients $v_{n,\rm c/s}$ in a single harmonic.
In fact, only the modes with an initial contribution to $\d E/\d y$ and $\{r^2\}$ resp.\ some $\varepsilon_{n,\rm c/s}$ contribute linearly to $\d N_\textrm{ch}/\d\eta$ and $[p_\textrm{T}]$ resp.\ $v_{n,\rm c/s}$. 

For a given mode, the linear-response coefficient for $\d N_\textrm{ch}/\d\eta$ has the same sign as that for $\d E/\d y$, while the coefficient for $[p_\textrm{T}]$ has the opposite sign to that for $\{r^2\}$.%
\footnote{The sign of a single $L_{\alpha,l}$ is not really meaningful, since it changes when replacing $\Psi_l$ by $-\Psi_l$. However the relative sign of two coefficients $L_{\alpha,l}$, $L_{\beta,l}$ is not affected by this change.} 
Note that the latter result also holds for initial states from the Saturation model, but not the former (see Fig.~\ref{fig:L_IS_FS_Saturation}). 
Regarding the anisotropic-flow harmonics, the $2\times2$-structure observed in the initial state is translated by the evolution into the final state. 
Yet one sees that the response becomes increasingly weaker for higher harmonics (with $n\geq 3$), which reflects the well-known effect of viscosity, which damps more strongly the finer spatial structures. 
This viscous damping also explains the comparatively weaker final-state response of the modes with higher $l$, which as discussed in Sec.~\ref{subsec:classification_modes} typically have more structure along the radial direction.

In general the sign of the linear-response coefficient for a flow harmonic $v_{n,\rm c/s}$ is the same as that of the corresponding $\varepsilon_{n,\rm c/s}$. 
However, this does not hold for $n=1$, where the response coefficients of $\varepsilon_1$ and $v_1$ have opposite signs (Fig.~\ref{fig:L_IS_FS_Glauber}). 
Indeed, the sign of $\varepsilon_1$, with its peculiar $r^3$ weight, yields the sign of $v_1$ at high $p_\textrm{T}$, which is opposite (to fulfill transverse momentum conservation) to the sign of $v_1$ at low $p_\textrm{T}$, which is that reflected in the momentum-integrated $v_1$~\cite{Teaney:2010vd}. 

Regarding the quadratic response coefficients in the top right panel of Fig.~\ref{fig:Q_IS_FS_Glauber}, they are typically smaller than the linear ones, but there are also more sizable coefficients than for initial-state observables. 
Thus, at quadratic order all modes seem to be contributing to the charged multiplicity $\d N_\textrm{ch}/\d\eta$ (negatively) and to the average transverse momentum $[p_\textrm{T}]$ (positively). 
This confirms the departure from linearity seen on a few modes in Sec.~\ref{subsec:linearitycheck} precisely for these two observables (top right panel of Fig.~\ref{fig:Linearity_test_Glauber}). 
That all modes contribute to $\d N_\textrm{ch}/\d\eta$ and $[p_\textrm{T}]$ at quadratic order should also be contrasted with the fact that at linear order only the radially symmetric modes contribute to these observables. 
It means that the first nonzero contribution of most modes to multiplicity and average momentum is at quadratic order. 
One can however see that the $\{Q_{\alpha,ll}\}$ for $\d N_\textrm{ch}/\d\eta$ and $[p_\textrm{T}]$ are of order $10^{-3}$ or smaller, i.e., change the value of the corresponding observable given by the average state by a similar relative amount, and that they generally decrease in absolute value with increasing $l$. 

Other nonzero coefficients $\{Q_{\alpha,ll}\}$ are those that represent the quadratic response of anisotropic flow in the harmonic $2n$ of an initial eccentricity in the $n$-th harmonic, $v_{2n} \propto \varepsilon_n^2$. 
This quadratic response yields $v_{2n}$ values of order $10^{-3}$ or smaller, which in real events consisting in a mixture of many modes will be subleading compared with the linear response of $v_{2n}$ to modes with a nonzero $\varepsilon_{2n}$. 
Note that the absence of quadratic response for the odd harmonics $v_{2n+1}$ in general, which also holds for events at $b=9$~fm, is due to the definite parity in position space of the average states (even) and the modes in general (either even or odd): 
the quadratic response in a mode of given position-space parity will always be even, and cannot give rise to odd (position- or momentum-space) observables.

At finite impact parameter $b=9$~fm, the plots showing the response coefficients are significantly more busy than at $b=0$, due to the breaking of the system rotational symmetry, but the overall trends are similar. 
In particular, in the final state the effect of viscous damping for higher-lying fluctuation modes and on higher anisotropic-flow harmonics is still clearly present. 

Every mode now affects several characteristics in the initial state at linear order (Fig.~\ref{fig:L_IS_FS_Glauber} bottom left), which is reflected one-to-one in the linear-response coefficients of final-state observables (Fig.~\ref{fig:L_IS_FS_Glauber}, bottom right). 
Regarding these linear coefficients, two differences with those for collisions at $b=0$ appear. 
First, the opposite signs of the $\{L_{\alpha,l}\}$ for $\d E/\d y$ and $\{r^2\}$ --- and the corresponding correlation resp.\ anticorrelation between $\d N_\textrm{ch}/\d\eta$ and $[p_\textrm{T}]$ in the Glauber resp.\ Saturation (Fig.~\ref{fig:L_IS_FS_Saturation}, bottom right) model --- is no longer systematic. 
Second, the $2\times 2$-structure of the eccentricities and flow harmonics has disappeared: 
the participant-plane angles $\Phi_n$ tend to align either along the impact-parameter direction (i.e., the $x$ axis), resulting in a finite $\varepsilon_{n,\rm c}$ and $\varepsilon_{n,\rm s}=0$, or at $\pi/2n$ from the $x$-axis, yielding $\varepsilon_{n,\rm c}=0$ and a finite $\varepsilon_{n,\rm s}$.

Although the final state in the right panel seems to mirror exactly the initial state on the left-hand side, one can find a couple of interesting differences for modes $l=58$ and 59. 
Mode $l=58$ resp.\ 59 has a visible $v_{1,\rm s}$ resp.\ $v_{1,\rm c}$, although the corresponding eccentricity $\varepsilon_{1,\rm s}$ resp.\ $\varepsilon_{1,\rm c}$ is very small. 
The explanation of this apparent inconsistency is that the coefficients $L_{v_1,l}$ of these modes measure the contribution to $v_1$ at order $c_l$ of events of the form $\bar{\Psi}+c_l\Psi_l$, i.e., consisting of the mode and the average initial state. 
Such events have a very small $\varepsilon_1$ (of the order of $10^{-3}$ for $c_l=1$) but significant $\varepsilon_2$ --- that of $\bar{\Psi}$ --- and $\varepsilon_3$ --- coming from $\Psi_l$. 
In the evolution, the ellipticity and triangularity of the initial geometry interfere and yield the nonlinear dynamical response
\begin{equation}
v_1(\bar{\Psi}+c_l\Psi_l) \propto \varepsilon_2(\bar{\Psi})_{}\varepsilon_3(c_l\Psi_l)
\end{equation}
already at linear order in $c_l$.\footnote{The notation $\varepsilon_3(c_l\Psi_l)$ is slightly inaccurate and should rather read $\varepsilon_3(\bar{\Psi}+c_l\Psi_l)$, since the energy density inserted in definition~\eqref{eq:eccentricities2} is $\bar{\Psi}+c_l\Psi_l$, but conveniently emphasizes that only $c_l\Psi_l$ contributes to $\varepsilon_3$.}
Note that this nonlinear dynamical response $v_{|n\pm 2|}(\bar{\Psi}+c_l\Psi_l) \propto \varepsilon_2(\bar{\Psi})_{}\varepsilon_n(c_l\Psi_l)$ at order $c_l$ is actually present for all modes at $b=9$~fm, but it cannot be pinpointed for most of the modes because they usually have all even or odd eccentricities. 

Looking at the quadratic response coefficients for initial-state characteristics (bottom left of Fig.~\ref{fig:Q_IS_FS_Glauber}), there is again some sizable nonlinear response for $\{r^2\}$, which is present for modes which show a significant linear response for both $\{r^2\}$ and $\d E/\d y$. 
In contrast with $b=0$, also the quadratic coefficients for some of the eccentricities --- mostly $\varepsilon_{2,\rm c}$ and $\varepsilon_{4,\rm c}$ --- are now sizable. 
For initial states from the Glauber model, this holds for modes which also show a finite linear-response coefficient for the same eccentricity; but in the case of the Saturation model, almost all modes have nonzero $Q_{\alpha,ll}$ for $\varepsilon_{2,\rm c}$ and $\varepsilon_{4,\rm c}$ (Fig.~\ref{fig:Q_IS_FS_Saturation} bottom left), even those with a small corresponding $L_{\alpha,l}$ (Fig.~\ref{fig:L_IS_FS_Saturation} bottom left). 
These nonzero quadratic coefficients are caused by the denominators in definition~\eqref{eq:eccentricities2} and the finite $\varepsilon_{2,\rm c}$ or $\varepsilon_{4,\rm c}$ of the average state. 

For final-state observables, a first salient feature is that all modes seem to have a negative resp.\ positive quadratic-response coefficient for $\d N_\textrm{ch}/\d\eta$ resp.\ $[p_\textrm{T}]$, as already seen at $b=0$.\footnote{Contrary to the linear coefficients, $Q_{\alpha,ll}$ is unchanged under the change $\Psi_l\to -\Psi_l$ and thus its sign is meaningful.} While most modes do not modify the total energy of the system, they do change the energy density profile, with regions with more energy than the average state and others with less energy.
The negative $Q_{\alpha,ll}$ for multiplicity can then be understood from the empirical scaling behavior~\cite{Giacalone:2019ldn}
\begin{equation}
\frac{\textrm{d}N_\textrm{ch}}{\textrm{d}\eta} \propto \int_\textbf{x}{e(\textbf{x})^{\frac{2}{3}}},
\label{eq:Nchrelationtoenergy}
\end{equation}
according to which a change in the initial energy density $e(\textbf{x})$ results in a less-than-linear change of the multiplicity. By considering the effects of fluctuations around the average background, $e(\textbf{x})=\bar{e}(\textbf{x})+c_{l} \delta e_{l}(\textbf{x})$, one finds that
\begin{equation}
\left\langle \frac{\textrm{d}N_\textrm{ch}}{\textrm{d}\eta} \right\rangle \propto \int_\textbf{x}\bar{e}(\textbf{x})^{\frac{2}{3}} \left[1 - \frac{1}{9} \sum_{l}\frac{\delta e_{l}(\textbf{x}) \delta e_{l}(\textbf{x})}{\bar{e}(\textbf{x})^2} \right],
\label{eq:Nchrelationtoenergy2}
\end{equation}
where we used Eqs.~\eqref{eq:<c_l>=0} and \eqref{eq:<c_lc_l'>}. Due to the exponent $2/3$ in Eq.~\eqref{eq:Nchrelationtoenergy}, the fluctuation-induced correction to the multiplicity in Eq.~\eqref{eq:Nchrelationtoenergy2} is always negative, indicating for the multiplicity $Q_{\alpha,ll}<0$ for all modes. 

Assuming that this local change of energy density results in a corresponding modification of the number of locally emitted particles with a scaling law similar to Eq.~\eqref{eq:Nchrelationtoenergy}, with an exponent smaller than~1, leads after integrating over the whole system to a less-than-linear fluctuation-induced modification of multiplicity, i.e., a negative $Q_{\alpha,ll}$. 

In turn, the anticorrelation between $\d N_\textrm{ch}/\d\eta$ and $[p_\textrm{T}]$ can be attributed to the fact that most modes do not modify the total energy of the system, so that a decrease in the multiplicity has to be accompanied by an increase of their average transverse momentum. 
What is, however, nontrivial is that this also holds for modes that change the total system energy --- but these modes also modify the mean square radius of the initial state and affect $\d N_\textrm{ch}/\d\eta$ and $[p_\textrm{T}]$ at linear order, so that disentangling all effects is beyond the scope of the present paper.\footnote{In a forthcoming paper we shall consider events with a fixed multiplicity, i.e., a given centrality, instead of fixed impact parameter. This will possibly facilitate the discussion, in addition to being closer to the experimental setup.} 

Eventually, the quadratic coefficients for $v_{2,\rm c}$ and $v_{4,\rm c}$ are sizable for a large number of modes, while the coefficients for the other flow harmonics are significantly smaller. 
The quadratic contributions to $v_{2,\rm c}$ and $v_{4,\rm c}$ have different origins, which are difficult to disentangle.
Thus, there is the linear response $v_{n,c}\propto \varepsilon_{n,c}$ to an initial eccentricity $\varepsilon_{2,c}$, $\varepsilon_{4,c}$ which is already quadratic in $c_l$ (see bottom-left panel of Fig.~\ref{fig:Q_IS_FS_Glauber}). 
Then, the modes with an initial $\varepsilon_1(\Psi_l)$ (either $\varepsilon_{1,\rm c}$ or $\varepsilon_{1,\rm s}$) can dynamically give rise to a quadratic $v_{2,c}\propto\varepsilon_1^2$, but also, due to the interference with the ellipticity of the average state to a $v_{4,c}(\bar{\Psi}+c_l\Psi_l) \propto \varepsilon_{2,c}(\bar{\Psi})_{}\varepsilon_1(c_l\Psi_l)^2$. 
Such a term also contributes to $v_{2,c}$. 
More generally $v_{2,c}$ and $v_{4,c}$ also have contributions of the form $\varepsilon_{2,c}(\bar{\Psi})_{}\varepsilon_1(\Psi_l)_{}\varepsilon_3(\Psi_l)$ or $\varepsilon_{2,c}(\bar{\Psi})_{}\varepsilon_3(\Psi_l)^2$ (for modes with odd eccentricities) or $\varepsilon_{2,c}(\bar{\Psi})_{}\varepsilon_2(\Psi_l)^2$ (for modes with even eccentricities).
As a final example of dynamical nonlinear response, let us mention mode $l=14$, which has both an $\varepsilon_{1,\rm s}$ (and higher sine odd harmonics) and an $\varepsilon_{2,\rm c}$: 
these give rise in the evolution to a $v_{1,\rm s} \propto \varepsilon_{2,c} \varepsilon_{1,s}$, and also to a $v_{3,\rm s}$, at quadratic order in $c_l$.\footnote{For mode $l=15$, the quadratic $v_{1,\rm s}$ comes from $\varepsilon_{2,\rm c}$ and $\varepsilon_{3,\rm s}$.}

\subsection{Anisotropic flow response to the initial eccentricities}
\label{subsec:response_coeff}

A number of studies, within either fluid dynamics --- ideal or dissipative --- or kinetic transport theory, have demonstrated the existence of simple relationships between initial-state eccentricities and the anisotropic flow harmonics in the final state. 
Considering the absolute values $\varepsilon_n$ and $v_n$ [Eqs.~\eqref{eq:eccentricities1}, \eqref{eq:eccentricities2}, \eqref{eq:flow_harmonics}], and restricting oneself to small eccentricities, one finds that the $n$th flow harmonic receives on the one hand a linear\footnote{In collisions with a large $\varepsilon_2$, an extra cubic term $\propto \varepsilon_2^3$ was found to contribute to $v_2$~\cite{Noronha-Hostler:2015dbi}.} contribution from $\varepsilon_n$~\cite{Ollitrault:1992bk,Alver:2010gr,Teaney:2010vd,Gardim:2011xv,Borghini:2010hy,Niemi:2012aj,Plumari:2015cfa}
\begin{equation}
v_n = {\cal K}_{n,n} \varepsilon_n ,
\label{eq:vn_vs_eps_n}
\end{equation}
and on the other hand contributions from eccentricities in other harmonics. 
In the simplest case, the latter are of the form~\cite{Borghini:2005kd,Gardim:2011xv,Teaney:2012ke,Niemi:2012aj}
\begin{equation}
v_n = {\cal K}_{n,mp} \varepsilon_m\varepsilon_p
\label{eq:vn_vs_eps_m,p}
\end{equation}
with $|m\pm p| = n$. 
We presently wish to discuss how such behaviors appear in our mode-by-mode analysis. 

Indeed, when discussing the response coefficients for anisotropic flow in the previous Section, we related them several times to the initial eccentricities of the system. 
At $b=9$~fm, the deformation of the average state $\bar{\Psi}$ with sizable values of $\varepsilon_2$ and  $\varepsilon_4$ (see Table~\ref{tab:average_state_eccentricities}) and the fact that most fluctuation modes have several nonzero $\varepsilon_m$ makes it difficult to isolate the influence of each individual eccentricity on a given flow harmonic, be it at linear or quadratic order in $c_l$. 

\begin{figure*}[!t]
	\includegraphics[width=0.495\linewidth]{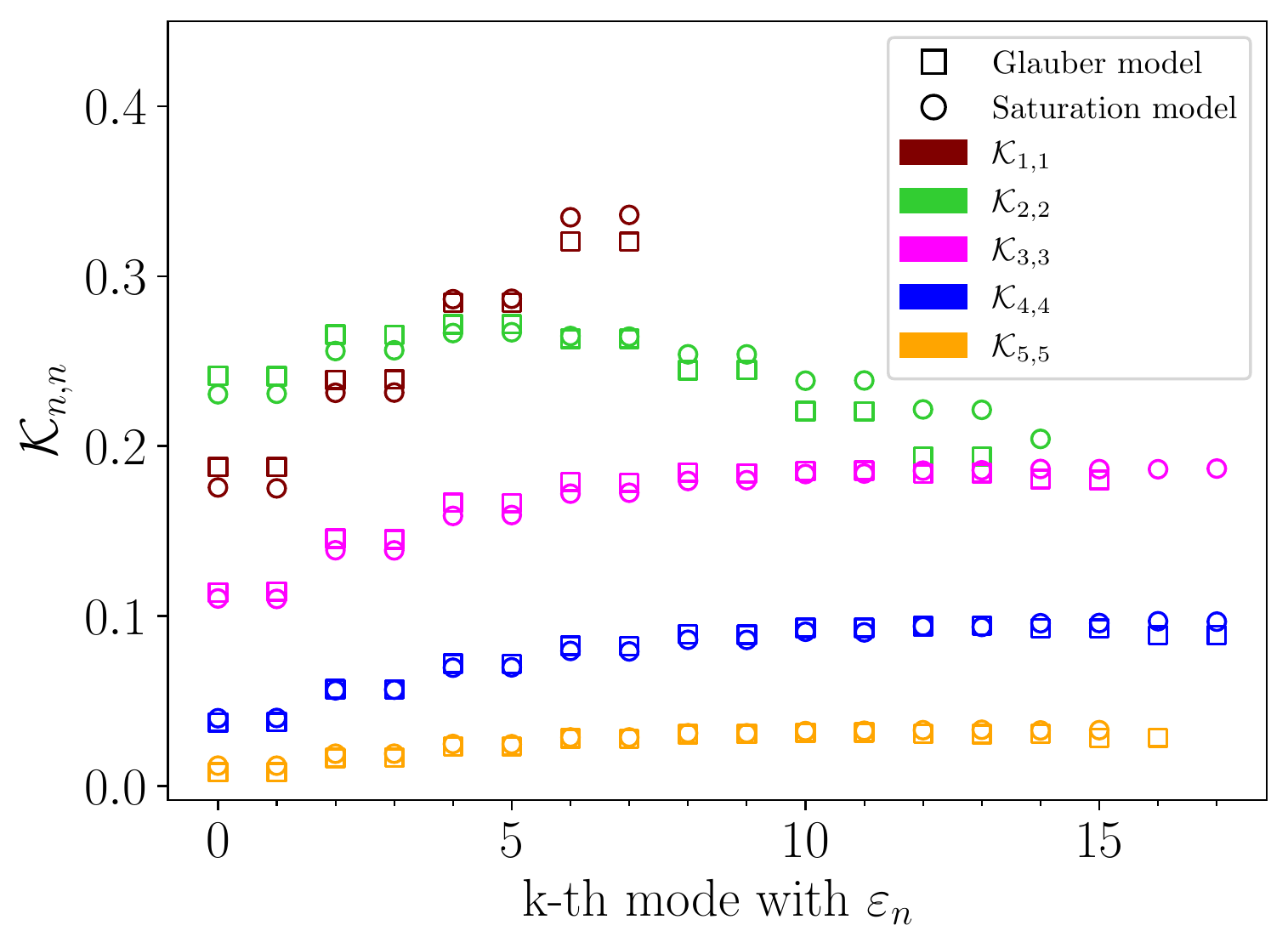}
	\includegraphics[width=0.495\linewidth]{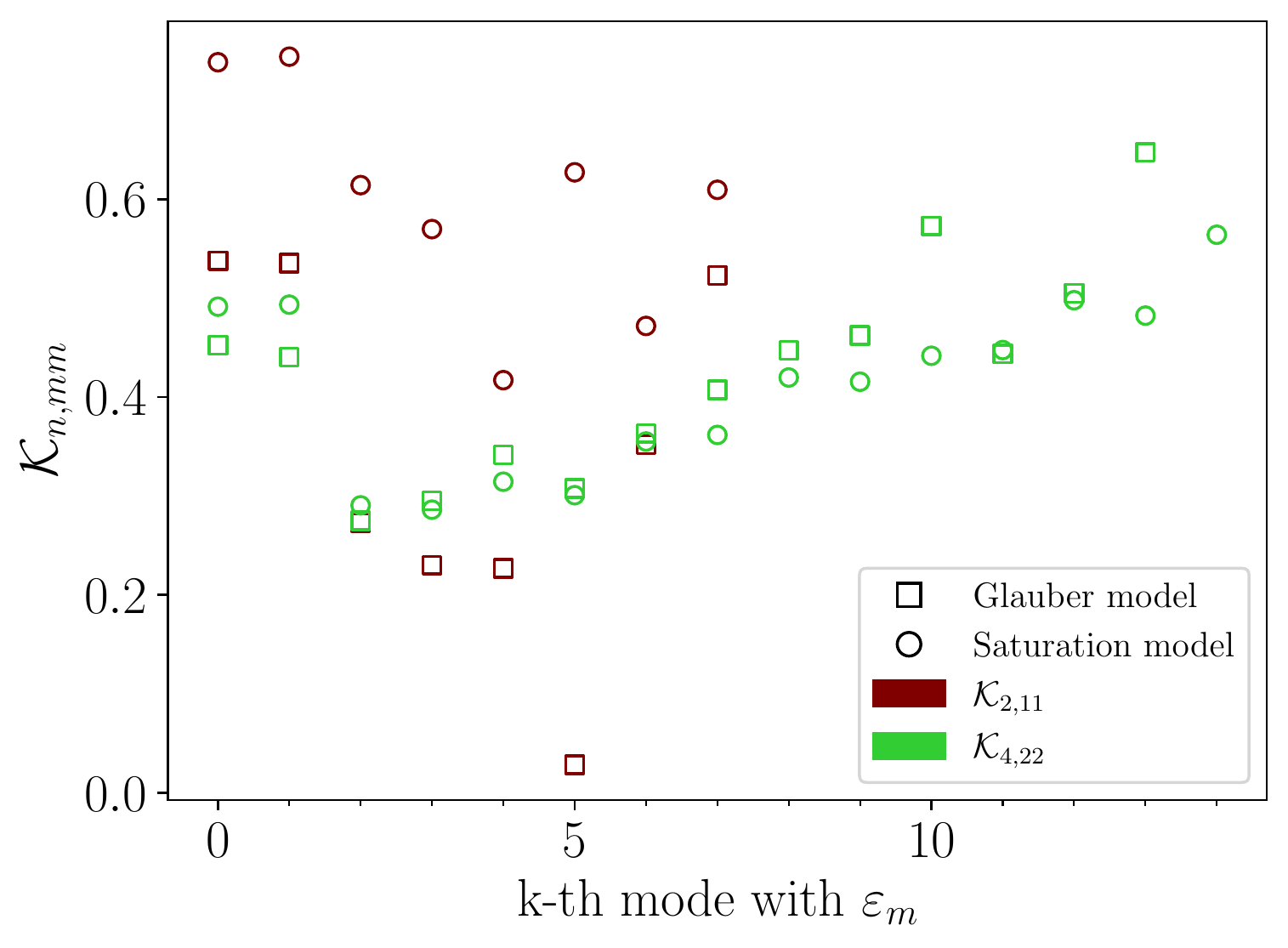}
	\vspace{-7mm}
	\caption{Linear flow-response coefficient $\mathcal{K}_{n,n}$ (left) and quadratic-response coefficient $\mathcal{K}_{n,mm}$ (right) with $n=2m$ for modes with $l<256$ for collisions at $b=0$ within the Glauber (squares) and Saturation (circles) models.}
	\label{fig:Kappab0}
\end{figure*}

In contrast, the situation at $b=0$ is cleaner: since the average state is radially symmetric, the flow response for an event of the form $\bar{\Psi}+c_l\Psi_l$ is entirely due to the asymmetry of the mode $\Psi_l$, which is what contributes to (the numerator of) the eccentricities $\varepsilon_m(\bar{\Psi}+c_l\Psi_l)$.
More precisely, we mostly encounter two cases. 
First, a linear response of the form~\eqref{eq:vn_vs_eps_n}, which manifests itself as a nonzero coefficients $L_{v_n,l}$, i.e., as a linear response in $c_l$, for a mode $\Psi_l$ with an initial $\varepsilon_n$. 
For such modes we compute
\begin{equation}
{\cal K}_{n,n} \equiv \lim_{c_l \to 0} \frac{v_n(\bar{\Psi}+c_l\Psi_l)}{\varepsilon_n(\bar{\Psi}+c_l\Psi_l)} = \frac{L_{v_{n},l}}{L_{\varepsilon_{n},l}},
\label{eq:kappadefinition}
\end{equation}
which is shown in the left panel of Fig.~\ref{fig:Kappab0}, where $k$ labels the $k$-th mode $\Psi_l$ with $l<256$ with an $\varepsilon_n\geq 0.01$.\footnote{The mode number $l$ may differ between the two models.} 
Second, one can identify a number of modes with an initial $\varepsilon_m$ with $m=1$ or 2 that gives rise to a final $v_n$ with $n=2m$, corresponding to Eq.~\eqref{eq:vn_vs_eps_m,p} with $p=m$. 
For those modes (again with $l<256$), characterized by a sizable $Q_{v_n,ll}$ (see top right panel of Fig.~\ref{fig:Q_IS_FS_Glauber}), we show 
\begin{equation}
{\cal K}_{n,mm} \equiv \lim_{c_l \to0} \frac{v_n(\bar{\Psi}+c_l\Psi_l)}{\varepsilon_m(\bar{\Psi}+c_l\Psi_l)^2} = \frac{Q_{v_n,ll}}{L_{\varepsilon_{m},l}^2}
\quad\text{with }n=2m
\label{eq:kappadefinitionmm}
\end{equation}
in the right panel of Fig.~\ref{fig:Kappab0}. 
We also noted in Sec.~\ref{subsec:lin_quad_response} the presence of modes with several sizable eccentricities, giving a ``mixed'' nonlinear response~\eqref{eq:vn_vs_eps_m,p} with $m\neq p$. 
For instance, for mode $l=40$ in the Glauber model one could compute a ${\cal K}_{3,21}$, which we did not do. 
By construction, the coefficients ${\cal K}_{n,n}$ and ${\cal K}_{n,mm}$ from Eqs.~\eqref{eq:kappadefinition} and \eqref{eq:kappadefinitionmm} are positive. 
However we noted in the previous section that the integrated $v_1$ has an opposite sign to $\varepsilon_1$, i.e., ${\cal K}_{1,1}$ should be negative~\cite{Teaney:2012ke}. 

The coefficients ${\cal K}_{n,n}$ in the left panel of Fig.~\ref{fig:Kappab0} clearly decrease with $n$, which can be attributed to viscous damping in the evolution. 
Yet this is less visible for the quadratic coefficients ${\cal K}_{2n,nn}$ on the right, although there is only a small number of modes with a sizable dipole deformation $\varepsilon_1$. 
One again sees that the modes come in doublets, as already seen for their eccentricities (Fig.~\ref{fig:mode_eccentricities} left) or in the $2\times 2$ structure of the linear-response coefficients $L_{\alpha,l}$ (top panels of Figs.~\ref{fig:L_IS_FS_Glauber} and \ref{fig:L_IS_FS_Saturation}).

Strikingly, the response coefficients ${\cal K}_{n,n}$ are very similar for modes from either initial-state model, although the corresponding modes have different radial profiles. 
This also holds for ${\cal K}_{4,22}$, but not for ${\cal K}_{2,11}$, where the fluctuation modes from the Saturation model with an $\varepsilon_1$ trigger a stronger $v_2$ response than those from the Glauber model.
For $n=3$, 4, 5, ${\cal K}_{n,n}$ seems to remain almost constant as the mode number $l$ increases, while ${\cal K}_{2,2}$ may show a decreasing trend, although rather mild. 
In fact, one could expect that fluctuation modes with higher $l$ should be more damped by viscous effects, since they show increasingly finer structure along the radial direction, but this is not obvious from this analysis. 

This approximate uniformity with increasing $l$ allows us to compare the mode-by-mode coefficients for initial states of the form $\bar{\Psi}+c_l\Psi_l$ with the values found from full event-by-event simulations.
Roughly speaking, the values of ${\cal K}_{n,n}$ shown in Fig.~\ref{fig:Kappab0} are in the same ballpark than those reported in the literature:
\begin{itemize}
\item ${\cal K}_{2,2} \simeq 0.2$--0.3 is in good agreement with the findings in hydrodynamical simulations with different setups~\cite{Niemi:2012aj,Noronha-Hostler:2015dbi,Qian:2016fpi,Rao:2019vgy,Liu:2018hjh} and in kinetic transport simulations~\cite{Wei:2018xpm,Plumari:2015cfa,Roch:2020zdl}.

\item ${\cal K}_{3,3} \simeq 0.1$--0.2 fits most values in the literature~\cite{Niemi:2012aj,Noronha-Hostler:2015dbi,Liu:2018hjh,Roch:2020zdl} apart from Ref.~\cite{Plumari:2015cfa}, which finds values larger by about a factor 2.

\item ${\cal K}_{4,4}\simeq 0.05$--0.1 is comparable to the results in Refs.~\cite{Niemi:2012aj,Qian:2016fpi,Liu:2018hjh}, although somewhat higher values were found in fluid-dynamical~\cite{Teaney:2012ke} simulations or kinetic theory simulations~\cite{Plumari:2015cfa,Kurkela:2020wwb}  with $\eta/s=1/4\pi$. 

\item In turn, ${\cal K}_{5,5}\simeq 0.008$--0.03 is twice larger than in Ref.~\cite{Teaney:2012ke}, but matches the value in Ref.~\cite{Liu:2018hjh}.

\item For the last linear coefficient ${\cal K}_{1,1}$ the four twofold-degenerate values shown in Fig.~\ref{fig:Kappab0} are too few to draw conclusions, yet we find a decent agreement with the results presented in Ref.~\cite{Teaney:2010vd}, while the value in Ref.~\cite{Teaney:2012ke} is roughly a factor of two smaller.
\end{itemize}

As regards the quadratic-response coefficients, we are not aware of any value of ${\cal K}_{2,11}$ in the literature, while ${\cal K}_{4,22}\simeq 0.3$--0.6 as shown in the right panel of Fig.~\ref{fig:Kappab0} is significantly larger than the value of 0.1 reported in Ref.~\cite{Liu:2018hjh}.

\section{Fluctuations and correlations of observables}
\label{sec:fluctuations_observables}

\begin{table*}[!b]
\caption{\label{tab:average_state_observables} Sample mean $\expval{O_\alpha}$ (using 8192 simulations), average-event value $O_\alpha(\bar{\Psi})$, and estimate~\eqref{eq:<O_alpha>} of the average value of initial and final-state observables for both models at $b=0$ and $b=9$~fm. 
The uncertainties given for $\expval{O_\alpha}$ are the standard errors on the mean.}
\begin{ruledtabular}
\begin{tabular}{c|cccc|cccc}
& $\frac{\d E}{\d y}$ (GeV) & $\lbrace r^2\rbrace$ (fm$^2$) & $\varepsilon_{2,\rm c}$ & $\varepsilon_{4,\rm c}$ & $  \frac{\d N_\textrm{ch}}{\d\eta}  $ & $[p_\textrm{T}]$ (MeV/$c$) & $v_{2,\mathrm{c}}$ ($\%$) & $v_{4,\mathrm{c}}$ (\textperthousand) \\ \hline
Glauber $b=0$  & &  &  & & &  &  &\\
sample mean  & $7247 \pm 3$ & $16.30 \pm 0.01$ & - & - &  $2203 \pm 1 $ & $792.9 \pm 0.2$ & - & - \\
average event &  7248  & 16.29 & - & - & 2257 & 783.9 & - & - \\
Eq.~\eqref{eq:<O_alpha>} & 7248 & 16.31 & - & - & 2233 & 796.9 & - & - \\
\hline
Glauber  $b=9$~fm  & &  &  & & &  &  & \\
sample mean & $1668\pm 2$ & $8.69 \pm 0.02$ & $0.299\pm 0.002$ & $-0.111 \pm 0.002$ & $597 \pm 1$ & $773.8 \pm 0.3$  & $5.9 \pm 0.03 $ & $3.7 \pm 0.1$\\
average event & 1667  &  8.72 & 0.296 & $-0.094$  & 636 & 746.3 & 6.1  & 4.3 \\
Eq.~\eqref{eq:<O_alpha>} & 1667 & 8.72 & 0.302 & $-0.111$ & 569 & 758.8 & 3.9 & 3.3 \\
\hline
Saturation  $b=0$  &  &  &  & & &  &  &\\
sample mean & $6468 \pm 2$ & $17.01 \pm 0.01$ & - & - & $2071 \pm 1$ & $779.1 \pm 0.1$ & - & - \\
average event  & 6468 & 17.00 & - & - & 2119 & 769.0  & -  & - \\
Eq.~\eqref{eq:<O_alpha>} & 6468 & 17.01 & - & - & 2080 & 772.3 & - & - \\
\hline
Saturation  $b=9$ fm & &  &  & & &  &  &\\
sample mean & $1384 \pm 2$ &$ 7.94 \pm 0.01$ & $0.410 \pm 0.001 $ & $-0.220 \pm 0.002$ & $493 \pm 1$ & $794.3 \pm 0.3$ & $8.2\pm 0.03$  &  $6.7\pm 0.1$\\
average event   & 1386 &  7.95 & 0.406 &$-0.201$ & 537 & 753.9  & 8.5 & 7.1 \\
Eq.~\eqref{eq:<O_alpha>} & 1386 & 7.95 & 0.410 & $-0.219$ & 432 & 803.5 & 6.4 & 9.9 \\
\end{tabular}
\end{ruledtabular}
\end{table*}

Now that we have established the mode decomposition and the framework to compute observables the linear and quadratic response of observables to the fluctuations, we will compare the results of this mode-by-mode approach to statistical averages of event-by-event simulations. Naturally, for a sample of events in a given class --- in the present study, at a given impact parameter and for one of the two initial-state models we consider --- the various system characteristics (initial-state geometry, final-state multiplicity and anisotropic flow) generally vary event-by-event, unless it defines the event class. 
One can thus consider the average value of each observable and the statistics of its fluctuations, in particular the variance, as well as the covariance between the fluctuations of different observables. 

Below we discuss how the statistics of observables can be assessed in the mode-by-mode approach via the decomposition in an average event and fluctuation modes introduced in Sec.~\ref{sec:stat_caracterization_IS}. To compare with the event-by-event approach, we also simulated the event-by-event evolution with K\o MP\o ST and MUSIC (see Sec.~\ref{subsec:time_evol}) for 8192 random events for each of our initial-state models and impact-parameter values. 
For these, we computed the average values (Sec.~\ref{subsec:prediction_observables}) of the observables introduced in Sec.~\ref{subsec:observables}, as well as their variances and some of their covariances (Sec.~\ref{subsec:variances_covariances}), which we relate to their values as calculated within the fluctuation-mode decomposition. 
Eventually, in Sec.~\ref{subsec:propability_distributions} we develop a framework for making predictions for the joint probability distribution of observables, under the assumption of Gaussian statistics of the coefficients $c_l$, with variances and covariances determined within the mode-by-mode decomposition.  
This approach is then applied to eccentricities and anisotropic-flow coefficients.

\subsection{Average values of observables}
\label{subsec:prediction_observables}

Based on the event-by-event simulations of the 8192 events simulated in each class, i.e., for collisions either at $b=0$ or $b=9$~fm with initial states from either the Glauber or the Saturation model, we display in Table~\ref{tab:average_state_observables} the ``sample mean,'' i.e., the mean value averaged over the sample, of a few system characteristics, namely the
total energy, mean square radius, eccentricities $\varepsilon_{2,\rm c}$ and $\varepsilon_{4,\rm c}$ in the initial state,  charged multiplicity, average transverse momentum, flow coefficients $v_{2,\rm c}$ and $v_{4,\rm c}$ in the final state. We also provide the results for the mode-by-mode approach, where the rows ``average event'' refer to the value of each of these observables computed in the average state $\bar{\Psi}$, while the row labeled ``Eq.~\eqref{eq:<O_alpha>}'' includes the effects of fluctuations up to quadratic order.

As could be  anticipated, in the isotropic case $b=0$ the eccentricities and anisotropic-flow coefficients have very small mean values (below $10^{-3}$) which we do not report.
For the initial-state characteristics, the values $O_\alpha(\bar{\Psi})$ computed in the average state are in very nice agreement with the sample-mean values $\expval{O_\alpha}$, except for $\varepsilon_{4,\rm c}$.
Going to the final-state observables, there are now sizable differences between the event-averaged and average-event values, especially for the charged multiplicity and $v_{4,\rm c}$. 
In particular for the former, this difference is related to the strong nonlinearities observed in Sec.~\ref{sec:response}.
Indeed, Eq.~\eqref{eq:<O_alpha>} shows how the quadratic-response coefficients $Q_{\alpha,ll}$ contribute to the average value of observable $O_\alpha$ at order ${\cal O}(c_l^2)$.
Including this term modifies significantly the predicted average value of a few observables (see Table~\ref{tab:average_state_observables}): 
charged multiplicity and $[p_\textrm{T}]$ in general, and at finite impact parameter $\varepsilon_{4,\rm c}$, $v_{2,\rm c}$, and $v_{4,\rm c}$. 
In both models, at $b=0$ the shift induced by the coefficients $\{Q_{\alpha,ll}\}$ brings the mode-by-mode average value of observables closer to the sample value, although not enough for $\d N_\textrm{ch}/\d\eta$. 
For collisions at $b=9$~fm, the estimate~\eqref{eq:<O_alpha>} of the average value yields a nice description of initial-state characteristics, and in the final state it improves the agreement of $[p_\textrm{T}]$ with the sample value. 
However the effect on $\expval{\d N_\textrm{ch}/\d\eta}$ or $\expval{v_{2,\rm c}}$ are of the proper sign, but much too large in absolute value. 
The presence of such sizable contributions is consistent with the lower right panels of Fig.~\ref{fig:Q_IS_FS_Glauber} or \ref{fig:Q_IS_FS_Saturation}. 
Yet we do not really understand why the contribution $\frac{1}{2}\sum_lQ_{v_{2,c},ll}$ should degrade the rather good agreement of $v_{2,c}(\bar{\Psi})$ with the value from the event sample, even though it is clear that higher-order terms (in $c_l$) can contribute --- which we did not attempt to estimate in the present study. 

\begin{figure*}[!htb]
\includegraphics[width=0.495\linewidth]{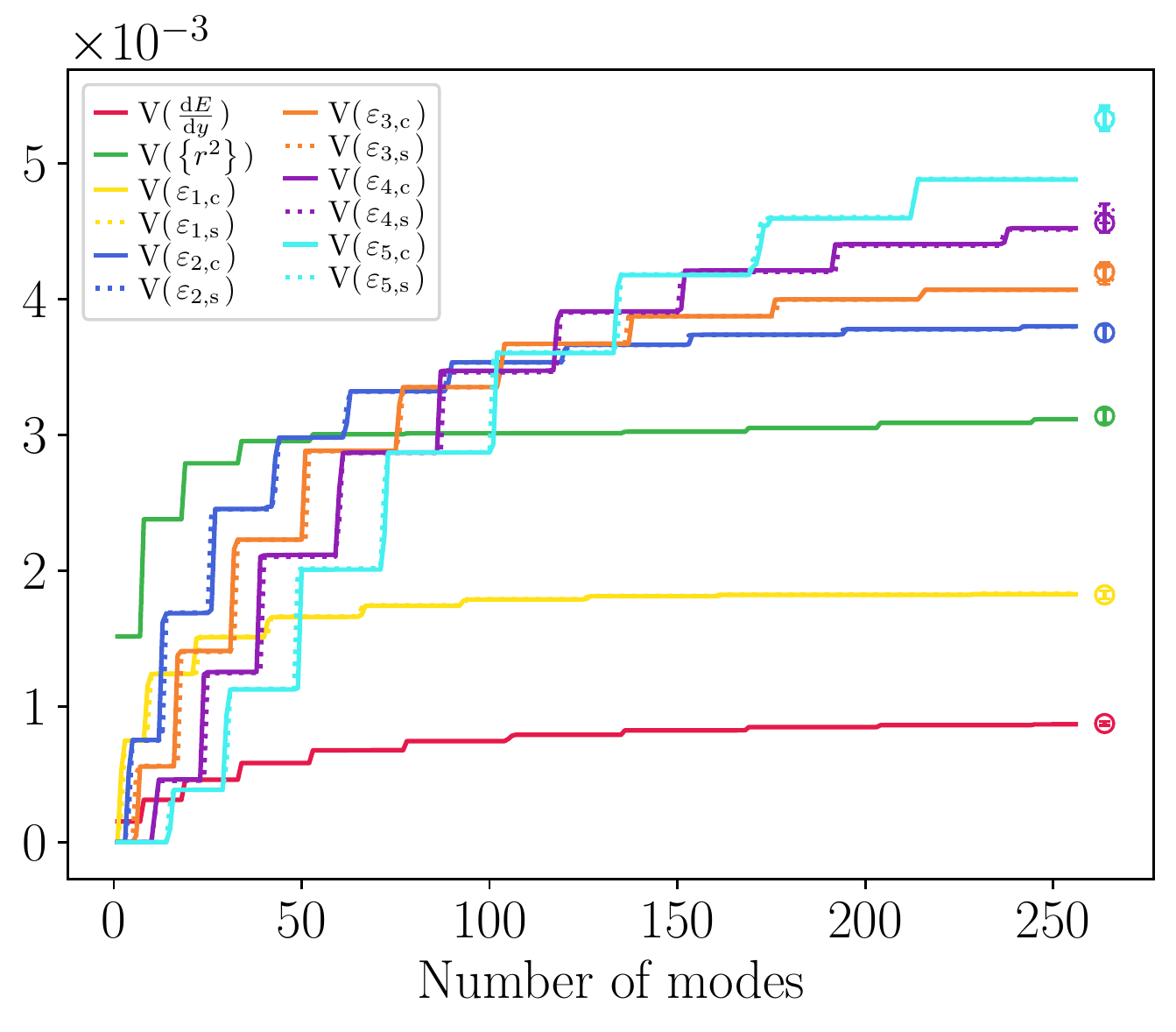}
\includegraphics[width=0.495\linewidth]{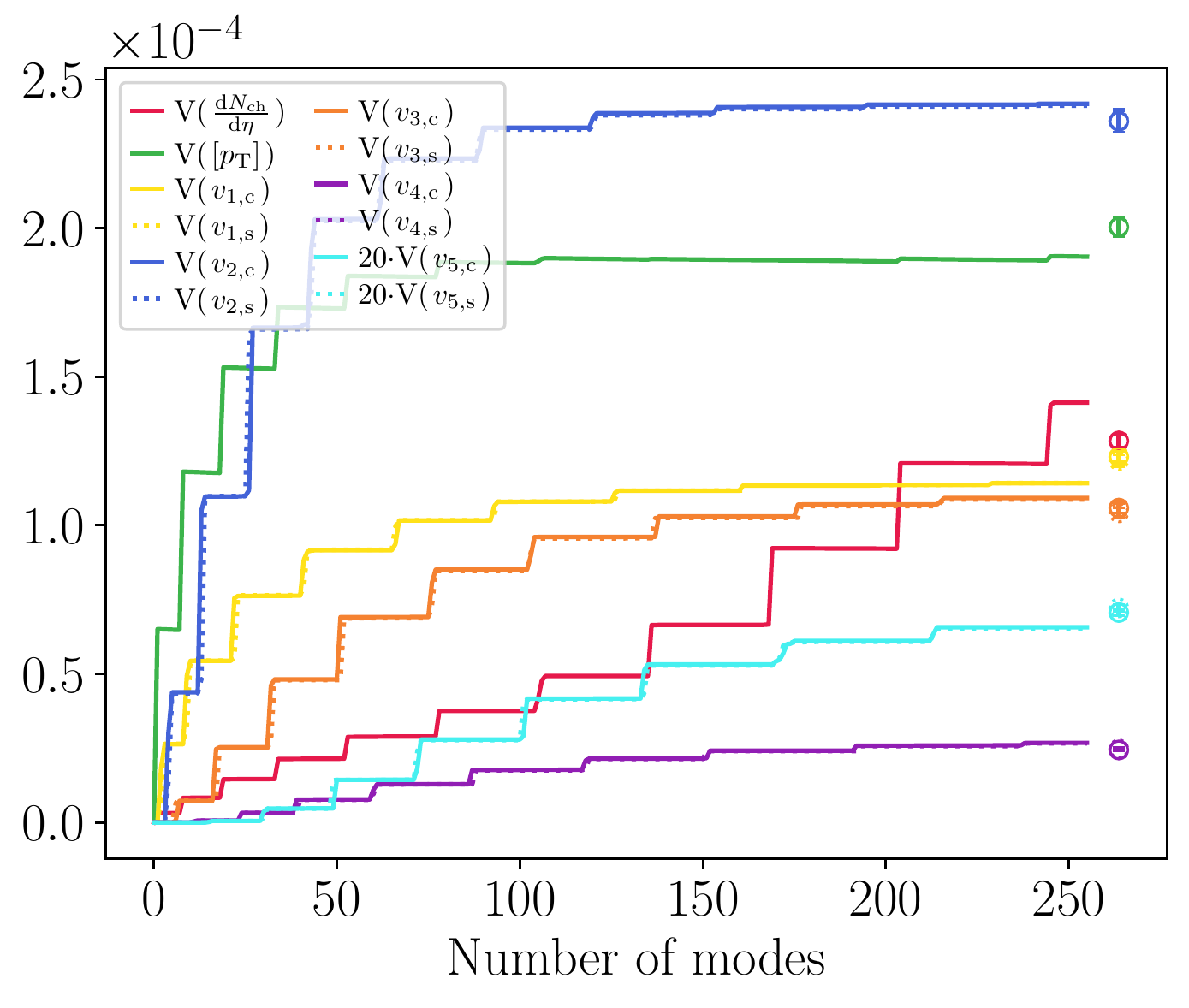}
\includegraphics[width=0.495\linewidth]{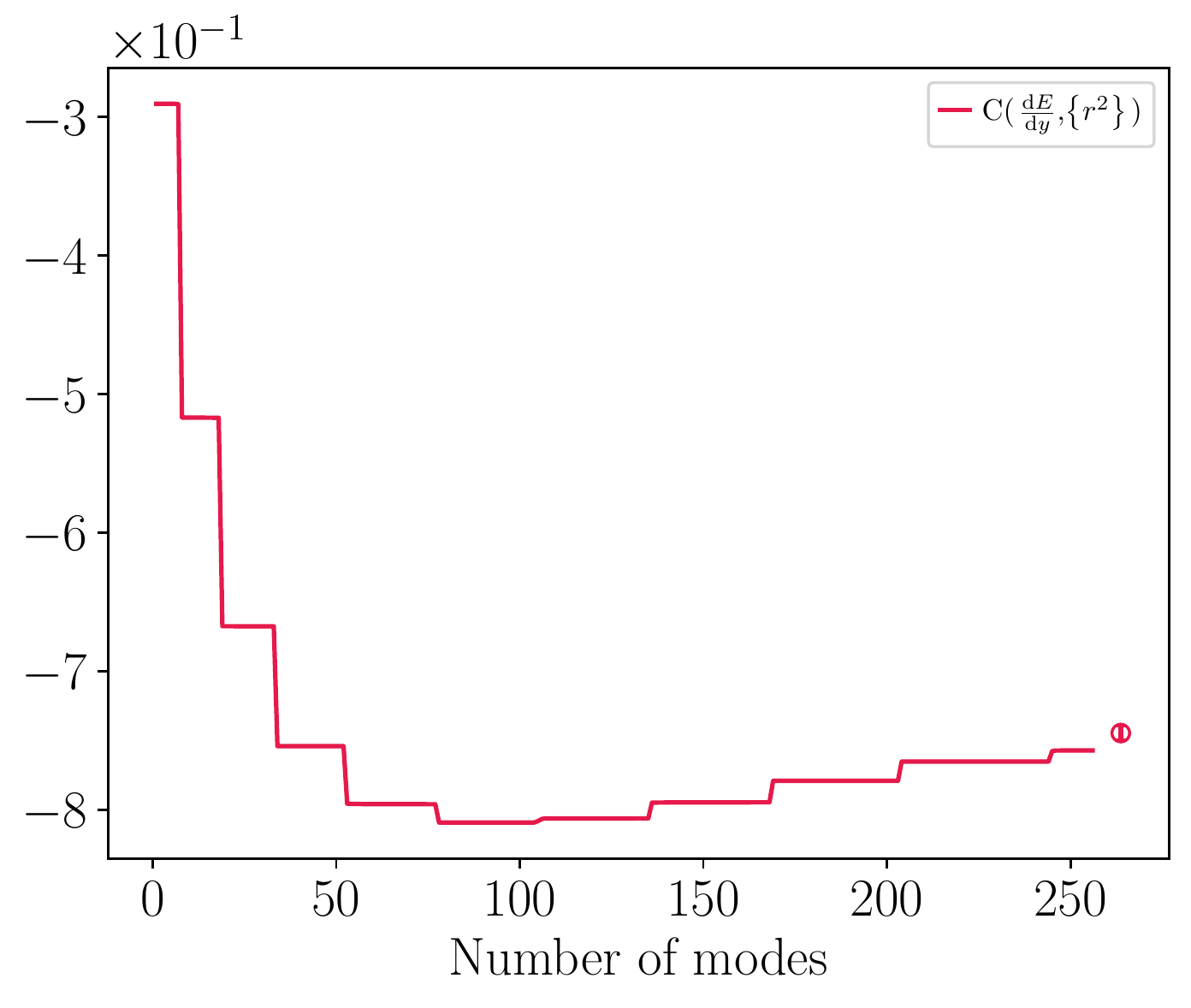}
\includegraphics[width=0.495\linewidth]{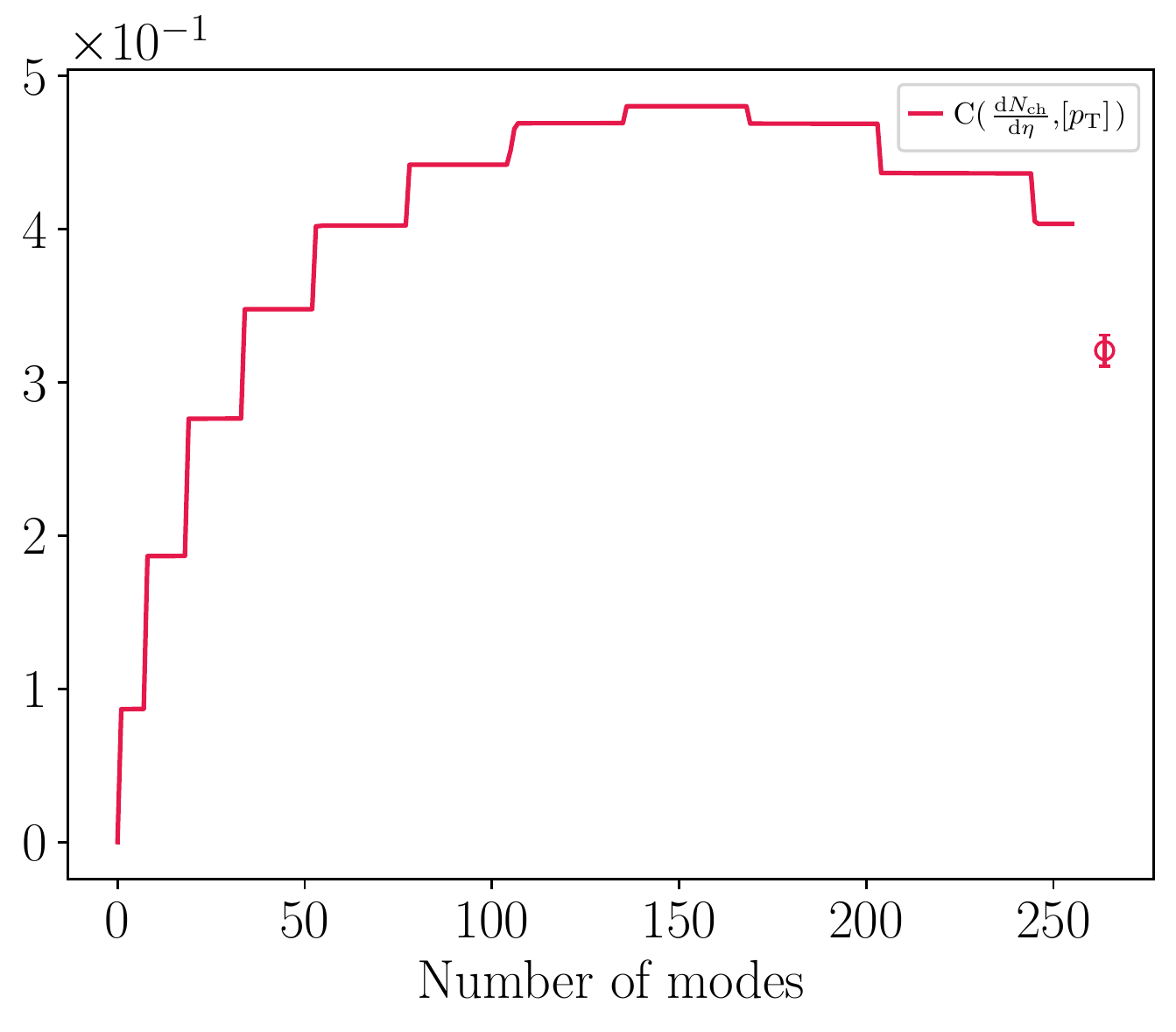}
\vspace{-7mm}
\caption{Variances (top) and correlation coefficients (bottom) of initial-state (left) and final-state observables (right) for collisions at $b=0$ in the Glauber model. 
The circles next to the right edge of each panel give the values computed from the random sample of 8192 events. 
The full lines show the quantities computed with Eqs.~\eqref{V(O_a)} and \eqref{eq:pearson_coeff}, including the number of modes given by the abscissa for the sums in V and the numerator of C. 
The sums in the denominator of C run over 256 modes.
The variances of $\d E/\d y$, $\{r^2\}$, $\d N_\textrm{ch}/\d\eta$, and $[p_\textrm{T}]$ are divided by the corresponding mean values.}
\label{fig:co_variances_convergence_Glauber_b0_IS_FS_modes_and_sampled}
\end{figure*}

\begin{figure*}[!htb]
\includegraphics[width=0.495\linewidth]{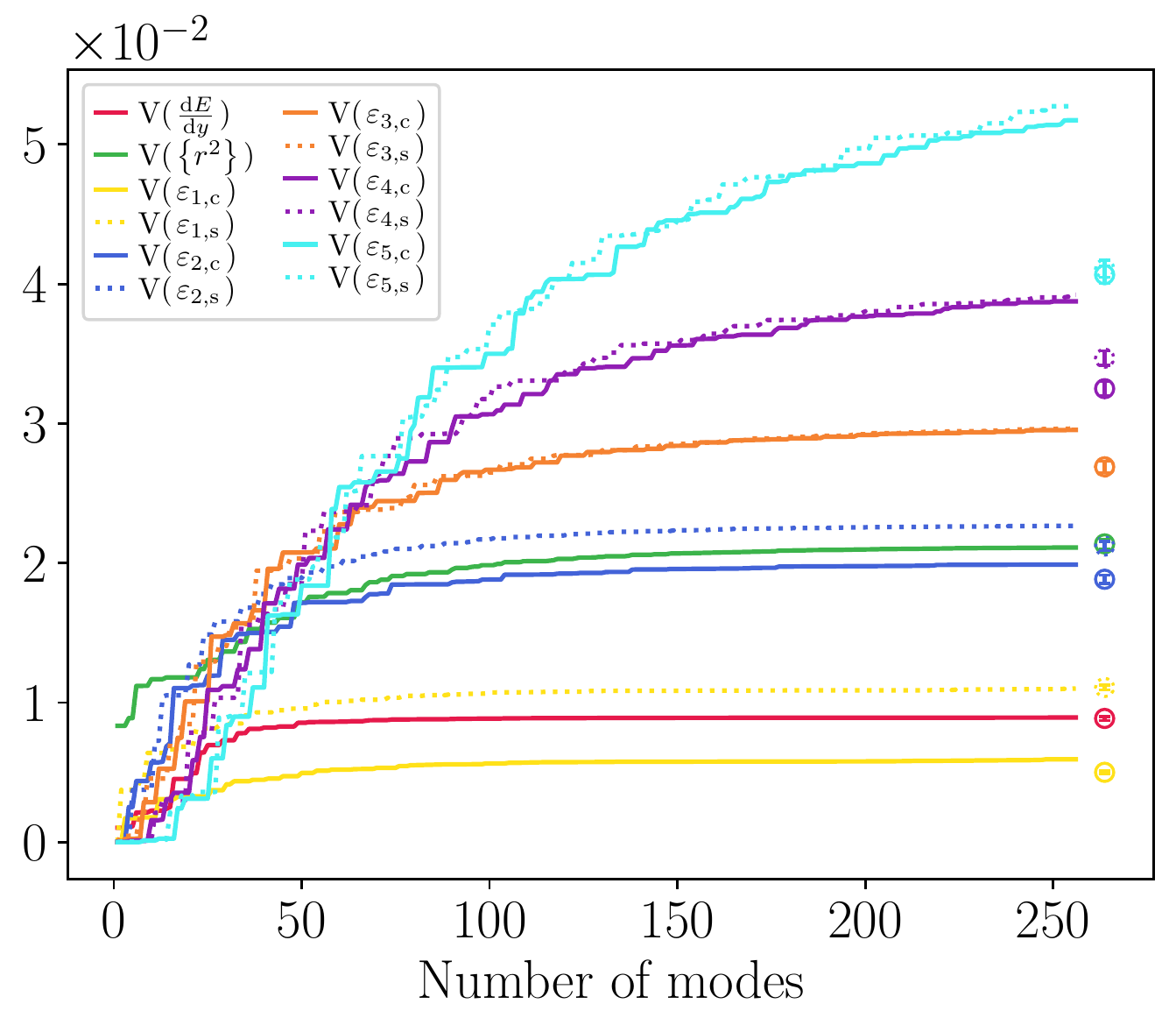}
\includegraphics[width=0.495\linewidth]{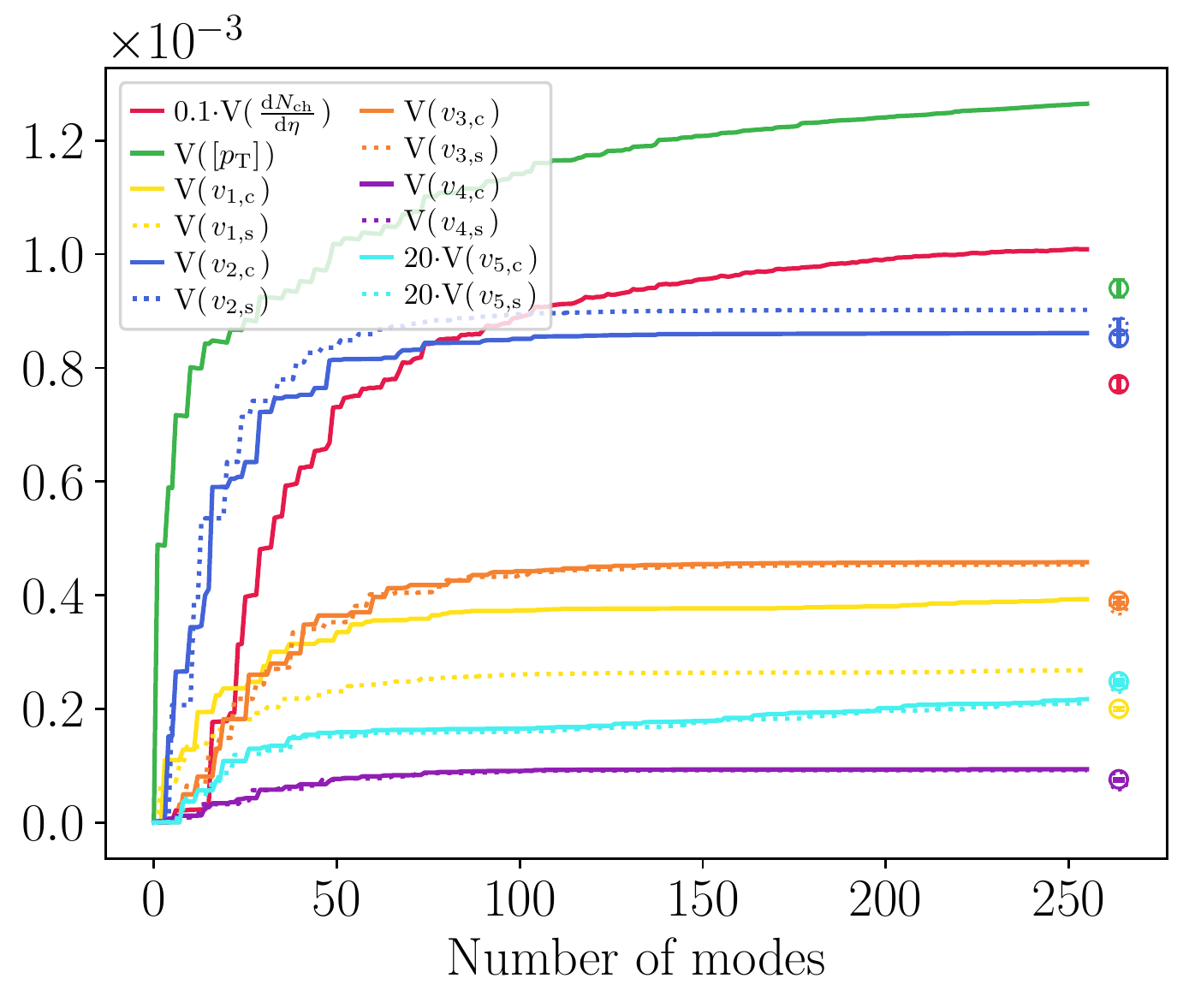}
\includegraphics[width=0.495\linewidth]{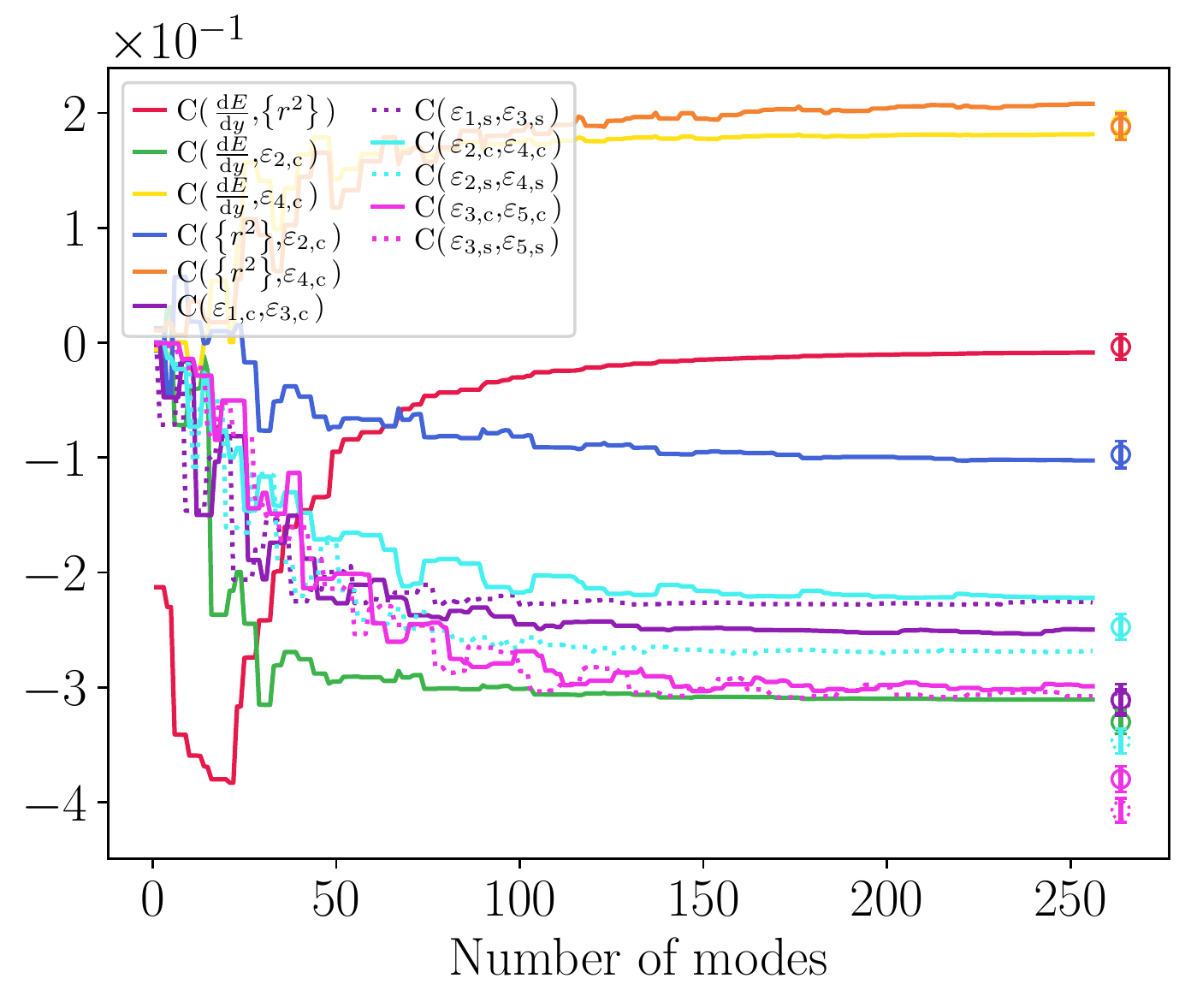}
\includegraphics[width=0.495\linewidth]{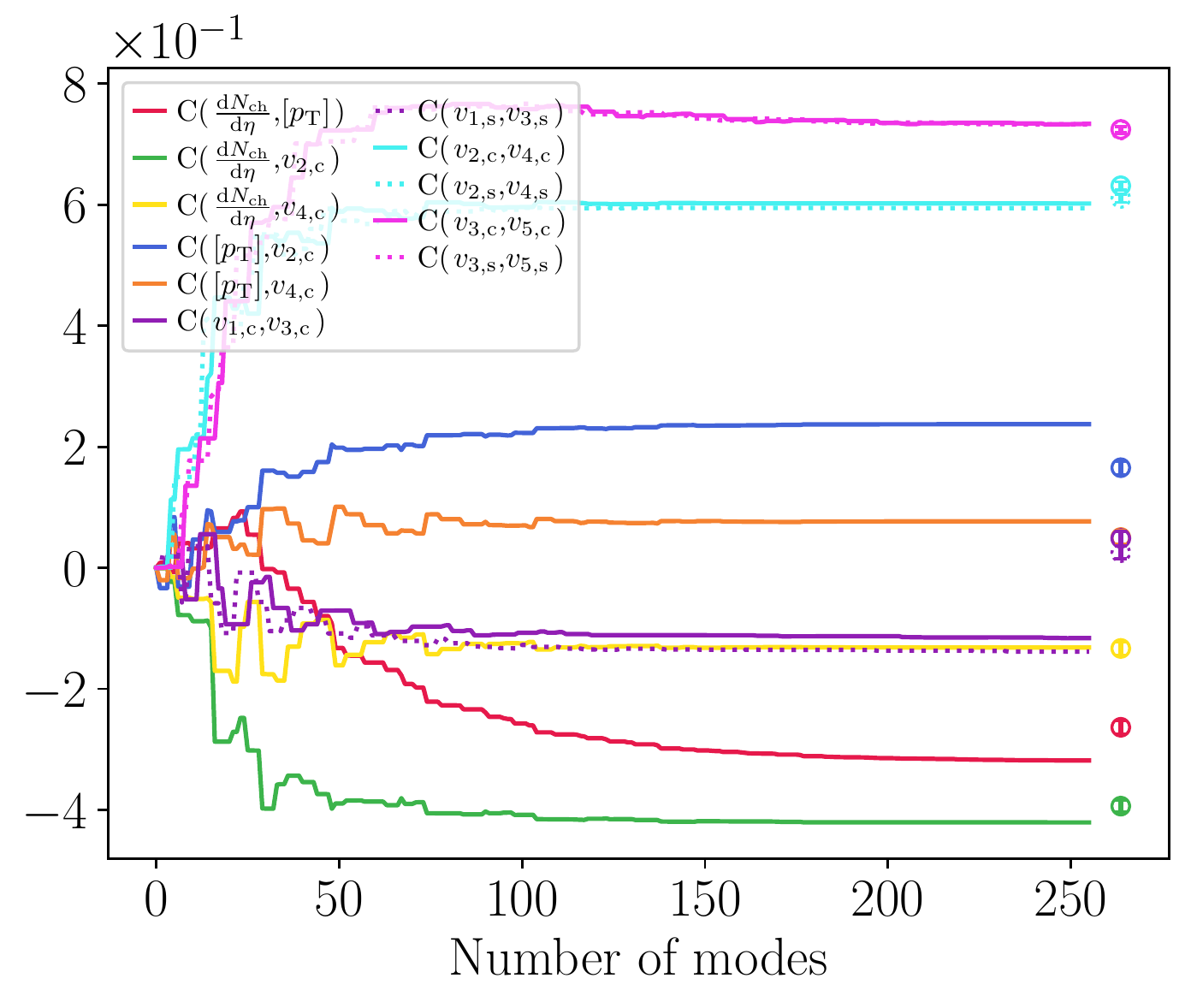}
\vspace{-7mm}
\caption{Same as Fig.~\ref{fig:co_variances_convergence_Glauber_b0_IS_FS_modes_and_sampled} for collisions at $b=9$~fm within the Glauber model.}
\label{fig:co_variances_convergence_Glauber_b9_IS_FS_modes_and_sampled}
\end{figure*}

\subsection{Variances and covariances}
\label{subsec:variances_covariances}

From the 8192 random events one can compute the sample standard deviation of every observable $O_\alpha$ about its mean value $\expval{O_\alpha}$.
On the other hand, in our dynamical setup\footnote{Using fluctuating fluid dynamics or early-time evolution would yield a further source of fluctuations.} the dispersion of the values taken by $O_\alpha$ event by event arises solely from the fluctuations in the initial state. 
One may hope to capture the fluctuations of observables with our mode decomposition, and in fact Eq.~\eqref{eq:covariance} gives the covariance between the fluctuations of two observables at order $c_l^2$. 
Setting $\alpha=\beta$ thus yields the variance of the fluctuations of $O_\alpha$
\begin{equation}
\label{V(O_a)}
\mathrm{V}(O_\alpha) \equiv \expval{(O_\alpha-\expval{O_\alpha})^2} \simeq 
\sum_l L_{\alpha,l}^2,
\end{equation}
involving the linear-response coefficients $L_{\alpha,l}$. 
In the top panels of Figs.~\ref{fig:co_variances_convergence_Glauber_b0_IS_FS_modes_and_sampled} and \ref{fig:co_variances_convergence_Glauber_b9_IS_FS_modes_and_sampled}, we show these variances (full lines) as function of the number of modes over which the sum runs.
We also show as circles close to the right edge of each panel the sample variances of the observables computed from the 8192 events, with error bars given by the standard uncertainty on the variance estimate. 

\begin{figure*}[!t]
\includegraphics[width=0.495\linewidth]{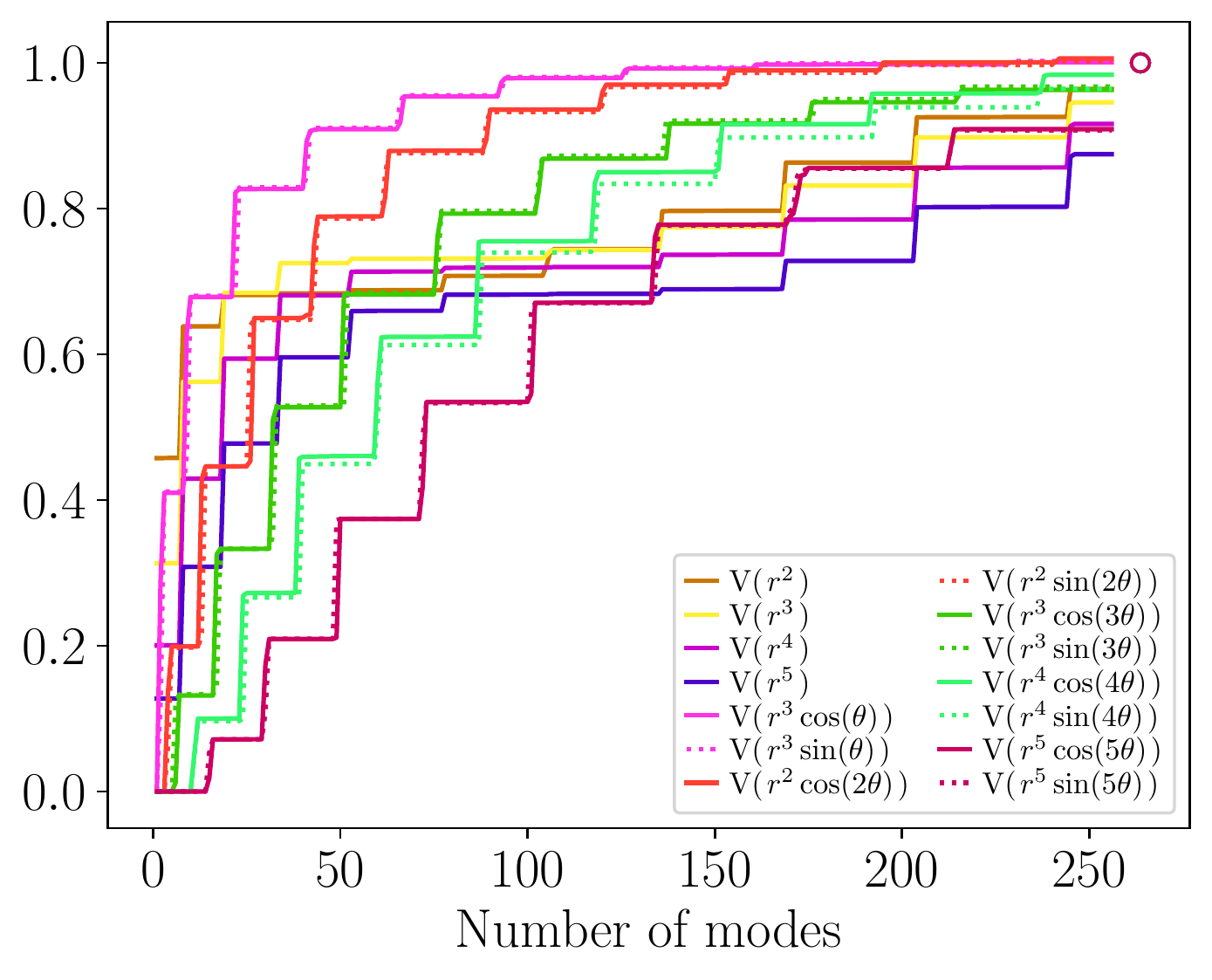}
\includegraphics[width=0.495\linewidth]{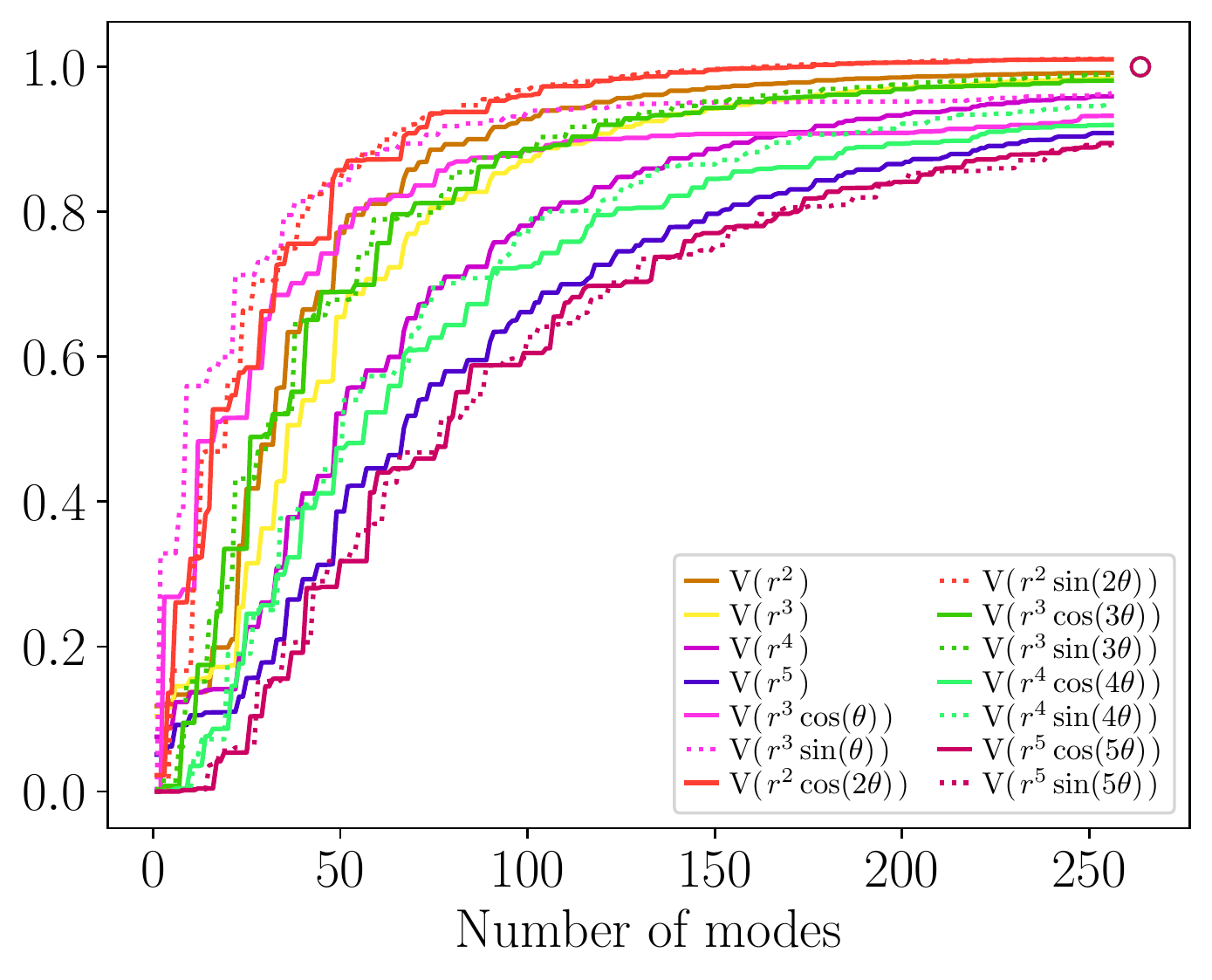}
\vspace{-7mm}
\caption{Variances of the numerators and denominators of the initial-state eccentricities~\eqref{eq:eccentricities1} for collisions at $b=0$ (left) and $b=9$ fm (right) in the Glauber model. 
The variances are computed with Eq.~\eqref{V(O_a)}, including the number of modes given by the abscissa for the sum, and divided by the sample variances from the 8192 random events, so that the open symbols at a value of~1 stand for these sample variances.}
\label{fig:variances_convergence_Glauber_IS_NumDenom}
\end{figure*}

From the covariance~\eqref{eq:covariance} one can also derive the Pearson correlation coefficient of two observables
\begin{equation}
\mathrm{C}(O_\alpha,O_\beta) \equiv 
\frac{\expval{(O_\alpha-\expval{O_\alpha})(O_\beta-\expval{O_\beta})}}{\sqrt{\textrm{V}(O_\alpha)\textrm{V}(O_\beta)}}
\simeq \frac{\sum_l L_{\alpha,l} L_{\beta,l}}{\sqrt{\sum_{k,k'}L_{\alpha,k}^2 L_{\beta,k'}^2}}.
\label{eq:pearson_coeff}
\end{equation}
These coefficients are shown for various pairs of observables as full lines in the bottom panels of Figs.~\ref{fig:co_variances_convergence_Glauber_b0_IS_FS_modes_and_sampled} and \ref{fig:co_variances_convergence_Glauber_b9_IS_FS_modes_and_sampled} as a function of the number of modes used in the sum in the numerator. 
The sums in the denominator always include the first 256 modes. 
That is, the values of $\textrm{V}(O_\alpha)$, $\textrm{V}(O_\beta)$ used in the denominator of Eq.~\eqref{eq:pearson_coeff} are those reached at the end of the full lines in the upper panels of the figures.
Again, the circles close to the right edge of the panels are the correlation coefficients from the sample of 8192 events.

We first focus on the behavior of the variances, shown in Fig.~\ref{fig:co_variances_convergence_Glauber_b0_IS_FS_modes_and_sampled} for collisions at $b=0$ while Fig.~\ref{fig:co_variances_convergence_Glauber_b9_IS_FS_modes_and_sampled} shows the corresponding results at $b=9$~fm. Both are obtained in the Glauber model, while the corresponding plots with initial states from the Saturation model are shown in Figs.~\ref{fig:co_variances_convergence_Saturation_b0_IS_FS_modes_and_sampled} and \ref{fig:co_variances_convergence_Saturation_b9_IS_FS_modes_and_sampled} in Appendix~\ref{appendix:co_variances_Saturation}, with similar results.

Since $L_{\alpha,l}^2$ is non-negative, including more modes in the sum increases the value of $\mathrm{V}(O_\alpha)$. 
But since the magnitude of the response $L_{\alpha,l}$ typically decreases with $l$, each sum should hopefully converge, and indeed each variance seems to reach a maximum value. 
At $b=0$ (Fig.~\ref{fig:co_variances_convergence_Glauber_b0_IS_FS_modes_and_sampled}) and for initial-state characteristics (left), the visual impression is that each mode-by-mode variance tends towards the corresponding sample variance, except for $\varepsilon_5$ for which more modes would be needed.  
For initial and final state observables at zero impact parameter, the variance does not grow with each mode, but there are only a few increasingly rarer steps: e.g., for $l = 0, 7, 18, 33, 52, 77\dots$ in the case of the variances of $\dd E/\dd y$ or multiplicity. 
The reason is simply that these are the only ``radial modes,'' i.e., the only modes for which the linear-response coefficient $L_{\alpha,l}$ for $\dd E/\dd y$  has a sizable value, as shown in the top panels of Fig.~\ref{fig:L_IS_FS_Glauber}.
In the case of eccentricities, the successive steps of the variance have the same size for $\varepsilon_{n,\rm c}$ and $\varepsilon_{n,\rm s}$ in a given harmonic but occur at values of $l$ differing by one: 
this mirrors once again the existence of pairs of degenerate, rotated modes, which we have encountered several times. 
Due to rotational symmetry at $b=0$, the sample variance is then the same for $\varepsilon_{n,\rm c}$ and $\varepsilon_{n,\rm s}$. 

Generally, the features observed for the fluctuations of the eccentricities are also found for the variances of the anisotropic-flow harmonics in the final state at $b=0$ (top right panel of Fig.~\ref{fig:co_variances_convergence_Glauber_b0_IS_FS_modes_and_sampled}). 
However, a significant difference appears, namely some of the mode-by-mode variances computed with Eq.~\eqref{V(O_a)} are larger than the sample variances from the 8192 random events. 
For $v_2$, $v_3$ or $v_4$ this could perhaps be caused by downwards fluctuations of the variances in the event sample. 
However, this cannot ba a valid explanation for the variance of multiplicity: $\textrm{V}(\d N_\textrm{ch}/\d\eta)$ has clearly not yet reached its maximum when summing over 256 modes in Eq.~\eqref{V(O_a)}, yet it is already markedly larger than the value found in the event sample. 
Since we shall again encounter this mismatch at $b=9$~fm, we momentarily postpone its discussion. 

Indeed, it is striking that some high fluctuation modes --- for instance $\Psi_l$ with $l=203$ --- yield a large contribution to the variance of charged multiplicity, while this does not happen for the other observables. 
This can be attributed to the system viscosity (here shear viscosity, since the bulk viscosity is zero in our simulations), which converts the small scale radial ``ripples'' of the initial energy density profile into extra particles. 
To test that idea we performed simulations with different values of the shear viscosity to entropy ratio ($\eta/s\in\{ 0, 0.16, 0.32\}$).
These revealed that the radial fluctuation modes with small $l$, for example $l=0$, are significantly affected by viscosity, which diminishes the mode contribution to the multiplicity. 
This is however less the case of radial modes with a relatively large $l$,~e.g., $l=203$. 
We checked that for $l\geq 256$ the contribution of radial modes to $\textrm{V}(\d N_\textrm{ch}/\d\eta)$ decreases, which is to be expected since the mode eigenvalues keep decreasing. 
Therefore, given a sufficient number of modes, $\textrm{V}(\d N_\textrm{ch}/\d\eta)$ computed with Eq.~\eqref{V(O_a)} converges to a finite value; but it still cannot coincide with the sample variance since it is already larger after 256 modes. 

At zero impact parameter there are only two sizable covariances, or equivalently Pearson correlation coefficients, between the observables we consider, namely, between total energy and typical system size in the initial state, and between multiplicity and average momentum in the final state.\footnote{We leave aside correlations between an initial-state quantity and a final-state observable.}
In the bottom panels of  Fig.~\ref{fig:co_variances_convergence_Glauber_b0_IS_FS_modes_and_sampled} we see that the mode-by-mode correlation coefficient $\textrm{C}(\d E/\d y,\{r^2\})$ seems to tend towards its sample value after 256 modes. 
This may also be true of $\textrm{C}(\d N_\textrm{ch}/\d\eta,[p_\textrm{T}])$, although it is less clear. 
As mentioned above, the radial fluctuation modes with a rather high $l$ still contribute to the fluctuations of multiplicity.
In any case, one may note that the mode-by-mode evolution of the correlation functions does not behave monotonically when more modes are included in the sums in the numerator, i.e., in the covariances: some modes affect both observables in the same direction, leading to a positive correlation function, while in other modes the observables are anti-correlated. 

In events at $b=9$ fm (Fig.~\ref{fig:co_variances_convergence_Glauber_b9_IS_FS_modes_and_sampled}), the first salient feature is that the variances and the correlation functions are now affected by many more modes, instead of only a few ones at zero impact parameter. 
Visually the typical contribution of a mode looks smaller, but one should beware that the vertical scales of the plots differ from Fig.~\ref{fig:co_variances_convergence_Glauber_b0_IS_FS_modes_and_sampled}. 
That more modes contribute to the fluctuations is simply related to the fact that, due to the mixing with the anisotropic average state (cf. Sec.~\ref{subsec:results_stat_charact_IS}) every mode now affects multiple initial-state characteristics. 
This also explains why there are now several nonzero correlation functions in both initial and final states. 

\begin{figure*}[!t]
	\includegraphics[width=0.495\linewidth]{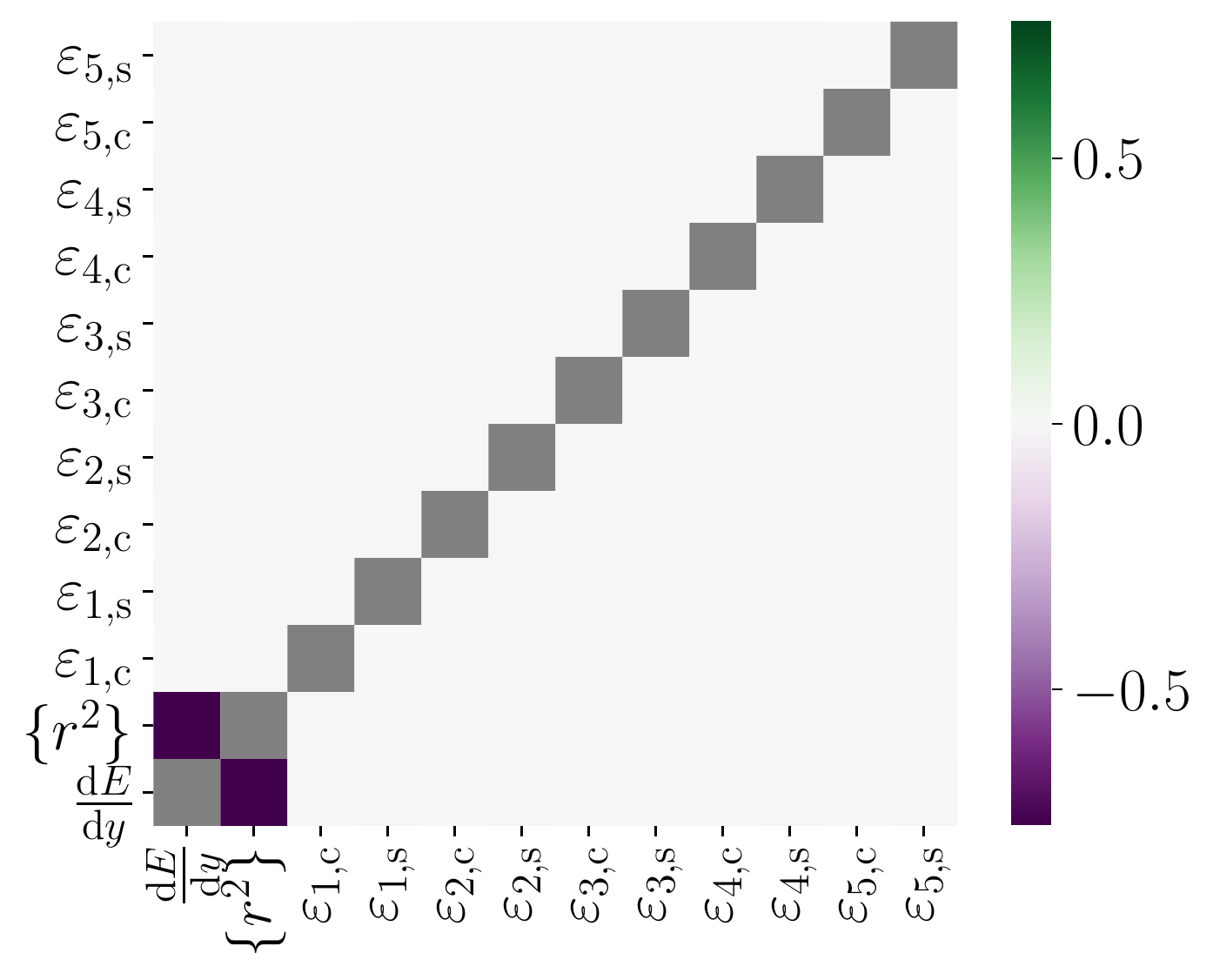}
	\includegraphics[width=0.495\linewidth]{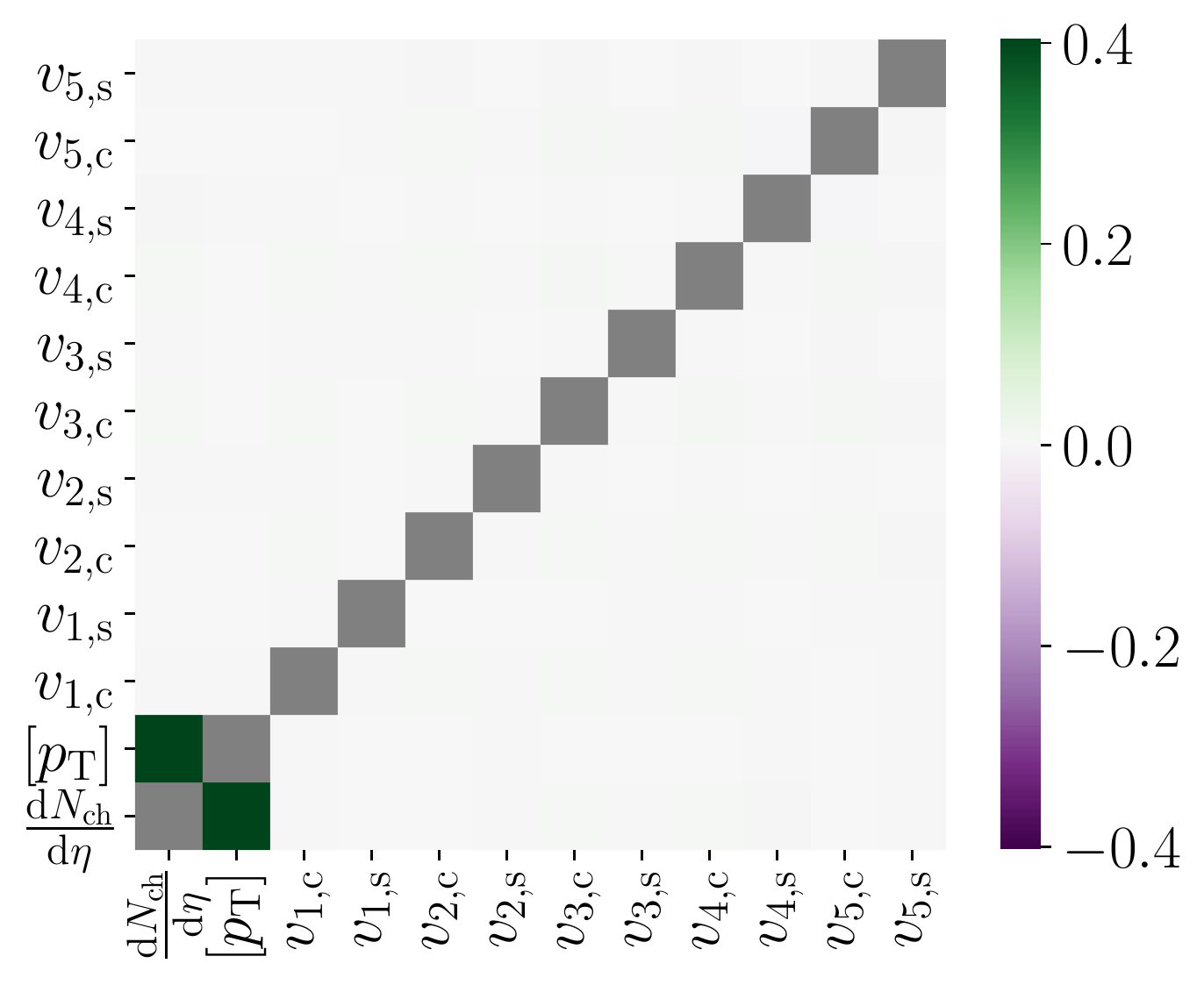}
	\includegraphics[width=0.495\linewidth]{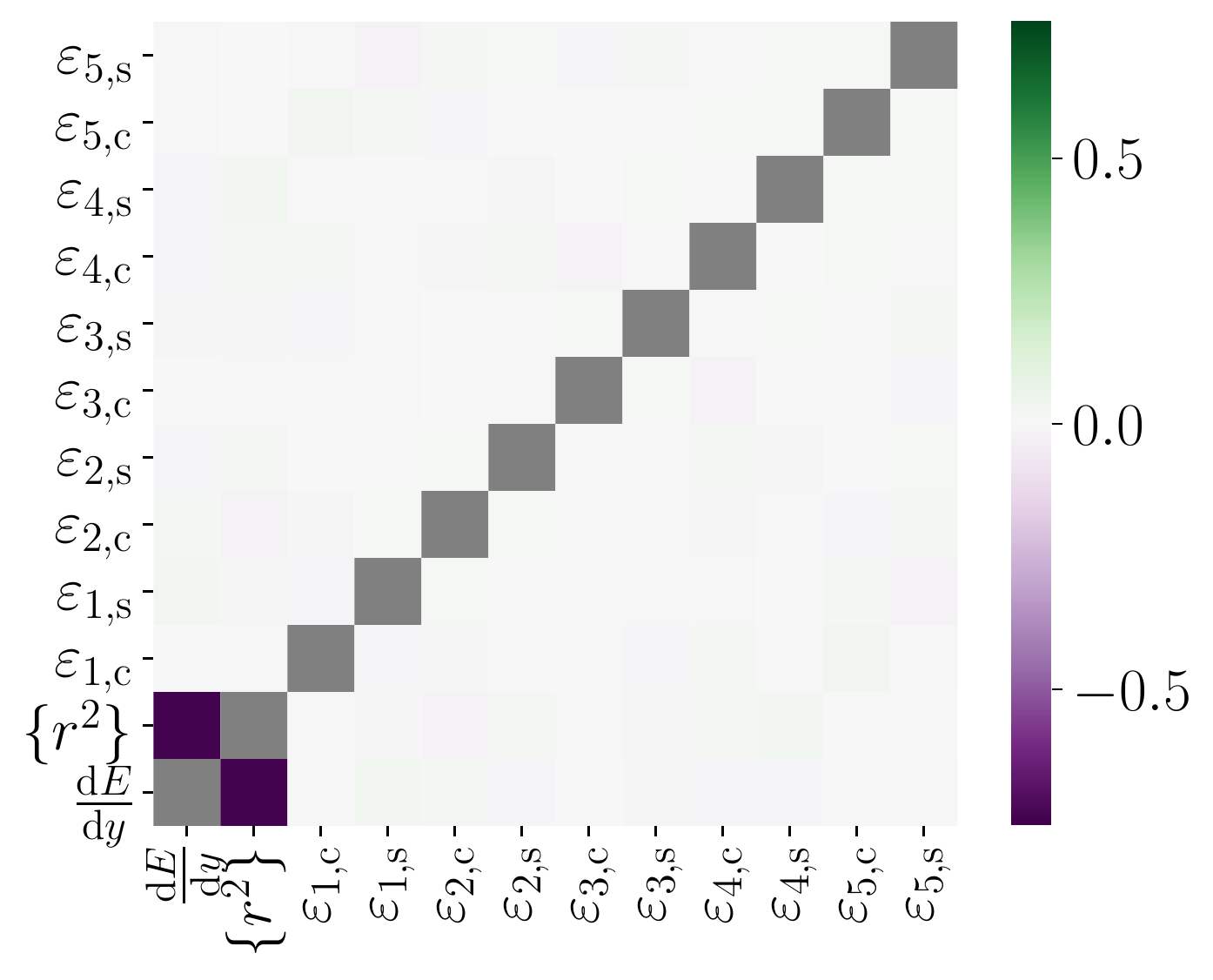}
	\includegraphics[width=0.495\linewidth]{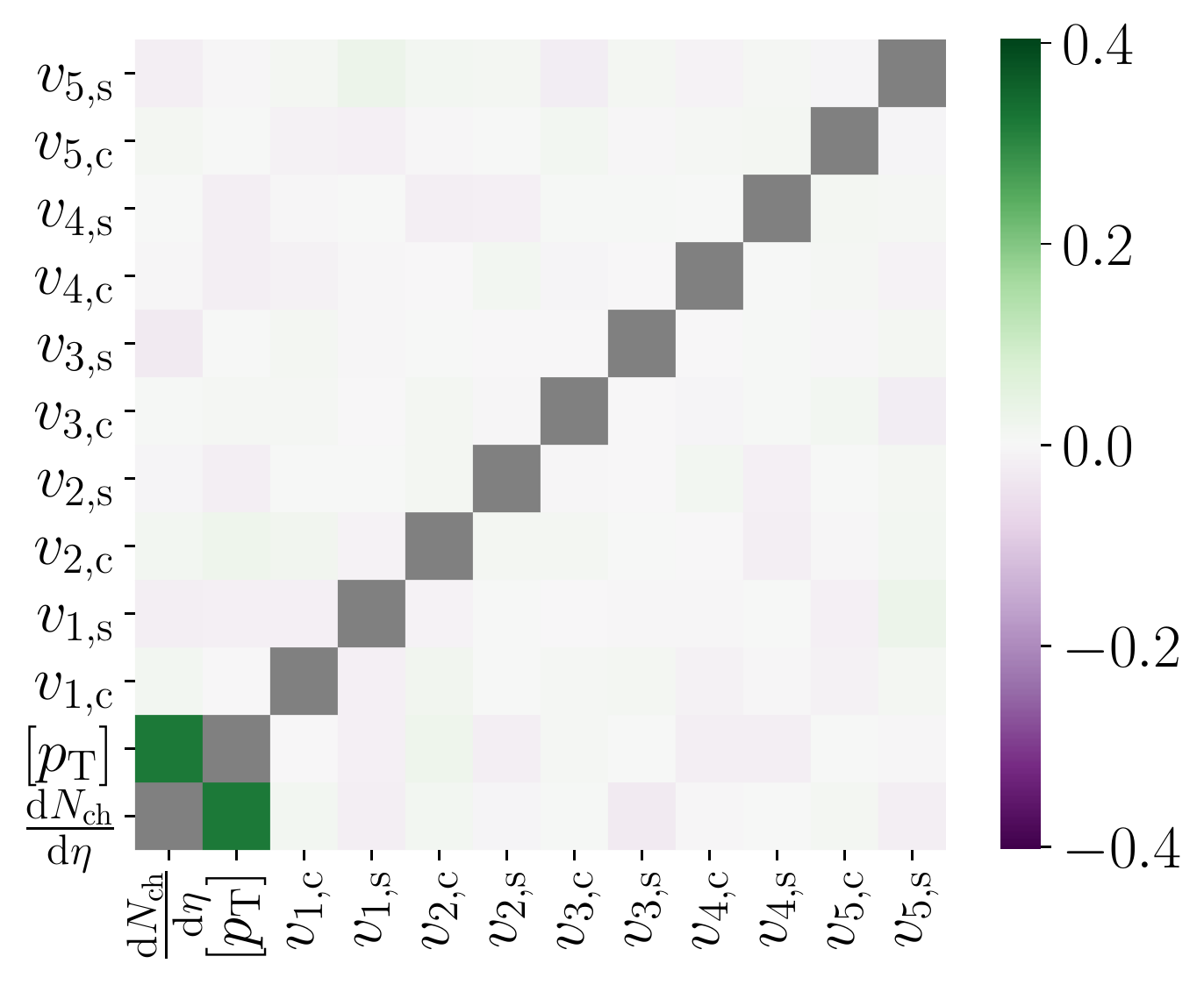}
	\vspace{-5mm}
	\caption{Correlation coefficients of initial state (left) and final state (right) observables for collisions at $b=0$ in the Glauber model.
		Top: values computed with 256 modes in the mode-by-mode approach [Eq.~\eqref{eq:pearson_coeff}]. 
		Bottom: values from the sample of 8192 random events. 
		Values on the diagonal, which by definition equal one, are shown as gray squares.}
	\label{fig:covariances_Glauber_b0_IS_FS_modes_and_sampled}
\end{figure*}

A second remark is that the parallel behavior of the variances of $\varepsilon_{n,\rm c}$ and $\varepsilon_{n,\rm s}$, and in the final state $v_{n,\rm c}$ and $v_{n,\rm s}$ is no longer observed, reflecting the breaking of rotational symmetry at finite impact parameter. 

Eventually, it is now clear that several variances computed within the mode-by-mode approach will not match the sample variance from our event-by-event simulations. 
In the top-left panel of Fig.~\ref{fig:co_variances_convergence_Glauber_b9_IS_FS_modes_and_sampled}, the values of $\textrm{V}(\varepsilon_n)$ with $n=2,3,4,5$ obtained with 256 modes are larger than the sample variance depicted by the circles. 
In the top right panel, the same holds for almost all final-state observables, especially $[p_\textrm{T}]$, $v_{1,\rm c/s}$ or $v_{3,\rm c/s}$. 

In the case of the initial-state characteristics, we found in Table~\ref{tab:average_state_observables} that the mode-by-mode approach provides a good estimate of the average value, especially when including the term in $Q_{\alpha,ll}$ in Eq.~\eqref{eq:<O_alpha>}. 
In contrast, Eq.~\eqref{V(O_a)} does not include these quadratic-response terms --- without contradiction, since both equations are valid at order ${\cal O}(c_l^2)$. 
Yet it hints at the origin of the mismatch between the mode-by-mode and sample values of the variances, since the eccentricities, which are defined as ratios with a moment of the energy density in the denominator, depend nonlinearly on the modes, i.e., on $c_l$. 
To test this idea, we took as observables $O_\alpha$ the numerators and denominators of the eccentricities~\eqref{eq:eccentricities1} alone, i.e., the integrals of $r^n$, $r^n\cos(n\theta)$, and $r^n\sin(n\theta)$ multiplied by the energy density. 
The mode-by-mode variances~\eqref{V(O_a)} of these quantities for initial states from the Glauber model, divided by the respective sample variances from the 8192 random events, are shown in Fig.~\ref{fig:variances_convergence_Glauber_IS_NumDenom}. 
The ratios of mode-by-mode over sample variances seem to converge to unity from below when more modes are accounted for in the calculation of $\textrm{V}(O_\alpha)$, showing that Eq.~\eqref{V(O_a)} provides a good estimate of the variance for characteristics that depend linearly on the modes. 
In turn, we can deduce that the lack of convergence of mode-by-mode variances to the sample values in Figs.~\ref{fig:co_variances_convergence_Glauber_b0_IS_FS_modes_and_sampled} and \ref{fig:co_variances_convergence_Glauber_b9_IS_FS_modes_and_sampled} means that nonlinear effects are present, already in the initial state, and naturally even more so in the final state.

At both impact parameters, the variance of $\varepsilon_{n,\rm c/s}$ increases with $n$. 
The reverse holds for the anisotropic flow harmonics, with $v_{2,\rm c/s}$ having the largest variance and $v_{5,\rm c/s}$ the smallest.\footnote{$v_1$ does not follow the trend, but the underlying physics is somewhat different, since global transverse-momentum conservation plays a crucial role for directed flow.}
This is consistent with the fact that the values of $v_n$ themselves follow the same hierarchy due to viscous damping.

\begin{figure*}[!htb]
\includegraphics[width=0.495\linewidth]{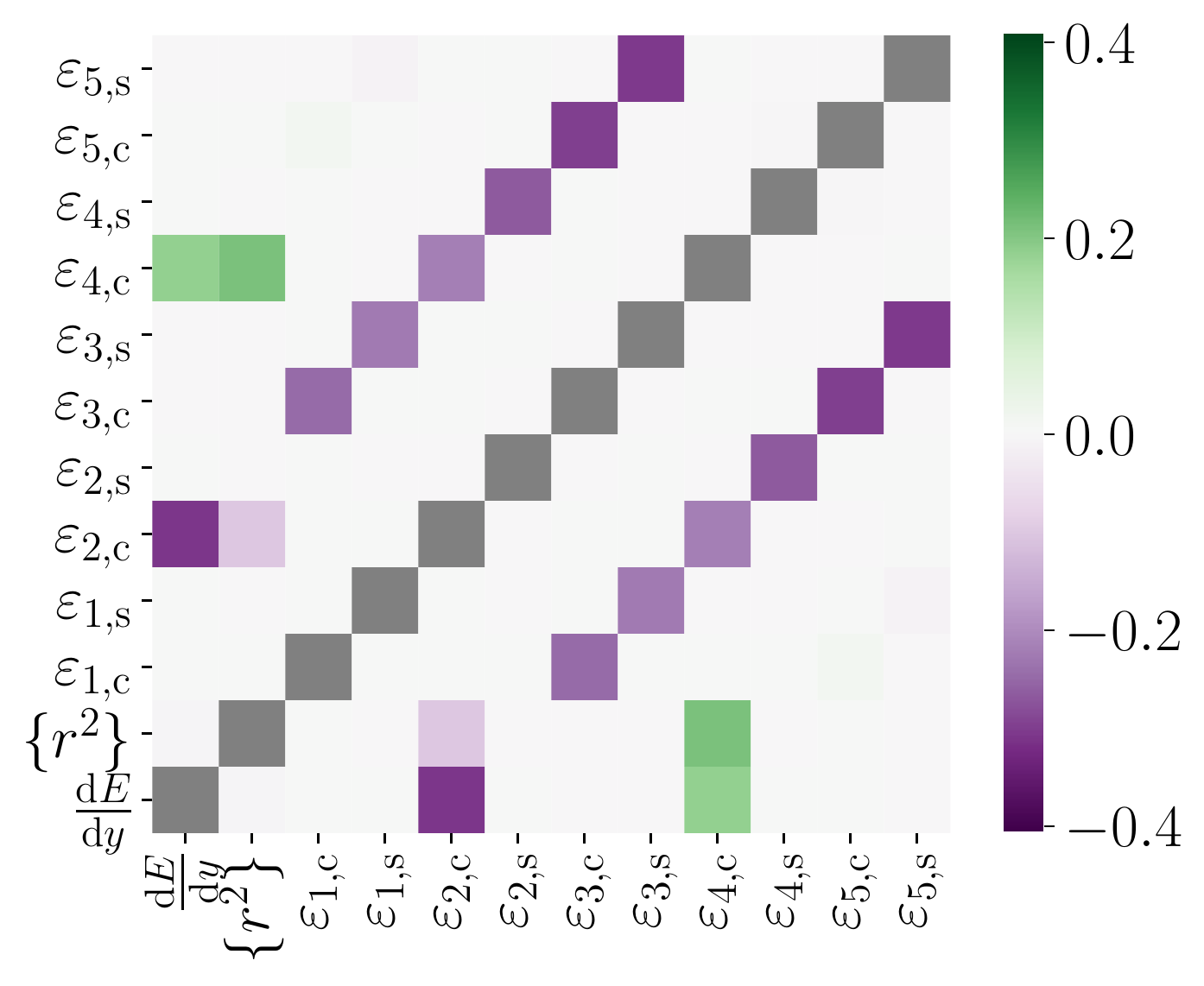}
\includegraphics[width=0.495\linewidth]{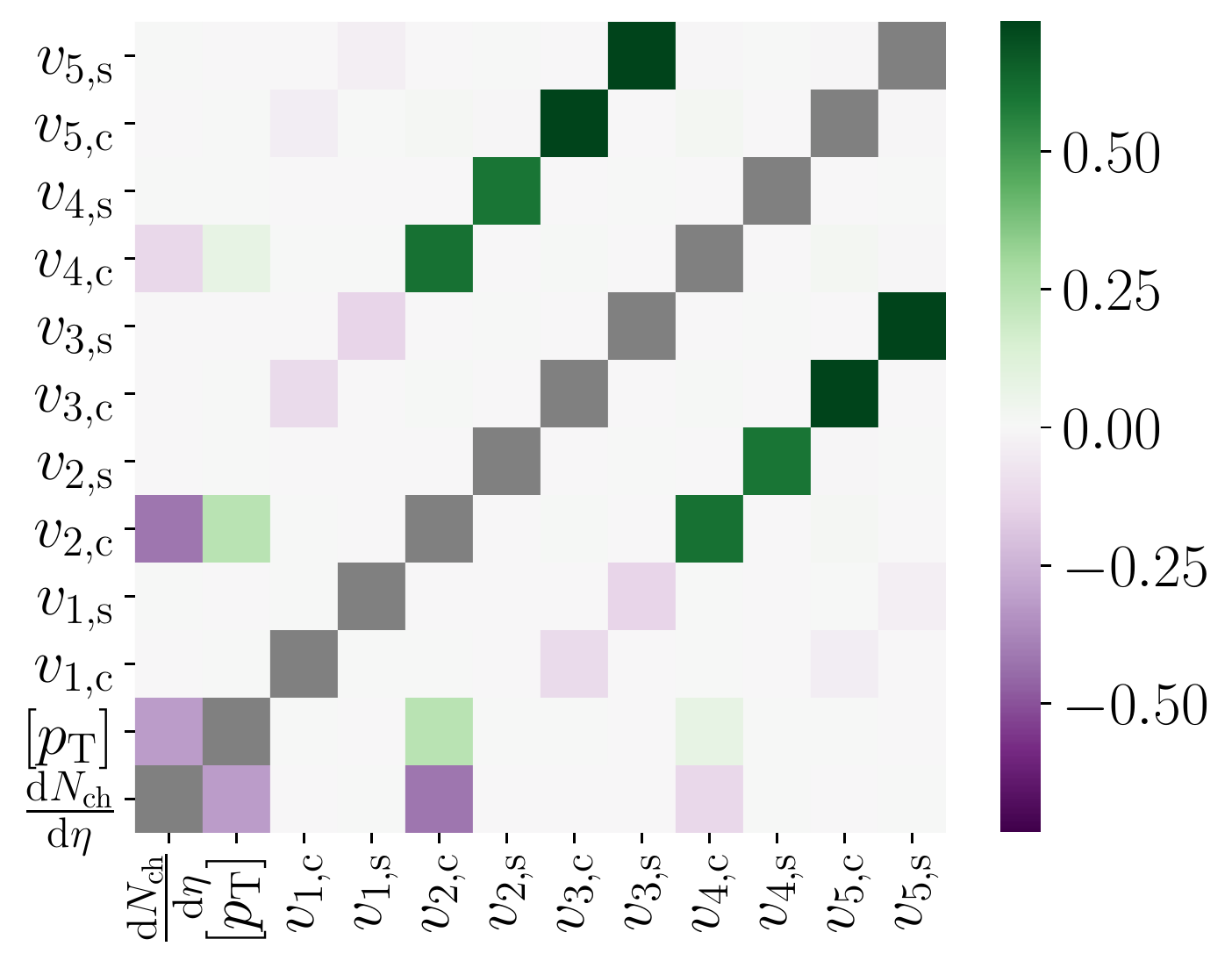}
\includegraphics[width=0.495\linewidth]{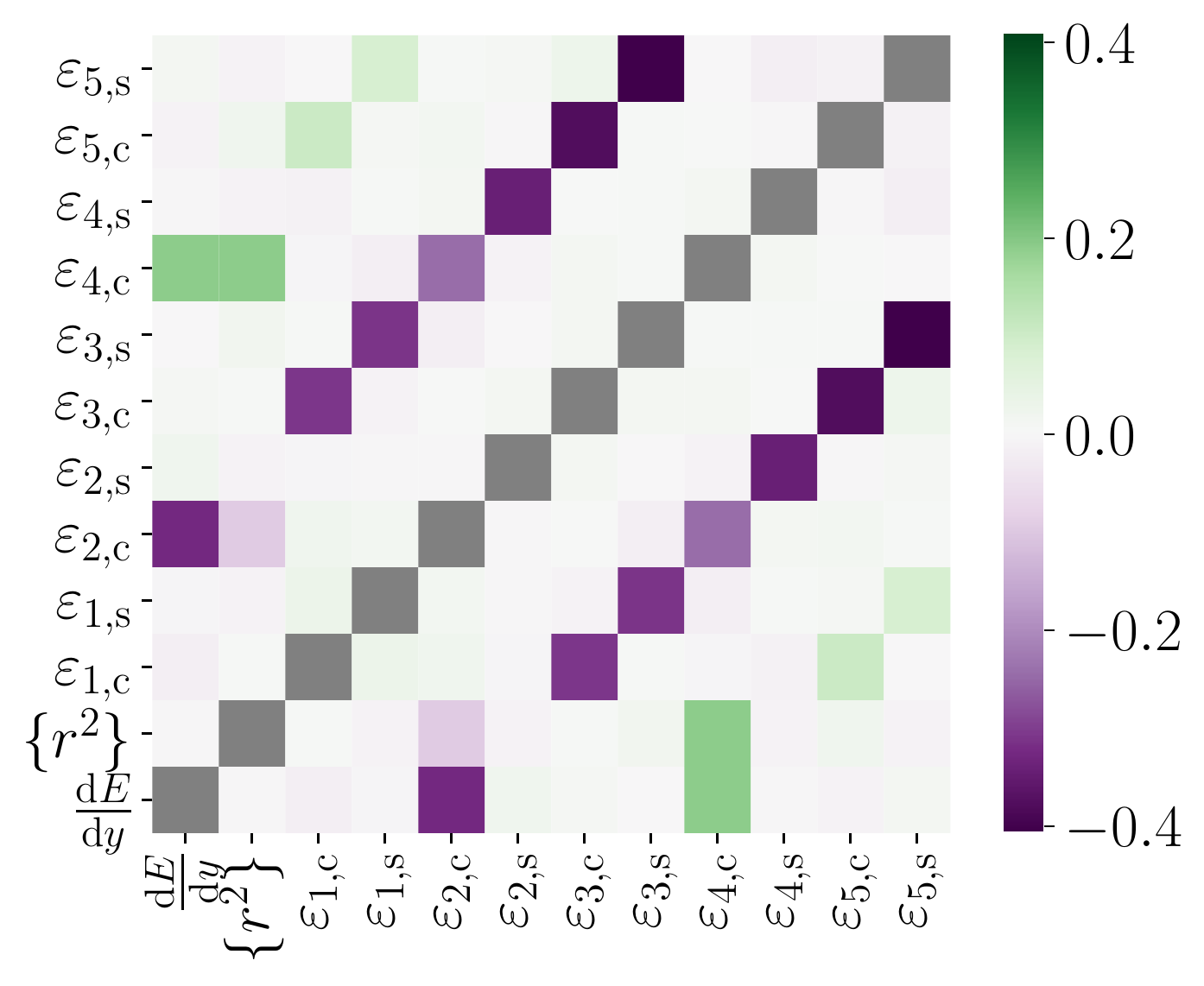}
\includegraphics[width=0.495\linewidth]{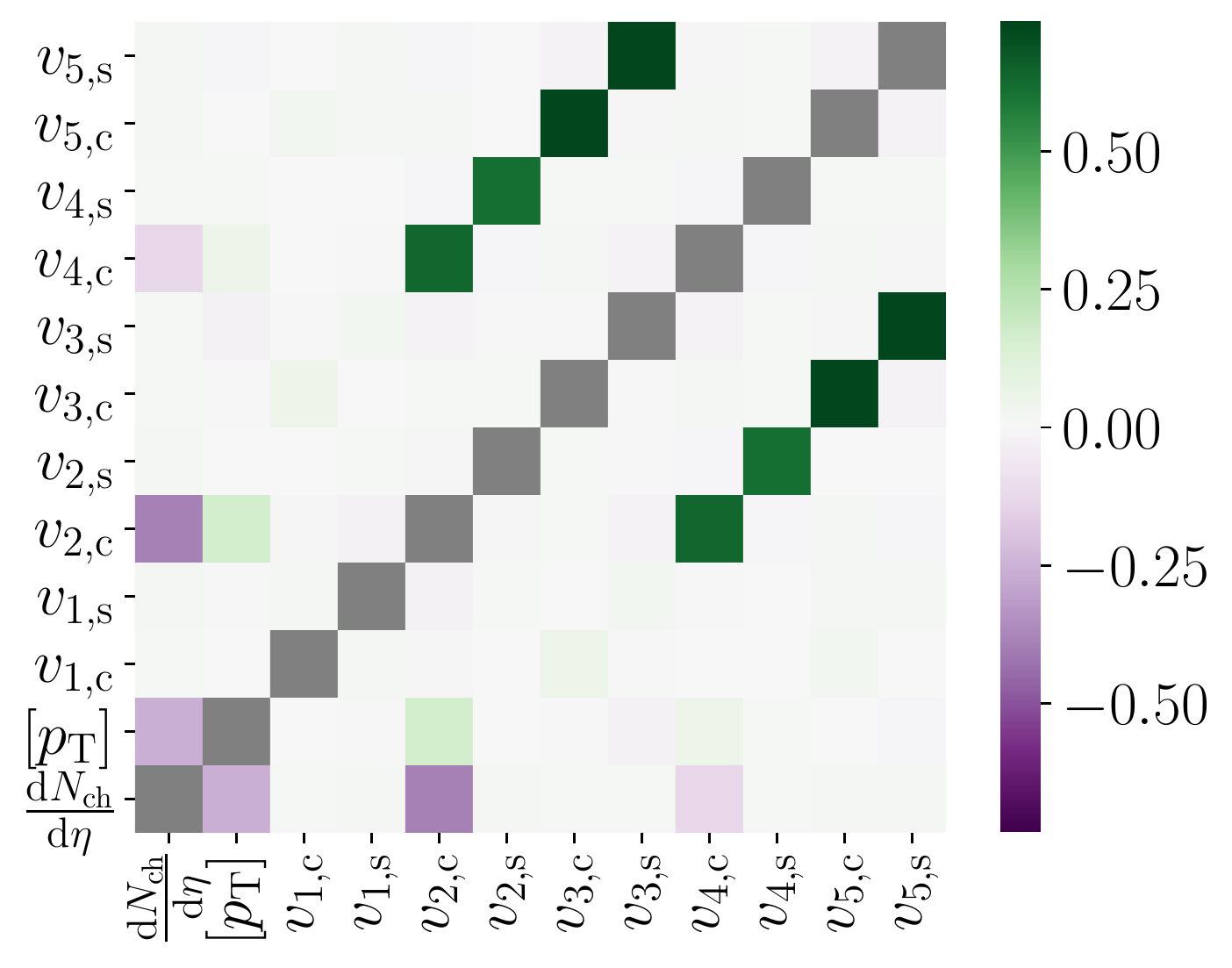}
\vspace{-5mm}
\caption{Same as Fig.~\ref{fig:covariances_Glauber_b0_IS_FS_modes_and_sampled} for collisions at $b=9$~fm within the Glauber model.}
\label{fig:covariances_Glauber_b9_IS_FS_modes_and_sampled}
\end{figure*}

Despite the discrepancy between mode-by-mode and sample variances, the correlation coefficients computed with the modes [Eq.~\eqref{eq:pearson_coeff}] or from the 8192 events are generally in rather decent agreement, see bottom panels of  Fig.~\ref{fig:co_variances_convergence_Glauber_b9_IS_FS_modes_and_sampled}.
In addition, most correlation functions seem to reach rather quickly a limiting value, namely after including about 150 modes, which rather contrasts the situation at $b=0$.  
This motivated us to provide a more direct comparison between the values computed with 256 modes of the correlation functions and the sample values, shown in Fig.~\ref{fig:covariances_Glauber_b0_IS_FS_modes_and_sampled} resp.\ Fig.~\ref{fig:covariances_Glauber_b9_IS_FS_modes_and_sampled} for collisions in the Glauber model at $b=0$ resp.\ $b=9$~fm, and in Figs~\ref{fig:covariances_Saturation_b0_IS_FS_modes_and_sampled} and \ref{fig:covariances_Saturation_b9_IS_FS_modes_and_sampled} for events with initial states from the Saturation model. 
A further advantage of these figures is that they also include correlation coefficients not shown in Figs.~\ref{fig:co_variances_convergence_Glauber_b0_IS_FS_modes_and_sampled} and \ref{fig:co_variances_convergence_Glauber_b9_IS_FS_modes_and_sampled}.
Here we merely describe the visible features, irrespective of any physical interpretation.

At vanishing impact parameter (Fig.~\ref{fig:covariances_Glauber_b0_IS_FS_modes_and_sampled}), we already saw that the only sizable correlation coefficients are between $\d E/\d y$ and $\{r^2\}$ in the initial state, and between charged multiplicity and average transverse momentum in the final state. 
These correlations are caused by the radially-symmetric fluctuation modes, which are the only ones that contribute to both pairs of observables.

In events at $b=9$~fm, a few clear trends appear: 
the cosine parts of even-order eccentricities resp.\ anisotropic-flow harmonics are (anti-)correlated with each other and with $\d E/\d y$ and $\{r^2\}$ resp.\ $\d N_\textrm{ch}/\d\eta$ and $[p_\textrm{T}]$, but not to the other observables.  
The cosine parts of odd-order eccentricities or anisotropic-flow harmonics are correlated with each other but not with the rest.  
Finally, the sine parts of eccentricities resp.\ flow coefficients only correlate to other $\varepsilon_{n,\rm s}$ resp.\ $v_{n,\rm s}$ in a harmonic $n$ with the same parity.

\subsection{Probability distributions and Gaussian statistics}
\label{section:gaussian_statistics}

We have seen that the probability distributions of the normalized coefficients $c_l$ of statistical fluctuations are close to being Gaussian, see Fig.~\ref{fig:cl_hist_Glauber}.
Assuming that they were in fact perfectly described by Gaussian statistics and that the fluctuations of a number of choice observables depend linearly on the fluctuations, it is straightforward to calculate model predictions of probability distributions for these observables. 
These predictions can serve as an approximation to the true statistical distributions of observables. 
In this section, we describe the mathematical framework for this approach ---  the details of the calculations are reported in Appendix~\ref{app:probability_distributions_Gfa} --- and present some results for the probability distributions of anisotropic flow coefficients.%

\subsubsection{Formalism}

It is clear that all observables cannot be assumed to respond linearly to the fluctuation modes. 
Indeed, many interesting characteristics of the system are by definition positive, such as, for example, for the eccentricities $\varepsilon_n=\sqrt{(\varepsilon_{n,\mathrm{c}})^2+(\varepsilon_{n,\mathrm{s}})^2}$, or they are (nonlinear) functions of other observables. 
With this caveat in mind, we postulate the existence of observables $O_\alpha$ whose fluctuations depend essentially linearly on the expansion coefficients $\{c_l\}$, i.e.,  such that Eq.~\eqref{Oseries} can be truncated at linear order:
\begin{equation}
O_\alpha(\{c_l\}) \simeq \bar{O}_\alpha + \sum_l L_{\alpha,l}c_l,
\label{eq:O_a_linear}
\end{equation}
where for brevity we denoted by $O_{\alpha}(\{c_l\})$ the value of an observable $O_{\alpha}$ evaluated at the (not necessarily physical) initial state $\bar{\Psi}+\sum_l c_l\Psi_l$ with a given set $\{c_l\}$.

Introducing the joint probability distribution $p(\{c_l\})$ of the expansion coefficients, the probability distribution of $O_\alpha$ reads
\begin{equation}
\label{eq:p_a(O_a)}
p_{\alpha}(O_\alpha) = \int\!\D c\,p(\{c_{l}\})\, \delta\big(O_\alpha-O_\alpha(\{c_{l}\})\big),
\end{equation}
where we introduced the shorthand notation
\begin{equation}
\label{eq:def_int(Dc)}
\int\!\D c \equiv \prod_l\int_{-\infty}^\infty \d c_l.
\end{equation}
Up to this point no assumption has been made in writing Eq.~\eqref{eq:p_a(O_a)}. 

By construction the coefficients $\{c_l\}$ are uncorrelated: the covariance of $c_l$ and $c_{l'}$ vanishes if $l\neq l'$, see Eq.~\eqref{eq:<c_lc_l'>}. 
Let us assume that they are statistically independent variables. 
In that case, the joint distribution factorizes into the product of single-variable probability distributions of the individual coefficients: 
\begin{equation}
p(\{c_l\}) = \prod_l p_l(c_l),
\end{equation}
where each distribution $p_l$ is centered and has a unit variance, Eqs.~\eqref{eq:<c_l>=0} and \eqref{eq:<c_lc_l'>}.
In addition, in this section we assume that each expansion coefficient $c_l$ is a Gaussian random variable:
\begin{equation}
\label{eq:p(c_l)_Gaussian}
p_l(c_l) \simeq \frac{1}{\sqrt{2\pi}} \exp\bigg(\!\!-\!\frac{c_l^2}{2}\bigg).
\end{equation}

In practical calculations, the decomposition in fluctuation modes can only involve a finite number of modes, so that the sum in Eq.~\eqref{eq:O_a_linear} or the product in Eq.~\eqref{eq:def_int(Dc)} are restricted to values $l<l_\textrm{max}$ with some $l_{\max}$. 
Accordingly, for an observable obeying relation~\eqref{eq:O_a_linear} and assuming that the $\{c_l\}$ are Gaussian, the probability distribution~\eqref{eq:p_a(O_a)} of $O_\alpha$ becomes
\begin{equation}
\label{eq:p_a(O_a)_Gaussian}
p_\alpha(O_\alpha) \simeq p^\textrm{G}_{\alpha}(O_\alpha) \equiv 
\frac{1}{\sqrt{2\pi_{} C_{\alpha\alpha}}}\exp\bigg[\!\!-\!\frac{(O_\alpha-\bar{O}_\alpha)^2}{2_{}C_{\alpha\alpha}}\bigg],
\end{equation}
where $C_{\alpha\alpha} \equiv \sum_l L_{\alpha,l}^2$ is the variance of $O_\alpha$, see also Eq.~\eqref{eq:covariance}.

More generally, one can obtain a similar result for the joint probability of $d$ observables $O_{\alpha_1}$, \dots, $O_{\alpha_d}$. 
In the Gaussian approximation, their joint probability distribution reads
\begin{equation}
\label{eq:p_a(O_a)_Gaussian_joint}
p^\textrm{G}_{\vec{\alpha}}(\{O_{\alpha_k}\}) = 
\frac{\exp\!\big[\!-\!\frac{1}{2}(O_{\alpha_i}-\bar{O}_{\alpha_i}) (\Sigma_{\Vec{\alpha}}^{-1})_{ij} (O_{\alpha_j}-\bar{O}_{\alpha_j})\big]}%
{\sqrt{(2\pi)^d \det (\Sigma_{\vec{\alpha}})}},
\end{equation}
with an implicit sum over $i$ and $j$ in the exponent. 
$\Sigma_{\vec{\alpha}}$ is the covariance matrix of the observables, with ($i,j$) entry
\begin{equation}
\label{eq:Sigma_a}
(\Sigma_{\vec{\alpha}})_{ij} \equiv \sum_l L_{\alpha_i,l}L_{\alpha_j,l} \,.
\end{equation}
For brevity, this entry will be denoted $(\Sigma_{\vec{\alpha}})_{ij} \equiv C_{\alpha_i\alpha_j}$.
Note that $l_{\max}$ does not appear explicitly in Eq.~\eqref{eq:p_a(O_a)_Gaussian_joint} [or Eq.~\eqref{eq:p_a(O_a)_Gaussian}], but it is hidden in the definition of the covariance matrix~\eqref{eq:Sigma_a}.

As could be anticipated, if the expansion coefficients $\{c_l\}$ on the basis of fluctuation modes are Gaussian-distributed, and if observables $\{O_{\alpha_k}\}$ depend linearly on these coefficients, then the joint probability distribution~\eqref{eq:p_a(O_a)_Gaussian_joint} of the observables is Gaussian.

For an observable of the form $O_\beta=\sqrt{O_{\alpha_1}^2+O_{\alpha_2}^2}$, the assumption of a Gaussian distribution is clearly not valid. 
But if it is (approximately) valid for $O_{\alpha_1}$ and $O_{\alpha_2}$, one can compute the probability distribution $p_\beta$ from the joint probability distribution $p^\textrm{G}_{\alpha_1,\alpha_2}$, starting from 
\begin{equation}
p_\beta^\textrm{G.f.a.}(O_\beta)=\!\int\!\d O_{\alpha_1}\!\int\!\d O_{\alpha_2}\,
p^\textrm{G}_{\alpha_1,\alpha_2}(O_{\alpha_1},O_{\alpha_2})\,
\delta\Big(O_\beta - \sqrt{O_{\alpha_1}^2+O_{\alpha_2}^2}\Big),\qquad
\end{equation}
where the superscript G.f.a.\ stands for ``Gaussian fluctuation approximation.''
As detailed in Appendix~\ref{app:probability_distributions_Gfa}, by plugging in the specific form of $p^\textrm{G}_{\alpha_1,\alpha_2}$ one can explicitly compute $p_\beta^\textrm{G.f.a.}$ and express it in terms of an infinite series of products of modified Bessel functions, whose arguments depend on the average values $\bar{O}_{\alpha_1}$, $\bar{O}_{\alpha_2}$ and the variances and covariances of $O_{\alpha_1}$ and $O_{\alpha_2}$, see Eq.~\eqref{eq:p_b(O_b)_Gaussian}.
If the average values $\bar{O}_{\alpha_1}$, $\bar{O}_{\alpha_2}$ vanish, the expression strongly simplifies and becomes
\begin{equation}
p_\beta^\textrm{G.f.a.}(O_\beta) = \frac{\Theta(O_\beta)\,O_\beta}{\sqrt{\det(\Sigma_{\alpha_1,\alpha_2})}} \exp 
\left[-\frac{O_\beta^2(C_{\alpha_2 \alpha_2 }+C_{\alpha_1 \alpha_1 })}{4\det(\Sigma_{\alpha_1,\alpha_2})}\right]  I_{0\!}\left(\frac{O_\beta^2\sqrt{(C_{\alpha_2\alpha_2}-C_{\alpha_1\alpha_1})^2+4C_{\alpha_1\alpha_2}}}{4 \det (\Sigma_{\alpha_1 \alpha_2})}\right)\!,
\label{eq:p_b(O_b)_Bessel-Gauss}
\end{equation}
where $\Theta$ denotes the Heaviside step function, since $O_\beta$ is clearly non-negative. 

Analogously, the joint probability distribution of two observables $O_\beta=\sqrt{O_{\alpha_1}^2+O_{\alpha_2}^2}$ and $O_\gamma=\sqrt{O_{\alpha_3}^2+O_{\alpha_4}^2}$ can be expressed as
\begin{equation}
p_{\beta,\gamma}^\textrm{G.f.a.}(O_\beta,O_\gamma) =
\Theta(O_\beta)\,O_\beta\, \Theta(O_\gamma)\,O_\gamma
\!\int_0^{2\pi}\!\d\phi \int_0^{2\pi}\!\d\psi\,
p^\textrm{G}_{\alpha_1,\alpha_2,\alpha_3,\alpha_4}(O_\beta\cos\phi, O_\beta\sin\phi, O_\gamma\cos\psi, O_\gamma\sin\psi),
\label{eq:joint_probab_dist}
\end{equation}
where the joint probability of the four observables $O_{\alpha_j}$ is assumed to be Gaussian. This equation is of the same form as (\ref{eq:pObeta_angular_integral}) for the case of a single observable of this type.
In general, i.e., when $O_{\alpha_3}$ or $O_{\alpha_4}$ have a nonzero covariance with $O_{\alpha_1}$ or $O_{\alpha_2}$, the integrals can no longer be computed as in the case of the single-variable distribution $p_\beta^\textrm{G.f.a.}$. 
Nevertheless, the distribution~\eqref{eq:joint_probab_dist} can still be evaluated numerically.

\subsubsection{Application to anisotropic flow}
\label{subsec:propability_distributions}

\begin{figure*}[!htb]
\includegraphics[width=0.495\linewidth]{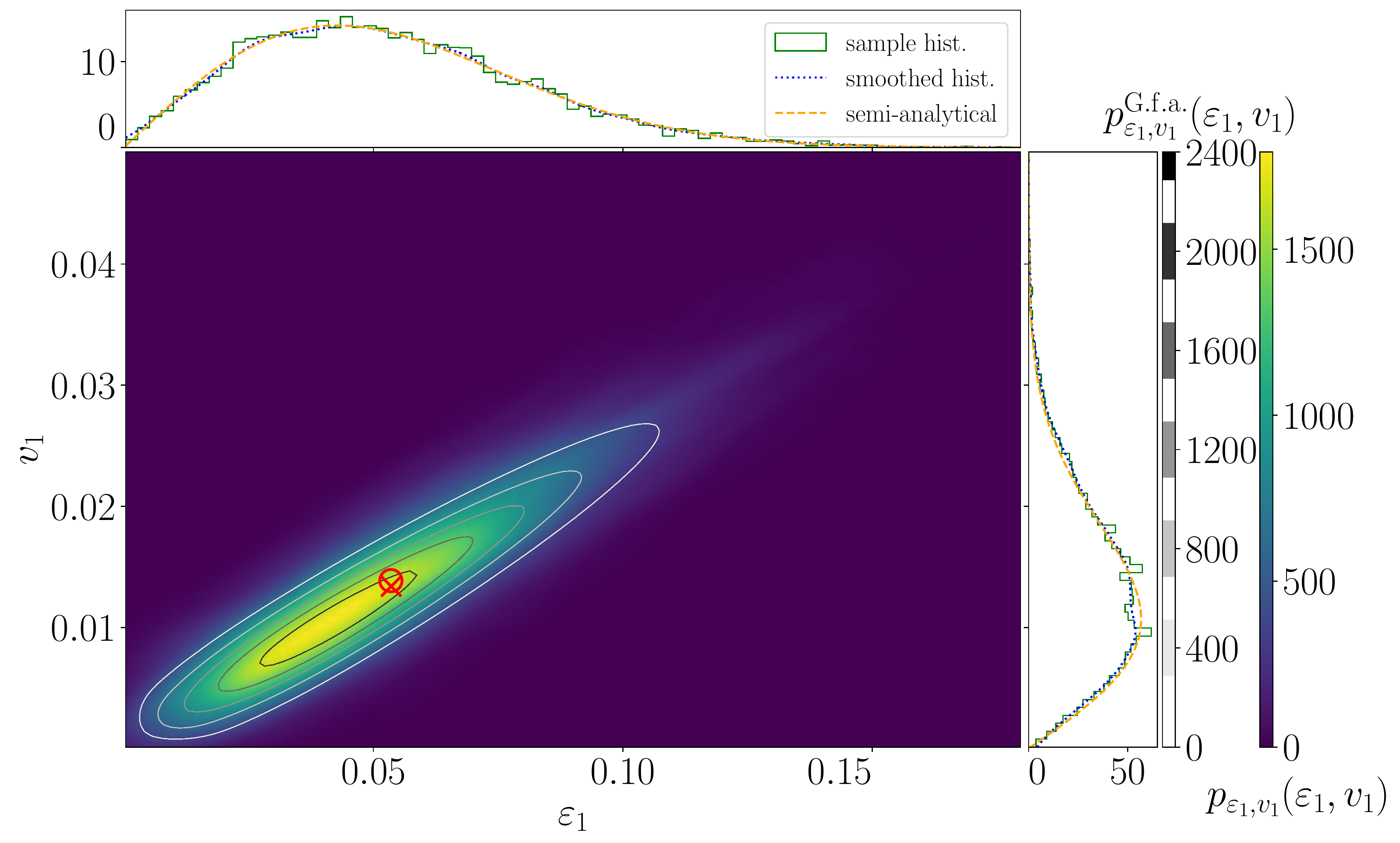}
\includegraphics[width=0.495\linewidth]{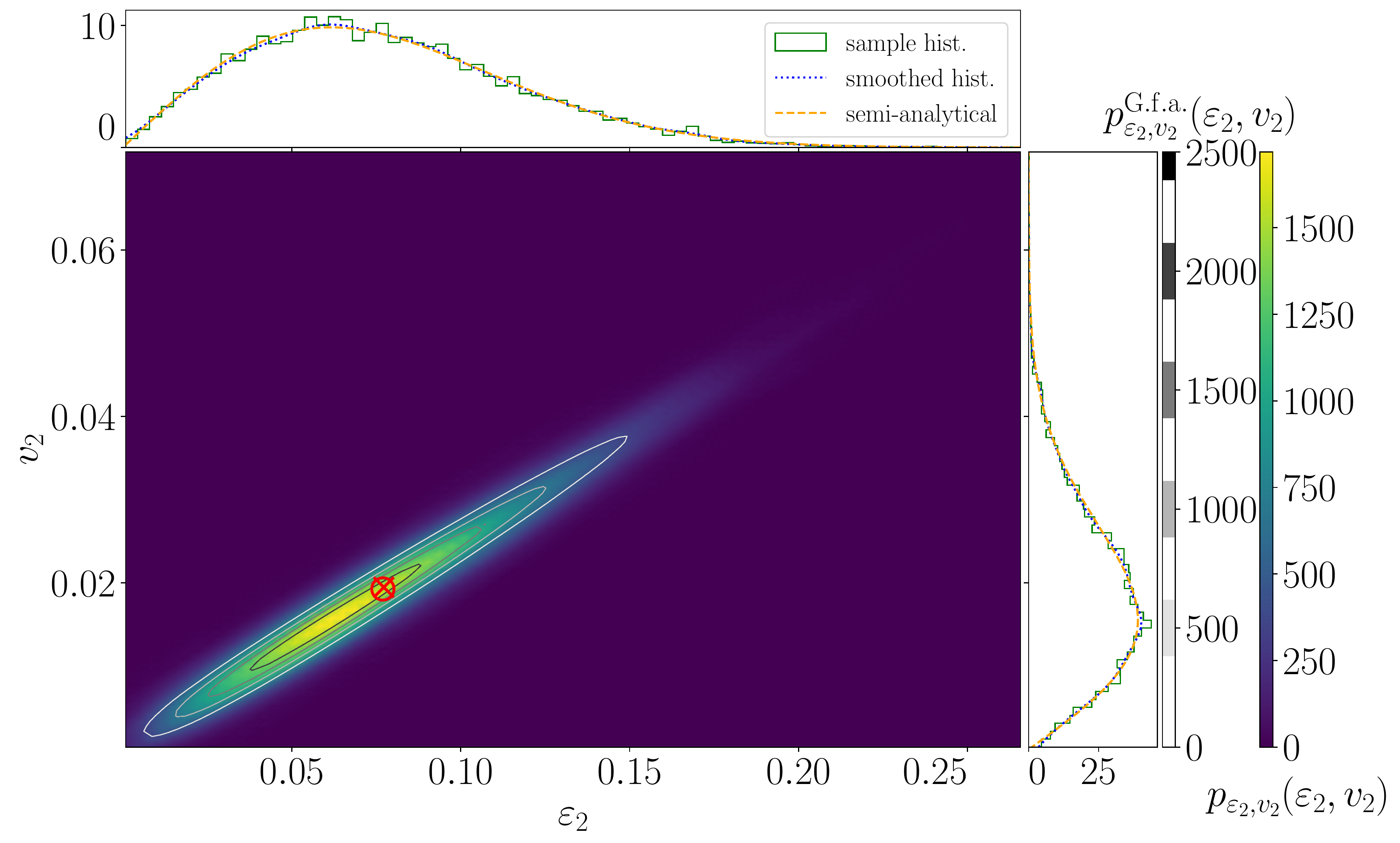}
\includegraphics[width=0.495\linewidth]{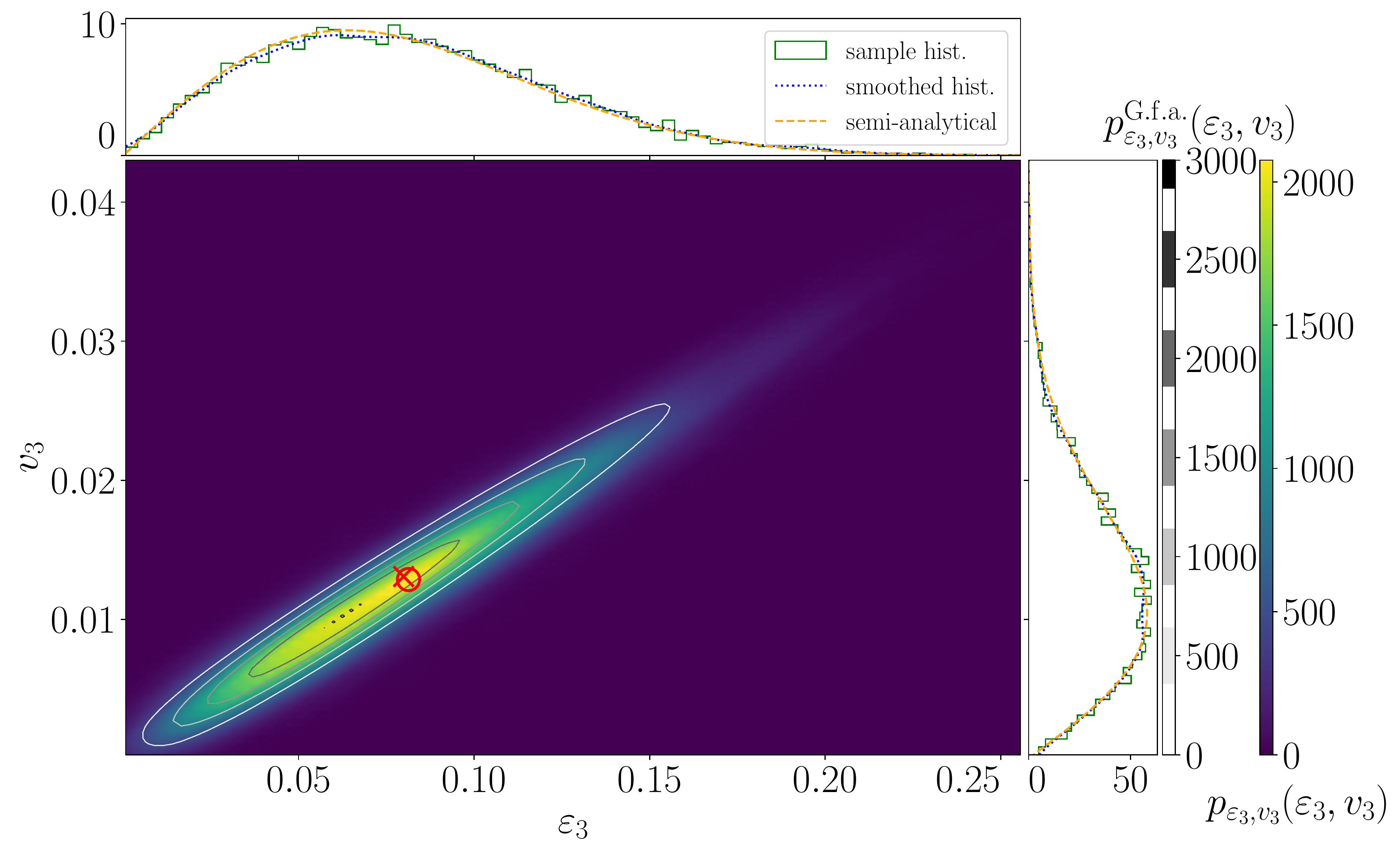}
\includegraphics[width=0.495\linewidth]{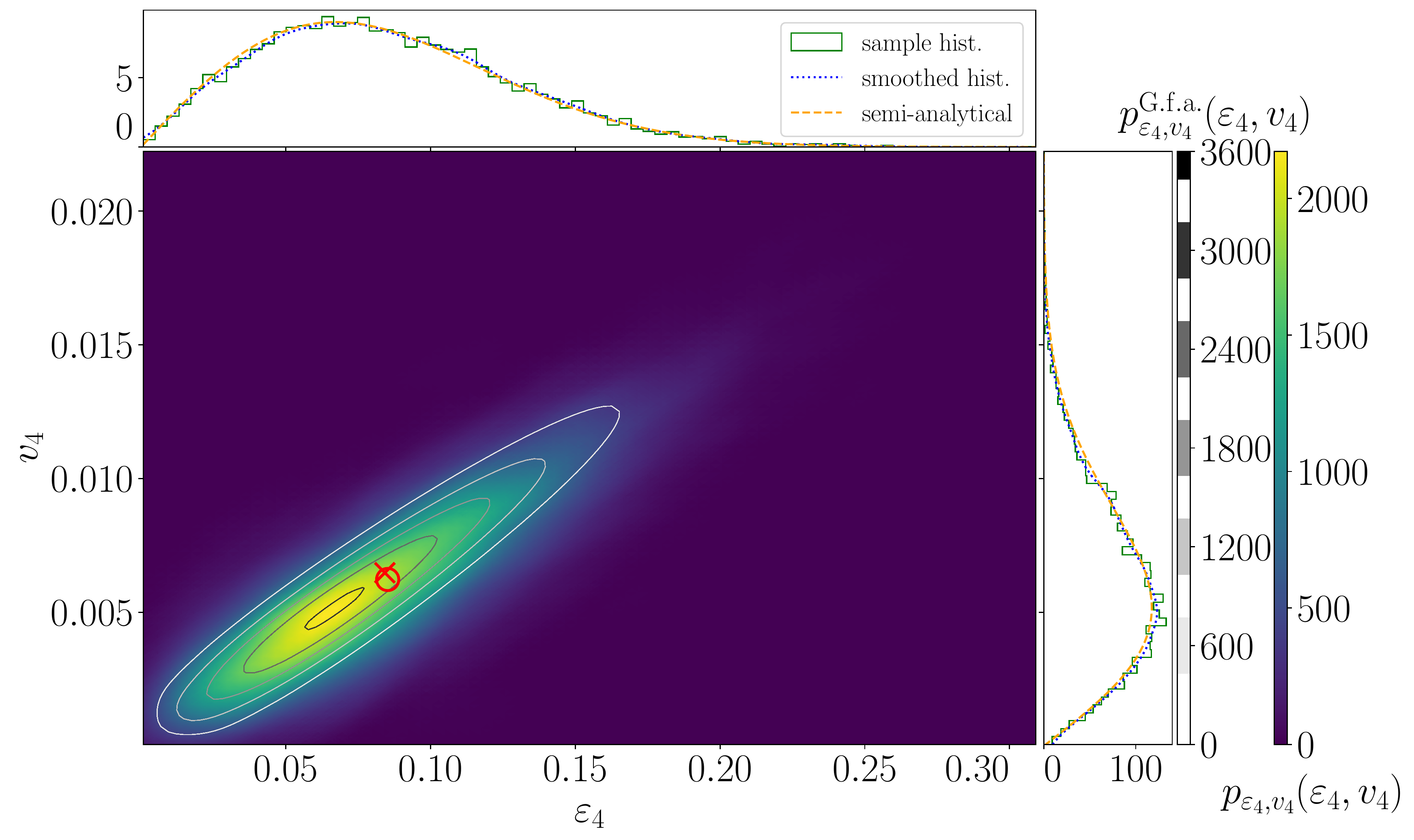}
\includegraphics[width=0.495\linewidth]{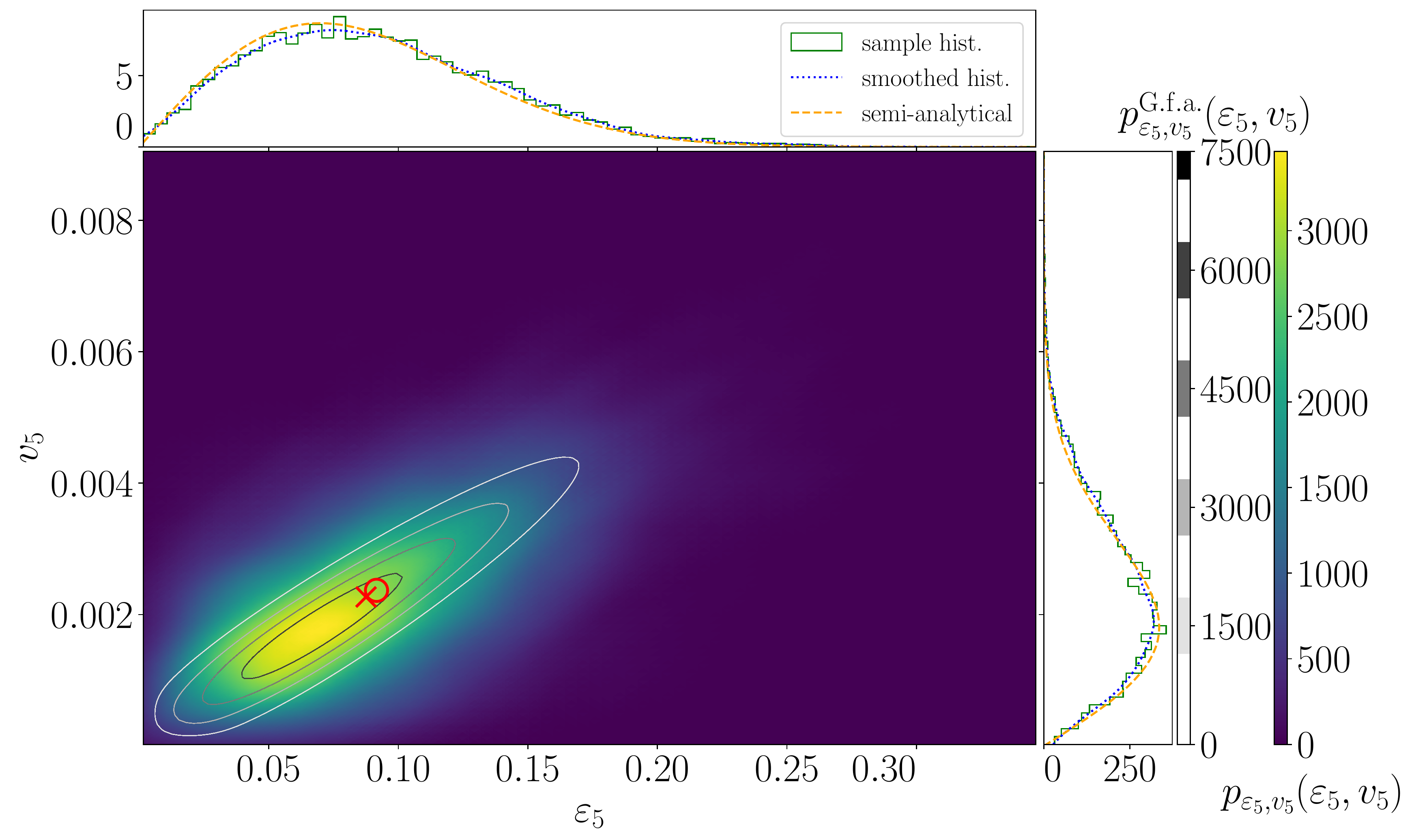}
\includegraphics[width=0.495\linewidth]{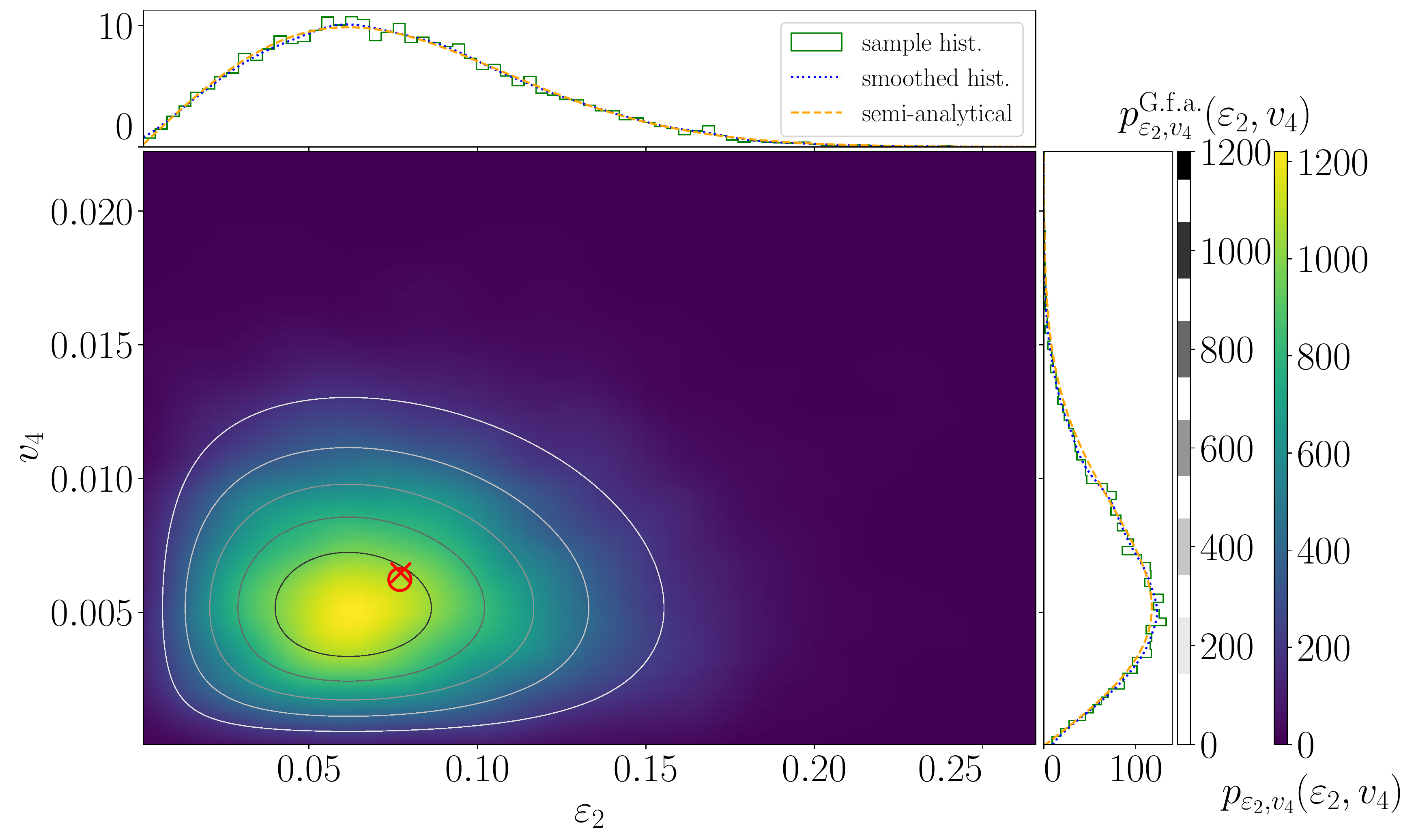}
\vspace{-5mm}
\caption{Joint probability distributions of $\varepsilon_n$ and $v_n$ for $n=1,2,3,4,5$ (top, middle, and bottom left panels) and $\varepsilon_2$, $v_4$ (bottom right) in collisions at $b=0$ within the Glauber model.
Contour lines: semi-analytical values from Eq.~\eqref{eq:joint_probab_dist}. 
Density plots: values from a sample of 8192 random events. 
Red cross resp.\ circle: mean value of the observables from the Gaussian-fluctuation calculations resp.\ the event sample. 
Top resp.\ right of each panel: probability distribution of $\varepsilon_n$ resp.\ $v_n$ (histogram: binned sample values; dotted: smoothing of the histogram with a Gaussian kernel density estimator; dashed line: semi-analytical values).}
\label{fig:2d_prob_dist_Glauber_b0}
\end{figure*}
\begin{figure*}[!htb]
\includegraphics[width=0.495\linewidth]{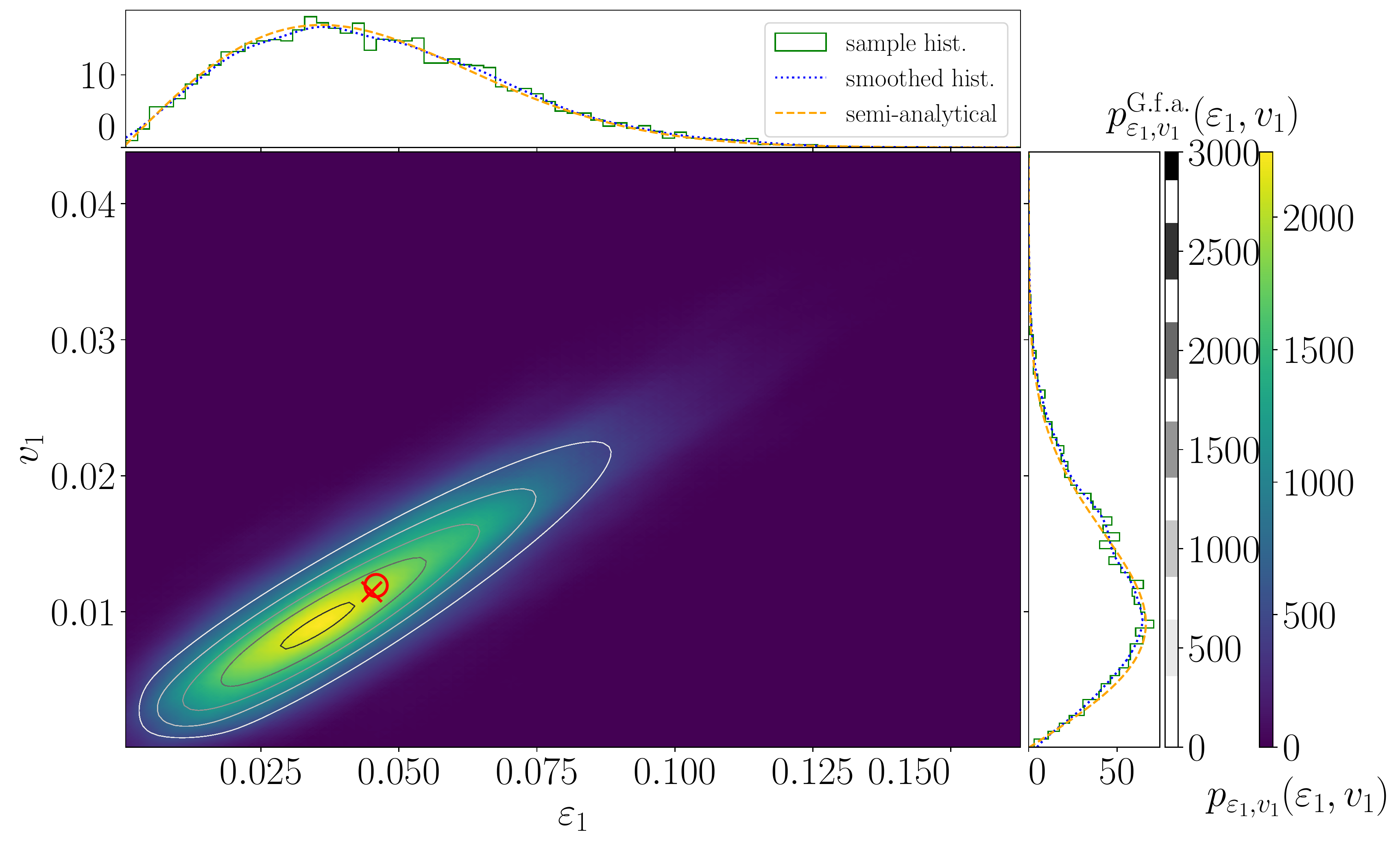}
\includegraphics[width=0.495\linewidth]{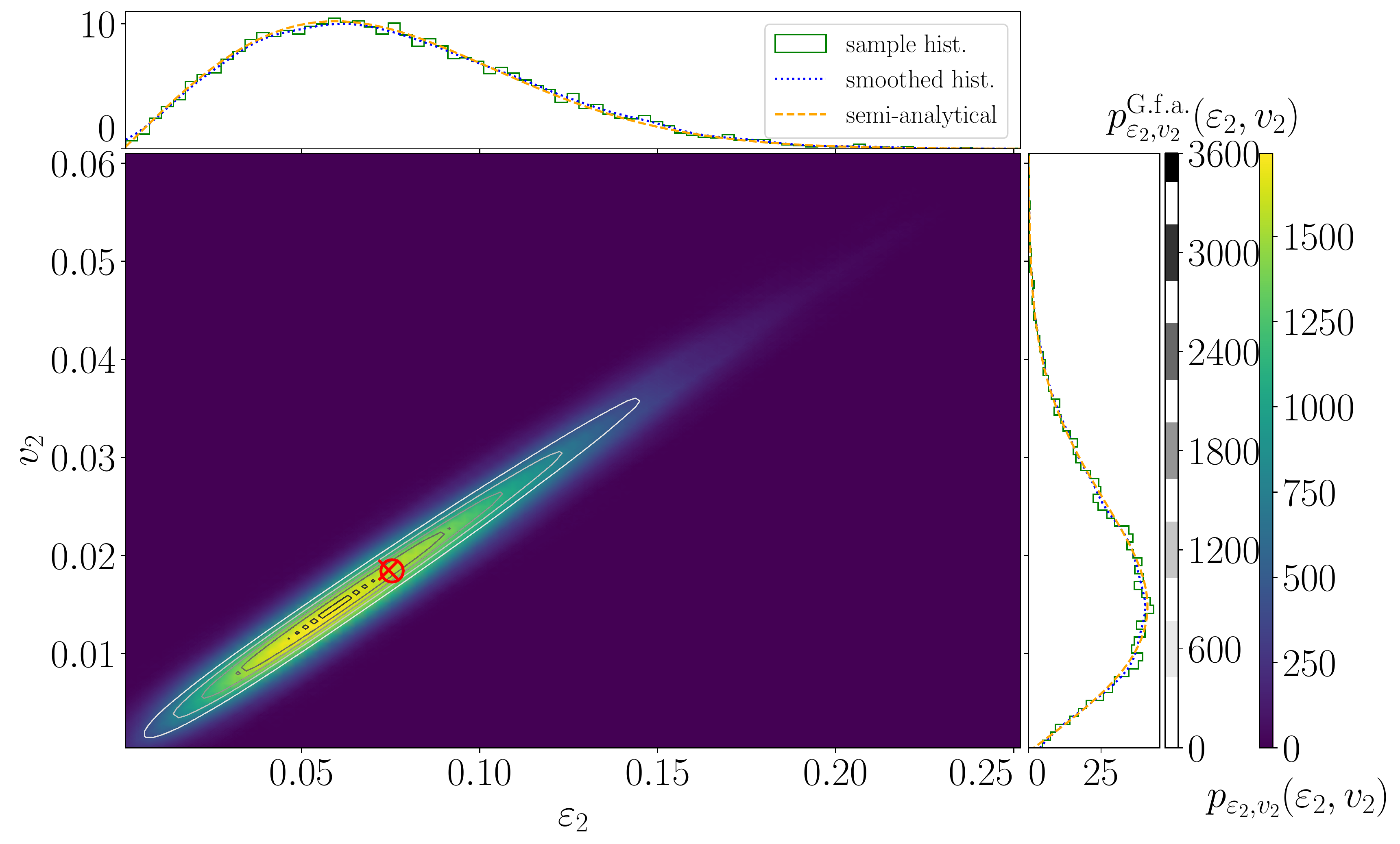}
\includegraphics[width=0.495\linewidth]{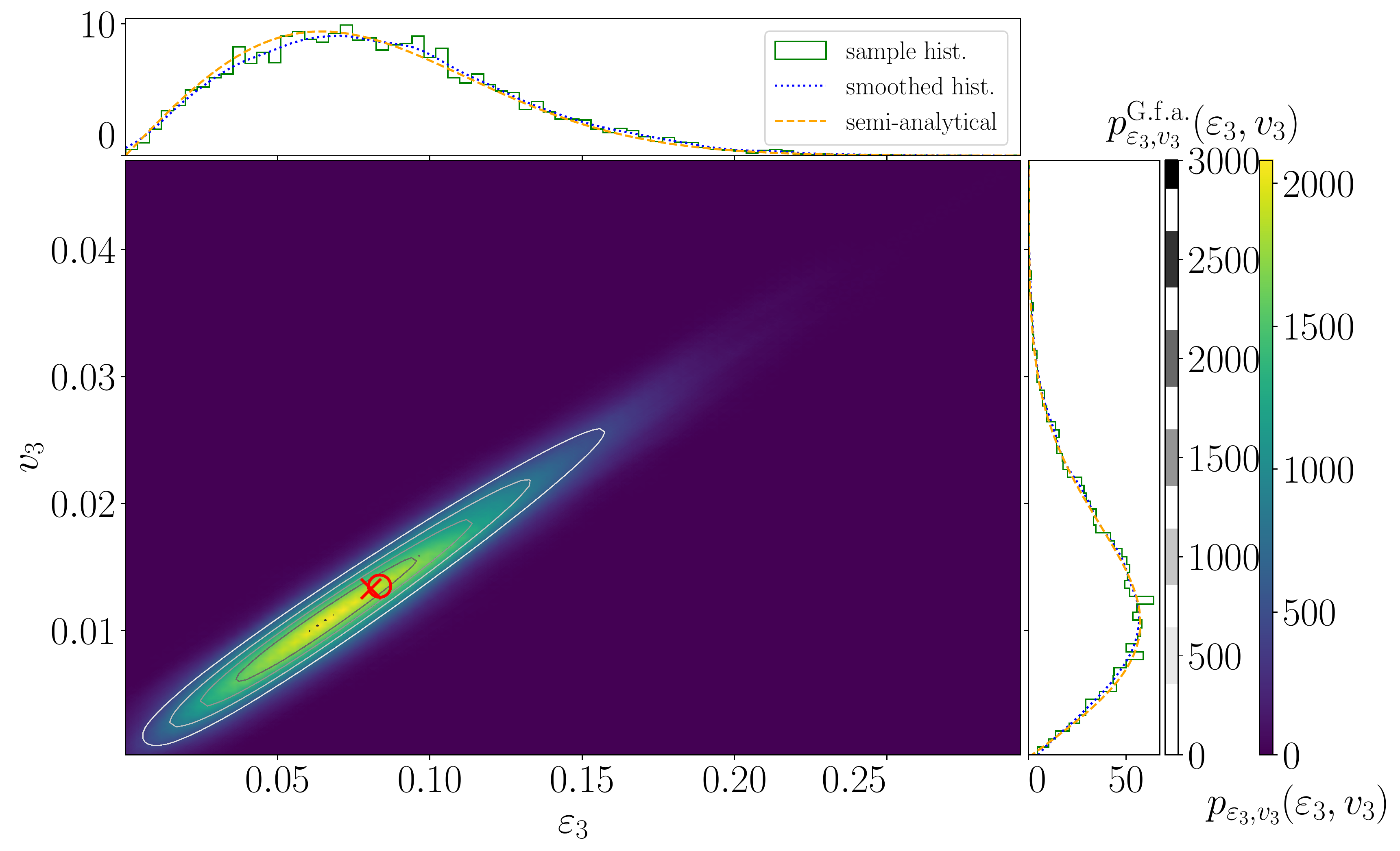}
\includegraphics[width=0.495\linewidth]{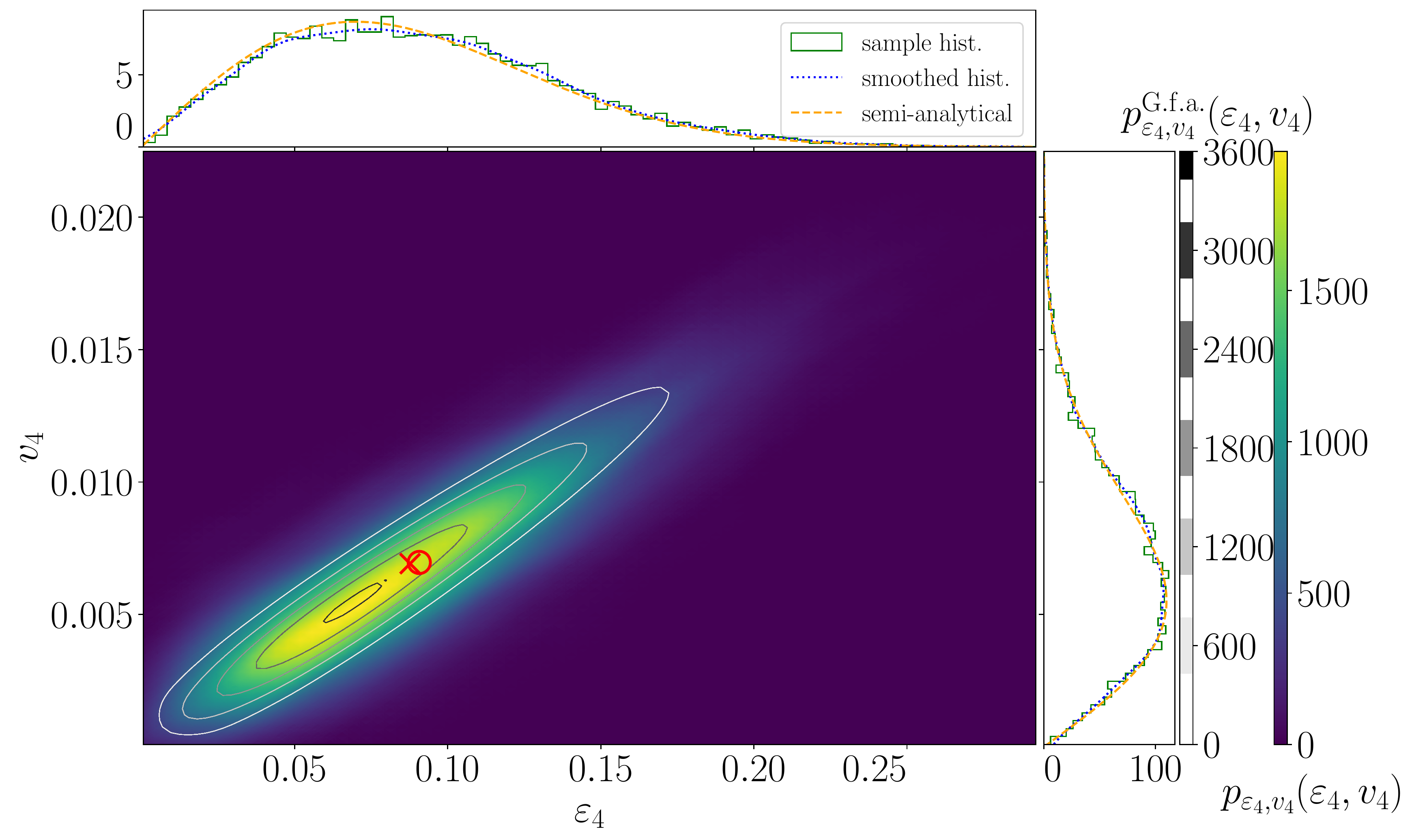}
\includegraphics[width=0.495\linewidth]{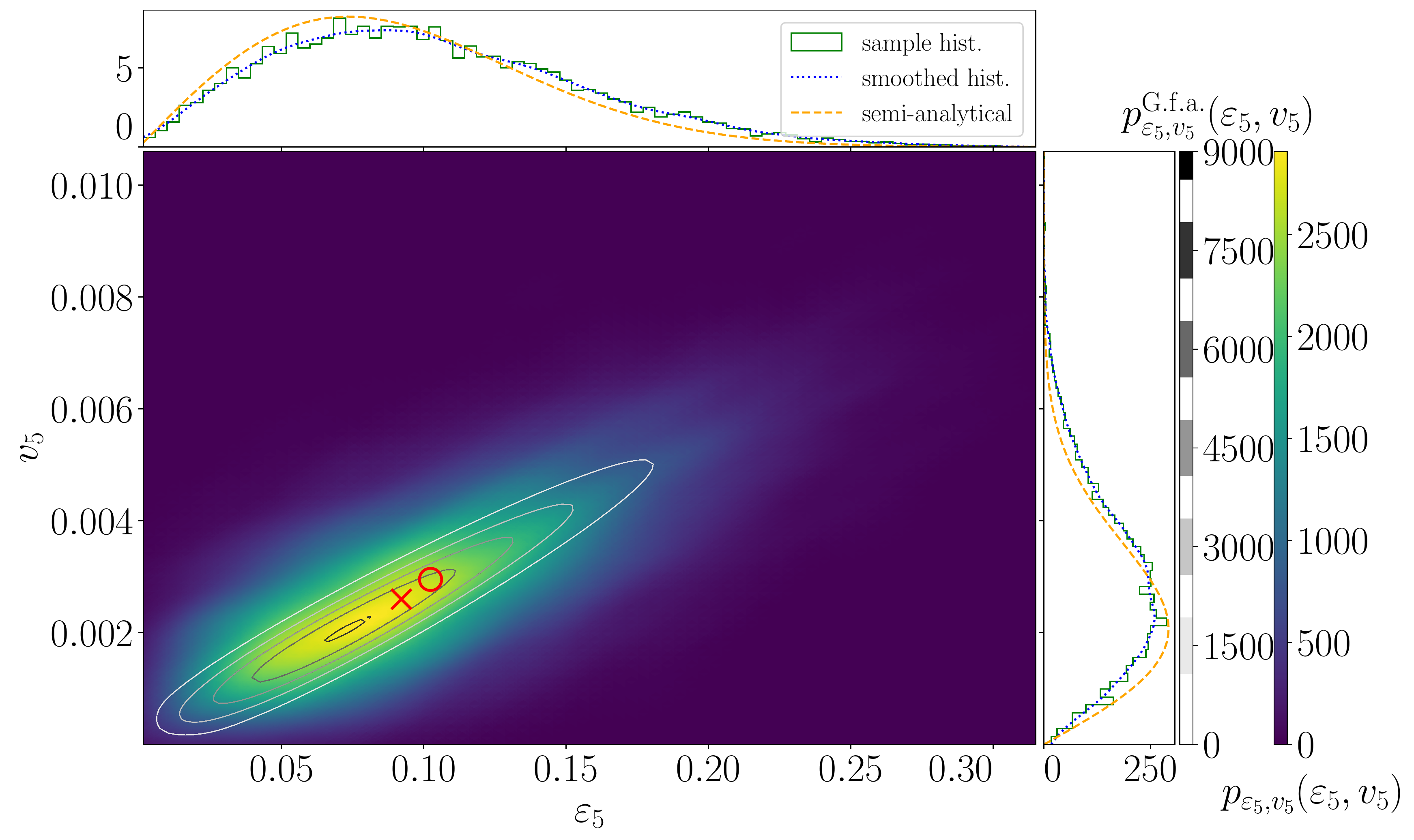}
\includegraphics[width=0.495\linewidth]{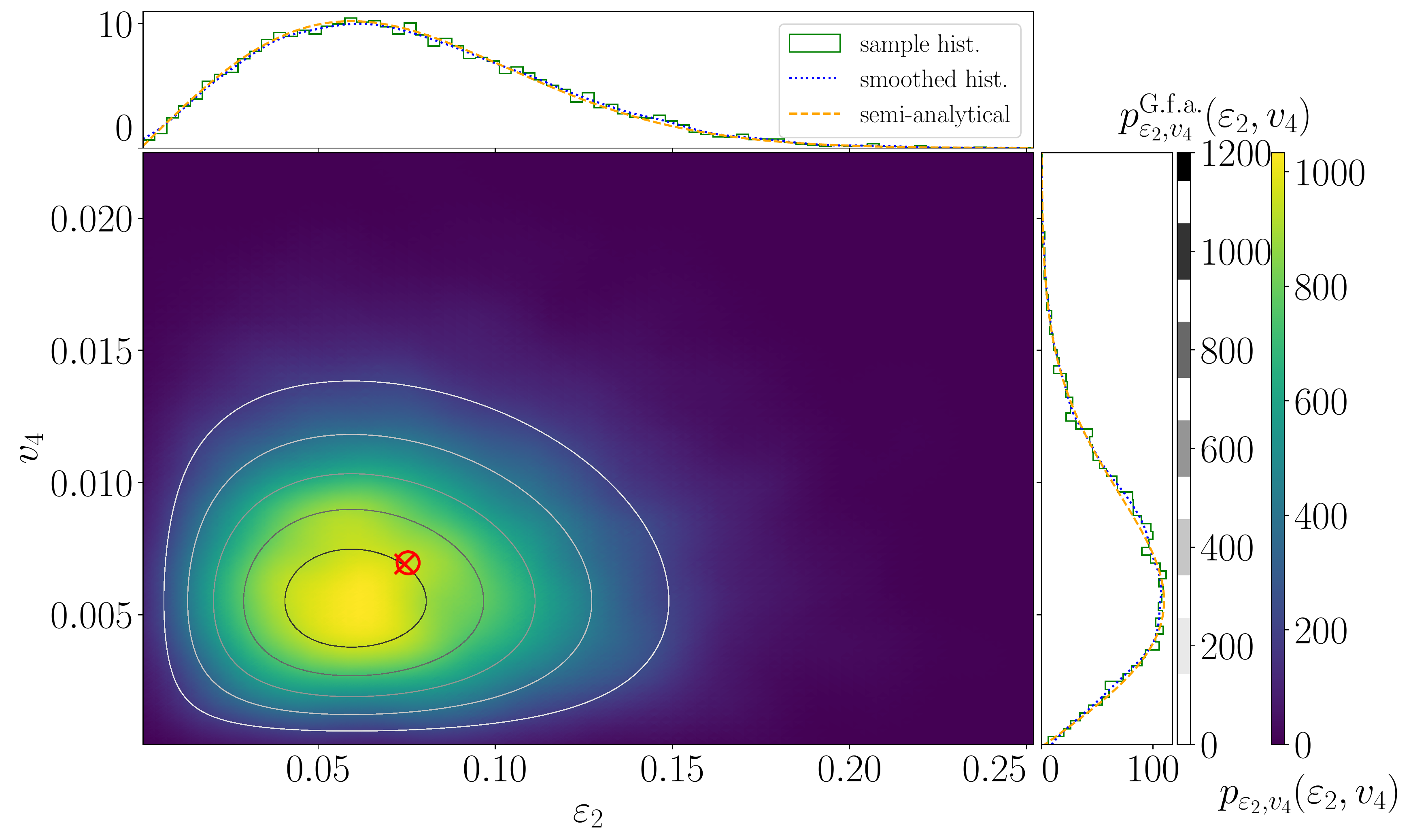}
\vspace{-5mm}
\caption{Same as Fig.~\ref{fig:2d_prob_dist_Glauber_b0} for collisions at $b=0$ within the Saturation model.}
\label{fig:2d_prob_dist_Saturation_b0}
\end{figure*}

We can now apply the formalism of the previous subsection to a few observables, and compute probability distributions of their fluctuations, using values of the (co)variances of observables determined with 256 fluctuation modes, i.e., as in Sec.~\ref{subsec:variances_covariances}.
These semi-analytical probability distributions are compared with those from a sample of 8192 dynamically evolved events (for each initial-state model and impact-parameter value). 

For the Pb-Pb collisions we consider in this paper, at a fixed impact parameter ranging up to 9~fm, the fluctuations of $\varepsilon_{n,\rm c}$ and $\varepsilon_{n,\rm s}$ can to a good approximation be assumed to be a two-dimensional Gaussian~\cite{Voloshin:2007pc,PHOBOS:2007vdf,Qiu:2011iv}, although this does not satisfy the actual constraint $\varepsilon_n\equiv\sqrt{(\varepsilon_{n,\mathrm{c}})^2+(\varepsilon_{n,\mathrm{s}})^2}\leq 1$.
At $b=0$, the values of the eccentricities in the average state $\bar{\Psi}$ vanish, $\bar{\varepsilon}_{n,\rm c/s} = 0$, so that the probability distribution of the modulus $\varepsilon_n$ is given by Eq.~\eqref{eq:p_b(O_b)_Bessel-Gauss}, namely a Bessel--Gaussian distribution~\cite{Voloshin:2007pc}. 

In turn, when all eccentricities are small, i.e., in our case in collisions at $b=0$, each flow coefficient $v_{n,\rm c/s}$ is approximately proportional to the corresponding eccentricity $\varepsilon_{n,\rm c/s}$
\begin{equation}
v_{n,\rm c/s} \propto \varepsilon_{n,\rm c/s},
\label{eq:vn_epsn_linear}
\end{equation}
which for $n=2$ and 3 also holds pretty well at $b=9$~fm.
For such a linear dynamical response, the fluctuations of $v_{n,\rm c/s}$ follow those of $\varepsilon_{n,\rm c/s}$, i.e., they are approximately Gaussian. 
Accordingly, the joint probability distribution of the four observables $\varepsilon_{n,\rm c}$, $\varepsilon_{n,\rm s}$, $v_{n,\rm c}$, $v_{n,\rm s}$ is itself Gaussian --- with a few large covariances ---, and one can estimate the joint probability distribution of $\varepsilon_n$ and $v_n\equiv\sqrt{(v_{n,\mathrm{c}})^2+(v_{n,\mathrm{s}})^2}$ using Eq.~\eqref{eq:joint_probab_dist}.

\begin{figure*}[!htb]
\includegraphics[width=0.495\linewidth]{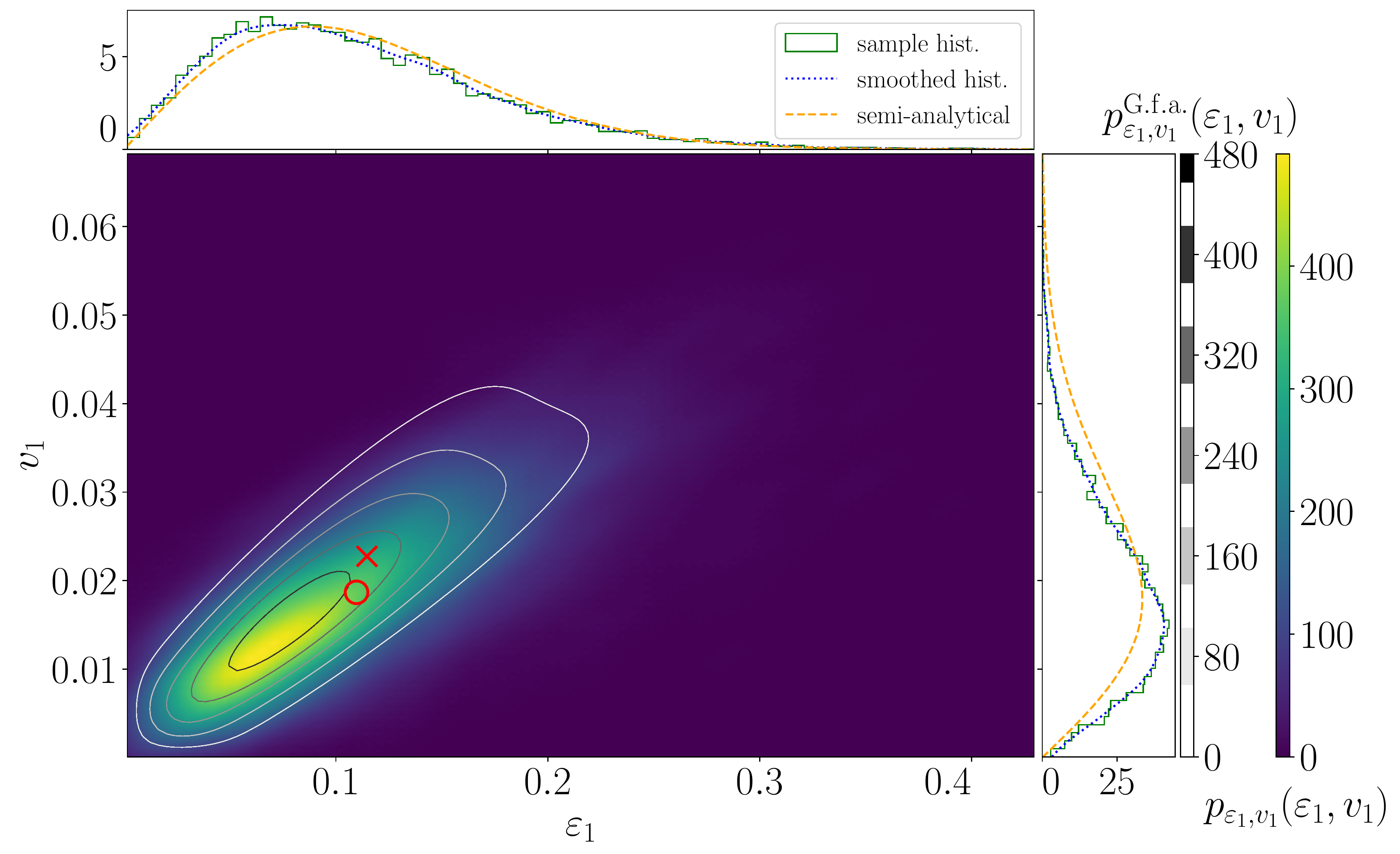}
\includegraphics[width=0.495\linewidth]{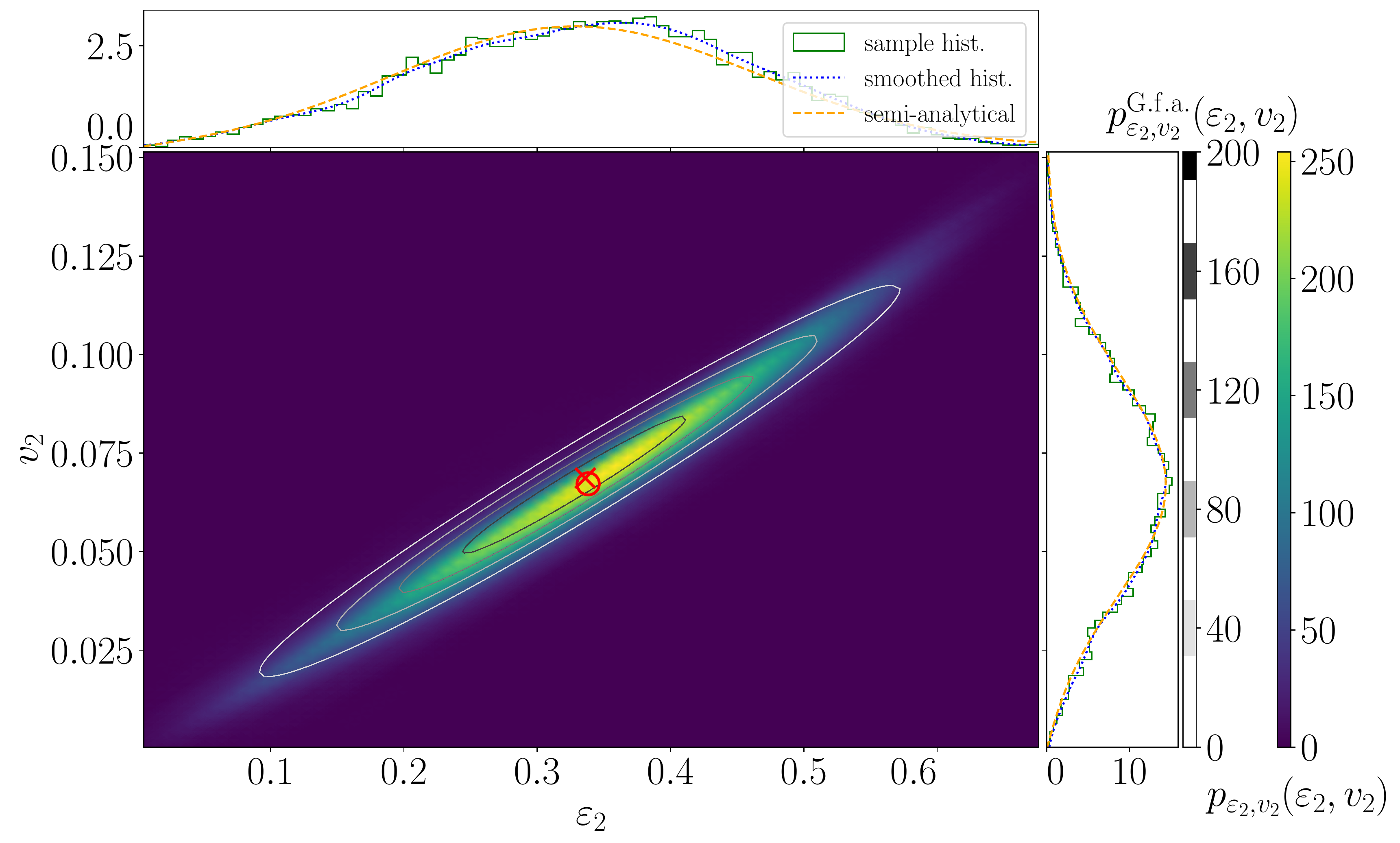}
\includegraphics[width=0.495\linewidth]{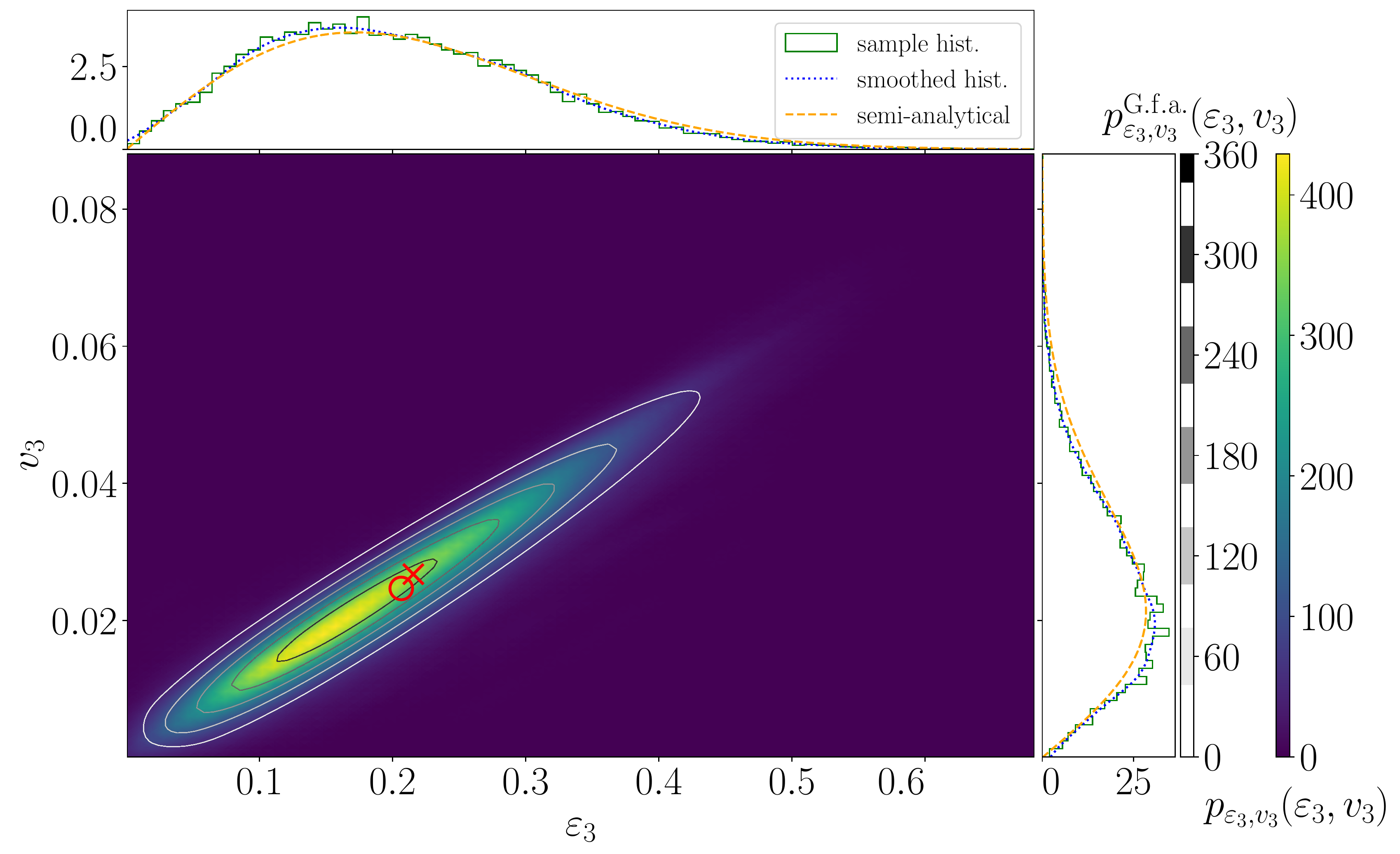}
\includegraphics[width=0.495\linewidth]{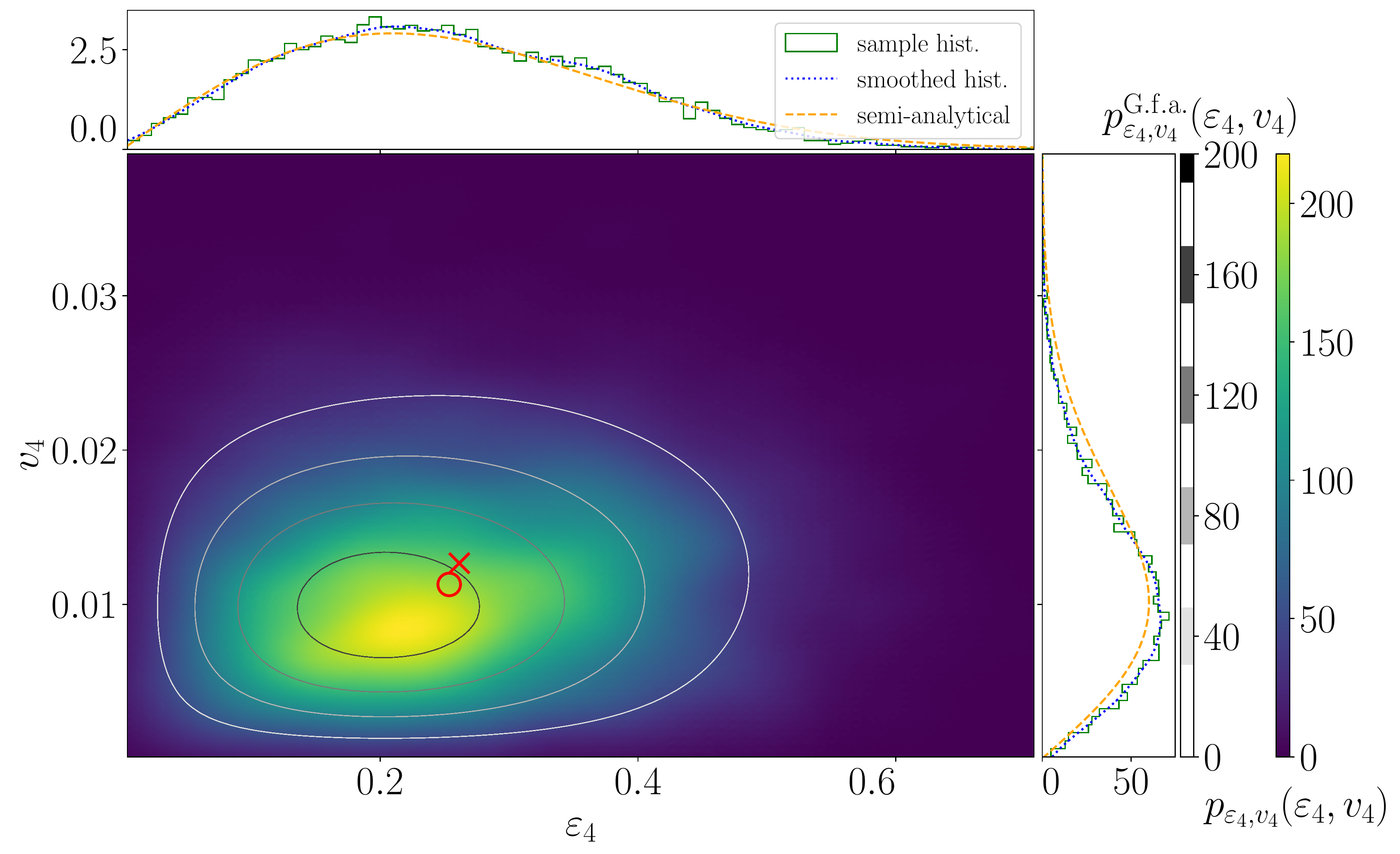}
\includegraphics[width=0.495\linewidth]{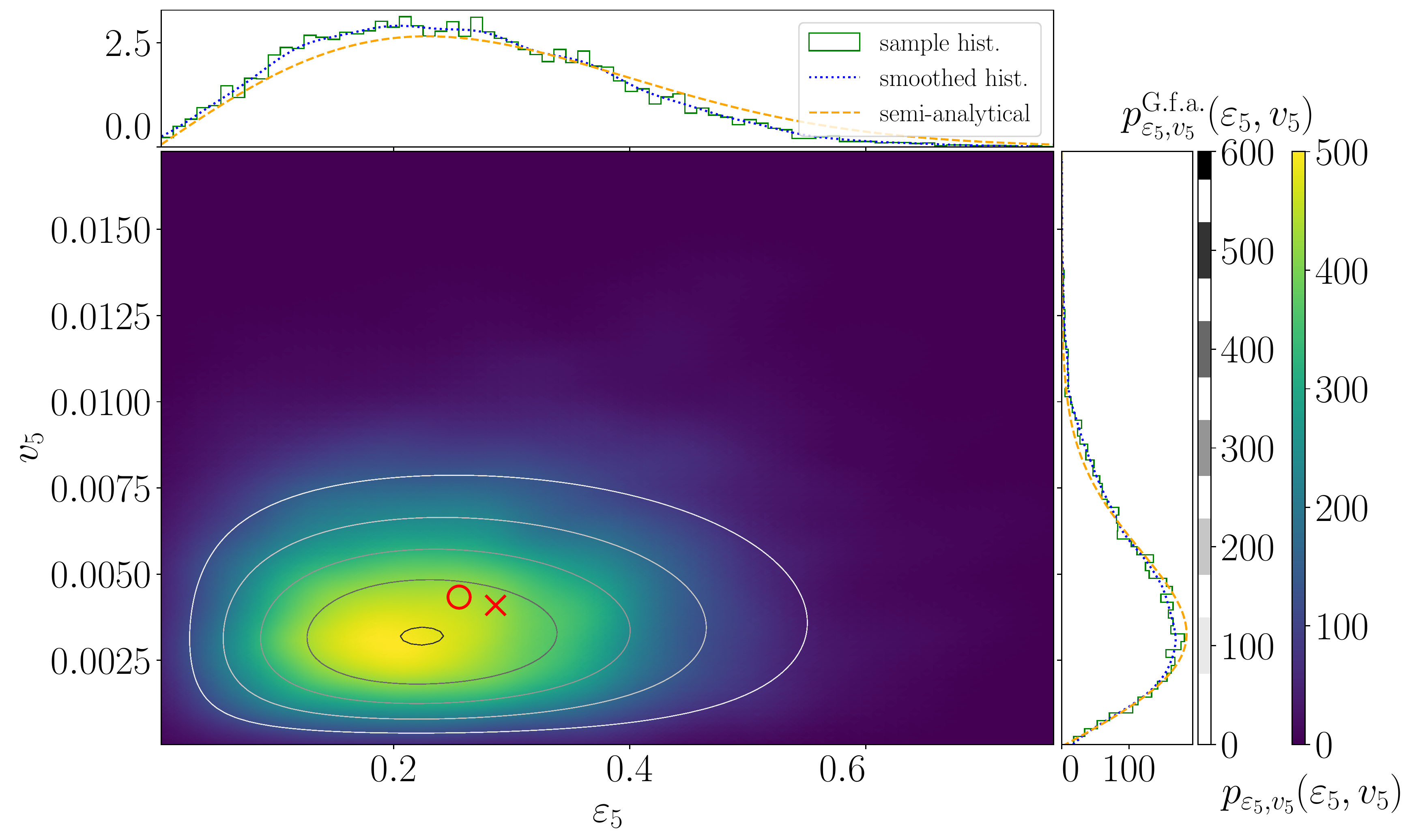}
\includegraphics[width=0.495\linewidth]{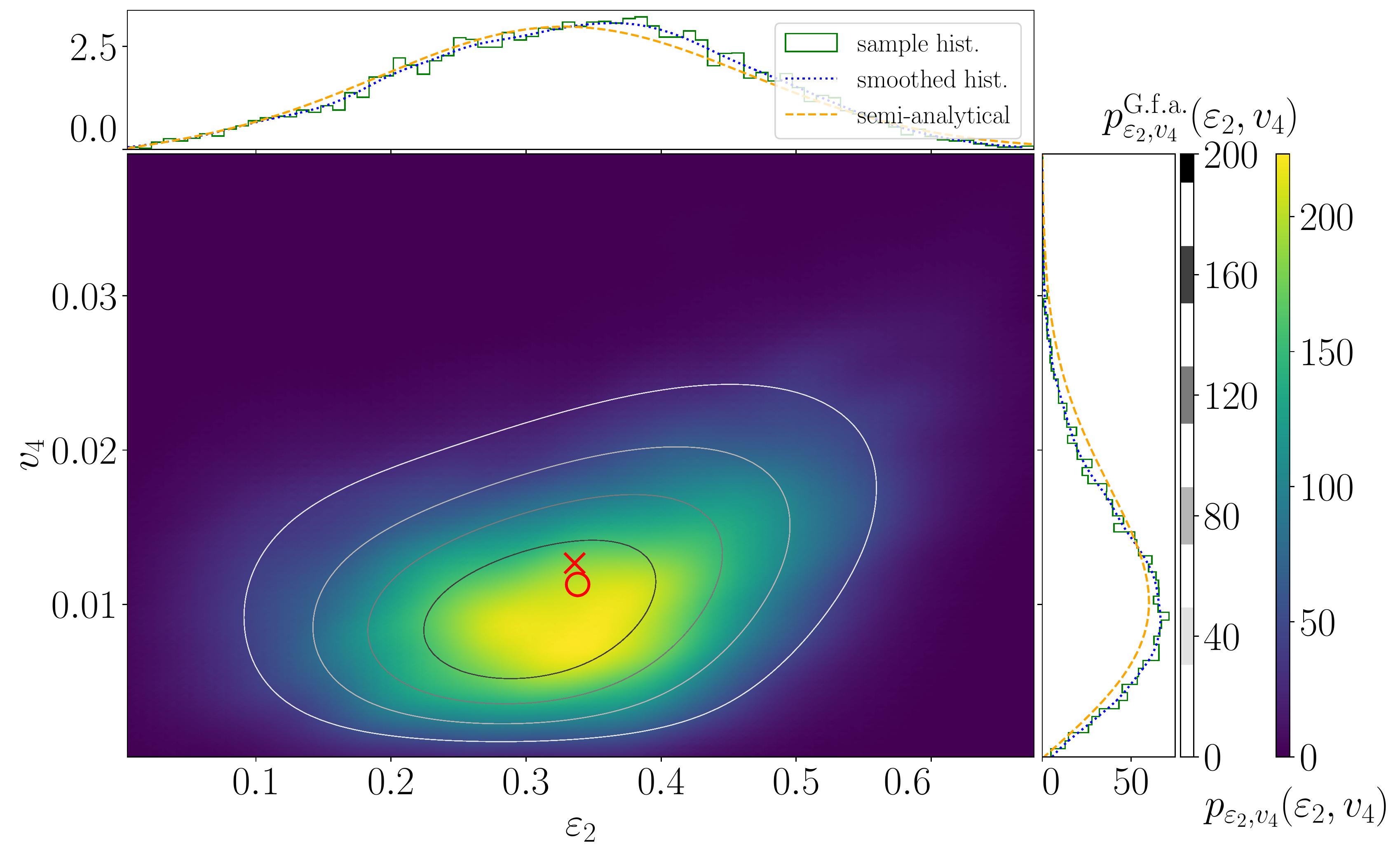}
\vspace{-5mm}
\caption{Same as Fig.~\ref{fig:2d_prob_dist_Glauber_b0} for collisions at $b=9$~fm within the Glauber model.}
\label{fig:2d_prob_dist_Glauber_b9}
\end{figure*}

\begin{figure*}[!htb]
\includegraphics[width=0.495\linewidth]{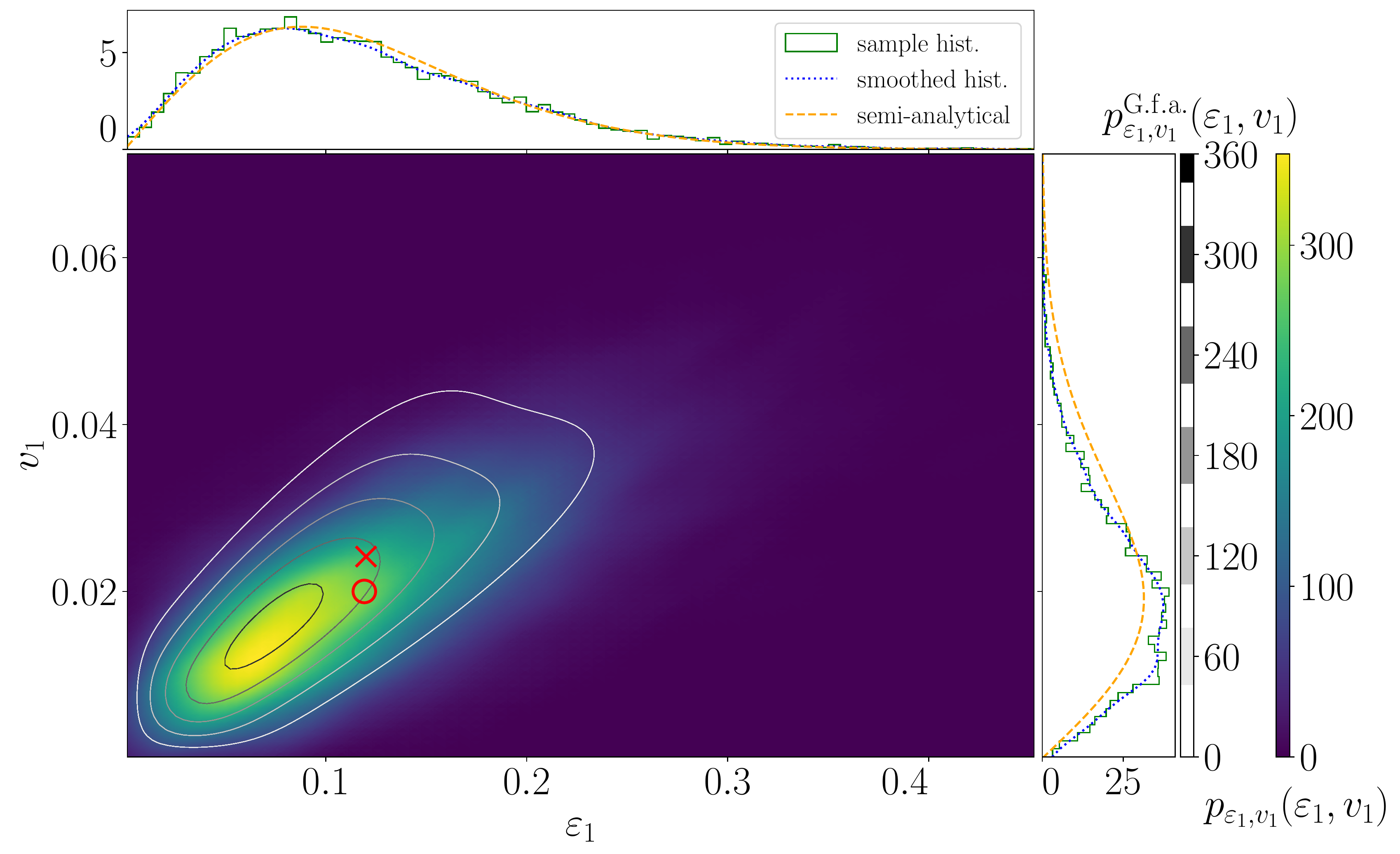}
\includegraphics[width=0.495\linewidth]{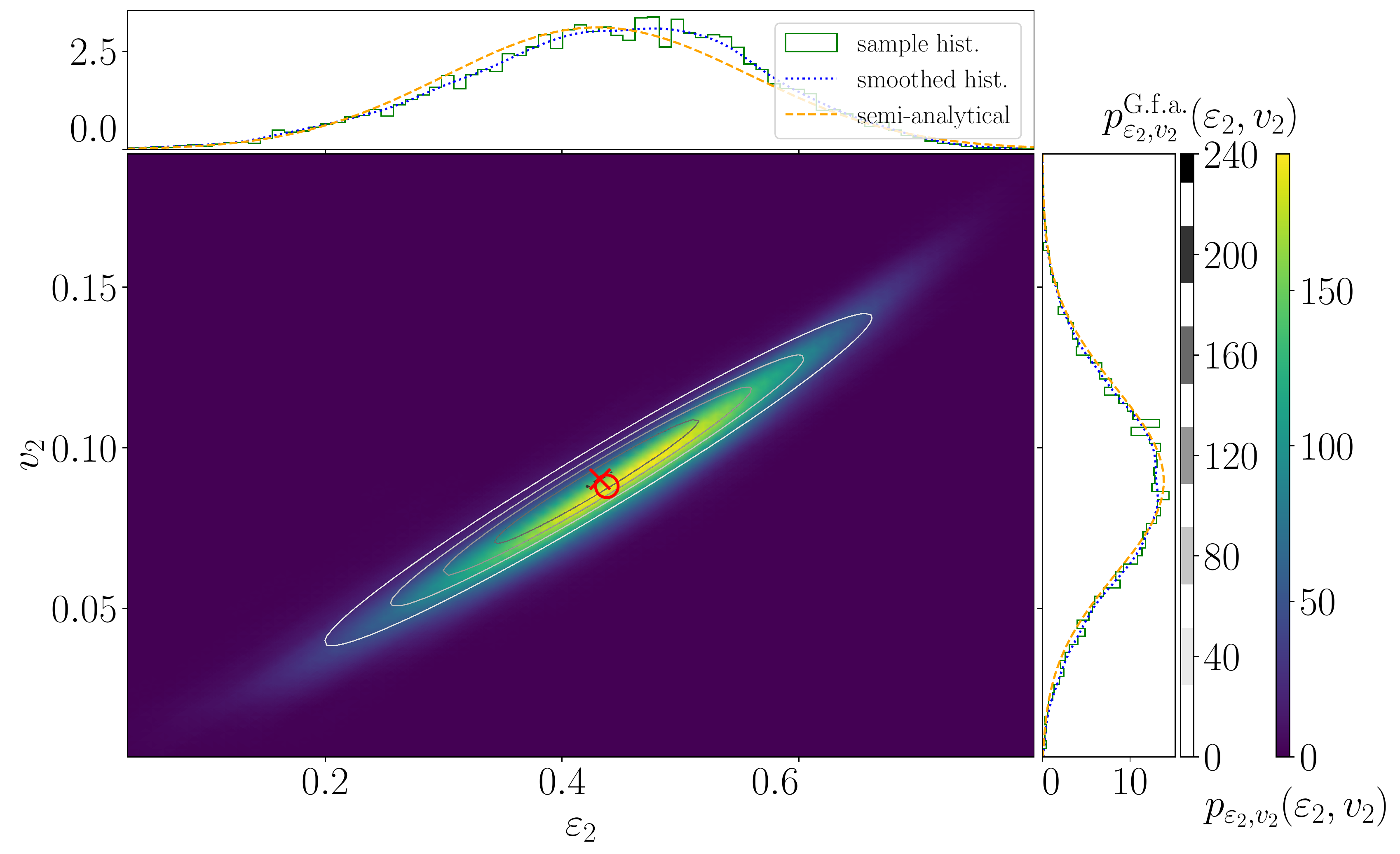}
\includegraphics[width=0.495\linewidth]{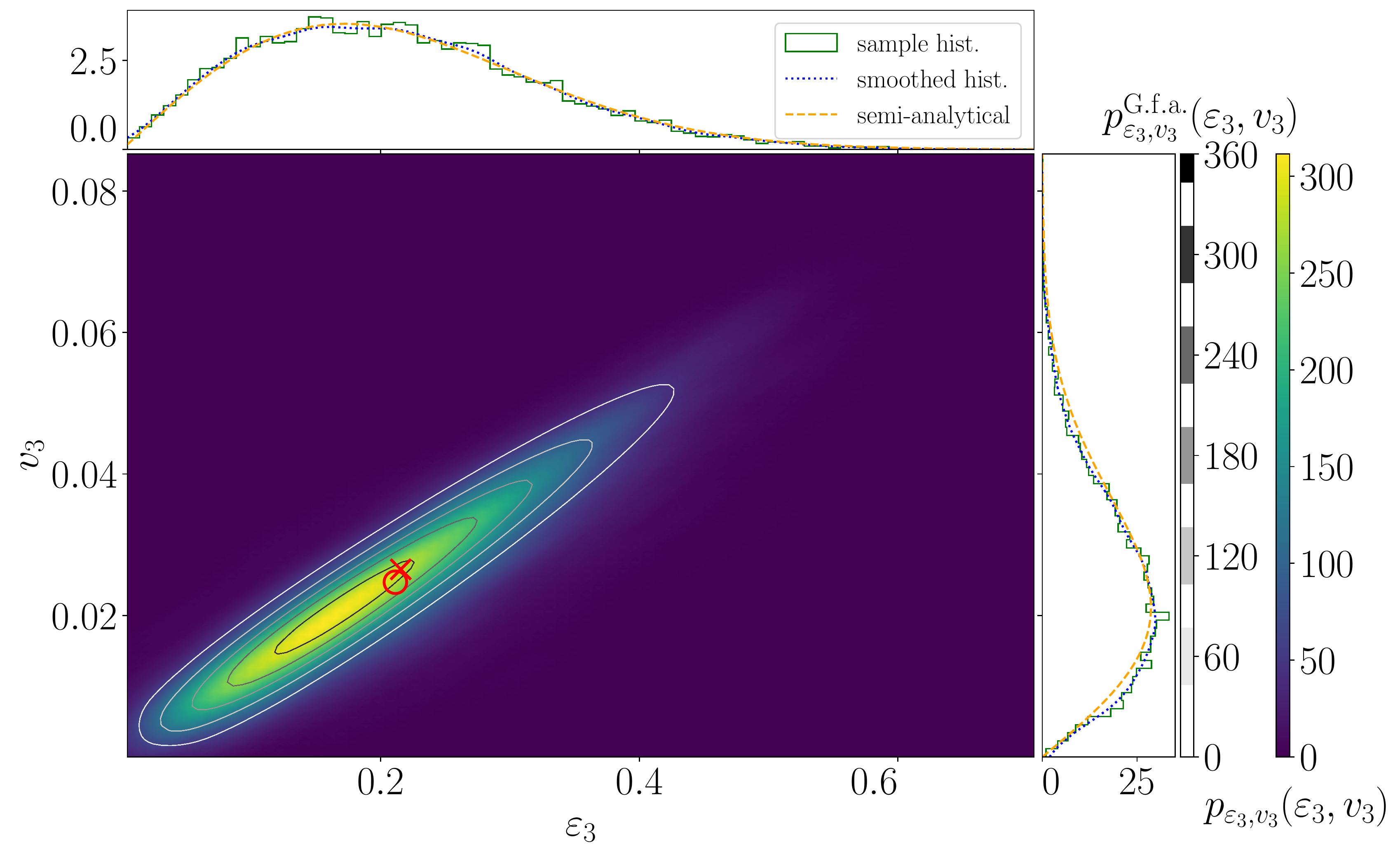}
\includegraphics[width=0.495\linewidth]{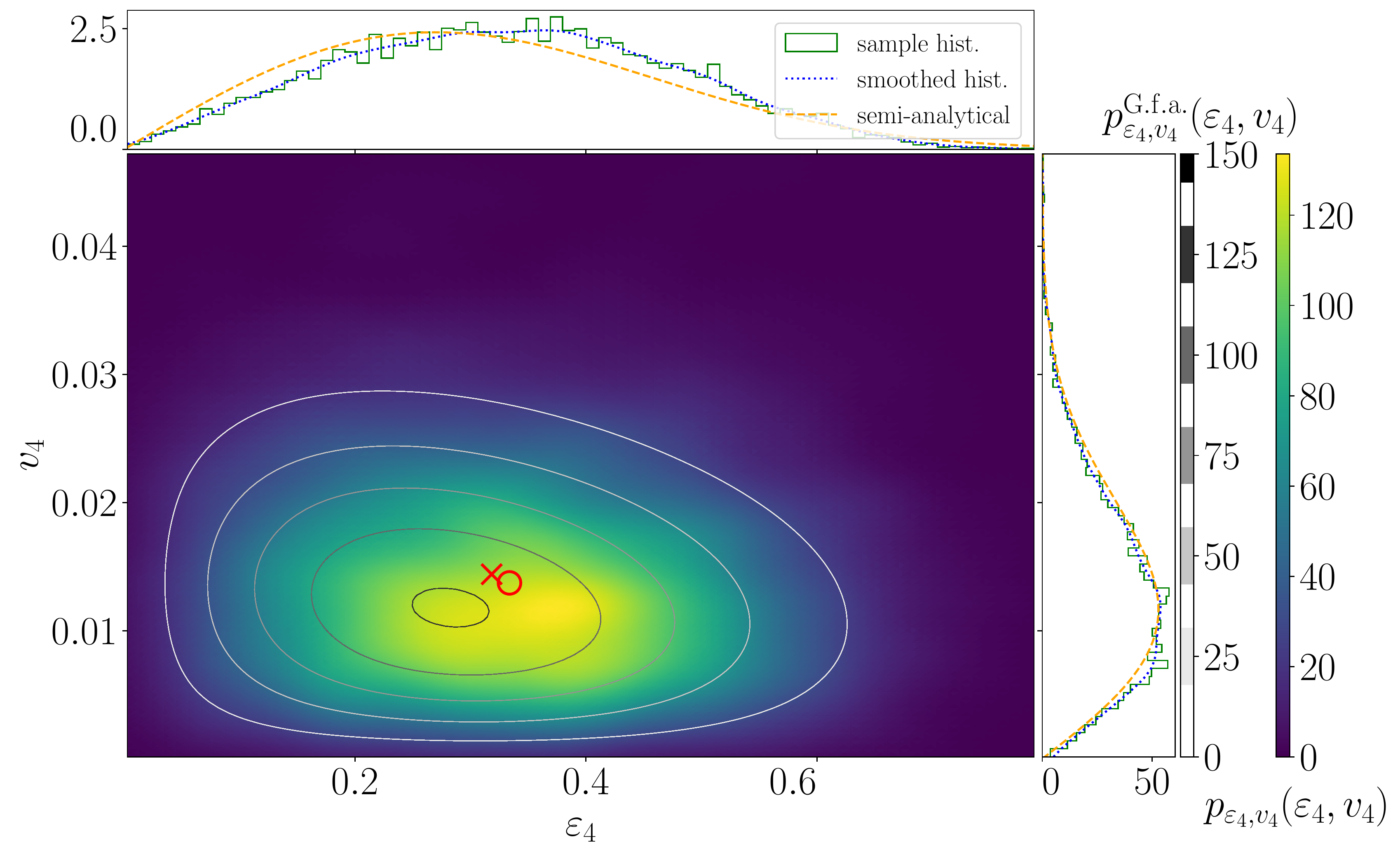}
\includegraphics[width=0.495\linewidth]{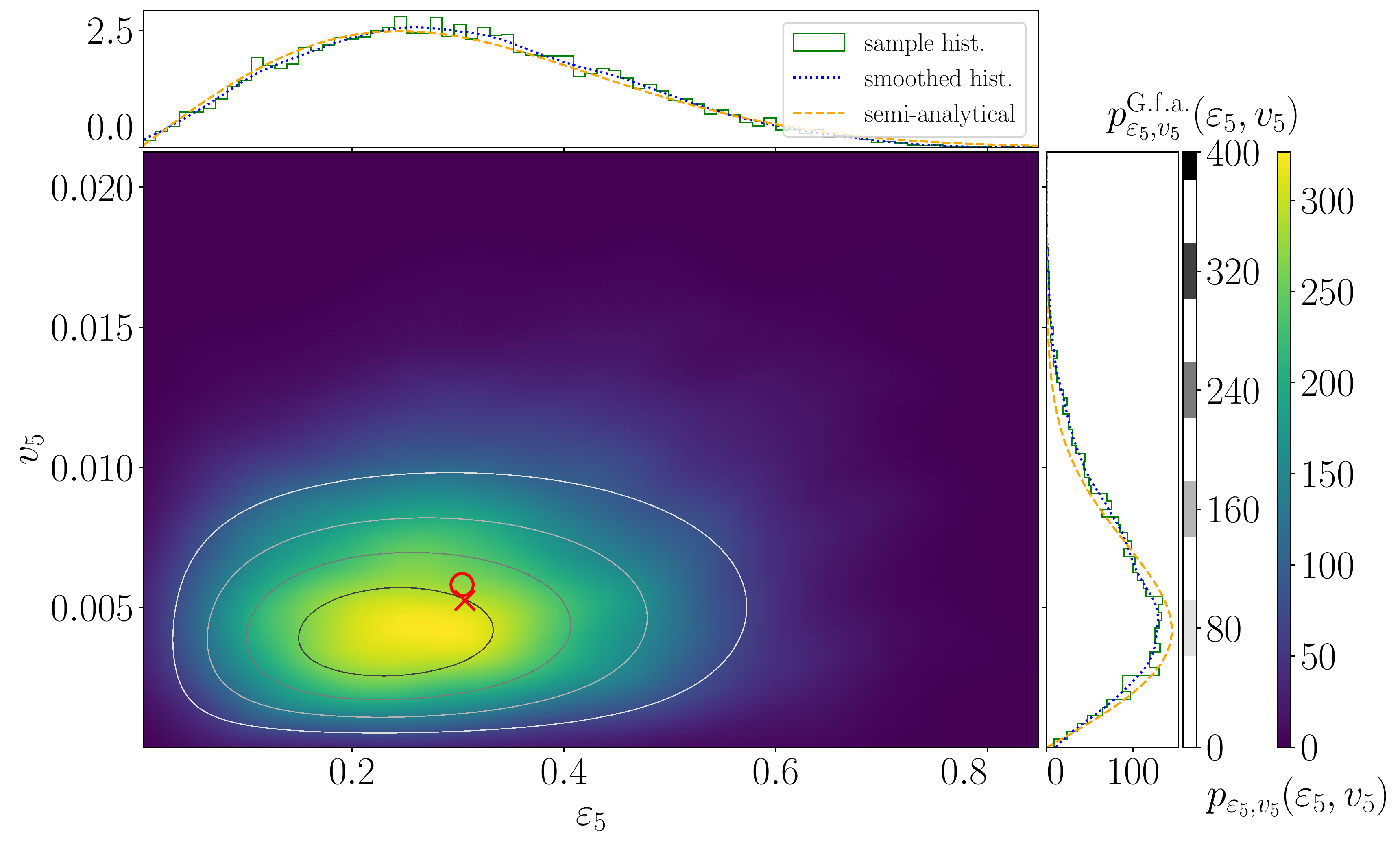}
\includegraphics[width=0.495\linewidth]{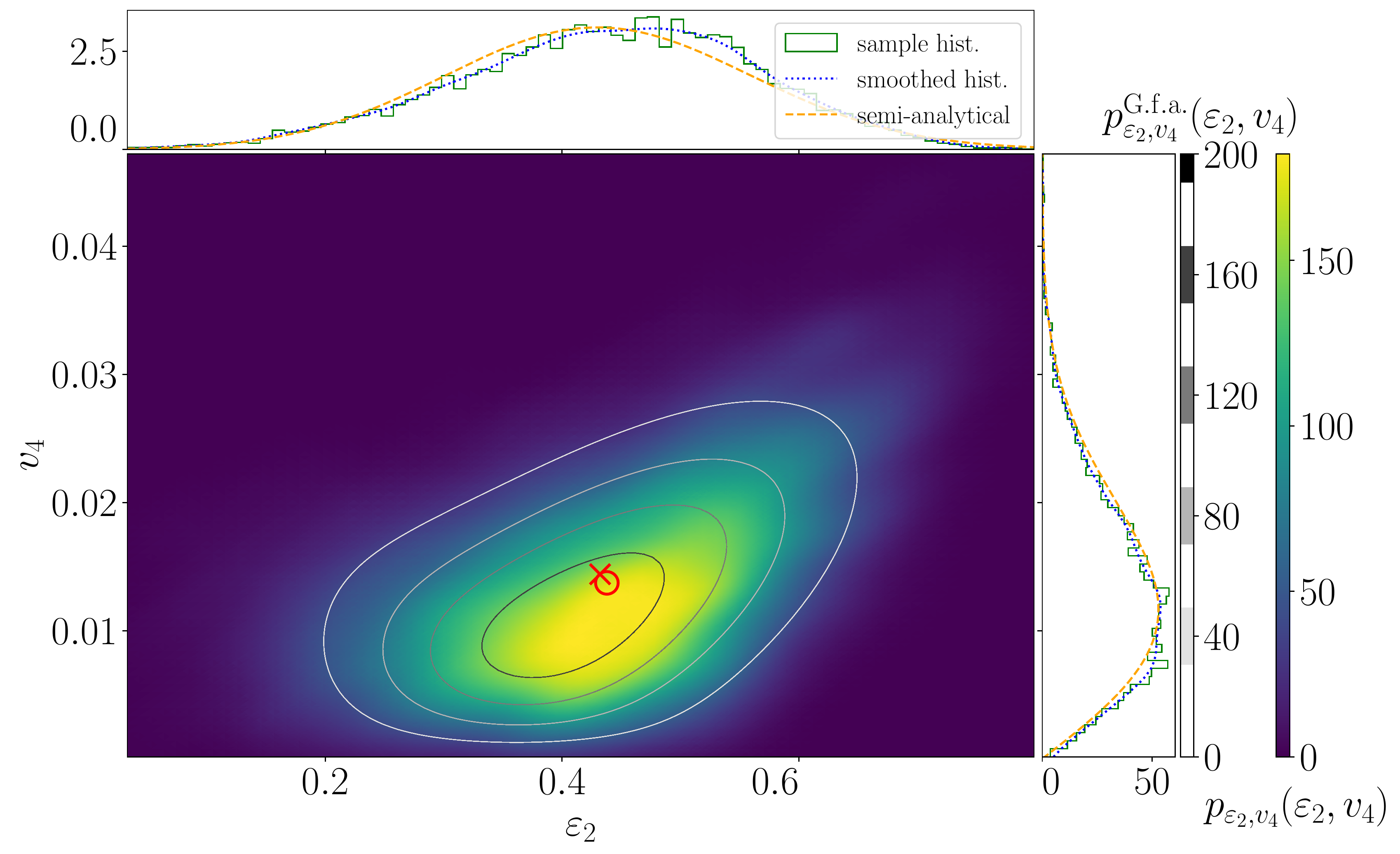}
\vspace{-5mm}
\caption{Same as Fig.~\ref{fig:2d_prob_dist_Glauber_b0} for collisions at $b=9$~fm within the Saturation model.}
\label{fig:2d_prob_dist_Saturation_b9}
\end{figure*}

More generally, we show in Figs.~\ref{fig:2d_prob_dist_Glauber_b0}--\ref{fig:2d_prob_dist_Saturation_b9} the joint probability distributions, either computed with Eq.~\eqref{eq:joint_probab_dist} or obtained from a sample of 8192 random events, of the observables $(\varepsilon_n,v_n)$ with $n\in\lbrace 1,\dots,5\rbrace$ and of $(\varepsilon_2,v_4)$, for both initial-state models at $b=0$ and 9~fm.
In every panel, we indicate with a cross resp.\ circle the average value of the two observables determined from the semi-analytical calculation resp.\ the event sample.
Above resp.\ right of each panel, we display the (marginal) distribution of the corresponding eccentricity resp.\ anisotropic-flow coefficient: 
the dashed curves stand for the semi-analytical results, the full-line histograms with visible bin widths are from the sampled events, while the dotted lines represent a Gaussian smoothing of these histograms. 
The first moments of these marginal distributions --- for the event samples and the calculations in the Gaussian fluctuation approximation --- are given in Table~\ref{tab:moments_eps_vn_distributions} in Appendix~\ref{app:moments_p(eps,v)}.


At vanishing impact parameter, the results from the Gaussian fluctuation approximation and the event sample are generally in excellent agreement, both in the Glauber (Figs.~\ref{fig:2d_prob_dist_Glauber_b0}) and Saturation (Fig.~\ref{fig:2d_prob_dist_Saturation_b0}) models. 
The only significant discrepancy is between the semi-analytical and sample mean values of $\varepsilon_5$ and $v_5$ in the Saturation model (red circle and cross in the bottom left panel of Fig.~\ref{fig:2d_prob_dist_Saturation_b0}).
This discrepancy can be related to the one already observed in the top panels of Fig.~\ref{fig:co_variances_convergence_Saturation_b0_IS_FS_modes_and_sampled} between the endpoint of the line(s) and the sample variances (represented by a circle) for $\varepsilon_{5,\rm c/s}$ or $v_{5,\rm c/s}$. 
Including more fluctuation modes in the calculation of the variances, which enter the semi-analytical estimate, would probably diminish the discrepancy. 

An additional observation at $b=0$ is that the dispersion away from the perfect proportionality~\eqref{eq:vn_epsn_linear} between $\varepsilon_n$ and $v_n$ seems to increase when going from $n=2$ to $n=5$. 
In contrast, the correlation between $\varepsilon_2$ and $v_4$ is much less marked: 
they rather seem to be uncorrelated, and the Gaussian fluctuation approximation captures remarkably well their joint probability because it factorizes into the product of their separate probability distributions, which are well reproduced. 

Going now to the results at $b=9$ presented in Figs.~\ref{fig:2d_prob_dist_Glauber_b9} and \ref{fig:2d_prob_dist_Saturation_b9} we find that the Gaussian fluctuation approach again provides in general a good approximation of the results from the event samples. 
However, deviations between the approaches are now more visible, starting with the average values of the observables. 
A first source of discrepancy is that the variances from the mode-by-mode approach sometimes miss the sample variances by a significant amount (see top panels of Fig.~\ref{fig:co_variances_convergence_Glauber_b9_IS_FS_modes_and_sampled} and \ref{fig:co_variances_convergence_Saturation_b9_IS_FS_modes_and_sampled}).
A second, possibly more important mismatch is that the linear approximation~\eqref{eq:vn_epsn_linear} is no longer always fulfilled: 
$v_3$ is still approximately proportional to $\varepsilon_3$, and this holds at the level of their sine and cosine parts. 
The proportionality between $v_2$ and $\varepsilon_2$ is no longer present at large $\varepsilon_2$, where a deviation appears, which was already reported in the literature~\cite{Noronha-Hostler:2015dbi}.
Eventually, nonlinear flow response is also present in the values of $v_1$, $v_4$, and $v_5$ found in the event sample. 
Indeed, the joint probability distribution of $\varepsilon_2$ and $v_4$ now has more structure than at $b=0$, demonstrating the existence of a correlation between them. 
However, the Gaussian fluctuation approximation does not include these nonlinearities, which for example spoil the assumption that the fluctuations of $v_{n,\rm c/s}$ are Gaussian. 
Nonetheless, it is worth emphasizing the results from the semi-analytical approach are not totally off, but yield a more than decent approximation.

\section{Summary and outlook}
\label{sec:conclusions}

In this paper we have introduced a general framework for characterizing the event-by-event fluctuations of the initial state of heavy-ion collisions predicted by a given model. 
Starting from a density-matrix formalism, the fluctuating initial states are written as the sum of an average event and a linear combination of uncorrelated fluctuation modes, Eq.~\eqref{eq:evt_vs_modes}, with expansion coefficients that are found to be almost Gaussian-distributed. 

For a set of observables --- both in the initial state and in the final state following a dynamical evolution with K{\o}MP{\o}ST and MUSIC --- we compared the mean values and (co)variances computed from a sample of events with those gained from a mode-by-mode calculation. 
In the mode-by-mode approach, we characterized the response of observables to the presence of a given mode by linear- and quadratic-response coefficients. 
The statistics of some observables (energy density, mean square radius, eccentricities $\varepsilon_{n,\rm c/s}$, to a large extent $v_{n,\rm c/s}$) is described very satisfactorily by the linear mode-by-mode response with the inclusion of a reasonable number of modes. 
This allows us in particular to predict the joint statistics of the eccentricities $\varepsilon_n$ and anisotropic flow coefficients $v_n$ in the mode-by-mode approach within a Gaussian fluctuation ansatz assuming a linear response. 
In contrast, charged multiplicity and to a lesser extent average transverse momentum are significantly nonlinear and may also possibly require more modes for a good description.\footnote{The issue with multiplicity may be mitigated if the initial state is defined in terms of entropy-density profiles, instead of energy densities as in the present paper.}

In the present paper we used initial states for Pb-Pb collisions at fixed impact parameter. 
This has the advantage that, at $b=0$, the colliding system is azimuthally symmetric in the transverse plane, yielding an average state and modes with a relatively simple structure: most fluctuation modes only have a single nonzero eccentricity $\varepsilon_n$, and accordingly in the final state only few sizable anisotropic flow harmonics $v_n$, $v_{2n}$, \dots. 
This simplifies the analysis of linear and nonlinear responses in the mode-by-mode evolution, in particular the calculation of response in the Gaussian-fluctuation approximation, and allowed us to validate our approach. 
In a forthcoming study, we shall consider the more experimentally relevant case of events within centrality classes.
Even then, it appears advantageous to consider a fixed orientation of the impact parameter, in order to  absorb the leading effects into an anisotropic average state, and thereby limit the contribution of fluctuations to observables. In this respect our approach is distinct from previous mode-by-mode studies \cite{Floerchinger:2013rya,Floerchinger:2013vua,Floerchinger:2013hza,Floerchinger:2013tya,Floerchinger:2014fta,Floerchinger:2018pje,Floerchinger:2020tjp}, where a rotationally symmetric average state was considered.
With the extra constraint of almost-fixed multiplicity, one can then meaningfully include further observables, and investigate (possibly with the help of a singular value decomposition) which modes are needed for which observables. 
This is a needed step towards the ultimate scope, which would be to discriminate between initial-state models. 
Here we considered two such models to test our method on different samples of initial states, but we did not attempt to really compare the models. 
Possibly the most prominent difference between them is the behavior of the eigenvalues of the density matrix of fluctuations, i.e., the relative importance of the contribution of modes to events. 

Quite naturally, our method can be extended to different colliding systems, in particular with deformed nuclei. 
It may also be interesting, although possibly costly, to investigate higher orders for the mode-by-mode response of observables, especially for those that are highly nonlinear.
The generality of our approach makes it also possible to consider more complicated initial states, like truly three-dimensional profiles or including conserved charges, although again at an increased computational cost. 

\begin{acknowledgments}
	We would like to thank Hannah Elfner, Giuliano Giacalone, Aleksas Mazeliauskas, Bjoern Schenke and Derek Teaney for valuable discussions.
	The authors acknowledge support by the Deutsche Forschungsgemeinschaft (DFG, German Research Foundation) through the CRC-TR 211 'Strong-interaction matter under extreme conditions' - project number 315477589 - TRR 211.
	Numerical simulations presented in this work were performed at the Paderborn Center for Parallel Computing (PC$^2$) and we gratefully acknowledge their support.
\end{acknowledgments}

\appendix

\section{Characteristics of the probability distributions of the expansion coefficients}
\label{appendix:moments_cl_coeff}

In this appendix we present quantitative measures of the probability distributions $p(c_l)$ of the expansion coefficients $\{c_l\}$. 
More specifically, we computed the average $\mu$, the variance $\sigma^2$, the skewness $\gamma_1$ and the excess kurtosis $\gamma_2$. 
By construction, $\mu$ resp.\ $\sigma^2$ should be close to 0 resp.\ 1, see Eqs.~\eqref{eq:<c_l>=0} resp.\ \eqref{eq:<c_lc_l'>}.
As in Sec.~\ref{subsec:results_stat_charact_IS}, these characteristics were obtained using 8192 randomly sampled events in both models and at both considered impact parameters.

\begin{figure*}[!htb]
\includegraphics[width=0.945\linewidth]{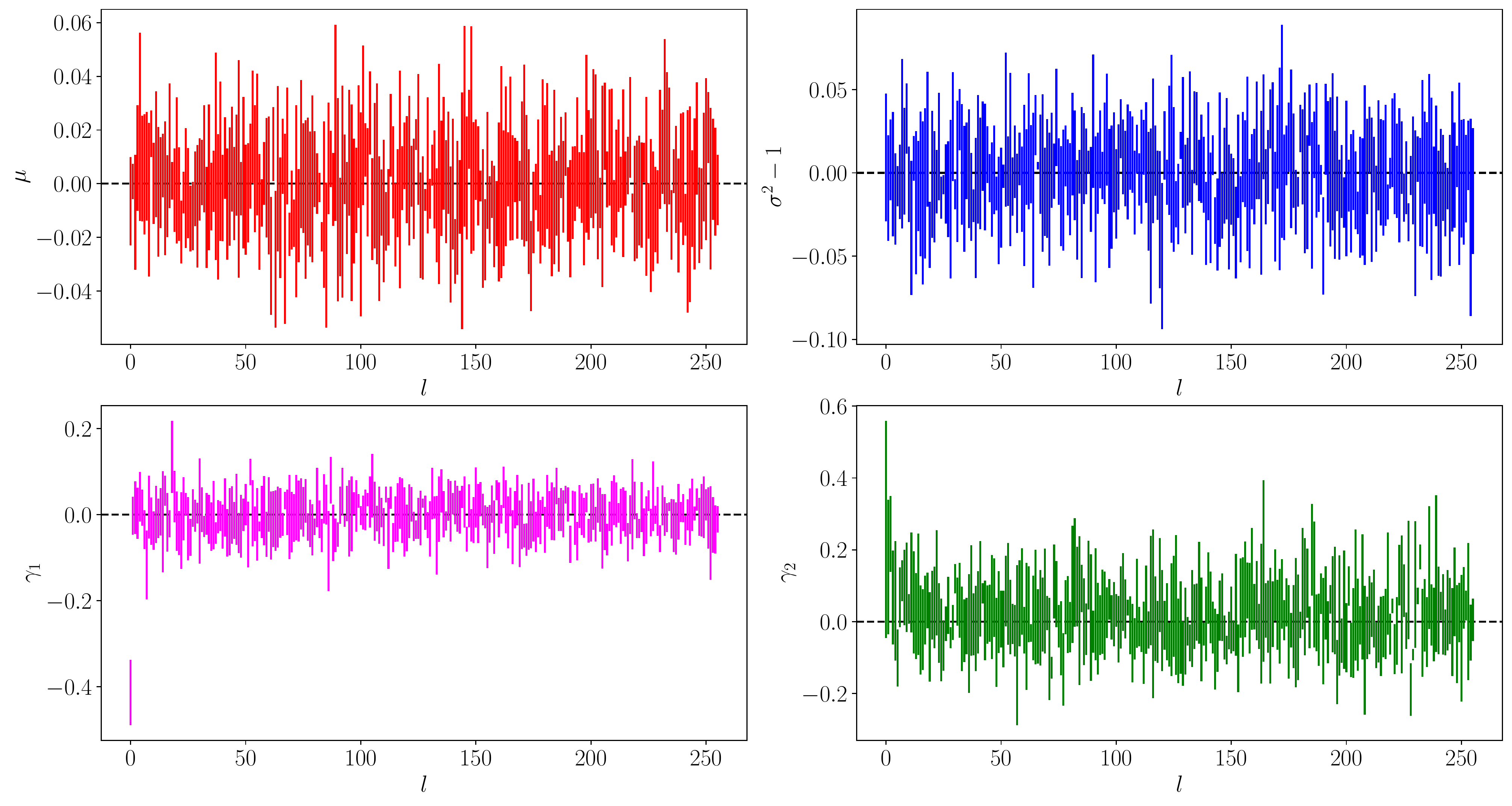}
\vspace{-3mm}
\caption{Average $\mu$, variance $\sigma^2$, skewness $\gamma_1$ and excess kurtosis $\gamma_2$ of the probability distributions of the expansion coefficients $c_l$ in the Glauber model at $b=0$ fm.}
\label{fig:cl_moments_Glauber_b0}
\end{figure*}
\begin{figure*}[!htb]
\includegraphics[width=0.945\linewidth]{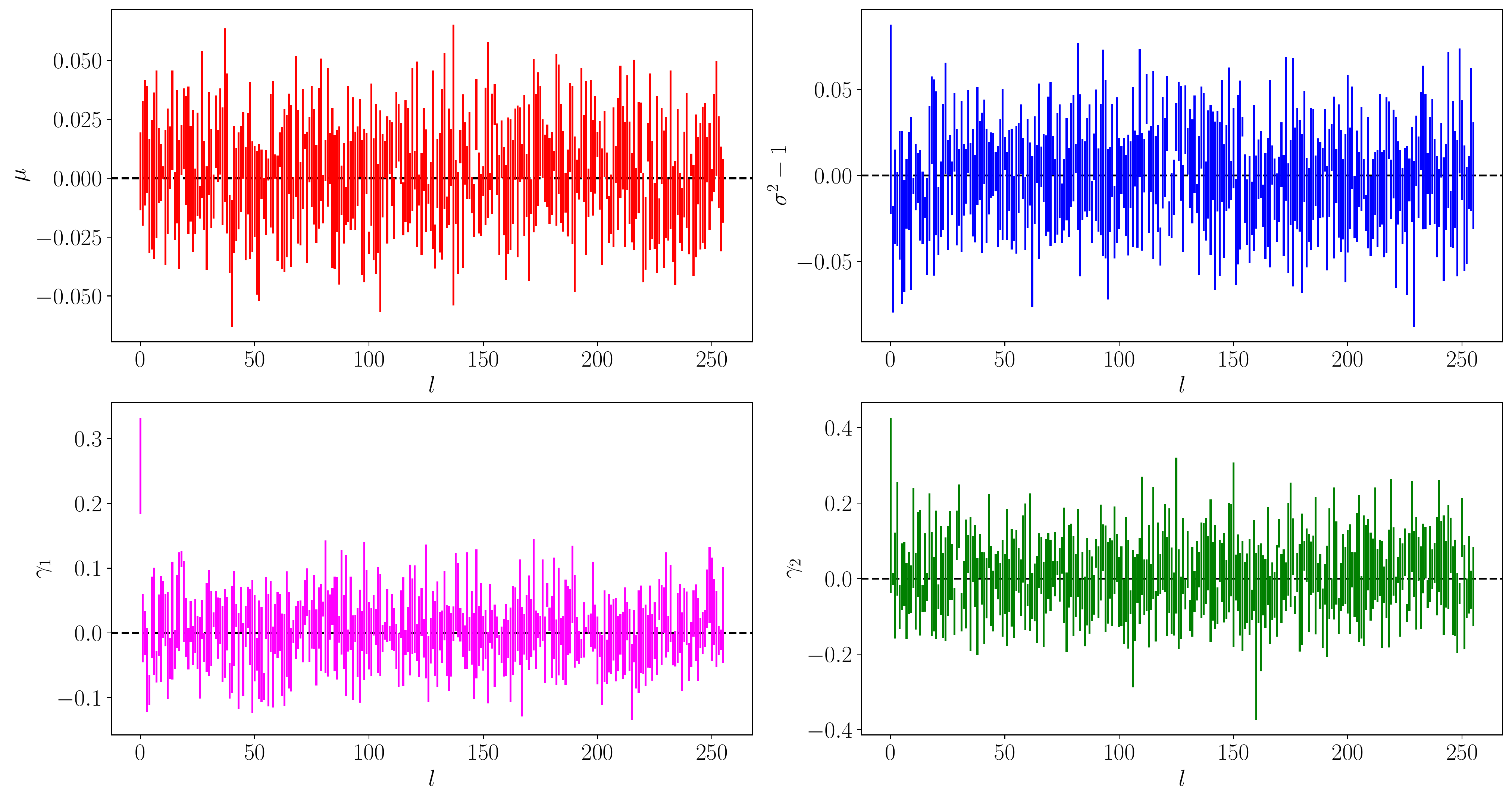}
\vspace{-3mm}
\caption{Average $\mu$, variance $\sigma^2$, skewness $\gamma_1$ and excess kurtosis $\gamma_2$ of the probability distributions of the expansion coefficients $c_l$ in the Saturation model at $b=0$ fm.}
\label{fig:cl_moments_Saturation_b0}
\end{figure*}
\begin{figure*}[!htb]
\includegraphics[width=0.945\linewidth]{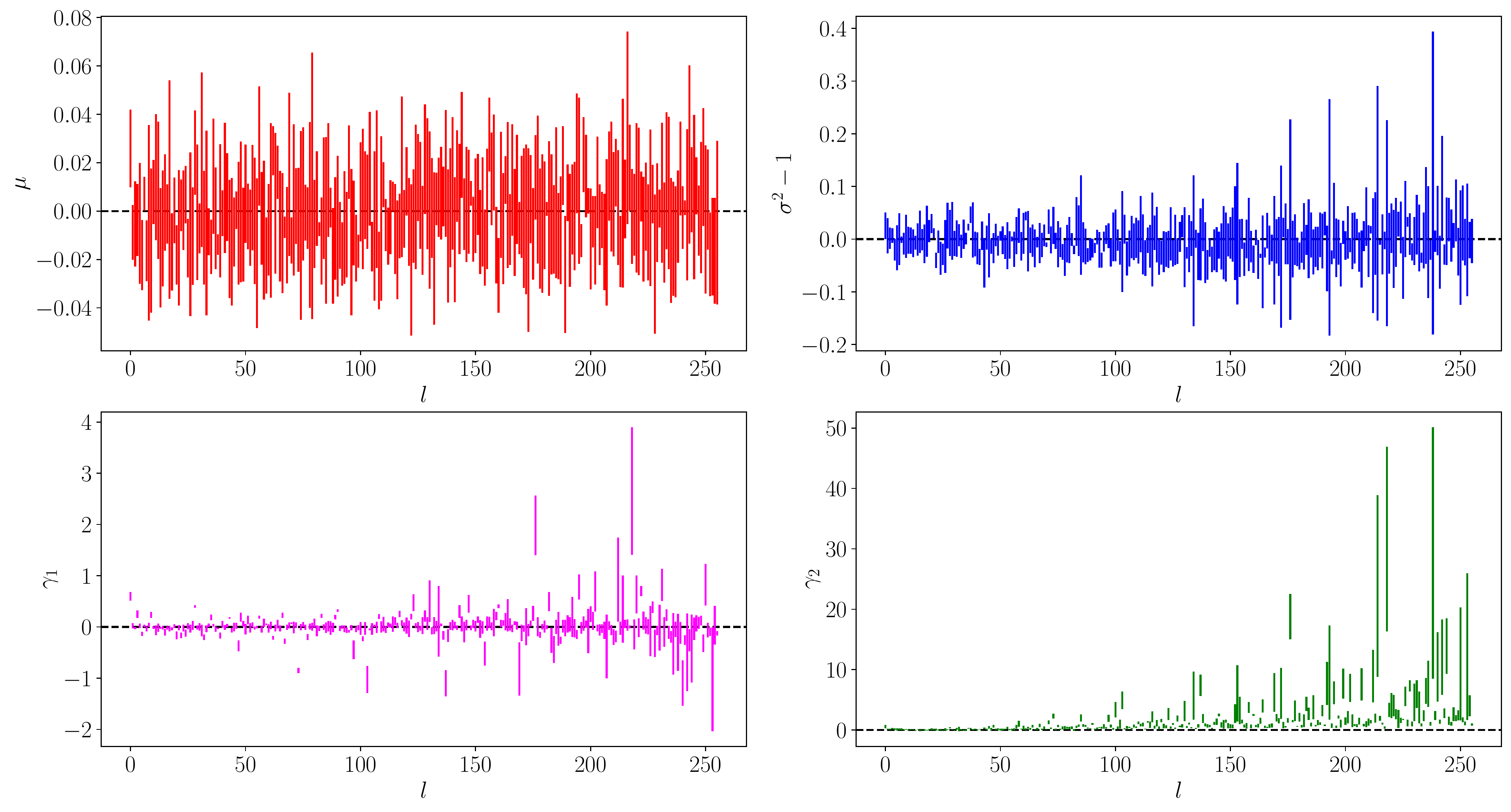}
\vspace{-3mm}
\caption{Average $\mu$, variance $\sigma^2$, skewness $\gamma_1$ and excess kurtosis $\gamma_2$ of the probability distributions of the expansion coefficients $c_l$ in the Glauber model at $b=9$ fm.}
\label{fig:cl_moments_Glauber_b9}
\end{figure*}
\begin{figure*}[!htb]
\includegraphics[width=0.945\linewidth]{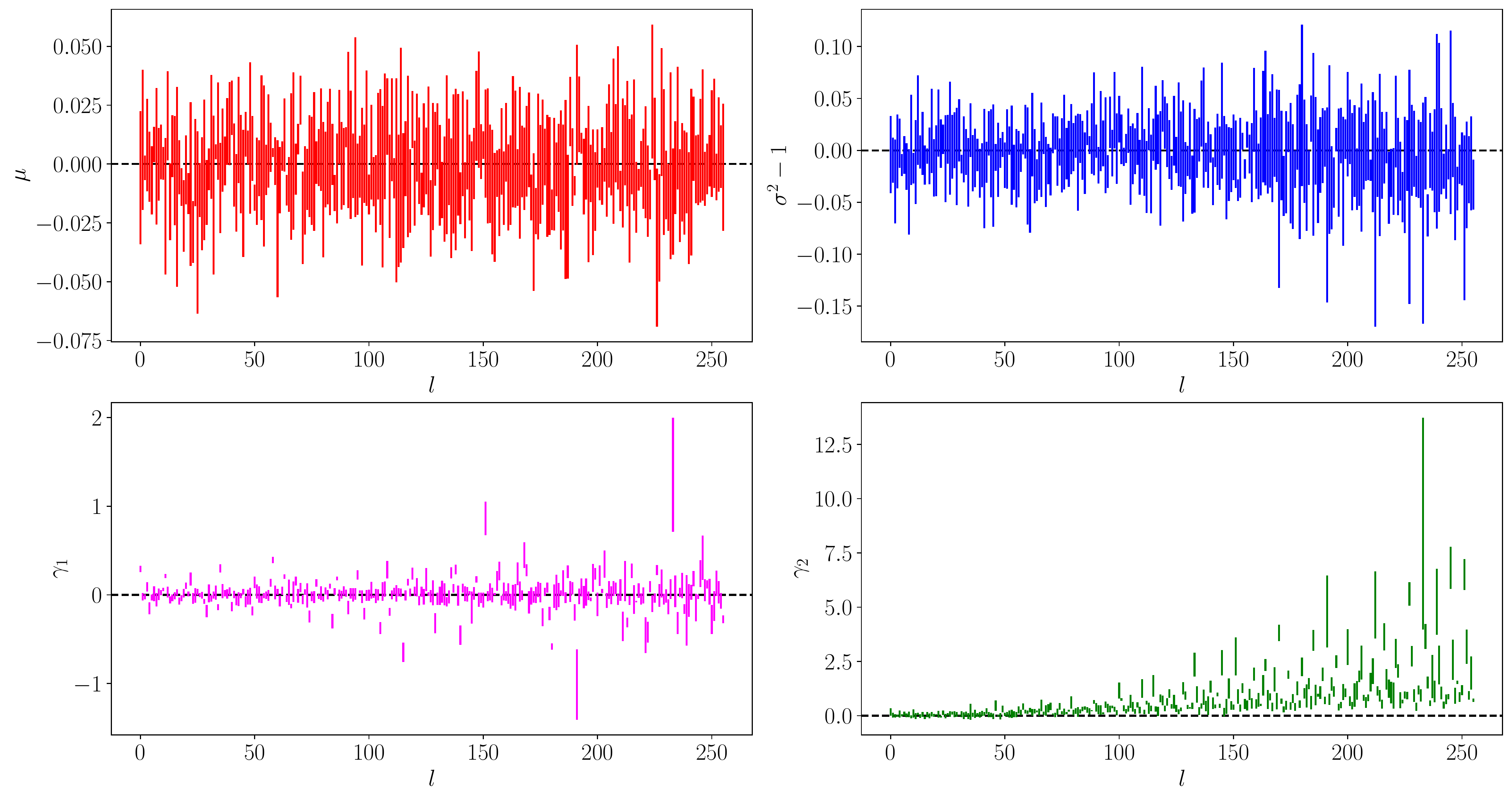}
\vspace{-3mm}
\caption{Average $\mu$, variance $\sigma^2$, skewness $\gamma_1$ and excess kurtosis $\gamma_2$ of the probability distributions of the expansion coefficients $c_l$ in the Saturation model at $b=9$ fm.}
\label{fig:cl_moments_Saturation_b9}
\end{figure*}

In Figs.~\ref{fig:cl_moments_Glauber_b0} and \ref{fig:cl_moments_Saturation_b0} we show the moments for events at zero impact parameter.
In both models the characteristics are consistent with those of a centered Gaussian distribution with unit variance, namely $\gamma_1=\gamma_2=0$, for almost all modes. 
The most notable exception is the mode with the largest relative weight, $l=0$, for which the distribution of the expansion coefficient $c_0$ has a sizable negative resp.\ positive skewness in the Glauber resp.\ Saturation model.%
\footnote{Closer inspection reveals that the skewness is also nonzero for a few further modes, in particular those with radial symmetry --- $l=7$, 18, 33, \ldots ---, although this is less visible.}
The asymmetry of the probability distribution signaled by this nonzero skewness is easily understood, as we now discuss on the example of the Saturation model. 
In that case the mode $l=0$ is positive at the center of the fireball (bottom left density plot in Fig.~\ref{fig:mode_examples}).
Consider the random event $\bar{\Psi} + c_0\Psi_0$ (and, for the sake of discussion, no other mode): 
the requirement that the energy density of the event should be non-negative everywhere, in particular at $r=0$, constrains the possible values of $c_0$, which cannot be too negative. 
In turn, $c_0$ cannot be too large a positive number either --- to ensure the positivity in regions with $r\approx 0.5R$ ---, yet that restriction is milder, in that it allows larger absolute values $|c_0|$. 
Overall, the distribution of $c_0$ can thus extend further towards positive values than towards negative ones: it has a longer tail on the right, i.e., precisely a positive skewness $\gamma_1$. 

In events at finite impact parameter ($b=9$ fm), for which the results are shown in Figs.~\ref{fig:cl_moments_Glauber_b9} and \ref{fig:cl_moments_Saturation_b9}, we find in both models larger deviations from the values for a Gaussian distribution. 
First, the skewness $\gamma_1$ of $p(c_l)$ departs considerably from zero for some modes, although without any clear trend. 
Second, the excess kurtosis $\gamma_2$ also deviates from zero, with a marked trend towards positive values that seems to increase with $l$.
Thus, the broken rotational symmetry at finite impact parameter leads to a slightly non-Gaussian probability distribution of the expansion coefficients $c_l$, especially for the higher modes.

\section{Fluctuation modes}
\label{appendix:mode_plots}

\begin{figure*}
\includegraphics[height=0.977\textheight]{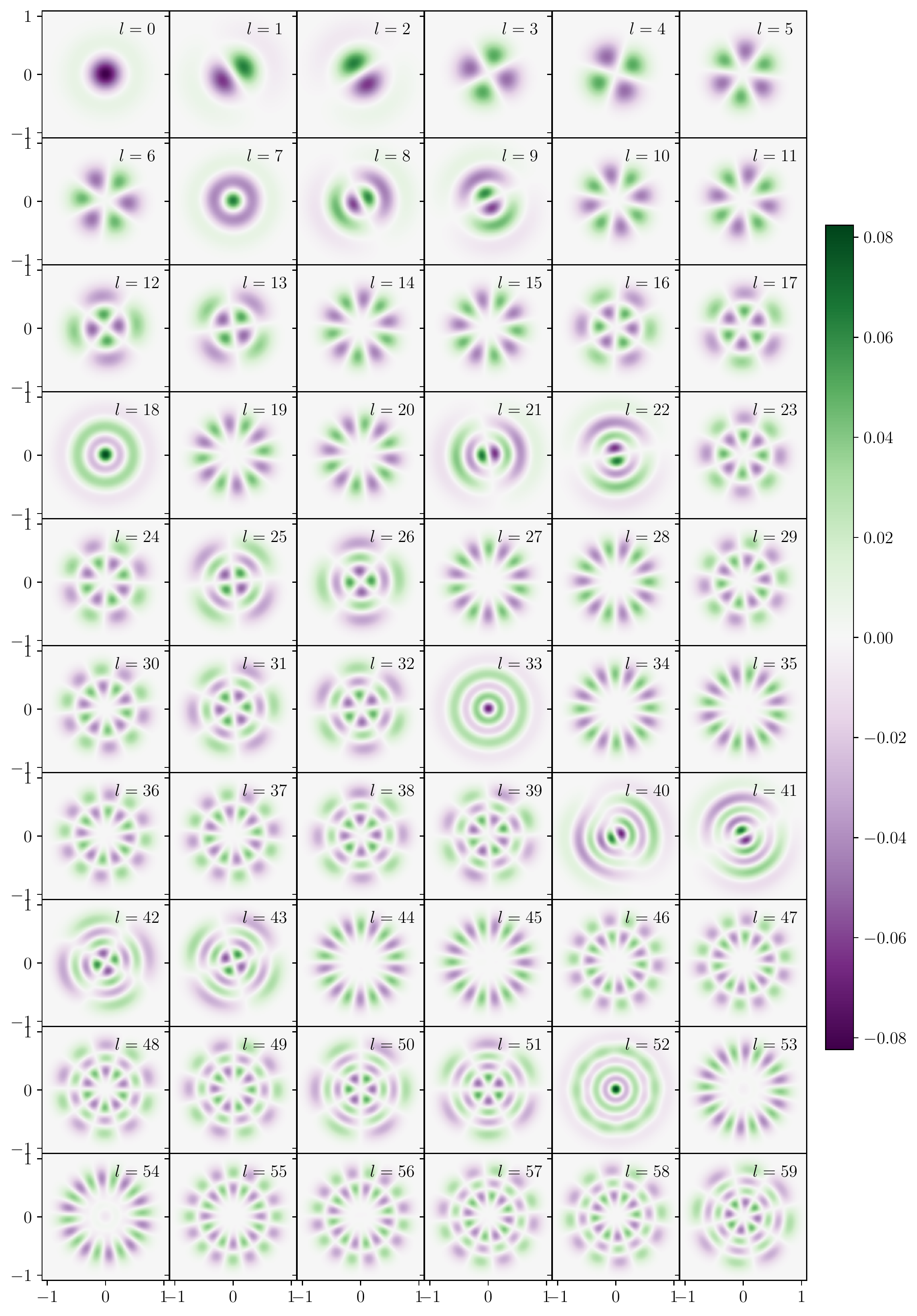}
\vspace{-3mm}
\caption{First 60 orthonormal eigenvectors for the Glauber model at $b=0$. Both axes are in units of $R$.}
\label{fig:60_modes_b0_Glauber}
\end{figure*}
\begin{figure*}
\includegraphics[height=0.977\textheight]{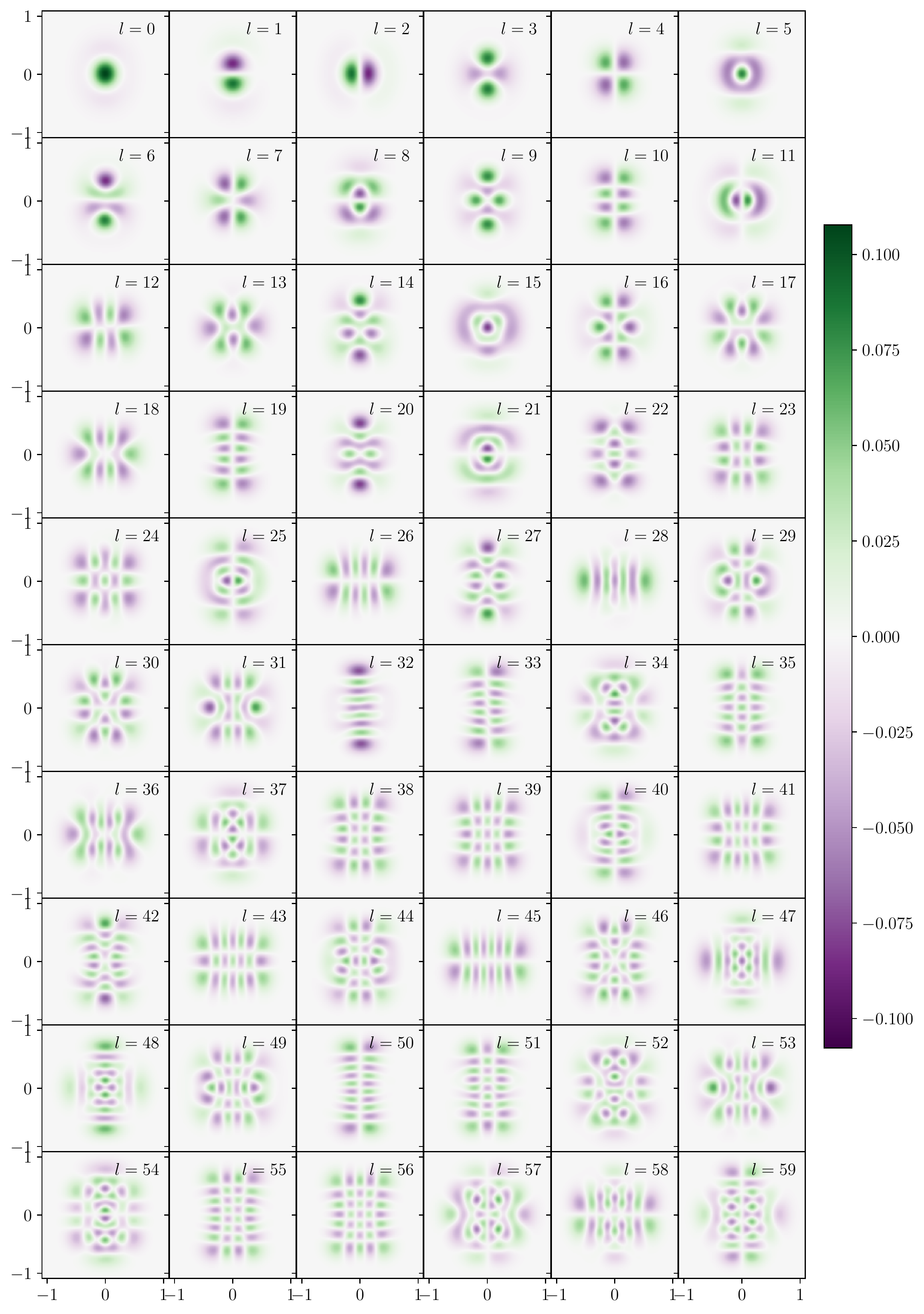}
\vspace{-3mm}
\caption{First 60 orthonormal eigenvectors for the Glauber model at $b=9$ fm. Both axes are in units of $R$.}
\label{fig:60_modes_b9_Glauber}
\end{figure*}
\begin{figure*}
\includegraphics[height=0.977\textheight]{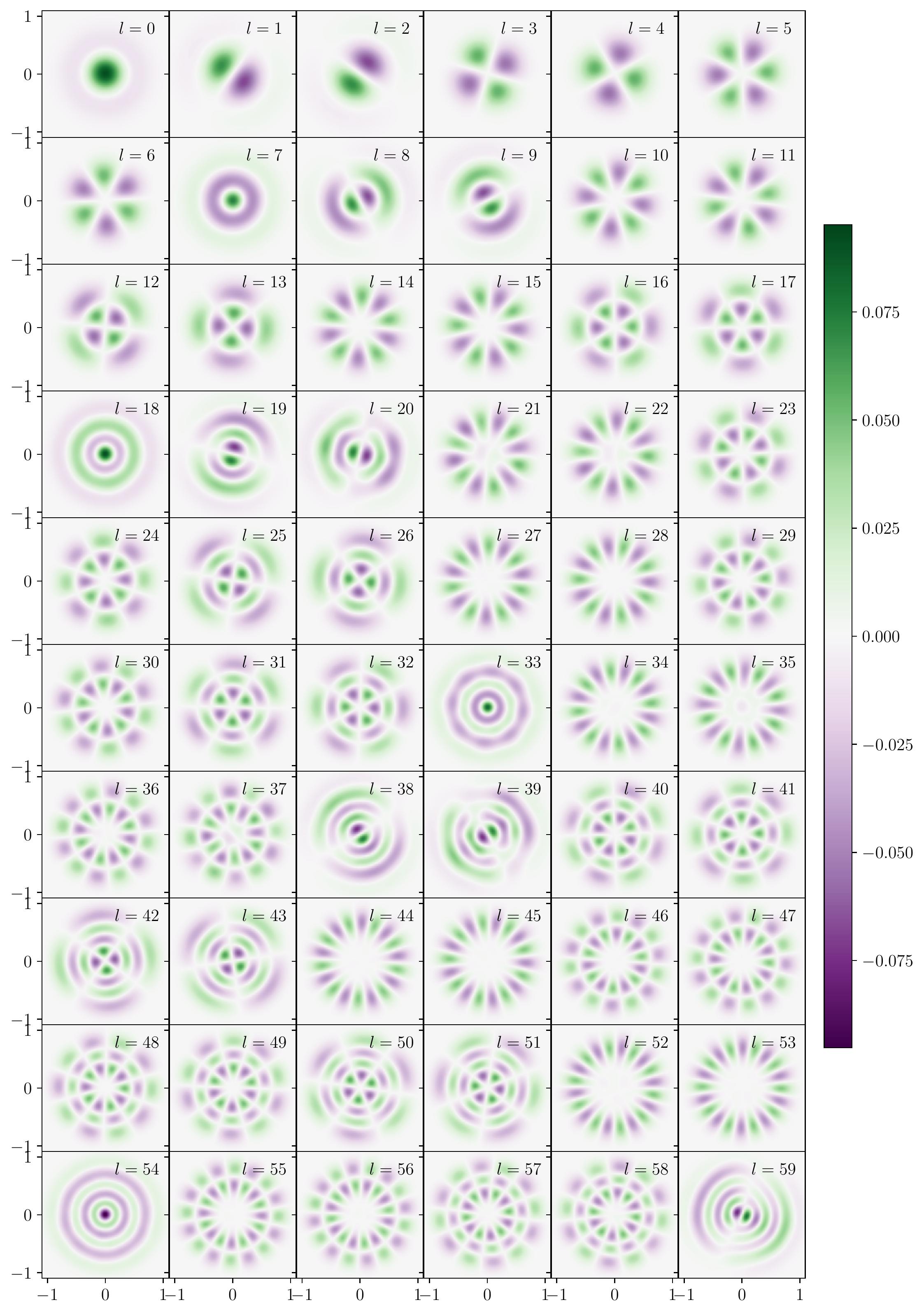}
\vspace{-3mm}
\caption{First 60 orthonormal eigenvectors for the Saturation model at $b=0$ fm. Both axes are in units of $R$.}
\label{fig:60_modes_b0_Saturation}
\end{figure*}
\begin{figure*}
\includegraphics[height=0.977\textheight]{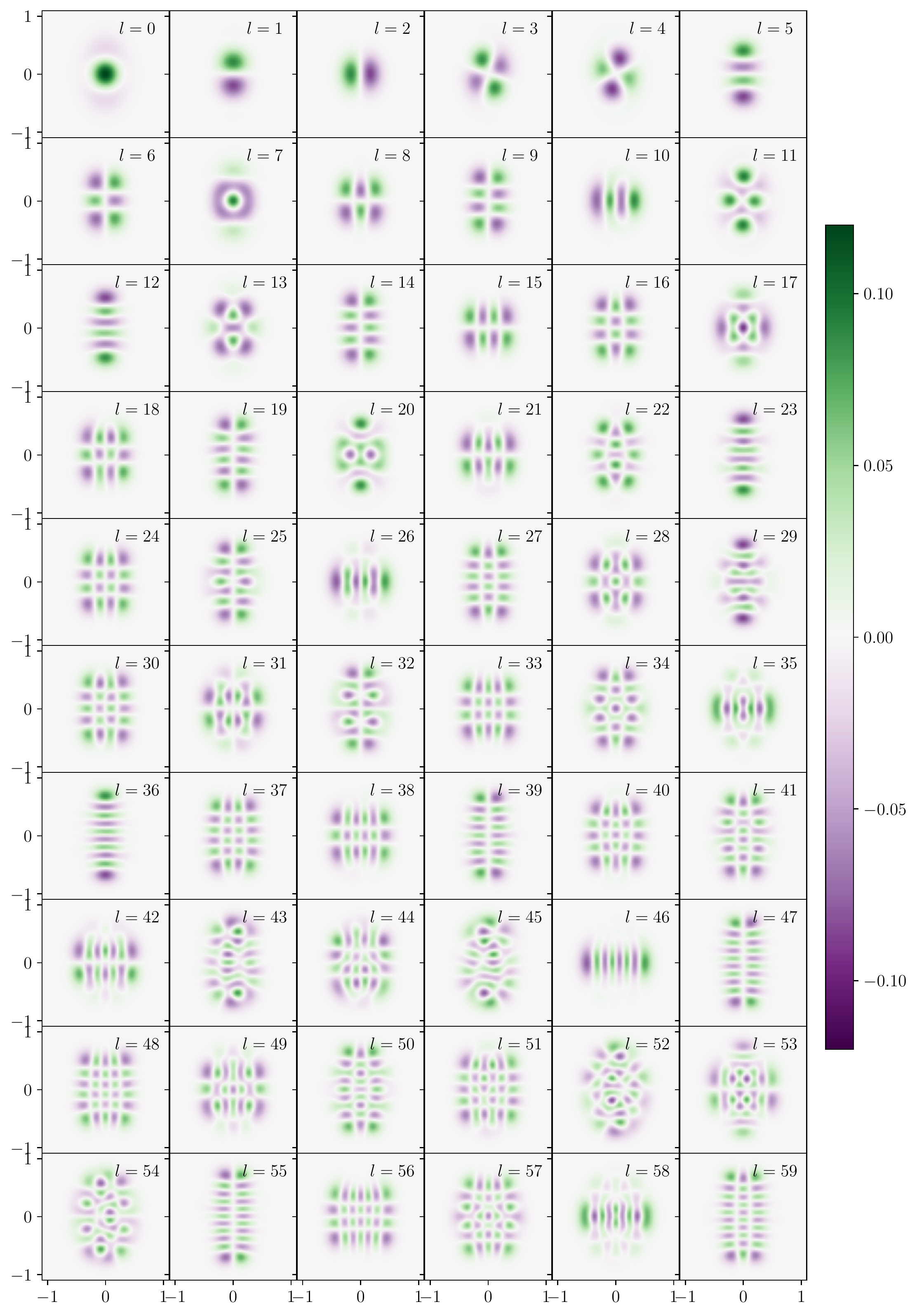}
\vspace{-3mm}
\caption{First 60 orthonormal eigenvectors for the Saturation model at $b=9$ fm. Both axes are in units of $R$.}
\label{fig:60_modes_b9_Saturation}
\end{figure*}

In this appendix we show the first 60 orthonormal eigenvectors of the density matrix of fluctuations~\eqref{eq:rho} for the initial state of Pb--Pb collisions at $\sqrt{\sNN}=5.02$~TeV within the Glauber model (Fig.~\ref{fig:60_modes_b0_Glauber} at $b=0$, Fig.~\ref{fig:60_modes_b9_Glauber} at $b=9$~fm) and the Saturation model (Fig.~\ref{fig:60_modes_b0_Saturation} at $b=0$, Fig.~\ref{fig:60_modes_b9_Saturation} at $b=9$~fm).
Note that these eigenvectors all have the same normalization, to allow a simpler comparison, while otherwise the norm of the successive fluctuation modes $\Psi_l$ decreases with increasing $l$. 

\begin{figure*}[!htb]
\includegraphics[width=0.495\linewidth]{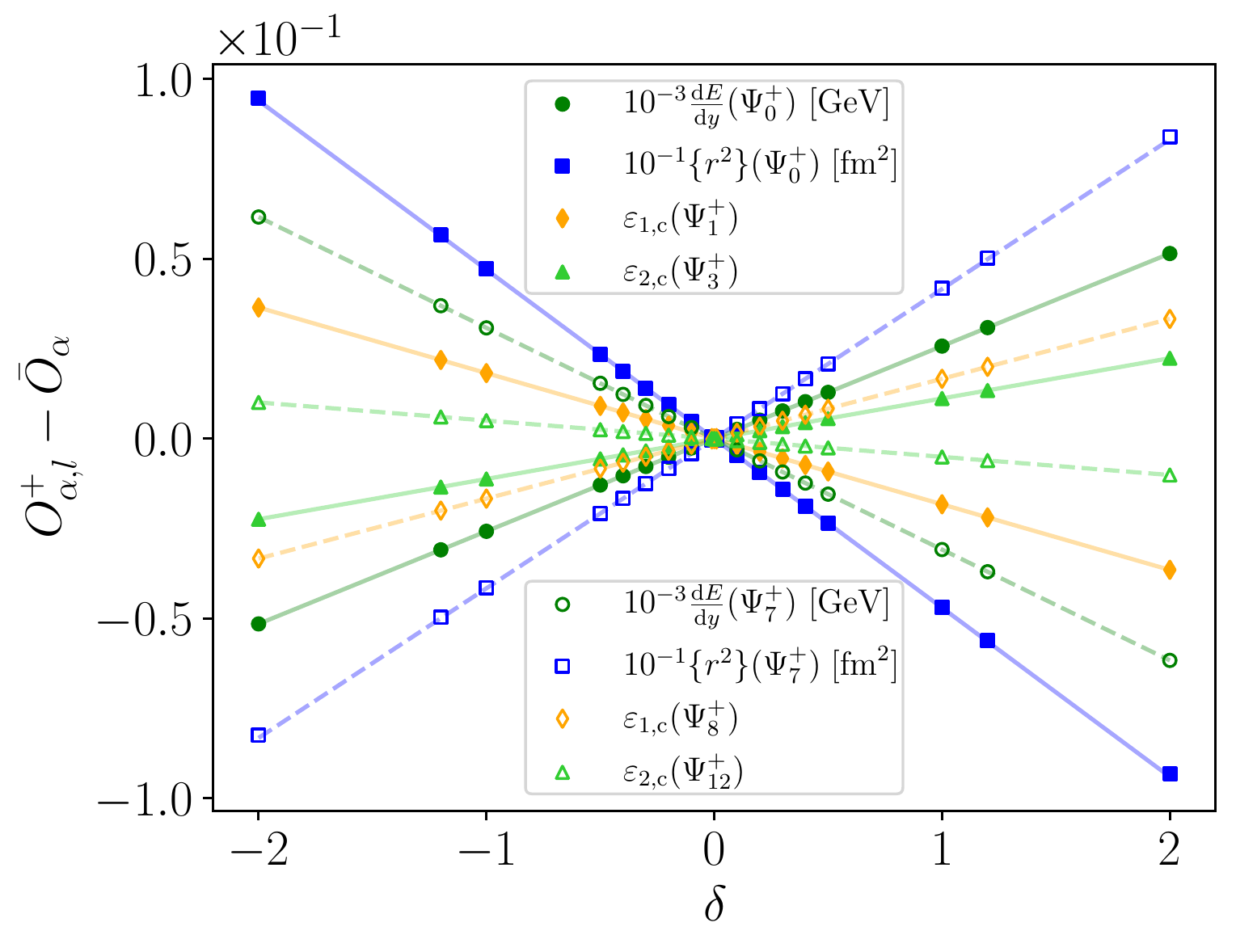}
\includegraphics[width=0.495\linewidth]{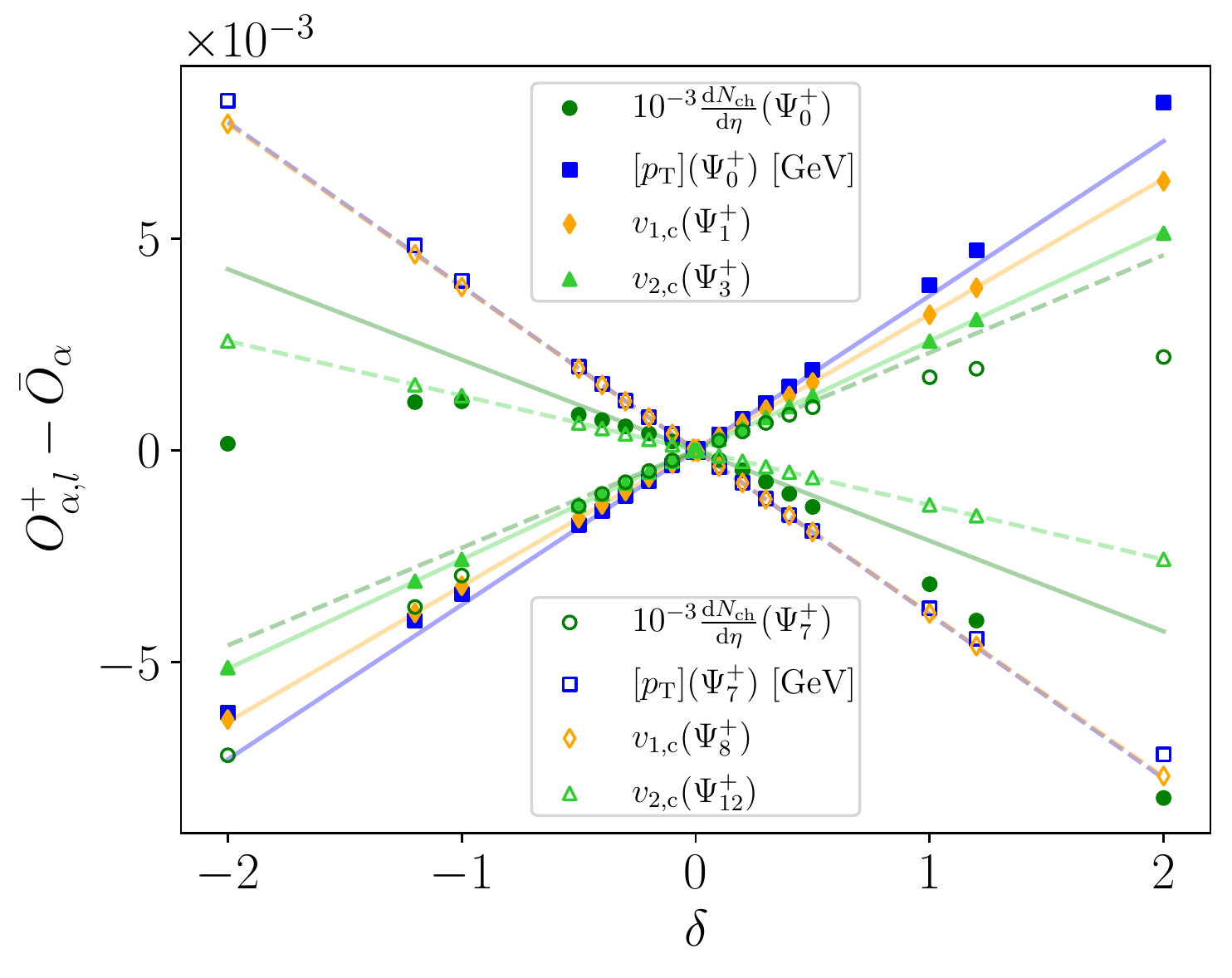}
\includegraphics[width=0.495\linewidth]{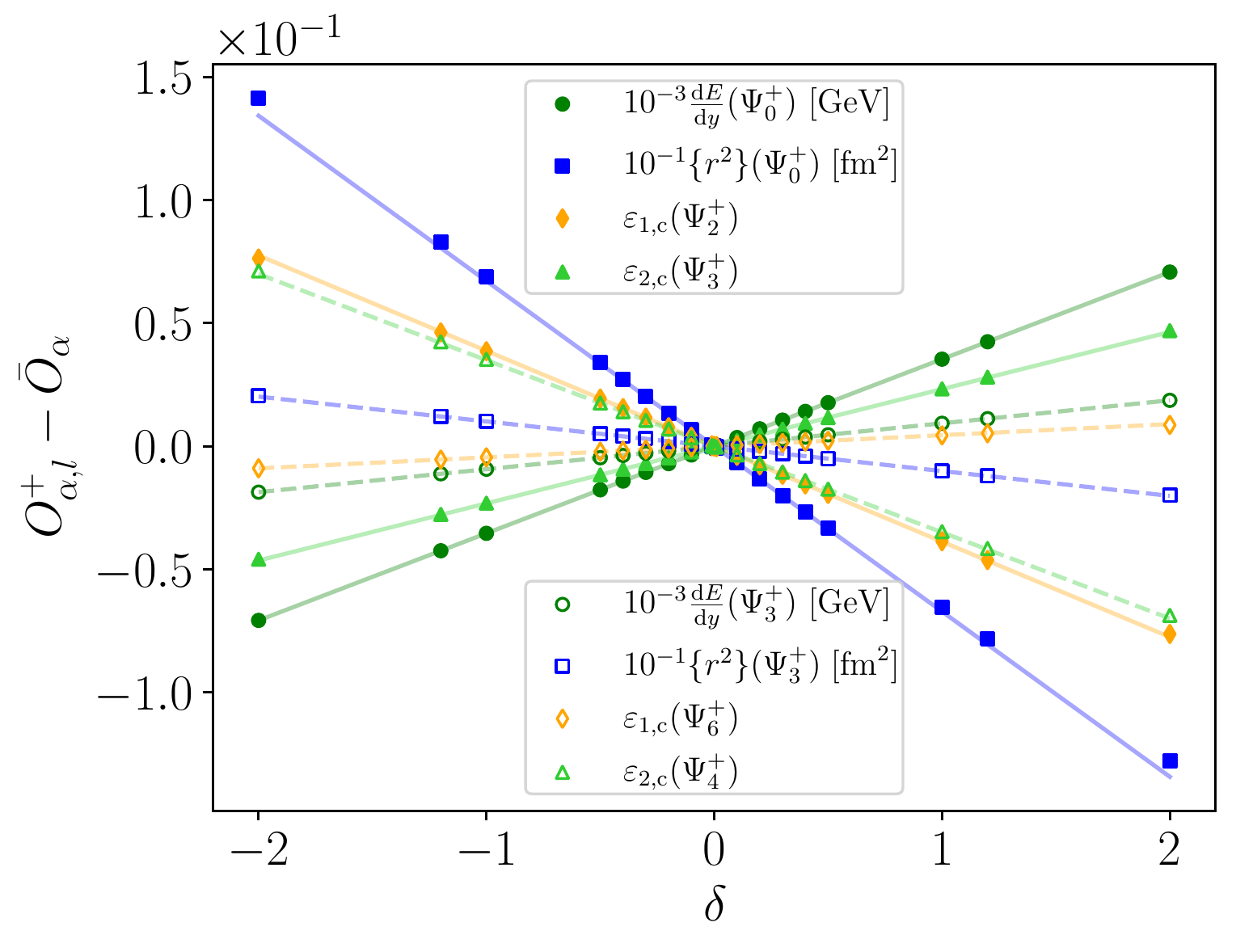}
\includegraphics[width=0.495\linewidth]{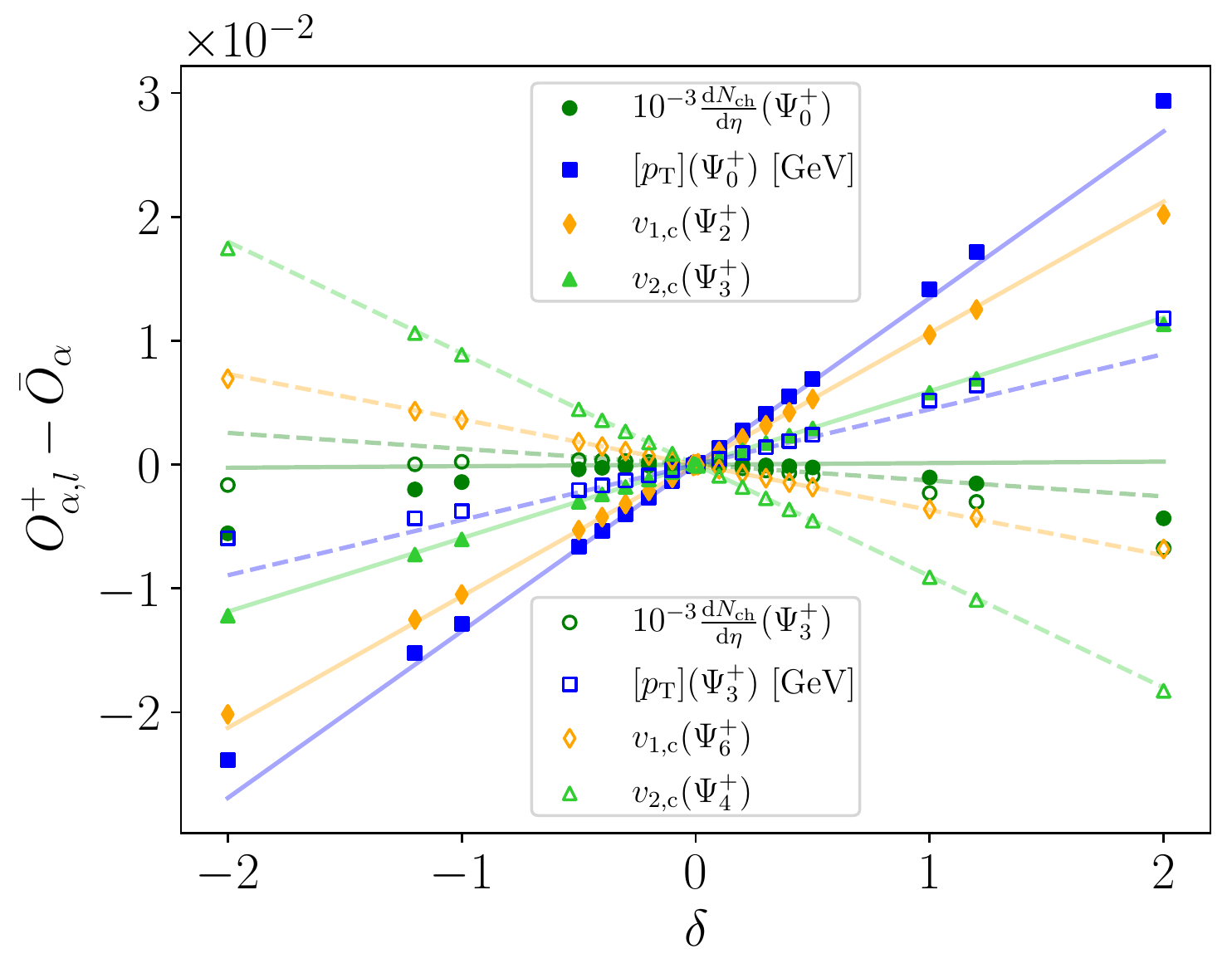}
\vspace{-5mm}
\caption{Variation with $\delta$ of $O^+_{\alpha,l}-\bar{O}_\alpha$ for initial-state (left) and final-state (right) observables, together with linear fits to the points with $|\delta|\leq 0.01$, using the Saturation model at $b=0$ (top) and $b=9$ fm (bottom)
Closed symbols and full lines correspond to the first modes contributing to the respective observable, while open symbols and dashed lines are for the second modes.}
\label{fig:Linearity_test_Saturation}
\end{figure*}

\section{Mode-by-mode response of observables in the Saturation model}
\label{appendix:response_saturation}

In this appendix we present the linear and quadratic mode-by-mode response of system observables for collisions with an initial state from the Saturation model, paralleling the results in the Glauber model of Sec.~\ref{subsec:lin_quad_response}.

\subsection{Linearity check}
\label{appendix:linearity_check}

In Fig.~\ref{fig:Linearity_test_Saturation} we show $O^+_{\alpha,l}-\bar{O}_\alpha \equiv O_\alpha(\bar{\Psi}+\delta\Psi_l)-O_\alpha(\bar{\Psi})$ for a number of initial-state and final-state observables $\{O_\alpha\}$ evaluated for a few modes, using various values of $\delta$ between $-2$ and 2.
On the left are collisions at zero impact parameter, on the right collisions at $b=9$~fm. 
This is similar to Fig.~\ref{fig:Linearity_test_Glauber}, with the difference that at $b=9$~fm we have chosen different fluctuation modes $\Psi_l$ for the eccentricities $\varepsilon_{n,\mathrm{c}}$, because the $l$th mode does not necessarily affect the same harmonics in the Glauber and Saturation models.

The results in the Saturation model are generally the same as in the Glauber model.
The only minor difference with Fig.~\ref{fig:Linearity_test_Glauber} regards the initial state at $b=0$, where the mean square radius $\{r^2\}$ looks more linear here than in the Glauber model.
Anticipating on what comes next, this can also be seen on the level of the quadratic-response coefficients in Fig.~\ref{fig:Q_IS_FS_Saturation} (top left panel), where no nonlinear response is visible for $\{r^2\}$, while small sizable coefficients can be spotted in the similar plot (Fig.~\ref{fig:Q_IS_FS_Glauber}) for the Glauber model.

\begin{figure*}[!t]
\includegraphics[width=0.495\linewidth]{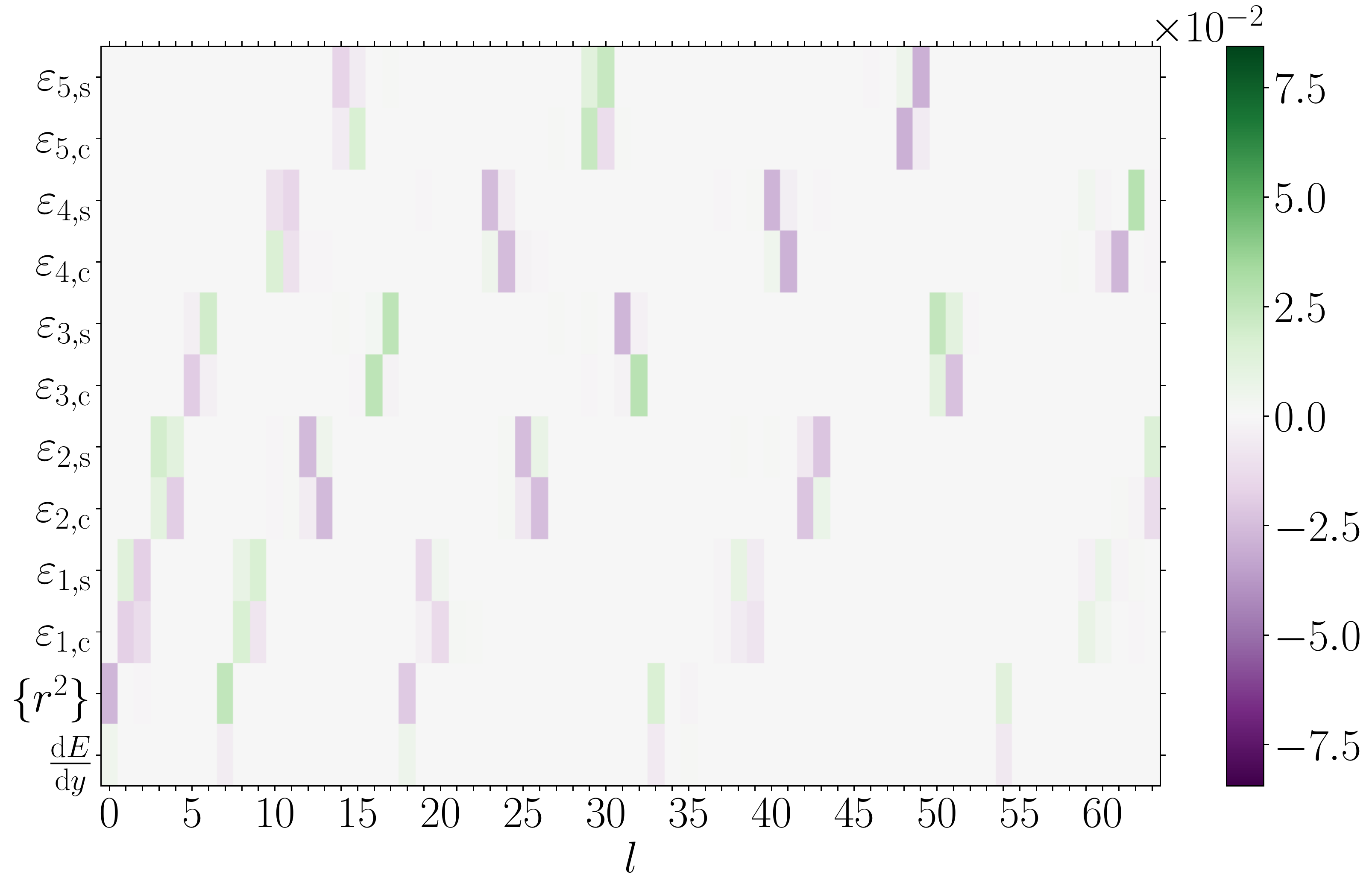}
\includegraphics[width=0.495\linewidth]{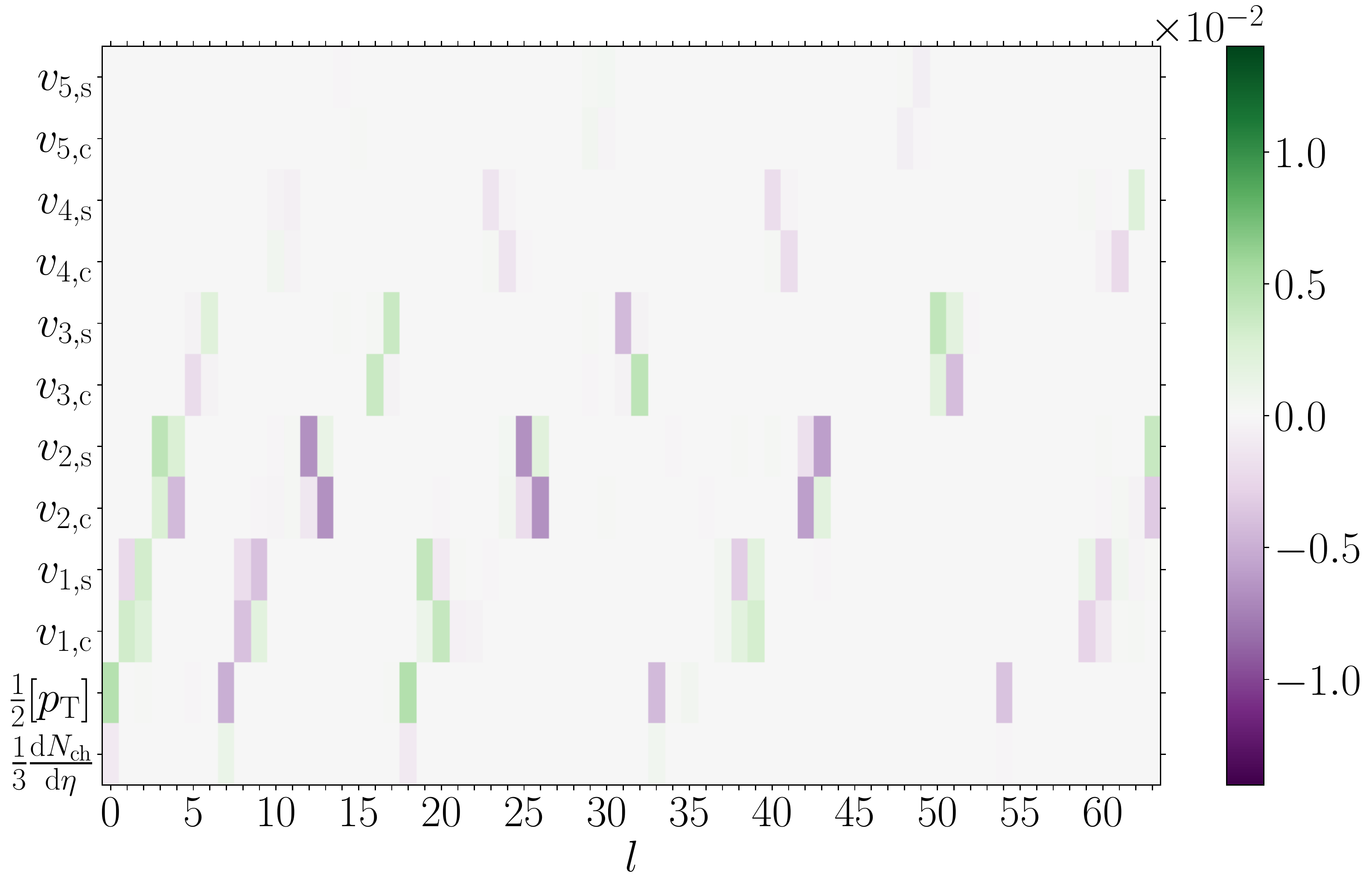}
\includegraphics[width=0.495\linewidth]{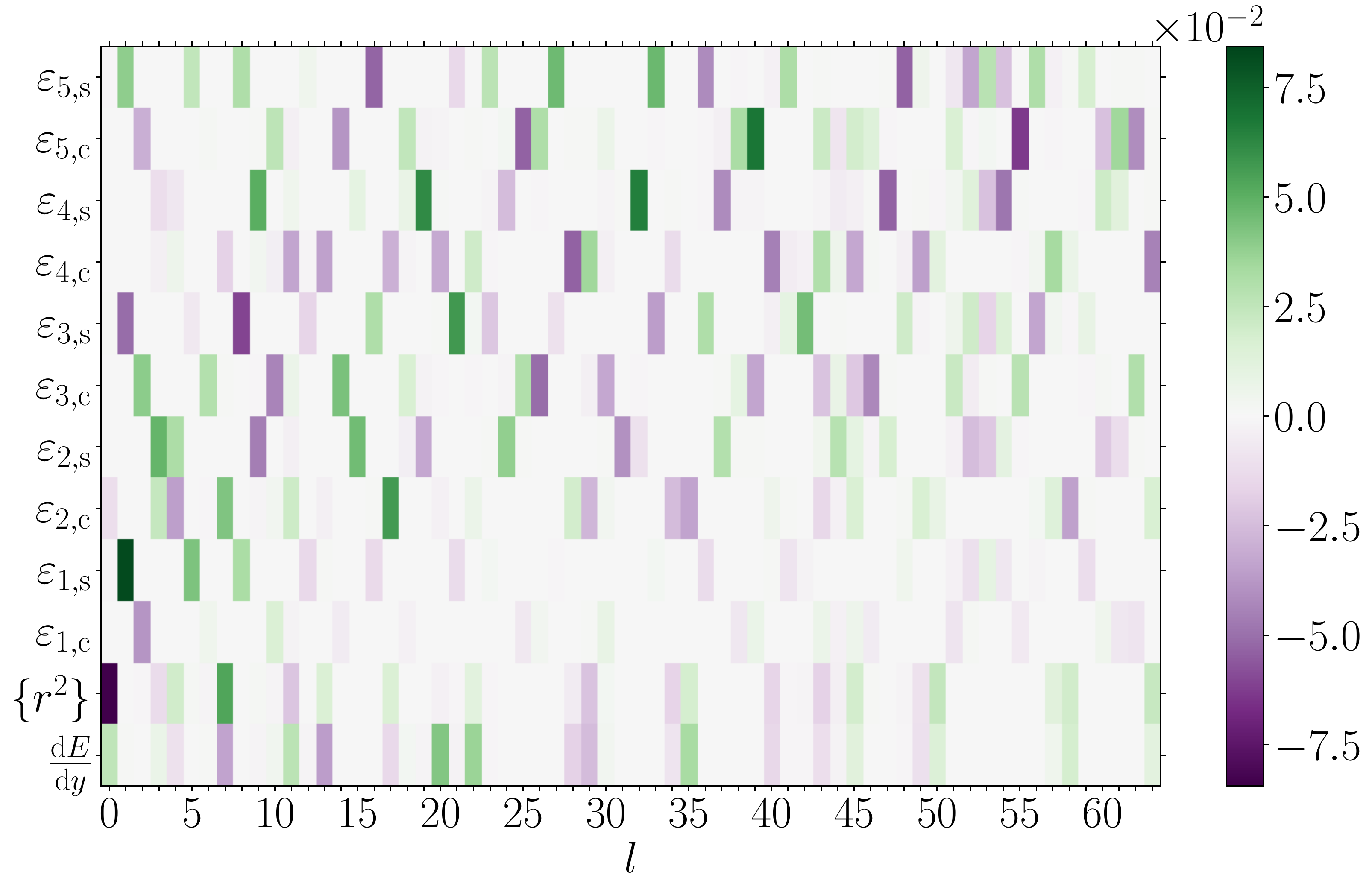}
\includegraphics[width=0.495\linewidth]{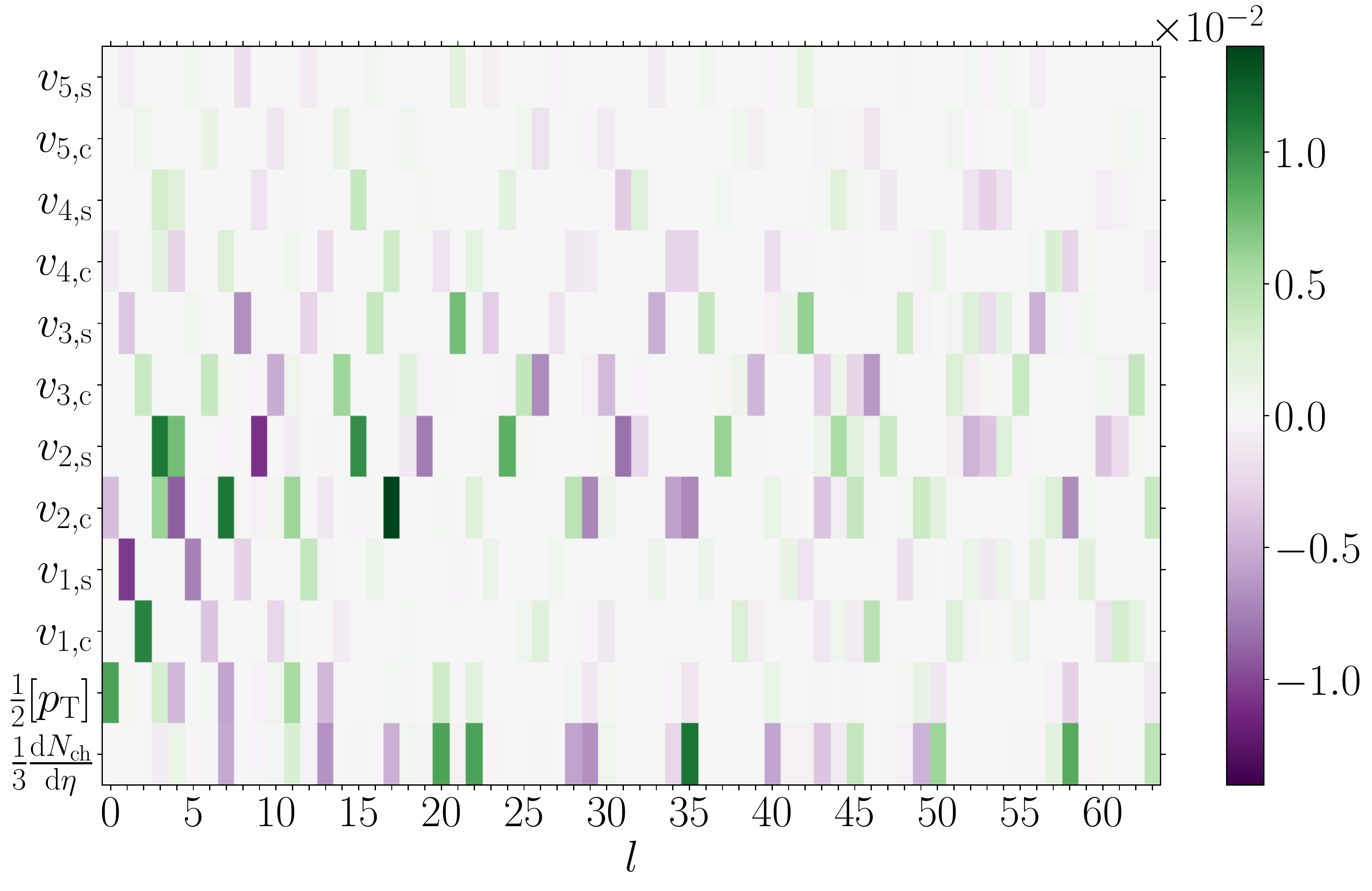}
\vspace{-5mm}
\caption{Linear-response coefficients $L_{\alpha,l}$ for the initial-state quantities (upper left) and final-state observables (upper right) at $b=0$ in the Saturation model. Bottom: same for $b=9$~fm. The dimensionful observables and multiplicity  have been normalized by $\bar{O}_\alpha$.}
\label{fig:L_IS_FS_Saturation}
\end{figure*}

\subsection{Linear- and quadratic-response coefficients}
\label{appendix:L_Q_Saturation}

\begin{figure*}[!htb]
\includegraphics[width=0.495\linewidth]{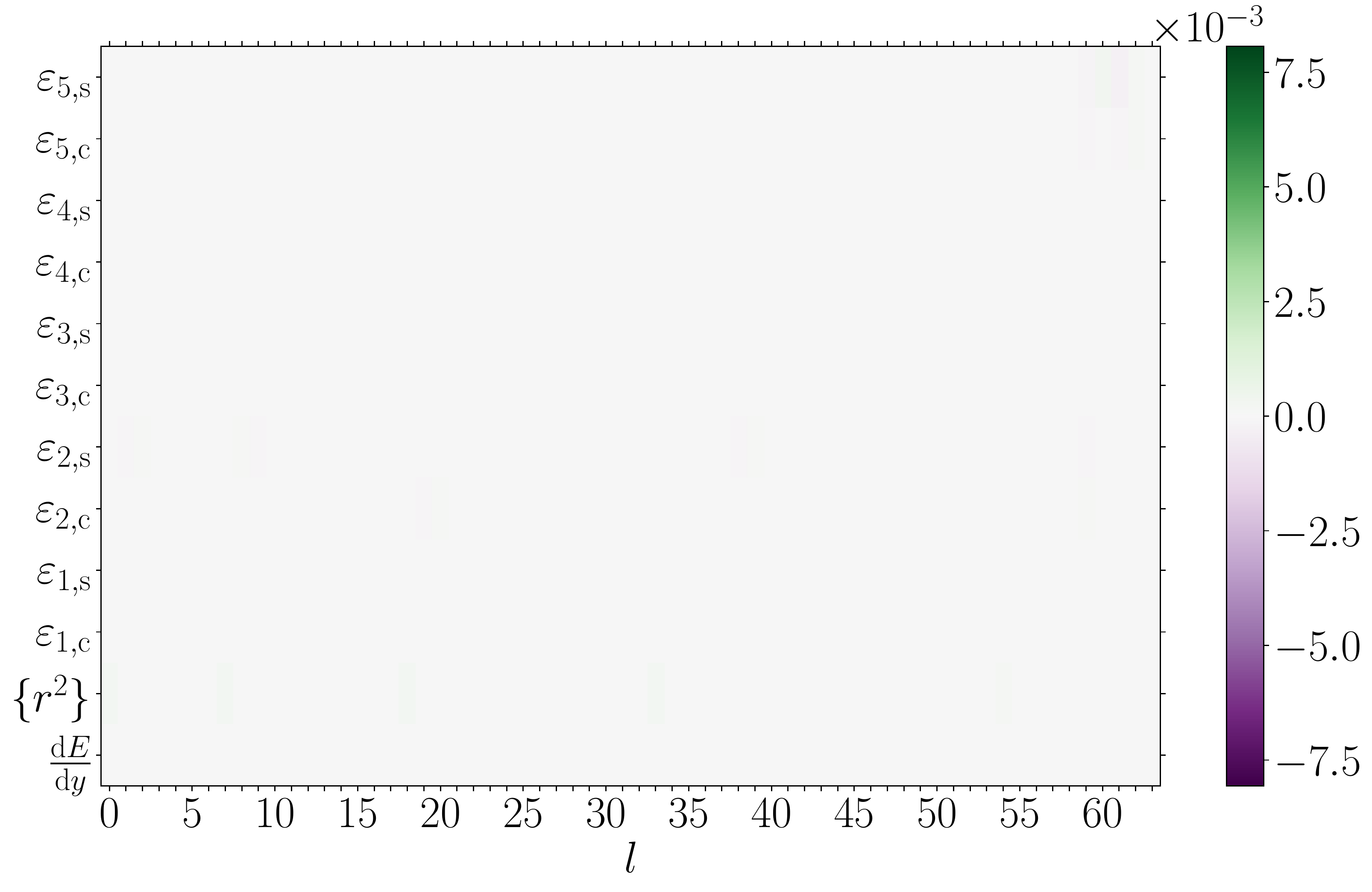}
\includegraphics[width=0.495\linewidth]{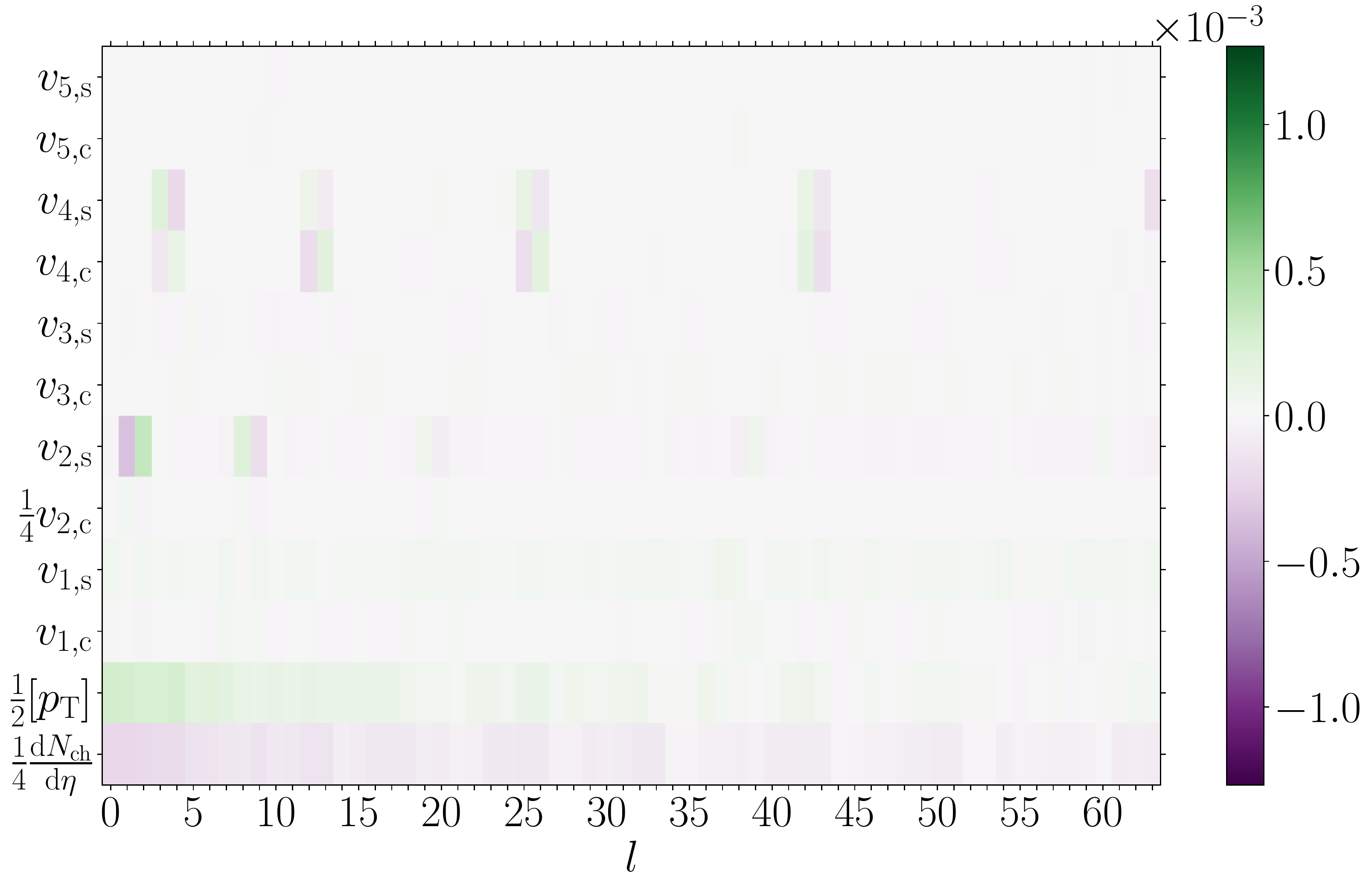}
\includegraphics[width=0.495\linewidth]{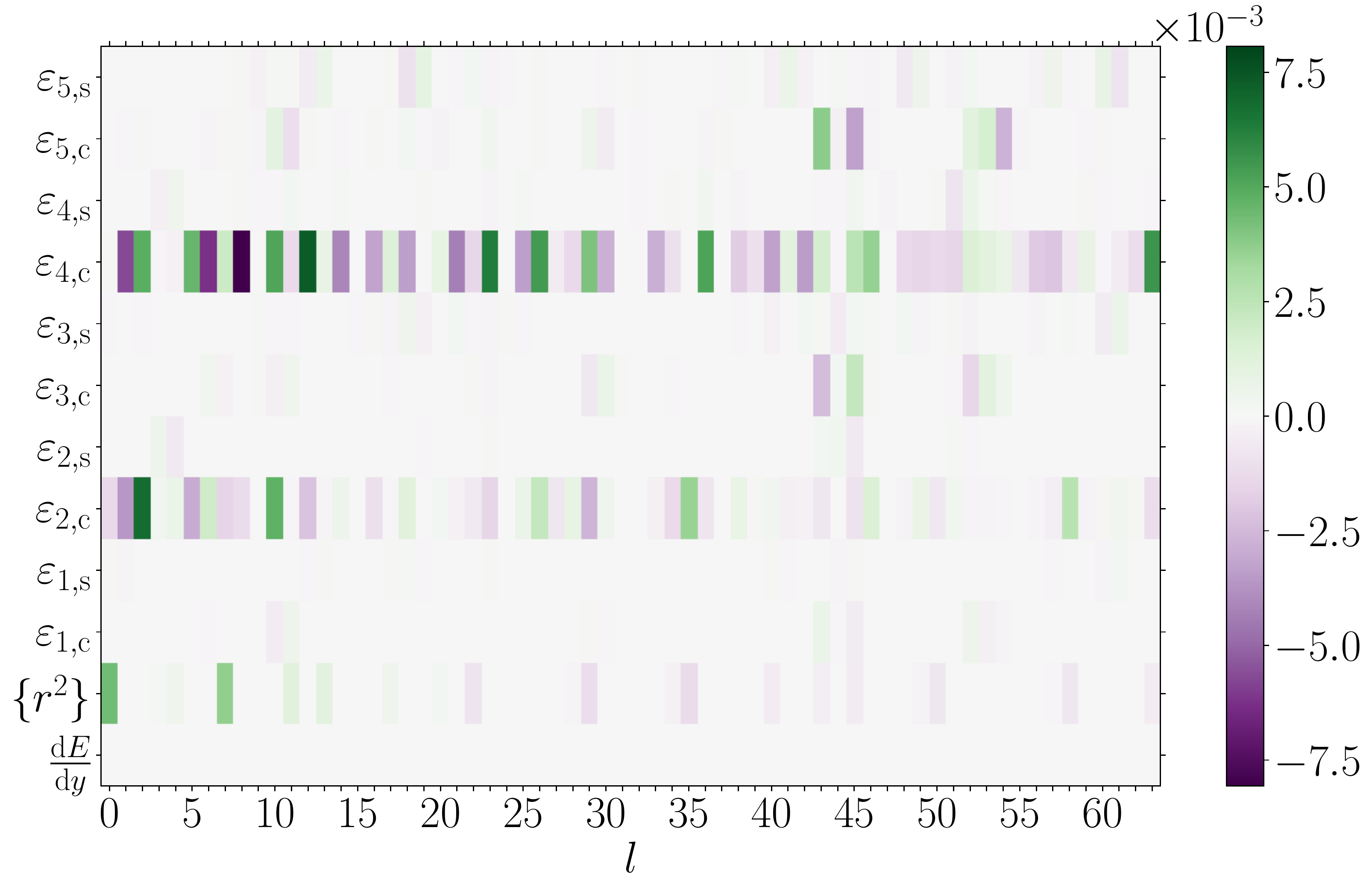}
\includegraphics[width=0.495\linewidth]{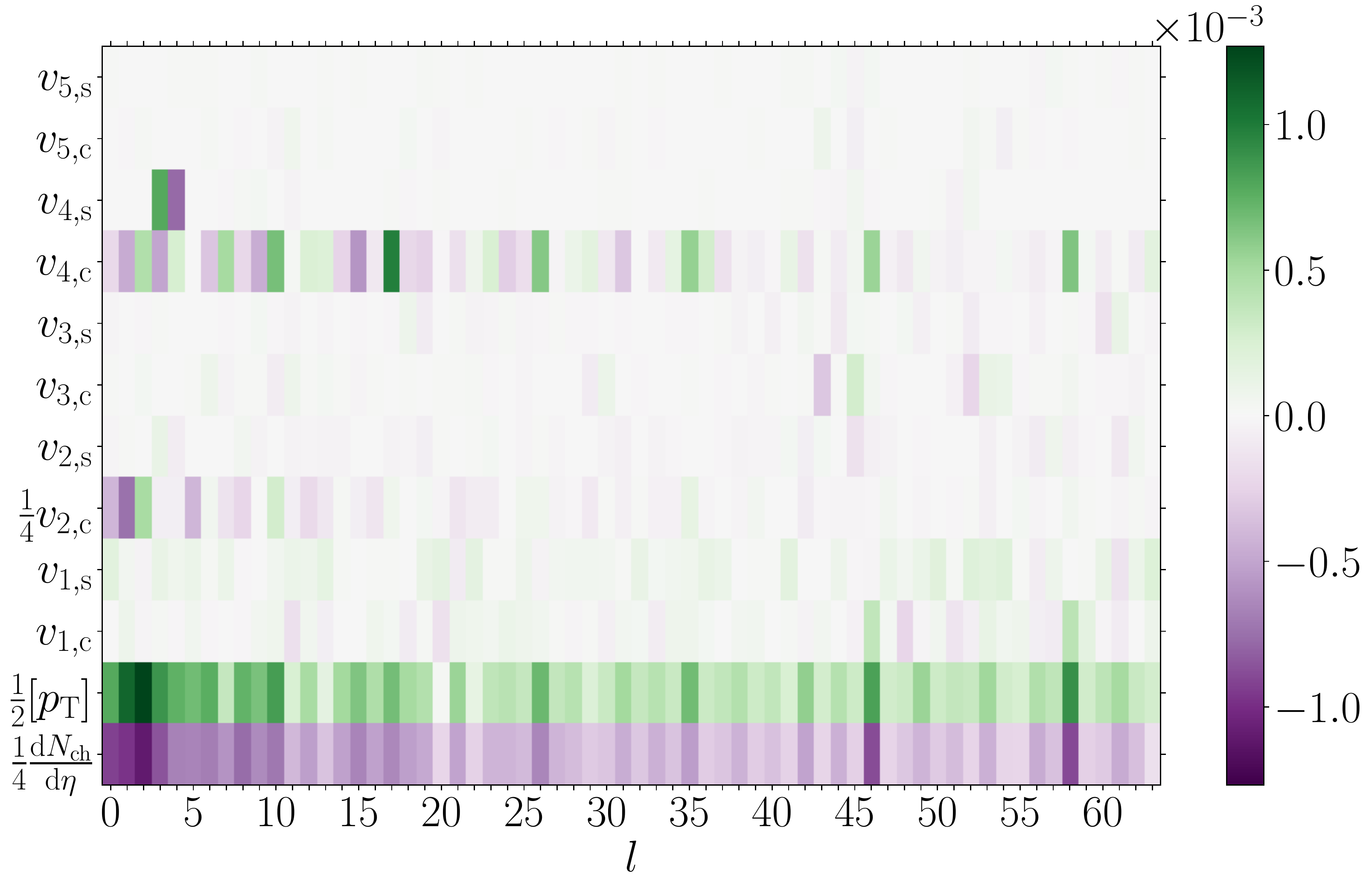}
\vspace{-5mm}
\caption{Quadratic-response coefficients $Q_{\alpha,ll}$ for the initial-state quantities (upper left) and final-state observables (upper right) at $b=0$ in the Saturation model. Bottom: same for $b=9$~fm. The dimensionful observables and multiplicity  have been normalized by $\bar{O}_\alpha$.}
\label{fig:Q_IS_FS_Saturation}
\end{figure*}

Figures~\ref{fig:L_IS_FS_Saturation} and \ref{fig:Q_IS_FS_Saturation} show the linear- and quadratic-response coefficients, respectively, of observables [Eq.~\eqref{Oseries}] for the fluctuation modes within the Saturation model. 
The results are generally qualitatively similar to those within the Glauber model (Figs.~\ref{fig:L_IS_FS_Glauber} and \ref{fig:Q_IS_FS_Glauber}), up to an important exception: 
Here the linear coefficients $L_{\alpha, l}$ of energy density and charged multiplicity in collisions at vanishing impact parameter (Fig.~\ref{fig:L_IS_FS_Saturation}, top panels) have opposite signs, while they have the same sign for events with initial states from the Glauber model.
Alternatively, the charged multiplicity $\d N_\textrm{ch}/\d\eta$ and the average transverse momentum $[p_\textrm{T}]$ at $b=0$ are negatively correlated in the Saturation model, while they are positively correlated in the Glauber model. 

Regarding the quadratic-response coefficients, an eye-catching result is the presence of sizable $Q_{\alpha,ll}$ for $v_{4,\rm s}$ in the modes $l=3$ and 4 at $b=9$~fm (bottom right panel of Fig.~\ref{fig:Q_IS_FS_Saturation}). 
These coefficients are readily explained, when one realizes that these two modes have both an $\varepsilon_{2,\rm c}(\Psi_l)$ and a $\varepsilon_{2,\rm s}(\Psi_l)$, as seen either in the first row of Fig.~\ref{fig:60_modes_b9_Saturation} --- these are the two modes that do not have the $x$ or $y$ direction as symmetry axis --- or in the bottom left panel of Fig.~\ref{fig:L_IS_FS_Saturation}. 
Accordingly, $v_{4,\rm s}(\bar{\Psi}+c_l\Psi_l)$ receives a contribution in $c_l^2\varepsilon_{2,\rm c}(\Psi_l)_{}\varepsilon_{2,\rm s}(\Psi_l)$. 
Consistently, there are also linear contributions $\propto c_l\varepsilon_{2,\rm c}(\bar{\Psi})_{}\varepsilon_{2,\rm c/s}(\Psi_l)$ to $v_{4,\rm c/s}(\bar{\Psi}+c_l\Psi_l)$, visible as linear-response coefficients in the bottom right panel of Fig.~\ref{fig:L_IS_FS_Saturation} for the same modes.

\section{Fluctuations and correlations of observables in the Saturation model}
\label{appendix:co_variances_Saturation}

\begin{figure*}[!htb]
\includegraphics[width=0.495\linewidth]{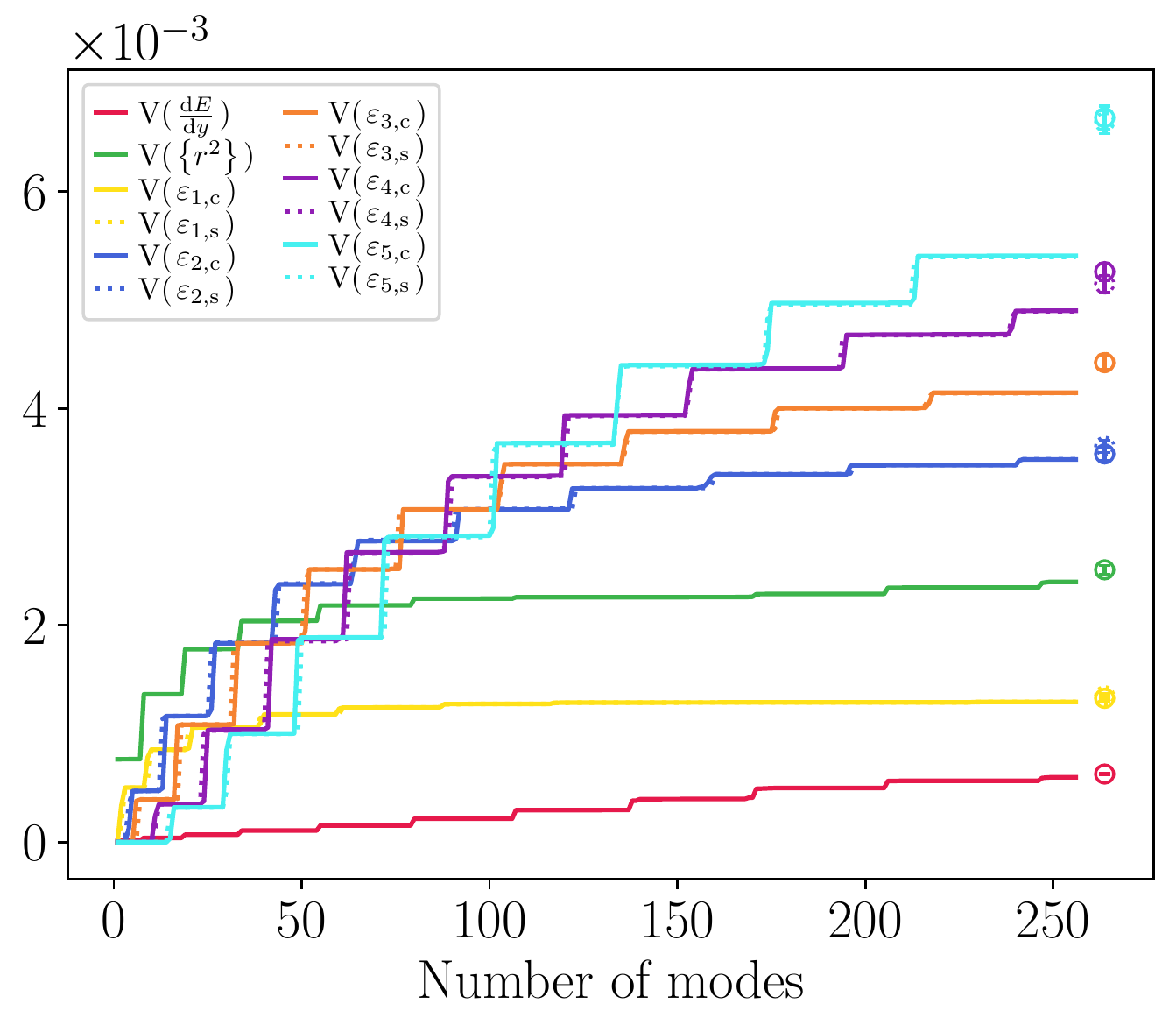}
\includegraphics[width=0.495\linewidth]{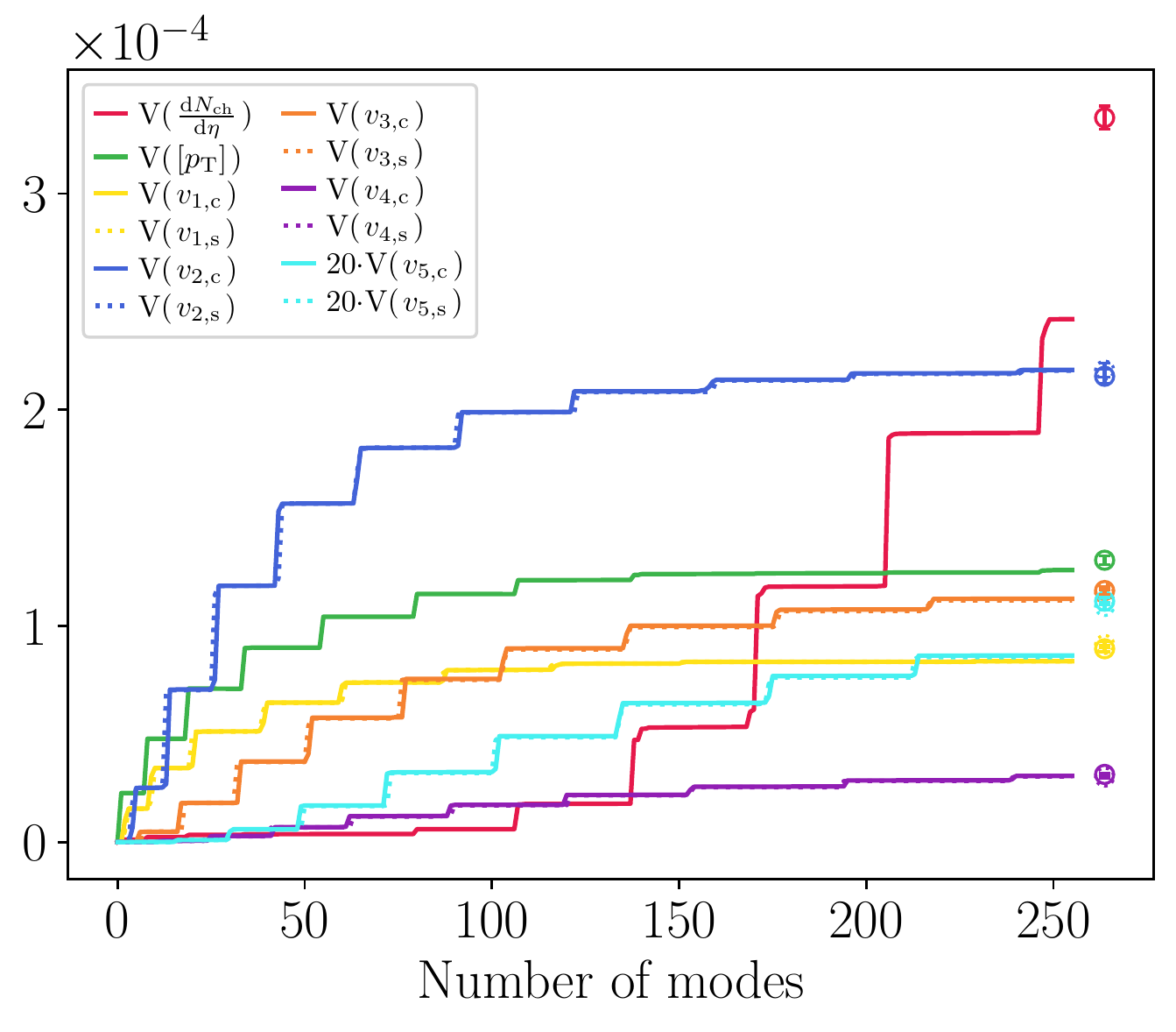}
\includegraphics[width=0.495\linewidth]{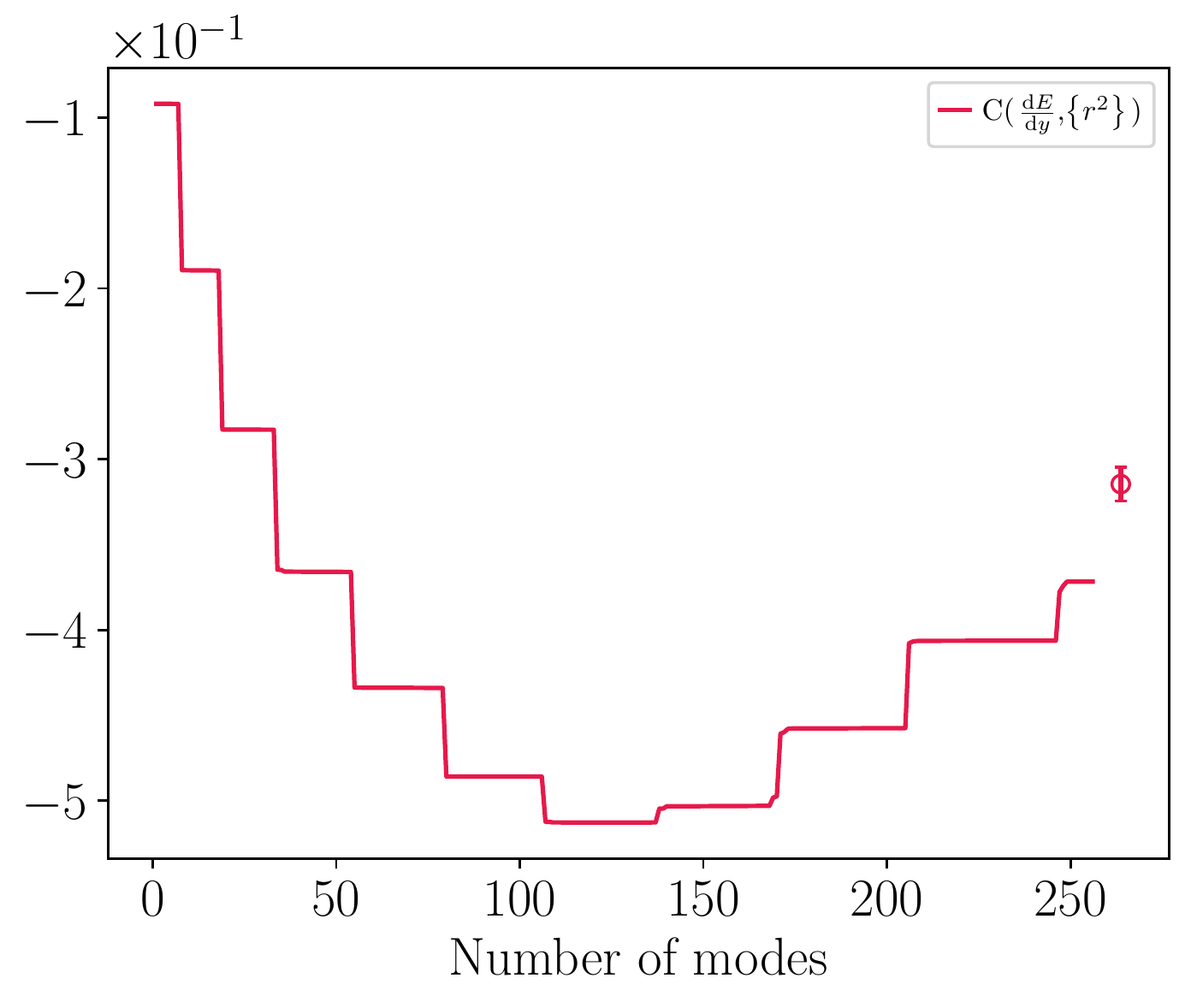}
\includegraphics[width=0.495\linewidth]{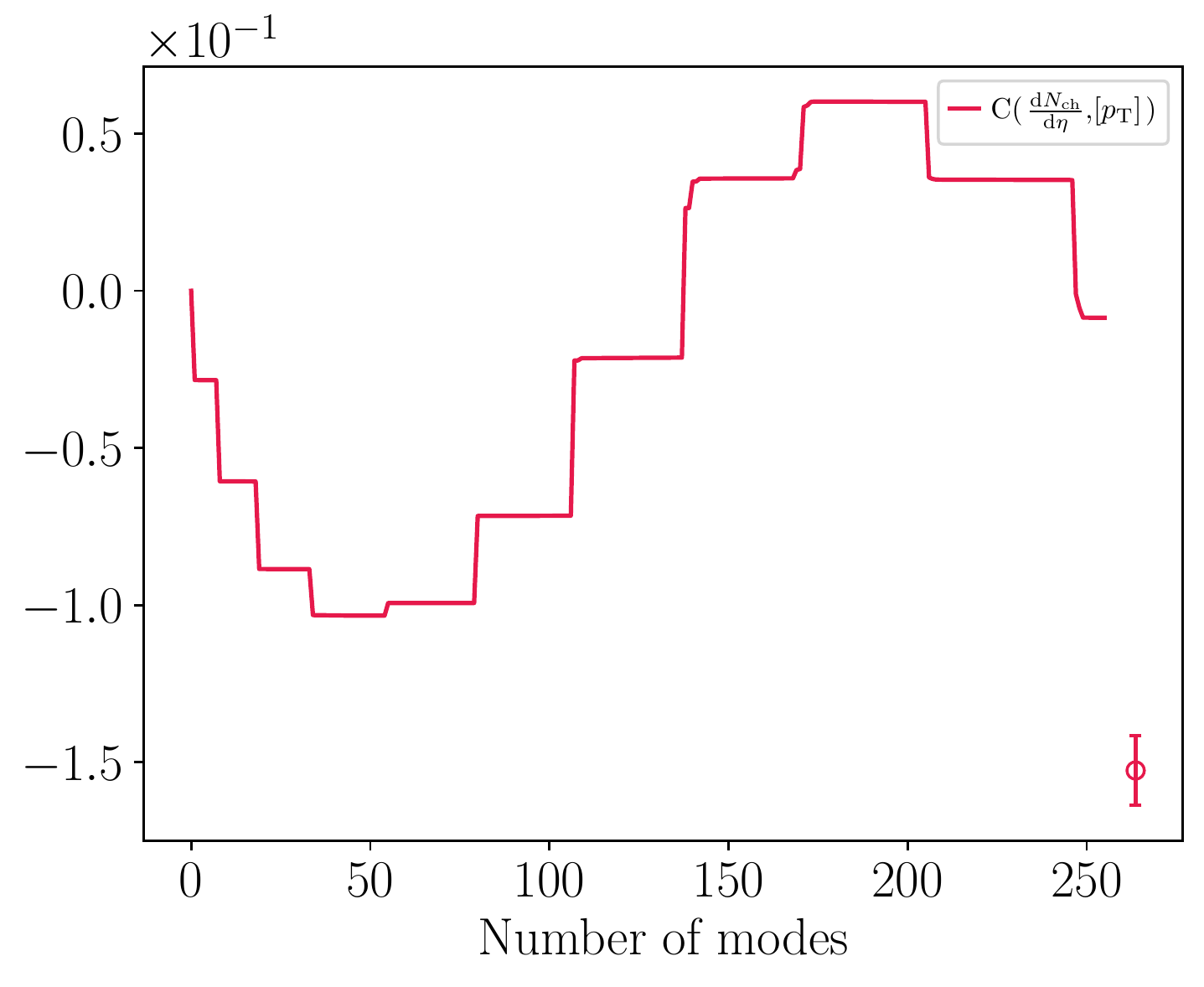}
\vspace{-7mm}
\caption{Variances (top) and correlation coefficients (bottom) of initial-state (left) and final-state observables (right) for collisions at $b=0$ in the Saturation model. 
The circles next to the right edge of each panel give the values computed from the random sample of 8192 events. 
The full lines show the quantities computed with Eqs.~\eqref{V(O_a)} and \eqref{eq:pearson_coeff}, including the number of modes given by the abscissa for the sums in V and the numerator of C. 
The sums in the denominator of C run over 256 modes.
The variances of $\d E/\d y$, $\{r^2\}$, $\d N_\textrm{ch}/\d\eta$, and $[p_\textrm{T}]$ are divided by the corresponding mean values.}
\label{fig:co_variances_convergence_Saturation_b0_IS_FS_modes_and_sampled}
\end{figure*}

\begin{figure*}[!htb]
\includegraphics[width=0.495\linewidth]{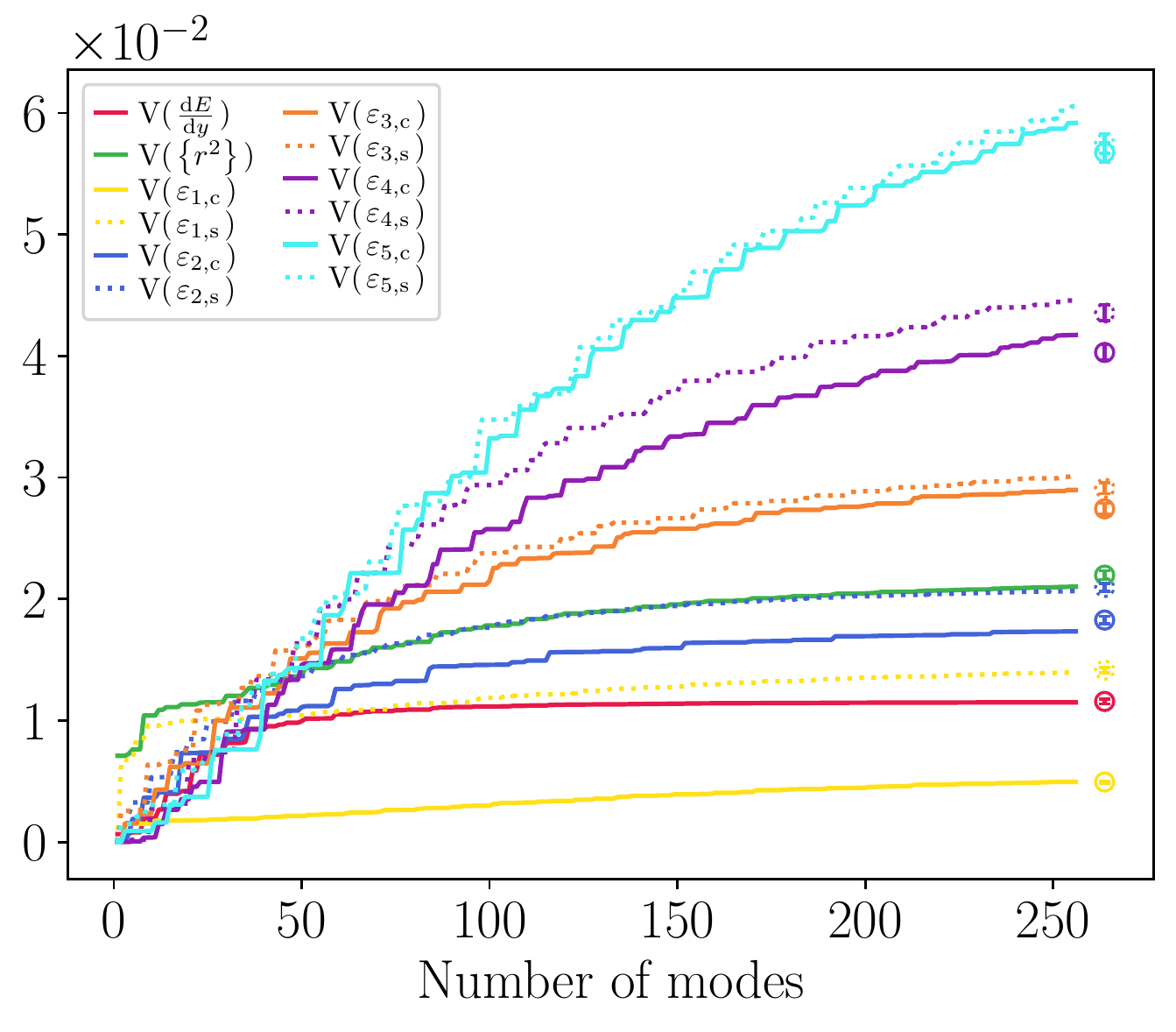}
\includegraphics[width=0.495\linewidth]{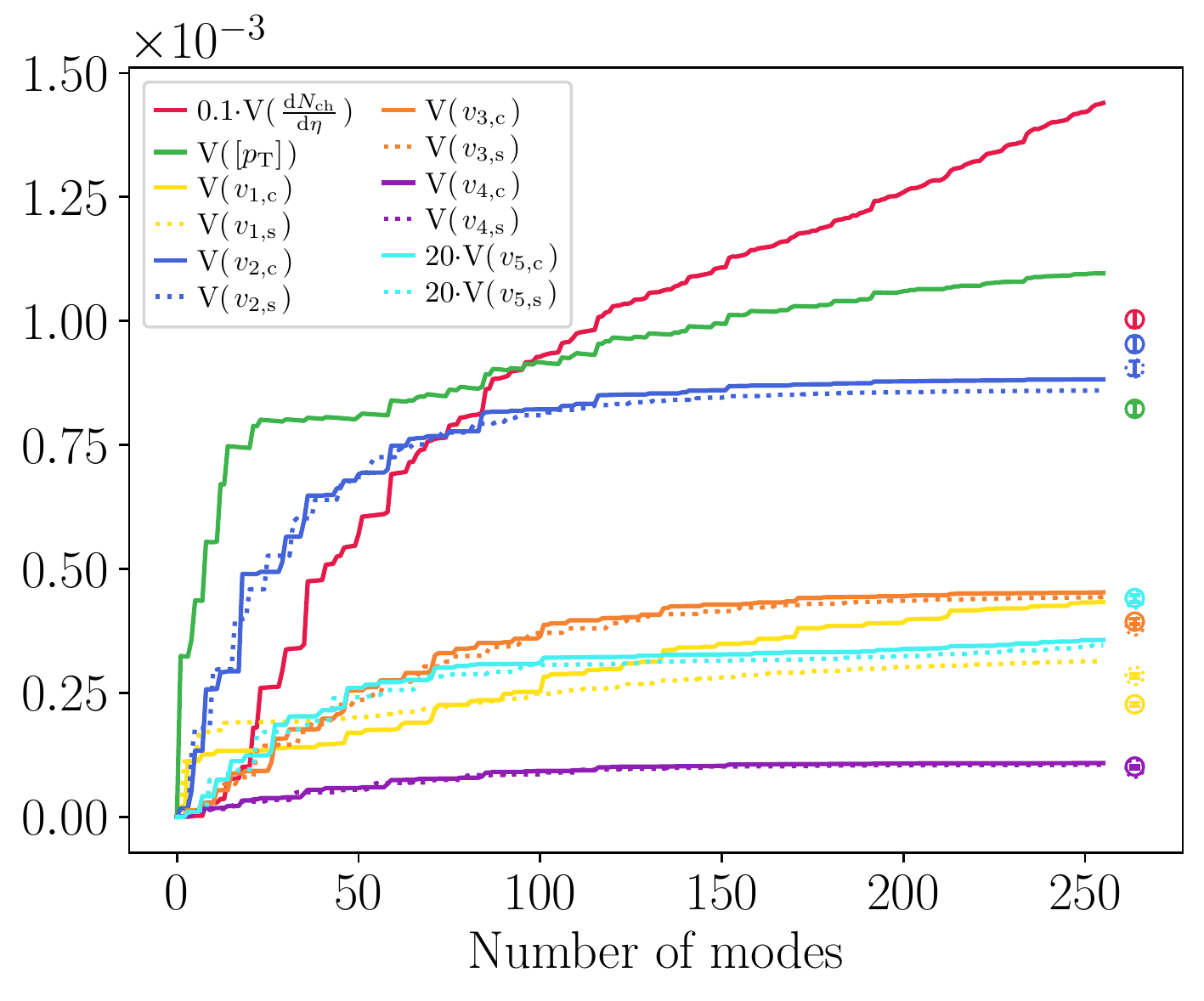}
\includegraphics[width=0.495\linewidth]{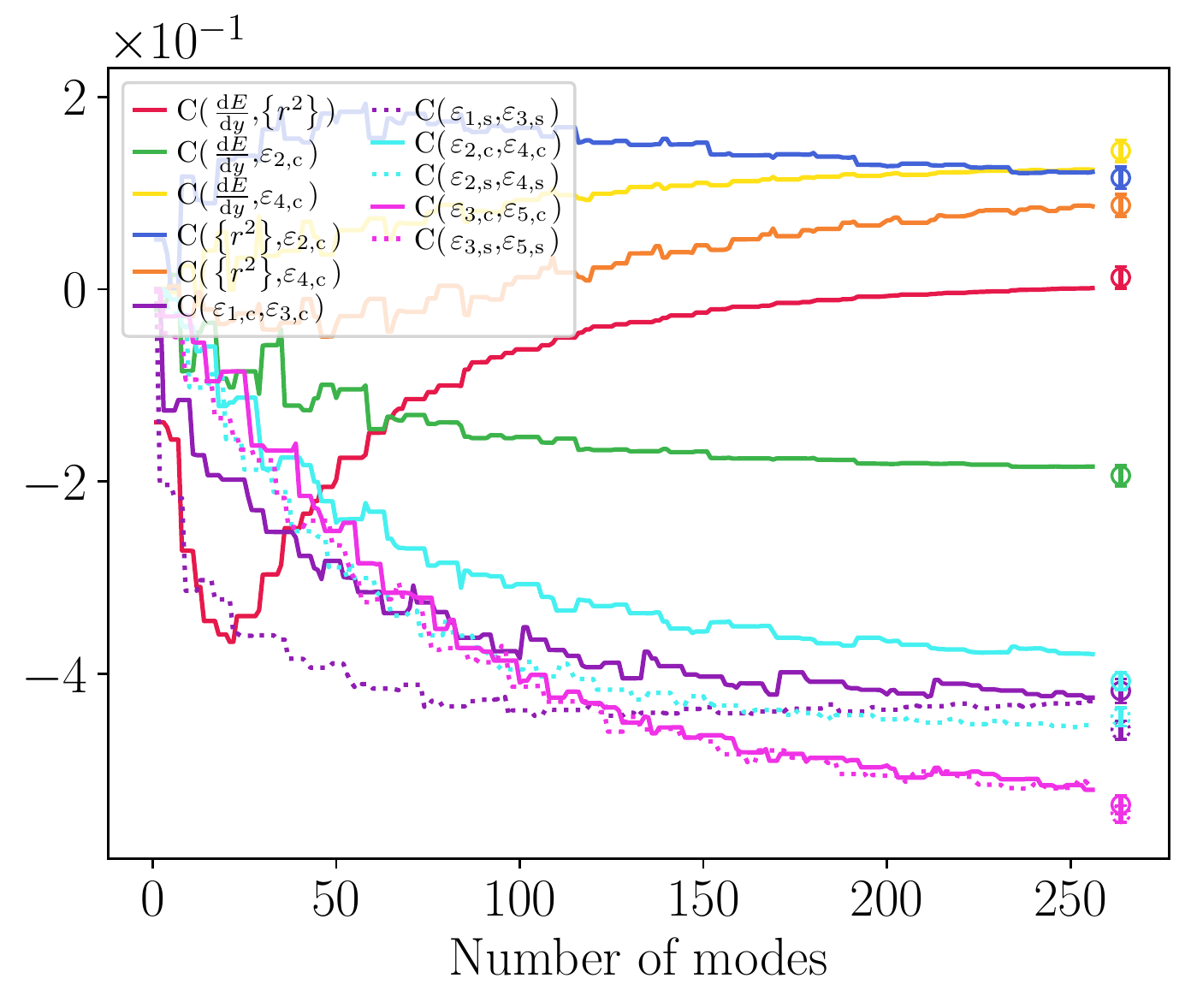}
\includegraphics[width=0.495\linewidth]{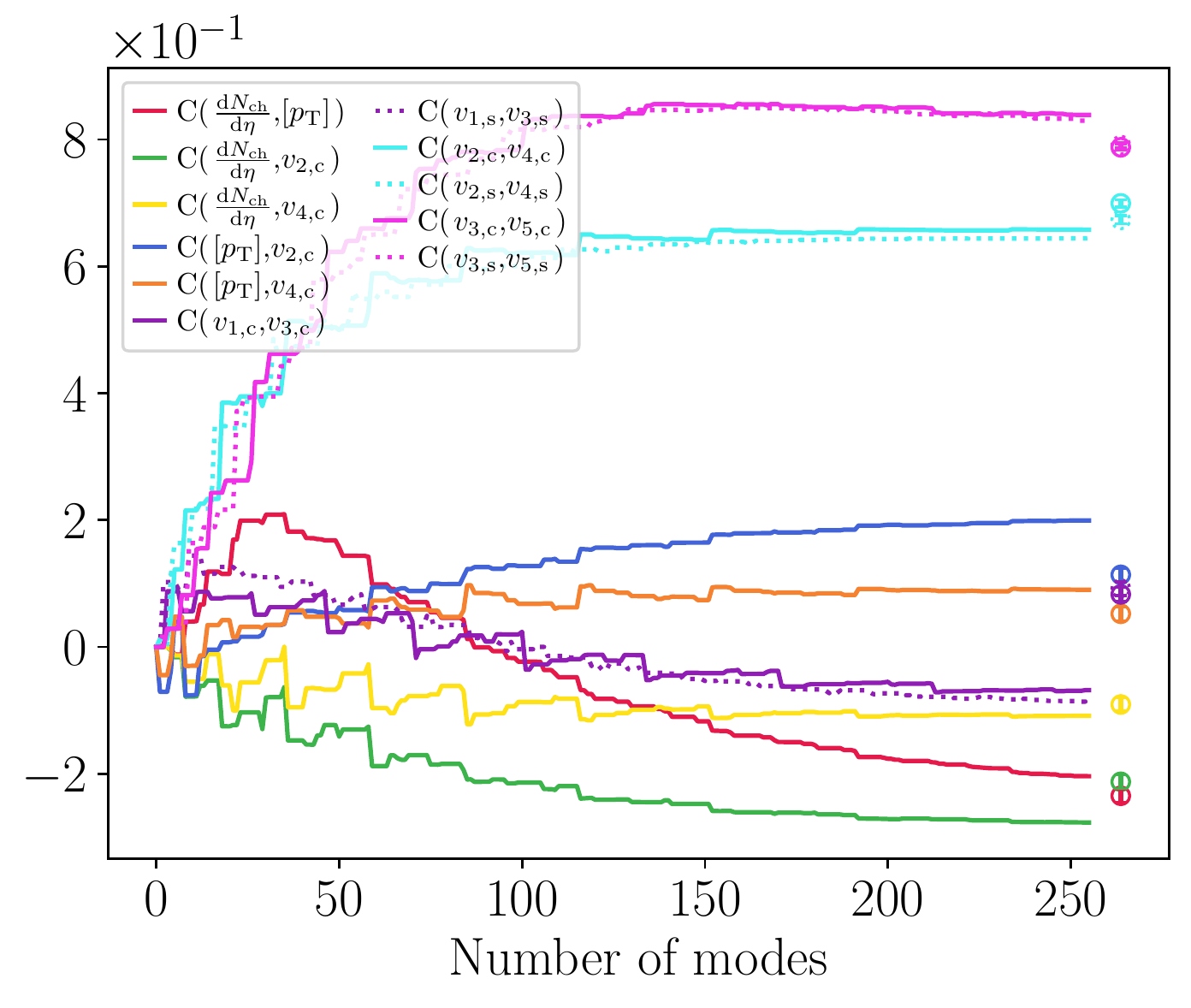}
\vspace{-7mm}
\caption{Same as Fig.~\ref{fig:co_variances_convergence_Saturation_b0_IS_FS_modes_and_sampled} for collisions at $b=9$~fm within the Saturation model.}
\label{fig:co_variances_convergence_Saturation_b9_IS_FS_modes_and_sampled}
\end{figure*}

In this appendix we present the variances and correlation coefficients of observables for events with initial states from the Saturation model, similar to Figs.~\ref{fig:co_variances_convergence_Glauber_b0_IS_FS_modes_and_sampled}--\ref{fig:covariances_Glauber_b9_IS_FS_modes_and_sampled}.
Figure~\ref{fig:co_variances_convergence_Saturation_b0_IS_FS_modes_and_sampled} is for collisions at $b=0$ and Fig.~\ref{fig:co_variances_convergence_Saturation_b9_IS_FS_modes_and_sampled} for events at $b=9$~fm.

As in Sec.~\ref{subsec:variances_covariances}, at $b=0$ the variances from the mode-by-mode approach tend towards the sample values obtained from 8192 random events for most observables except the charged multiplicity --- for which  modes with a high $l$ still yield large contributions to the variance --- and $\varepsilon_5$, which would necessitate a few more modes. 
Paralleling the nonconvergence of the variance $\d N_\textrm{ch}/\dd\eta$, its correlation coefficient with $[p_\textrm{T}]$ from the mode-by-mode approach is far from its sample value. 
In turn, at $b=9$~fm nonlinear effects spoil the agreement between mode-by-mode values and sample values. 

\begin{figure*}[!htb]
\includegraphics[width=0.495\linewidth]{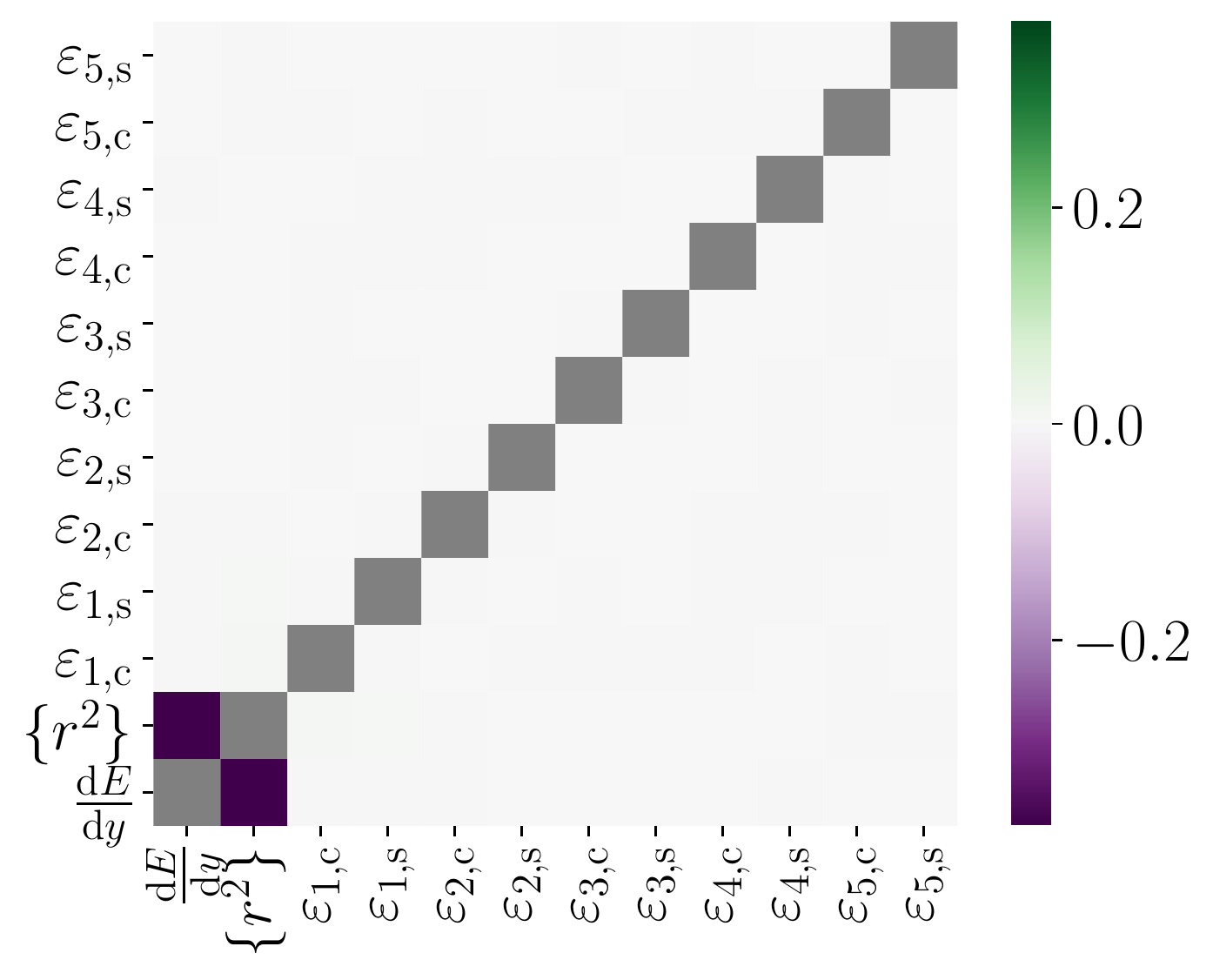}
\includegraphics[width=0.495\linewidth]{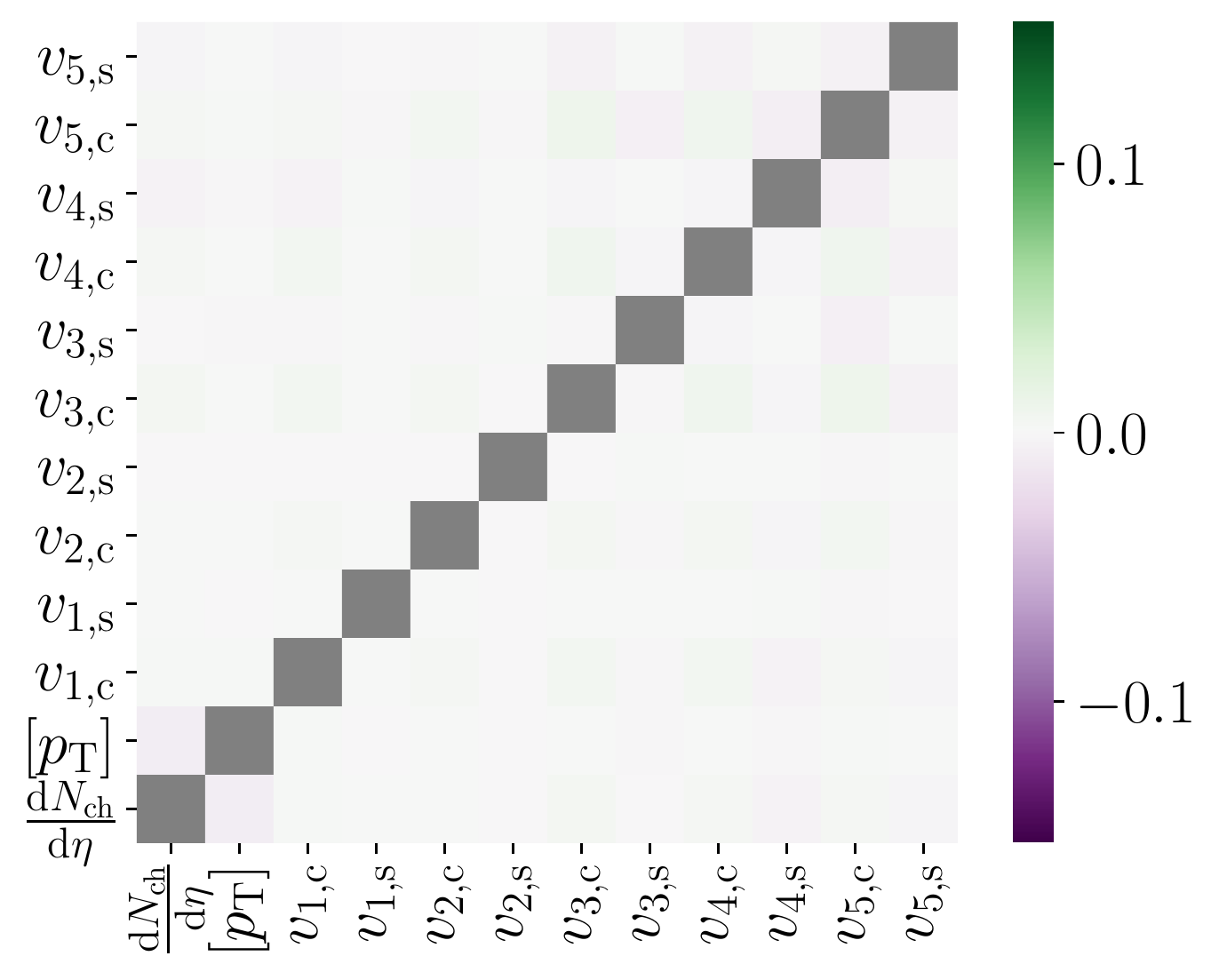}
\includegraphics[width=0.495\linewidth]{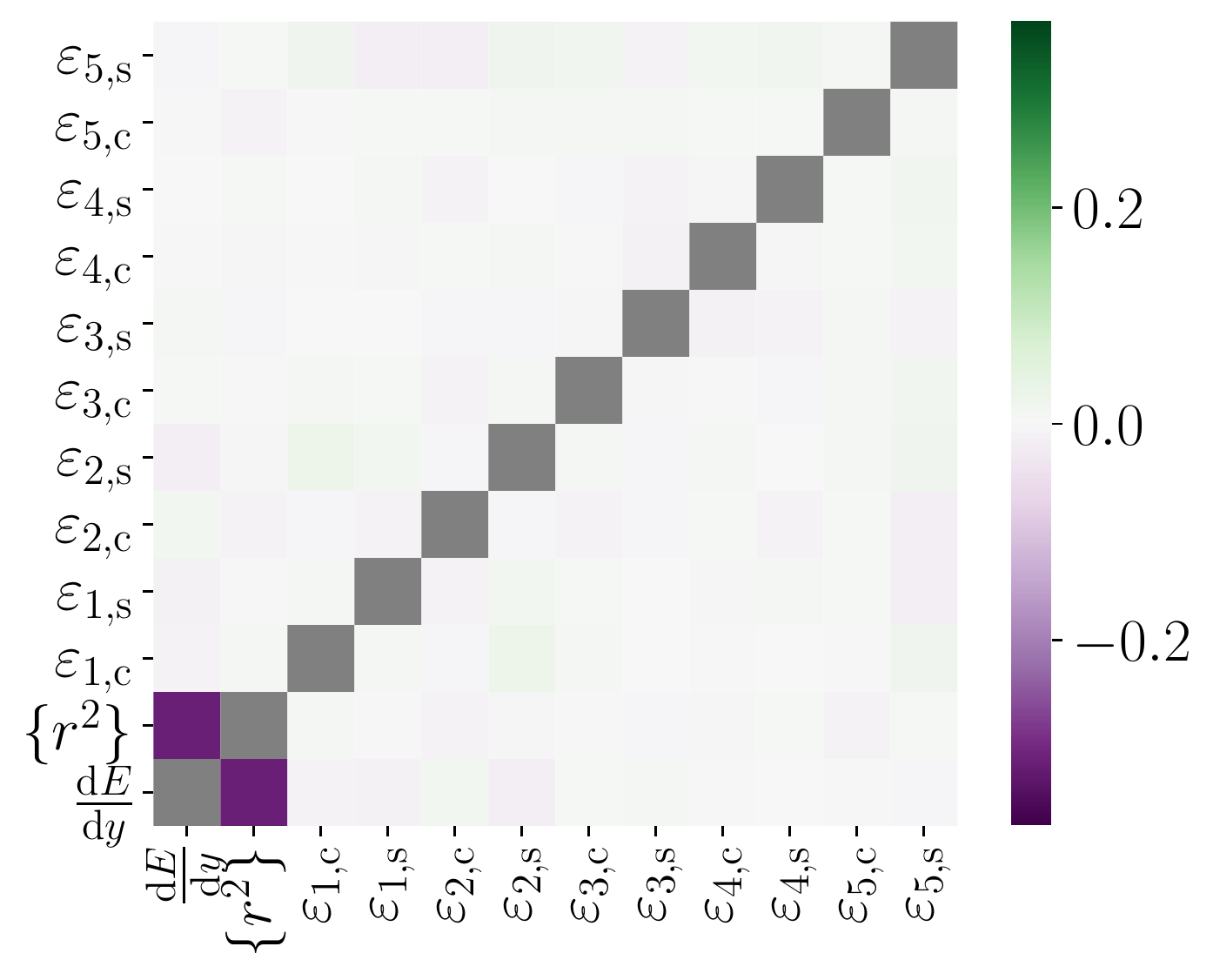}
\includegraphics[width=0.495\linewidth]{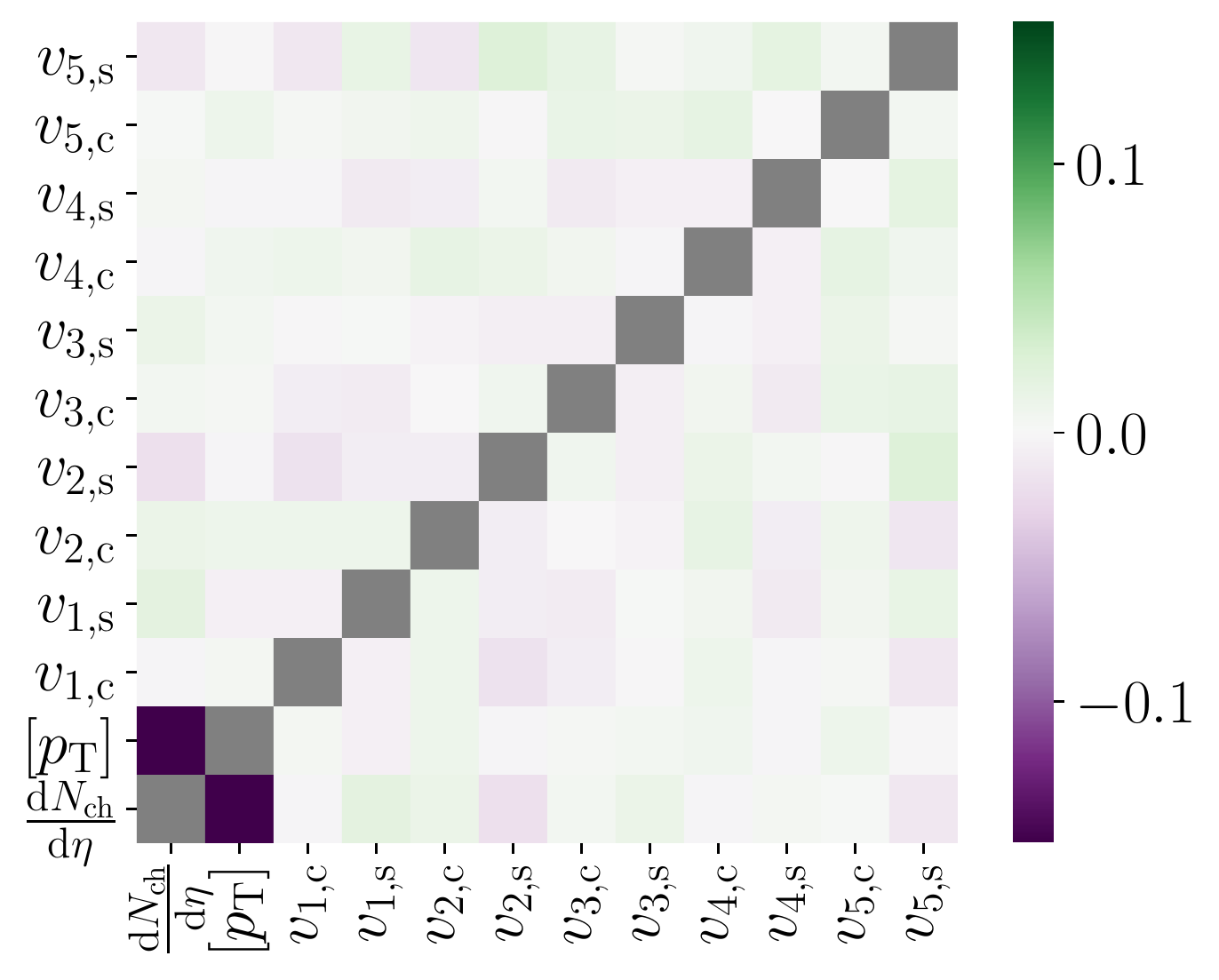}
\vspace{-5mm}
\caption{Correlation coefficients of initial-state (left) and final-state (right) observables for collisions at $b=0$ in the Saturation model.
Top: values computed with 256 modes in the mode-by-mode approach [Eq.~\eqref{eq:pearson_coeff}]. 
Bottom: values from the sample of 8192 random events. 
Values on the diagonal, which by definition equal one, are shown as gray squares.}
\label{fig:covariances_Saturation_b0_IS_FS_modes_and_sampled}
\end{figure*}

\begin{figure*}[!htb]
\includegraphics[width=0.495\linewidth]{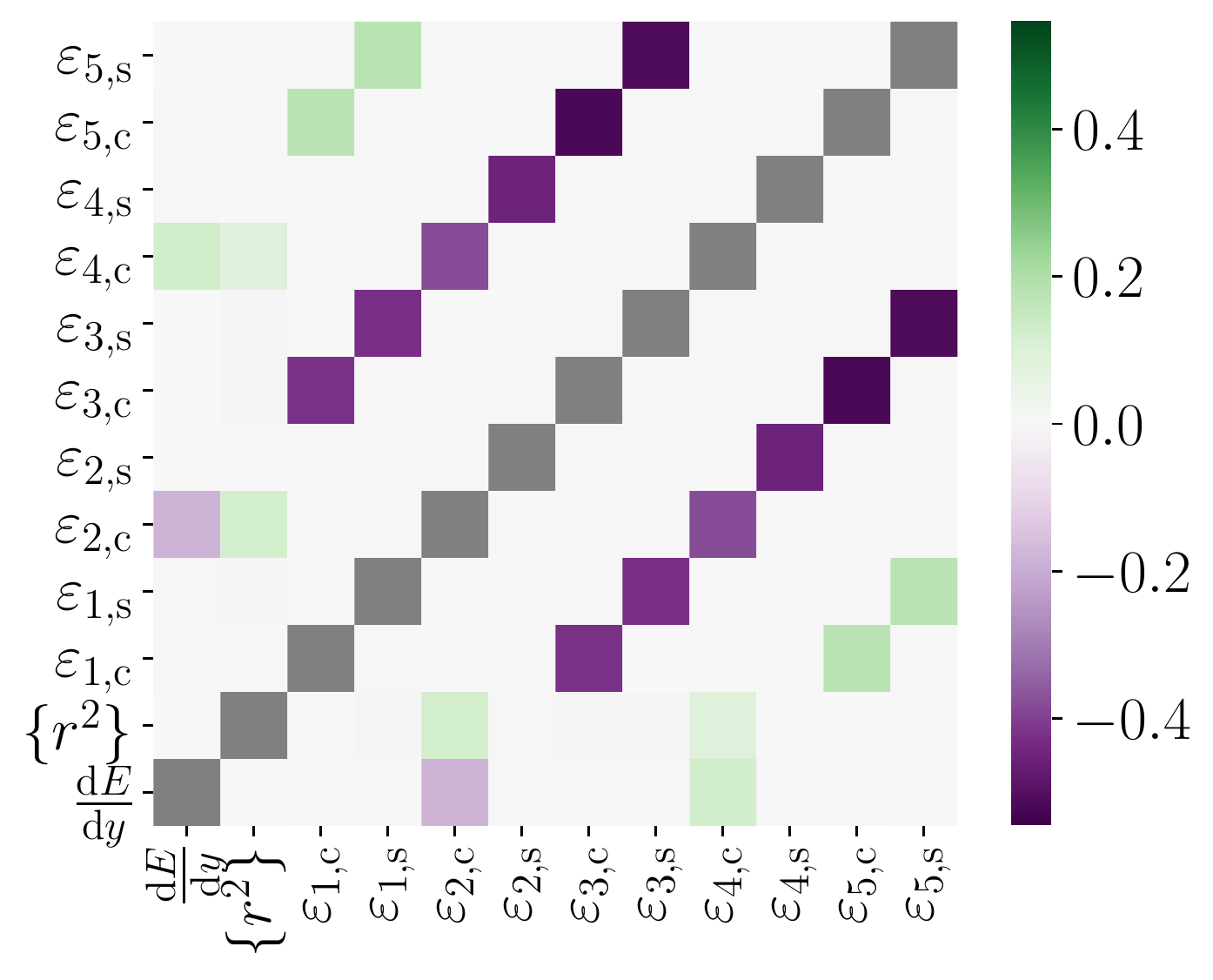}
\includegraphics[width=0.495\linewidth]{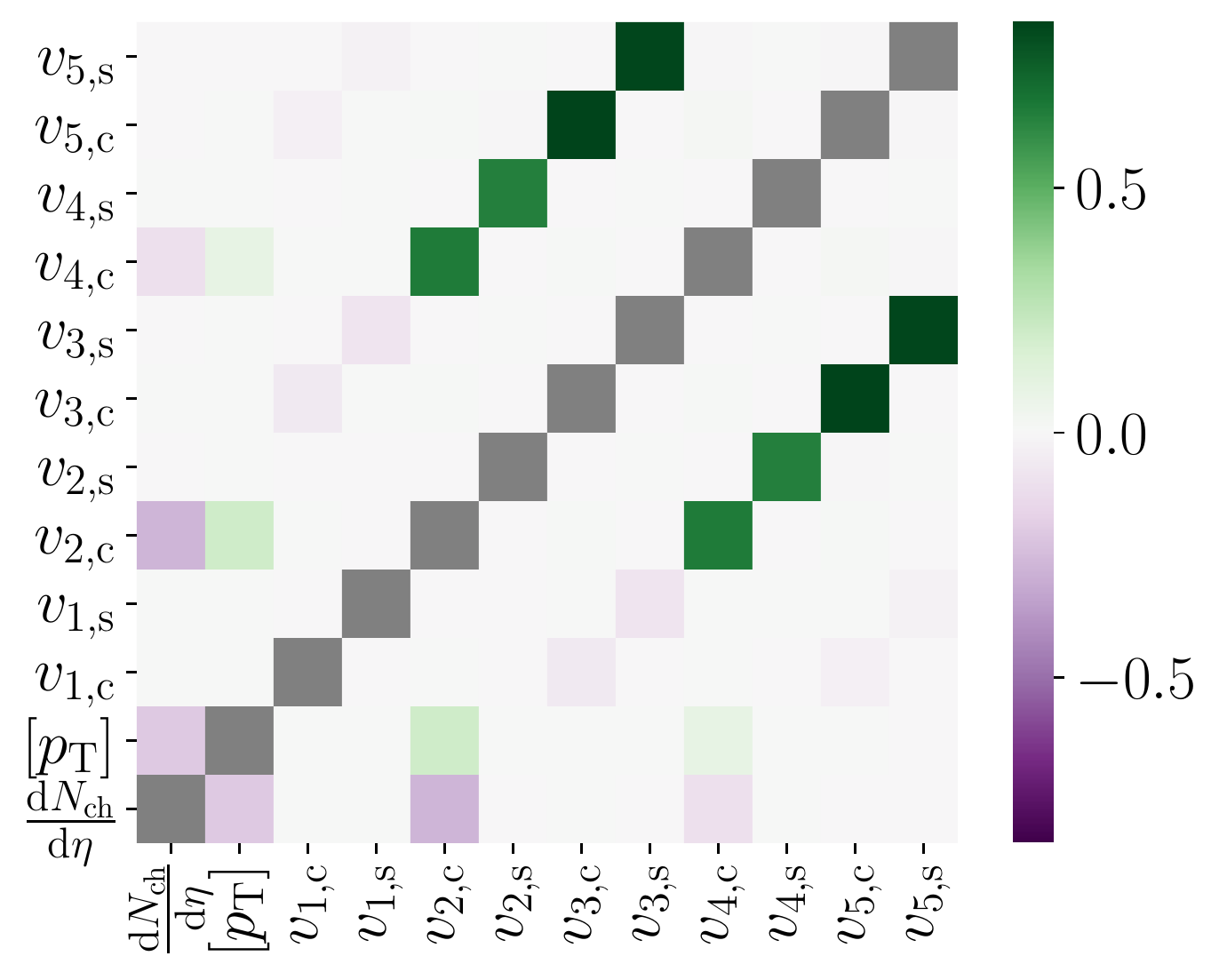}
\includegraphics[width=0.495\linewidth]{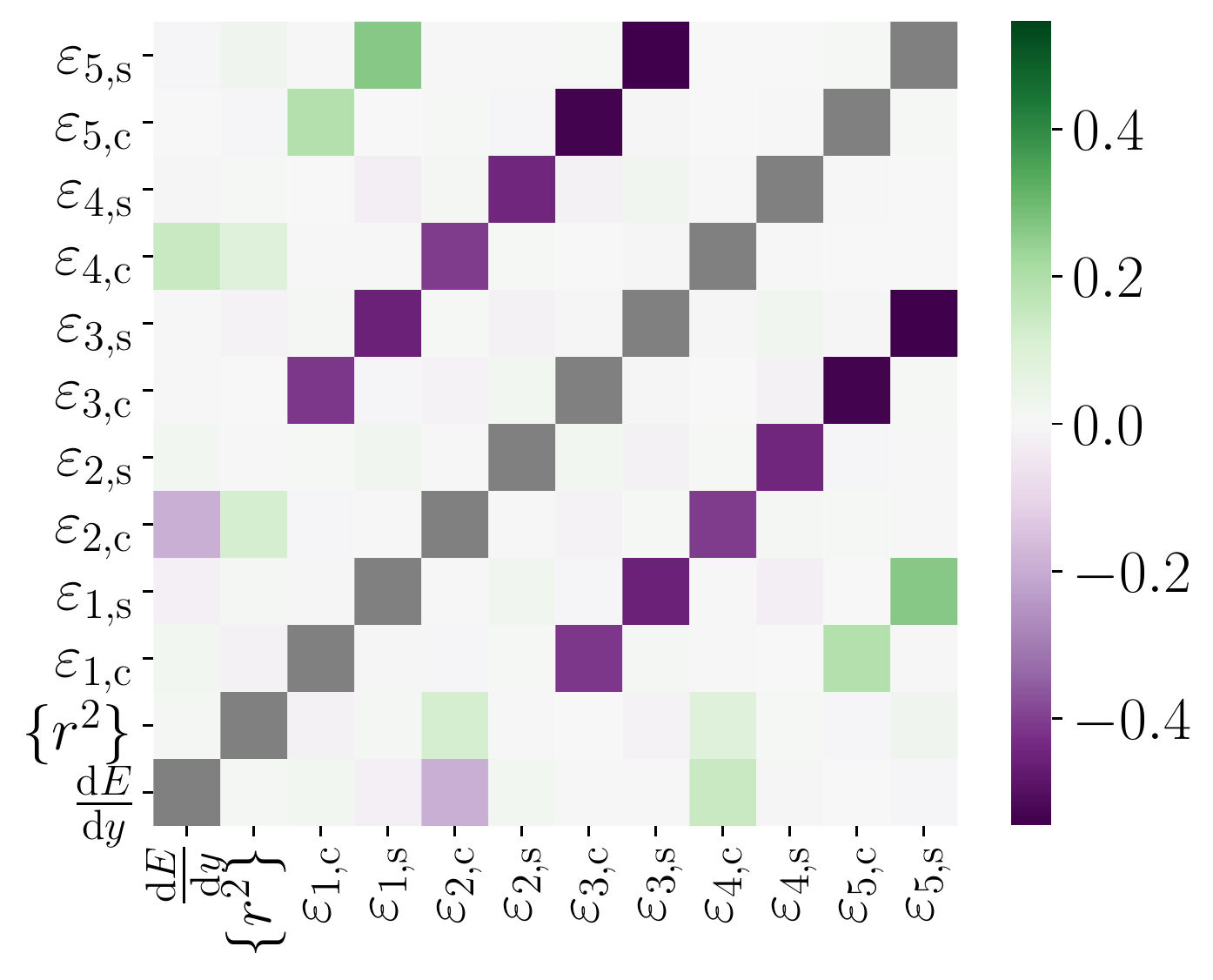}
\includegraphics[width=0.495\linewidth]{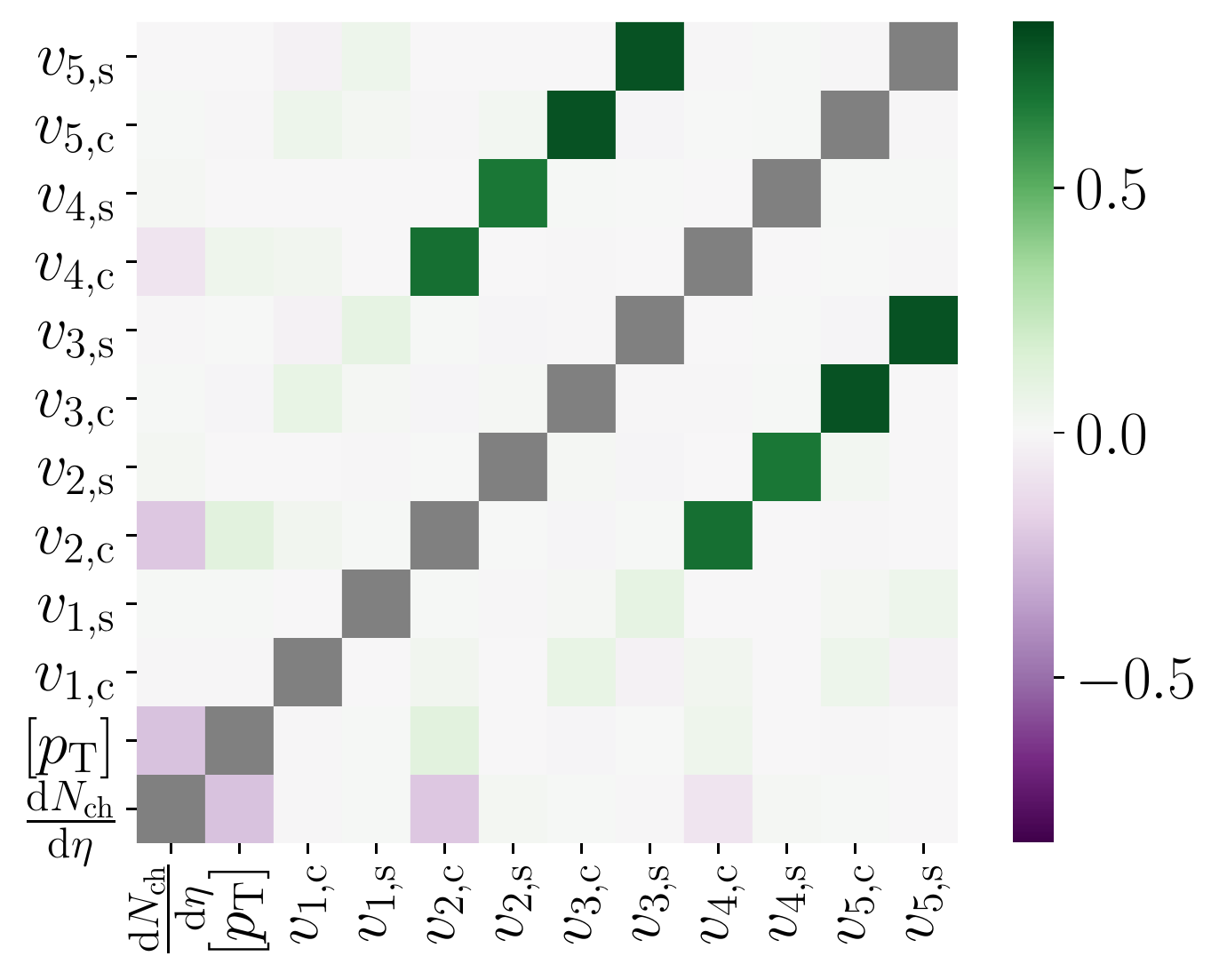}
\vspace{-5mm}
\caption{Same as Fig.~\ref{fig:covariances_Saturation_b0_IS_FS_modes_and_sampled} for collisions at $b=9$~fm in the Saturation model.}
\label{fig:covariances_Saturation_b9_IS_FS_modes_and_sampled}
\end{figure*}

In Figs.~\ref{fig:covariances_Saturation_b0_IS_FS_modes_and_sampled} and \ref{fig:covariances_Saturation_b9_IS_FS_modes_and_sampled} we show the values of the correlation coefficients, for events at $b=0$ and $b=9$~fm, respectively. 
For the initial-state observables (left panels) at $b=0$ and generally at $b=9$~fm the coefficients from the mode-by-mode approach computed with 256 modes (top panels) are in qualitative agreement with the sample values extracted from the 8192 random events (bottom panels). 
On the other hand, the large discrepancy between the mode-by-mode and sample values of $\textrm{C}(\d N_\textrm{ch}/\d\eta,[p_\textrm{T}])$ at $b=0$ is clearly reflected in the mismatch between the top and bottom panels on the right of Fig.~\ref{fig:covariances_Saturation_b0_IS_FS_modes_and_sampled}.
The difference comes from the limitation to order $c_l^2$ in the calculation of the mode-by-mode value. 
If we momentarily assume that all expansion coefficients $\{c_l\}$ are Gaussian-distributed and that they are statistically independent, one can compute all moments of their distributions.
Under this assumption, one finds that at order $c_l^4$ the covariance of two observables $O_\alpha$, $O_\beta$ receives a contribution proportional to the sum over $l$ of $Q_{\alpha,ll} Q_{\beta,ll}$.\footnote{There are also a contribution proportional to the sum of $Q_{\alpha,ll'} Q_{\beta,ll'}$ over pairs $l\neq l'$ as well as terms coming from pushing Eq.~\eqref{Oseries} to order $c_l^4$, which we did not attempt to estimate.}
Now for both models and all modes, the quadratic-response coefficients for $\d N_\textrm{ch}/\d\eta$ and $[p_\textrm{T}]$ have opposite sign, so that this contribution to $\textrm{C}(\d N_\textrm{ch}/\d\eta,[p_\textrm{T}])$ is negative.
In the Glauber model this contribution is much smaller than that in the numerator of Eq.~\eqref{eq:pearson_coeff} from the linear coefficients. 
But in the Saturation model they are actually of similar magnitude, which at least partly explains the strong difference observed in Figs.~\ref{fig:co_variances_convergence_Saturation_b0_IS_FS_modes_and_sampled} or~\ref{fig:covariances_Saturation_b0_IS_FS_modes_and_sampled}.

\section{Probability distributions of observables}
\label{appendix:probability_distributions}

\subsection{Probability distributions in the Gaussian fluctuation approximation}
\label{app:probability_distributions_Gfa}

In this appendix we detail the derivation of the results presented in Sec.~\ref{section:gaussian_statistics} on the probability distributions of observables in the linearized Gaussian approximation. 
Assuming that the expansion coefficients $c_l$ follow a centered Gaussian distribution with unit variance [Eq.~\eqref{eq:p(c_l)_Gaussian}] and that the response of an observable $O_\alpha$ to initial-state fluctuations is linear, the probability distribution for $O_\alpha$ is approximately given by
\begin{equation}
p^\textrm{G}_{\alpha}(O_\alpha) = \frac{1}{(2\pi)^{l_\textrm{max}/2}}
\int\!\D c\,\prod_l\textrm{e}^{-c_l^2/2}\, \delta\bigg(O_\alpha-\bar{O}_\alpha-\sum_l L_{\alpha,l}c_l\bigg),
\end{equation}
where $l_\textrm{max}$ is the number of fluctuation modes entering a practical calculation. 
Introducing a Fourier representation of the delta-distribution
\begin{equation}
\delta(x) = \int\!\frac{\d s}{2\pi}\,\textrm{e}^{\textrm{i}sx}\;,
\end{equation}
we can recast the integral over every $c_l$ into a complex Gaussian form
\begin{equation}
p^\textrm{G}_{\alpha}(O_\alpha) = 
\frac{1}{(2\pi)^{l_\textrm{max}/2}}\int\!\frac{\d s}{2\pi} \textrm{e}^{\textrm{i}s(O_\alpha -\bar{O}_\alpha )} 
\prod_l \int\!\d c_l\,\exp\bigg(\!\!-\!\frac{c_l^2}{2} - \textrm{i}_{}sL_{\alpha,l}c_l\bigg)
\end{equation}
that is then easily computed:
\begin{equation}
p^\textrm{G}_{\alpha}(O_\alpha) = 
\int\frac{\d s}{2\pi} \exp\bigg[\textrm{i}s(O_\alpha-\bar{O}_\alpha) - \frac{s^2 \sum_lL_{\alpha,l}^2}{2} \bigg] = 
\frac{1}{\sqrt{2\pi C_{\alpha\alpha}}} \exp\bigg[\!-\!\frac{(O_\alpha-\bar{O}_\alpha)^2}{2C_{\alpha\alpha}}\bigg]\;,
\end{equation}
where in the last step we introduced the diagonal coefficients $\alpha=\beta$ of the covariance matrix
\begin{equation}
\label{eq:C_ab_def}
C_{\alpha\beta} \equiv \sum_l L_{\alpha,l}L_{\beta,l}.
\end{equation}

This calculation is readily extended to that of the joint probability distribution of $d$ observables $O_{\alpha_1}$, \dots, $O_{\alpha_d}$.
Starting from 
\begin{equation}
p^\textrm{G}_{\vec{\alpha}}(\{O_{\alpha_k}\}) = 
\frac{1}{(2\pi)^{l_\textrm{max}/2}}\!\int\!\D c\,\prod_l\textrm{e}^{-c_l^2/2}\,
\prod_k
\delta\bigg(O_{\alpha_k} - \bar{O}_{\alpha_k} - \sum_l L_{\alpha_k,l}c_l\bigg)
\end{equation}
one finds
\begin{align}
p^\textrm{G}_{\vec{\alpha}}(\{O_{\alpha_k}\}) &= \frac{1}{(2\pi)^{l_\textrm{max}/2}}
\prod_k\!\int\!\frac{\d s_k}{2\pi}\,\textrm{e}^{\textrm{i}s_k(O_{\alpha_k} - \bar{O}_{\alpha_k})}
\prod_l \!\int\!\d c_l\,\exp\bigg(\!-\!\frac{c_l^2}{2} - \textrm{i} s_k L_{\alpha_k,l}c_l \bigg)  \cr
&= \prod_k\!\int\!\!\frac{\d s_k}{2\pi} \exp\bigg[ \textrm{i} s_k\big(O_{\alpha_k}-\bar{O}_{\alpha_k}\big)
-\frac{\sum_{j}s_j\left(\Sigma_{\vec{\alpha}\,}\right)_{jk}s_k}{2}\bigg],
\end{align}
where we have introduced the matrix
\begin{equation}
\left(\Sigma_{\vec{\alpha}}\right)_{jk} \equiv \sum_l L_{\alpha_j,l} L_{\alpha_k,l}.
\end{equation}
Eventually, one obtains
\begin{equation}
p^\textrm{G}_{\vec{\alpha}}(\{O_{\alpha_k}\}) = \frac{1}{\sqrt{(2\pi)^d \det (\Sigma_{\vec{\alpha}})}}
\exp\bigg[\!-\!\frac{1}{2}\sum_{i,j}\big({O}_{{\alpha_i}}-{\bar{O}}_{{\alpha_i}}\big) \left(\Sigma_{{\vec{\alpha}}}^{-1}\right)_{ij} \big({O}_{{\alpha_j}}-{\bar{O}}_{{\alpha_j}}\big)\bigg],
\label{eq:joint_gaussian_prob_dist}
\end{equation}
which shows that $\Sigma_{\vec{\alpha}}$ is in fact the covariance of the $k$ observables, whose entries we shall denote
\begin{equation}
(\Sigma_{\vec{\alpha}})_{ij} \equiv C_{\alpha_i\alpha_j}.
\end{equation}

To be able to perform further calculations without specifying the covariances, we will make use of Cramer's rule for the general expression of the $(i,j)$ entry of the inverse of a $n\times n$ matrix $A$:
\begin{equation}
(A^{-1})_{ij} = \frac{(-1)^{i+j}}{\det (A)} \det(A_{(j,i)}),
\end{equation}
with $A_{(j,i)}$ the $(n-1)\times (n-1)$ matrix obtained by deleting the $j$-th row and $i$-th column from $A$. 
Using this result for the elements of $\Sigma_{\vec{\alpha}}^{-1}$ in Eq.~\eqref{eq:joint_gaussian_prob_dist}, we find
\begin{equation}
p^\textrm{G}_{\vec{\alpha}}(\{O_{\alpha_k}\})  = \frac{1}{\sqrt{(2\pi)^d \det (\Sigma_{{\vec{\alpha}}})}}
\exp\bigg[\!-\!\frac{\sum_{i,j}(-1)^{i+j}(O_{\alpha_i}-\bar{O}_{\alpha_i})(O_{\alpha_j}-\bar{O}_{\alpha_j})\det (\Sigma_{{\vec{\alpha}},(j,i)})}{2 \det (\Sigma_{{\vec{\alpha}}})}\bigg],
\label{eq:joint_gaussian_prob_dist_cramer}
\end{equation}
where the sum in the numerator of the exponent runs over $i$ and $j$ between 1 and $k$. 

Let us now compute the probability distribution $p_\beta$ of an observable $O_\beta = \sqrt{O_{\alpha_1}^2+O_{\alpha_2}^2}$ with Gaussian distributed $O_{\alpha_1}$ and $O_{\alpha_2}$. 
The starting point is simply
\begin{equation}
p_\beta^\textrm{G.f.a.}(O_\beta) = \int\!\d O_{\alpha_1} \,\d O_{\alpha_2}\, p^\textrm{G}_{\alpha_1,\alpha_2}(O_{\alpha_1},O_{\alpha_2})\,
\delta\Big(O_\beta - \sqrt{O_{\alpha_1}^2+O_{\alpha_2}^2}\Big),
\end{equation}
with $p^\textrm{G}_{\alpha_1,\alpha_2}$ the (Gaussian) joint probability distribution of $O_{\alpha_1}$ and $O_{\alpha_2}$, while the superscript ``G.f.a.'' stands for ``Gaussian fluctuation approximation.'' 
To tackle the integral, which in principle runs over the whole two-dimensional plane spanned by $O_{\alpha_1}$ and $O_{\alpha_2}$, we switch to polar coordinates such that $O_{\alpha_1} = \rho\cos \phi$ and $O_{\alpha_2} = \rho\sin\phi$.
One can then perform the radial integral with the delta distribution:
\begin{equation}
p_\beta^\textrm{G.f.a.}(O_\beta) = 
\int_0^{2\pi}\!\d\phi \int_0^\infty \d\rho\,\rho\,p^\textrm{G}_{\alpha_1,\alpha_2}(\rho\cos\phi,\rho\sin\phi)\, \delta(O_\beta - \rho) =
\Theta(O_\beta)\,O_\beta \int_0^{2\pi}\!\d\phi\,p^\textrm{G}_{\alpha_1,\alpha_2}(O_\beta \cos\phi,O_\beta \sin\phi)
\label{eq:pObeta_angular_integral}
\end{equation}
with $\Theta$ the Heaviside step function.
Analogously, if we need the joint probability distribution of $O_\beta$ and $O_\gamma=\sqrt{O_{\alpha_3}^2+O_{\alpha_4}^2}$, where $O_{\alpha_3}$ and $O_{\alpha_4}$ are two Gaussian distributed observables, we can perform two substitutions with polar coordinates --- in the planes $(O_{\alpha_1},O_{\alpha_2})$ and $(O_{\alpha_3},O_{\alpha_4})$ ---, which yield
\begin{equation}
p_{\beta,\gamma}^\textrm{G.f.a.}(O_\beta,O_\gamma) =
\Theta(O_\beta)\,O_\beta\,\Theta(O_\gamma)\,O_\gamma \int_0^{2\pi}\!\d\phi \int_0^{2\pi}\!\d\psi\,
p^\textrm{G}_{\alpha_1,\alpha_2,\alpha_3,\alpha_4}(O_\beta\cos\phi, O_\beta\sin\phi, O_\gamma\cos\psi, O_\gamma\sin\psi).
\end{equation}
However, the remaining angular integrals are generally highly nontrivial, because the joint probability distribution $p^\textrm{G}_{\vec{\alpha}}$, even if it is Gaussian, still contains the correlation between the observables $\{O_{\alpha_j}\}$, which enter as covariances in the argument of the exponential function. 

In the case of the single-variable probability distribution $p_\beta^\textrm{G.f.a.}$, with only one angular integration, one can make further analytical progress with Eq.~\eqref{eq:pObeta_angular_integral} as follows. 
Inserting expression~\eqref{eq:joint_gaussian_prob_dist_cramer} of $p^\textrm{G}_{\alpha_1,\alpha_2}$ into Eq.~\eqref{eq:pObeta_angular_integral} yields
\begin{align*}
p_\beta^\textrm{G.f.a.}(O_\beta) = \frac{\Theta(O_\beta)~O_\beta}{\sqrt{\det (\Sigma_{\alpha_1,\alpha_2})}}\int_0^{2\pi} \frac{\d \phi}{2\pi} 
&\exp\left[-\frac{C_{\alpha_2 \alpha_2 } (O_\beta \cos \phi - \bar{O}_{\alpha_1})^2+C_{\alpha_1\alpha_1}(O_\beta \sin \phi - \bar{O}_{\alpha_2})^2}{2\det (\Sigma_{\alpha_1,\alpha_2})}\right]\\
&\times 
\exp\left[-\frac{2C_{\alpha_1\alpha_2}(O_\beta \cos \phi - \bar{O}_{\alpha_1})(O_\beta \sin \phi - \bar{O}_{\alpha_2})}{2\det (\Sigma_{\alpha_1,\alpha_2})}\right].
\end{align*}
In the numerator of the argument of the exponential functions, we gather all terms linear resp.\ quadratic in the trigonometric functions into single $\cos(\phi-\theta_1)$ resp.\ $\cos(2\phi-\theta_2)$-terms:
\begin{align}
C_{\alpha_2 \alpha_2}&(O_\beta\cos\phi - \bar{O}_{\alpha_1})^2 + C_{\alpha_1\alpha_1}(O_\beta\sin\phi - \bar{O}_{\alpha_2})^2 - 2C_{\alpha_1\alpha_2}(O_\beta\cos\phi - \bar{O}_{\alpha_1})(O_\beta\sin\phi - \bar{O}_{\alpha_2}) = \cr
&\rho_1 \cos (\phi - \theta_1) + \rho_2 \cos (2\phi - \theta_2) + \frac{1}{2}C_{\alpha_2 \alpha_2}O_\beta^2+\frac{1}{2}C_{\alpha_1 \alpha_1}O_\beta^2+ C_{\alpha_2 \alpha_2 } \bar{O}_{\alpha_1}^2+C_{\alpha_1 \alpha_1 } \bar{O}_{\alpha_2}^2-2C_{\alpha_1 \alpha_2 } \bar{O}_{\alpha_1}\bar{O}_{\alpha_2},\qquad
\end{align}
where we defined
\begin{align}
\rho_1\cos\theta_1 = 2 \big(C_{\alpha_1\alpha_2} \bar{O}_{\alpha_2} - C_{\alpha_2\alpha_2} \bar{O}_{\alpha_1}\big) O_\beta
\quad&,\quad
\rho_1\sin\theta_1 = 2 \big(C_{\alpha_1\alpha_2} \bar{O}_{\alpha_1} - C_{\alpha_1\alpha_1 } \bar{O}_{\alpha_2}\big) O_\beta \label{eq:rho1}\\
\rho_2\cos\theta_2 = \frac{1}{2}\big(C_{\alpha_2\alpha_2} - C_{\alpha_1 \alpha_1}\big) O_\beta^2
\quad&,\quad
\rho_2\sin\theta_2 = -C_{\alpha_1\alpha_2} O_\beta^2 \label{eq:rho2}.
\end{align}
This results in
\begin{align}
p_\beta^\textrm{G.f.a.}(O_\beta) = \frac{\Theta(O_\beta)~O_\beta}{\sqrt{\det (\Sigma_{\alpha_1,\alpha_2})}}  
&\exp\left[-\frac{C_{\alpha_2 \alpha_2}O_\beta^2+C_{\alpha_1 \alpha_1}O_\beta^2+ 2C_{\alpha_2 \alpha_2 } \bar{O}_{\alpha_1}^2+2C_{\alpha_1 \alpha_1 } \bar{O}_{\alpha_2}^2-4C_{\alpha_1 \alpha_2 } \bar{O}_{\alpha_1}\bar{O}_{\alpha_2}}{4\det (\Sigma_{\alpha_1,\alpha_2})}\right] \cr
&\times \int_0^{2\pi}\frac{\d\phi}{2\pi} \exp\left[-\frac{\rho_1\cos(\phi-\theta_1) + \rho_2\cos(2\phi-\theta_2)}{2\det(\Sigma_{\alpha_1,\alpha_2})}\right].
\label{eq:p(O_b)_middle-step}
\end{align}
To handle the terms in $\exp[\rho_j\cos(j\phi-\theta_j)/2\det(\Sigma_{\alpha_1,\alpha_2})]$, we follow the same trick as in Ref.~\cite{Borghini:2002vp} and write
\begin{equation}
\exp(z_j^*\textrm{e}^{\textrm{i}_{}j\phi} +z_{j\,}\textrm{e}^{-\textrm{i}_{}j\phi}) = \sum_{q=-\infty}^\infty\! \textrm{e}^{-\textrm{i}qj\phi} \bigg(\frac{z_j}{|z_j|}\bigg)^{\!\!q}I_q(2|z_j|),
\end{equation}
where $I_q$ is the modified Bessel of the first kind of order $q$, while
\begin{equation}
z_1 \equiv -\frac{\rho_1 e^\textrm{i}\theta_1}{4\det(\Sigma_{\alpha_1,\alpha_2})}
\quad,\quad
z_2 \equiv -\frac{\rho_2 e^\textrm{i}\theta_2}{4\det(\Sigma_{\alpha_1,\alpha_2})}.
\label{eq:z_j_vs_rho_j}
\end{equation}
This yields
\begin{equation}
\int_0^{2\pi}\frac{\d\phi}{2\pi}\exp\left[-\frac{\rho_1\cos(\phi-\theta_1) + \rho_2\cos(2\phi-\theta_2)}{2\det(\Sigma_{\alpha_1,\alpha_2})}\right] = 
\sum_{q_1,q_2=-\infty}^\infty\!\bigg(\frac{z_1}{|z_1|}\bigg)^{\!\!q_1} I_{q_1}(2|z_1|)
\bigg(\frac{z_2}{|z_2|}\bigg)^{\!\!q_2}I_{q_2}(2|z_2|)
\!\int_0^{2\pi}\frac{\d\phi}{2\pi}\,\textrm{e}^{-\textrm{i}(q_1+2q_2)\phi},
\end{equation}
so that the integral over $\phi$ is now trivial. 
Going back to Eq.~\eqref{eq:p(O_b)_middle-step}, we thus find 
\begin{align}
p_\beta^\textrm{G.f.a.}(O_\beta)&=
\frac{\Theta(O_\beta)~O_\beta}{\sqrt{\det(\Sigma_{\alpha_1,\alpha_2})}}  
\exp\left[-\frac{C_{\alpha_2 \alpha_2}O_\beta^2 + C_{\alpha_1\alpha_1}O_\beta^2 + 2C_{\alpha_2\alpha_2}\bar{O}_{\alpha_1}^2 + 2C_{\alpha_1\alpha_1}\bar{O}_{\alpha_2}^2 - 4C_{\alpha_1\alpha_2}\bar{O}_{\alpha_1}\bar{O}_{\alpha_2}}%
{4\det (\Sigma_{\alpha_1,\alpha_2})}\right]\cr
&\quad\times \sum_{q=-\infty}^\infty \bigg(\frac{z_1^2z_2}{|z_1^2z_2|}\bigg)^{\!\!q} I_{2q}(2|z_1|)I_{q}(2|z_2|). \label{eq:p_b(O_b)_Gaussian}
\end{align}
There are two special cases in which all terms in the sum over $q$ vanish but one. 
When $O_{\alpha_1}$ and $O_{\alpha_2}$ are independent ($C_{\alpha_1\alpha_2}=0$) and have the same variance ($C_{\alpha_1\alpha_1}=C_{\alpha_2\alpha_2}$), Eq.~\eqref{eq:rho2} yields $\rho_2 = 0$ and thus [Eq.~\eqref{eq:z_j_vs_rho_j}] $z_2=0$. 
In turn, in the case of vanishing average values $\bar{O}_1=\bar{O}_2=0$, Eqs.~\eqref{eq:rho1} and \eqref{eq:z_j_vs_rho_j} give $z_1=0$.
In either case, only the term $q=0$ in the sum in the second line of Eq.~\eqref{eq:p_b(O_b)_Gaussian} is nonzero, and the sum itself simply reduces to a single modified Bessel function $I_0$, resulting in a simple form for the probability distribution of $O_\beta$.

\subsection{Moments of the probability distributions of eccentricities and flow coefficients}
\label{app:moments_p(eps,v)}

In Table~\ref{tab:moments_eps_vn_distributions} we show the first moments of the probability distributions $p(\varepsilon_n)$ and $p(v_n)$ shown at the top and on the right of every panel in Figs.~\ref{fig:2d_prob_dist_Glauber_b0}--\ref{fig:2d_prob_dist_Saturation_b9}.
We give the average $\mu$, variance $\sigma^2$, skewness $\gamma_1$ and excess kurtosis $\gamma_2$ for the distributions computed within the Gaussian fluctuation approximation (G.f.a.)\ and for those obtained from the sampled events.

In collisions at vanishing impact parameter and for both initial-state models, these moments confirm the visual impression of Figs.~\ref{fig:2d_prob_dist_Glauber_b0} and \ref{fig:2d_prob_dist_Saturation_b0}, namely the very good agreement between the probability distributions from the event sample and those computed in the G.f.a., which are of the Bessel--Gaussian type. 
The only sizable discrepancies are for the excess kurtosis $\gamma_2$ (of basically all $\varepsilon_n$ and $v_n$) and to a lesser extent the variance of $\varepsilon_5$ in the Saturation model. 

At $b=9$~fm, the Gaussian fluctuation approximation still provides a very good description of the mean and variance of the distributions of $\varepsilon_n$ and $v_n$, although less good for the variances of $\varepsilon_4$ and $\varepsilon_5$~--- which is correlated to the mismatch seen for these observables in the top left panels of Figs.~\ref{fig:co_variances_convergence_Glauber_b0_IS_FS_modes_and_sampled} and \ref{fig:co_variances_convergence_Saturation_b0_IS_FS_modes_and_sampled} --- and for the mean value of $v_1$. 
Regarding the higher moments, the G.f.a.\ generally reproduces the skewness values of the event samples rather well --- with the important exception of $\varepsilon_2$, for which the wrong sign is predicted, and $\varepsilon_4$. 
Going to $\gamma_2$, larger departures between the event-sample and G.f.a.\ values occur more often. 
Part of the explanation may be that the (excess) kurtosis is more sensitive to the constraints $\varepsilon_n\leq 1$, $v_n\leq 1$, which are necessarily there in the event samples but not accounted for in the Gaussian-fluctuation approach. 

\begin{table*}
\caption{\label{tab:moments_eps_vn_distributions} Average $\mu$, variance $\sigma^2$, skewness $\gamma_1$ and excess kurtosis $\gamma_2$ of the probability distributions $p(\varepsilon_n)$ and $p(v_n)$ calculated with the mode-by-mode approach within the Gaussian fluctuation approximation (G.f.a.)\ and extracted from random samples of 8192 events.}
\begin{ruledtabular}
\begin{tabular}{c|c|c|c|c|c}
moment & $O_\alpha$ & Glauber $b=0$ & Glauber $b=9\;\mathrm{fm}$ & Saturation $b=0$ & Saturation $b=9\;\mathrm{fm}$ \\
\hline
$\mu^{\mathrm{G.f.a.}}$ & $\varepsilon_1$ (\%) $\ |\ $ $v_1$ (\%) & $5.36$ $\ |\ $  $1.34$ & $11.47$ $\ |\ $  $2.27$ & $4.51$ $\ |\ $  $1.15$ & $12.00$ $\ |\ $  $2.42$ \\
$\mu$ & $\varepsilon_1$ (\%) $\ |\ $ $v_1$ (\%) & $5.35$ $\ |\ $  $1.39$ & $10.97$ $\ |\ $  $1.87$ & $4.59$ $\ |\ $  $1.19$ & $11.92$ $\ |\ $  $2.00$\\
\hline
$\mu^{\mathrm{G.f.a.}}$ & $\varepsilon_2$ (\%) $\ |\ $ $v_2$ (\%) & $7.72$ $\ |\ $  $1.95$ & $33.60$ $\ |\ $  $6.87$ & $7.45$ $\ |\ $  $1.85$ & $43.21$ $\ |\ $  $9.02$\\
$\mu$ & $\varepsilon_2$ (\%) $\ |\ $ $v_2$ (\%) & $7.70$ $\ |\ $  $1.93$ & $33.82$ $\ |\ $  $6.72$ & $7.54$ $\ |\ $  $1.84$ & $43.78$ $\ |\ $  $8.80$\\
\hline
$\mu^{\mathrm{G.f.a.}}$ & $\varepsilon_3$ (\%) $\ |\ $ $v_3$ (\%) & $8.00$ $\ |\ $  $1.31$ & $21.57$ $\ |\ $  $2.68$ & $8.07$ $\ |\ $  $1.33$ & $21.55$ $\ |\ $  $2.65$\\
$\mu$ & $\varepsilon_3$ (\%) $\ |\ $ $v_3$ (\%) & $8.13$ $\ |\ $  $1.29$ & $20.65$ $\ |\ $  $2.47$ & $8.35$ $\ |\ $  $1.35$ & $21.13$ $\ |\ $  $2.47$\\
\hline
$\mu^{\mathrm{G.f.a.}}$ & $\varepsilon_4$ (\%) $\ |\ $ $v_4$ (\%) & $8.42$ $\ |\ $  $0.65$ & $26.14$ $\ |\ $  $1.27$ & $8.77$ $\ |\ $  $0.69$ & $31.84$ $\ |\ $  $1.44$\\
$\mu$ & $\varepsilon_4$ (\%) $\ |\ $ $v_4$ (\%) & $8.52$ $\ |\ $  $0.62$ & $25.36$ $\ |\ $  $1.13$ & $9.07$ $\ |\ $  $0.70$ & $33.38$ $\ |\ $  $1.37$\\
\hline
$\mu^{\mathrm{G.f.a.}}$ & $\varepsilon_5$ (\%) $\ |\ $ $v_5$ (\%) & $8.76$ $\ |\ $  $0.23$ & $28.64$ $\ |\ $  $0.41$ & $9.21$ $\ |\ $  $0.26$ & $30.64$ $\ |\ $  $0.53$\\
$\mu$ & $\varepsilon_5$ (\%) $\ |\ $ $v_5$ (\%) & $9.16$ $\ |\ $  $0.24$ & $25.55$ $\ |\ $  $0.43$ & $10.24$ $\ |\ $  $0.30$ & $30.38$ $\ |\ $  $0.58$\\
\hline\hline
$(\sigma^2)^\textrm{G.f.a.}$ & $\varepsilon_1$ (\textperthousand) $\ |\ $ $v_1$ (\textperthousand) & $0.78$ $\ |\ $  $0.05$ & $3.79$ $\ |\ $  $0.14$ & $0.55$ $\ |\ $  $0.04$ & $4.50$ $\ |\ $  $0.16$ \\
$\sigma^{2}$ & $\varepsilon_1$ (\textperthousand) $\ |\ $ $v_1$ (\textperthousand) & $0.79$ $\ |\ $  $0.05$ & $4.03$ $\ |\ $  $0.10$ & $0.57$ $\ |\ $  $0.04$ & $4.88$ $\ |\ $  $0.11$ \\
\hline
$(\sigma^2)^\textrm{G.f.a.}$ & $\varepsilon_2$ (\textperthousand) $\ |\ $ $v_2$ (\textperthousand) & $1.63$ $\ |\ $  $0.10$ & $17.05$ $\ |\ $  $0.73$ & $1.52$ $\ |\ $  $0.09$ & $16.33$ $\ |\ $  $0.82$ \\
$\sigma^{2}$ & $\varepsilon_2$ (\textperthousand) $\ |\ $ $v_2$ (\textperthousand) & $1.58$ $\ |\ $  $0.10$ & $15.12$ $\ |\ $  $0.71$ & $1.55$ $\ |\ $  $0.09$ & $15.54$ $\ |\ $  $0.83$ \\
\hline
$(\sigma^2)^\textrm{G.f.a.}$ & $\varepsilon_3$ (\textperthousand) $\ |\ $ $v_3$ (\textperthousand) & $1.75$ $\ |\ $  $0.05$ & $12.70$ $\ |\ $  $0.20$ & $1.78$ $\ |\ $  $0.05$ & $12.69$ $\ |\ $  $0.19$ \\
$\sigma^{2}$ & $\varepsilon_3$ (\textperthousand) $\ |\ $ $v_3$ (\textperthousand) & $1.76$ $\ |\ $  $0.04$ & $11.14$ $\ |\ $  $0.16$ & $1.88$ $\ |\ $  $0.05$ & $11.90$ $\ |\ $  $0.17$ \\
\hline
$(\sigma^2)^\textrm{G.f.a.}$ & $\varepsilon_4$ (\textperthousand) $\ |\ $ $v_4$ (\textperthousand) & $1.94$ $\ |\ $  $0.01$ & $18.53$ $\ |\ $  $0.04$ & $2.10$ $\ |\ $  $0.01$ & $25.29$ $\ |\ $  $0.06$ \\
$\sigma^{2}$ & $\varepsilon_4$ (\textperthousand) $\ |\ $ $v_4$ (\textperthousand) & $1.93$ $\ |\ $  $0.01$ & $15.12$ $\ |\ $  $0.03$ & $2.18$ $\ |\ $  $0.01$ & $21.00$ $\ |\ $  $0.05$ \\
\hline
$(\sigma^2)^\textrm{G.f.a.}$ & $\varepsilon_5$ (\textperthousand) $\ |\ $ $v_5$ (\textperthousand) & $2.10$ $\ |\ $  $0.001$ & $22.38$ $\ |\ $  $0.005$ & $2.32$ $\ |\ $  $0.002$ & $25.57$ $\ |\ $  $0.008$ \\
$\sigma^{2}$ & $\varepsilon_5$ (\textperthousand) $\ |\ $ $v_5$ (\textperthousand) & $2.27$ $\ |\ $  $0.002$ & $16.49$ $\ |\ $  $0.006$ & $2.85$ $\ |\ $  $0.002$ & $21.94$ $\ |\ $  $0.01$ \\
\hline\hline
$\gamma^{\mathrm{G.f.a.}}_1$ & $\varepsilon_1$ $\ |\ $  $v_1$ & $0.63$ $\ |\ $  $0.63$ & $0.72$ $\ |\ $  $0.67$ & $0.63$ $\ |\ $  $0.63$ & $0.84$ $\ |\ $  $0.66$\\
$\gamma_1$ & $\varepsilon_1$ $\ |\ $  $v_1$ & $0.65$ $\ |\ $  $0.64$ & $0.93$ $\ |\ $  $0.71$ & $0.61$ $\ |\ $  $0.60$ & $0.94$ $\ |\ $  $0.68$\\
\hline
$\gamma^{\mathrm{G.f.a.}}_1$ & $\varepsilon_2$ $\ |\ $  $v_2$ & $0.63$ $\ |\ $  $0.63$ & $0.18$ $\ |\ $  $0.19$ & $0.63$ $\ |\ $  $0.63$ & $0.05$ $\ |\ $  $0.07$\\
$\gamma_1$ & $\varepsilon_2$ $\ |\ $  $v_2$ & $0.61$ $\ |\ $  $0.64$ & $-0.07$ $\ |\ $  $0.12$ & $0.61$ $\ |\ $  $0.63$ & $-0.24$ $\ |\ $  $0.05$\\
\hline
$\gamma^{\mathrm{G.f.a.}}_1$ & $\varepsilon_3$ $\ |\ $  $v_3$ & $0.63$ $\ |\ $  $0.63$ & $0.63$ $\ |\ $  $0.63$ & $0.63$ $\ |\ $  $0.63$ & $0.63$ $\ |\ $  $0.63$\\
$\gamma_1$ & $\varepsilon_3$ $\ |\ $  $v_3$ & $0.59$ $\ |\ $  $0.60$ & $0.58$ $\ |\ $  $0.60$ & $0.64$ $\ |\ $  $0.65$ & $0.54$ $\ |\ $  $0.60$\\
\hline
$\gamma^{\mathrm{G.f.a.}}_1$ & $\varepsilon_4$ $\ |\ $  $v_4$ & $0.63$ $\ |\ $  $0.63$ & $0.62$ $\ |\ $  $0.62$ & $0.63$ $\ |\ $  $0.63$ & $0.50$ $\ |\ $  $0.60$\\
$\gamma_1$ & $\varepsilon_4$ $\ |\ $  $v_4$ & $0.61$ $\ |\ $  $0.62$ & $0.38$ $\ |\ $  $0.64$ & $0.60$ $\ |\ $  $0.60$ & $0.13$ $\ |\ $  $0.69$\\
\hline
$\gamma^{\mathrm{G.f.a.}}_1$ & $\varepsilon_5$ $\ |\ $  $v_5$ & $0.63$ $\ |\ $  $0.63$ & $0.63$ $\ |\ $  $0.63$ & $0.63$ $\ |\ $  $0.63$ & $0.62$ $\ |\ $  $0.63$\\
$\gamma_1$ & $\varepsilon_5$ $\ |\ $  $v_5$ & $0.62$ $\ |\ $  $0.60$ & $0.48$ $\ |\ $  $0.83$ & $0.61$ $\ |\ $  $0.60$ & $0.39$ $\ |\ $  $0.72$\\
\hline\hline
$\gamma^{\mathrm{G.f.a.}}_2$ & $\varepsilon_1$ $\ |\ $  $v_1$ & $0.25$ $\ |\ $  $0.25$ & $0.51$ $\ |\ $  $0.35$ & $0.25$ $\ |\ $  $0.25$ & $0.80$ $\ |\ $  $0.32$\\
$\gamma_2$ & $\varepsilon_1$ $\ |\ $  $v_1$ & $0.33$ $\ |\ $  $0.25$ & $1.01$ $\ |\ $  $0.41$ & $0.16$ $\ |\ $  $0.16$ & $0.99$ $\ |\ $  $0.38$\\
\hline
$\gamma^{\mathrm{G.f.a.}}_2$ & $\varepsilon_2$ $\ |\ $  $v_2$ & $0.25$ $\ |\ $  $0.25$ & $-0.18$ $\ |\ $  $-0.18$ & $0.25$ $\ |\ $  $0.25$ & $-0.06$ $\ |\ $  $-0.09$\\
$\gamma_2$ & $\varepsilon_2$ $\ |\ $  $v_2$ & $0.19$ $\ |\ $  $0.30$ & $-0.42$ $\ |\ $  $-0.33$ & $0.13$ $\ |\ $  $0.15$ & $-0.18$ $\ |\ $  $-0.17$\\
\hline
$\gamma^{\mathrm{G.f.a.}}_2$ & $\varepsilon_3$ $\ |\ $  $v_3$ & $0.25$ $\ |\ $  $0.25$ & $0.24$ $\ |\ $  $0.25$ & $0.25$ $\ |\ $  $0.25$ & $0.25$ $\ |\ $  $0.25$\\
$\gamma_2$ & $\varepsilon_3$ $\ |\ $  $v_3$ & $0.15$ $\ |\ $  $0.17$ & $0.11$ $\ |\ $  $0.14$ & $0.32$ $\ |\ $  $0.34$ & $-0.03$ $\ |\ $  $0.07$\\
\hline
$\gamma^{\mathrm{G.f.a.}}_2$ & $\varepsilon_4$ $\ |\ $  $v_4$ & $0.25$ $\ |\ $  $0.25$ & $0.21$ $\ |\ $  $0.23$ & $0.25$ $\ |\ $  $0.25$ & $-0.03$ $\ |\ $  $0.16$\\
$\gamma_2$ & $\varepsilon_4$ $\ |\ $  $v_4$ & $0.25$ $\ |\ $  $0.22$ & $-0.31$ $\ |\ $  $0.29$ & $0.18$ $\ |\ $  $0.20$ & $-0.55$ $\ |\ $  $0.38$\\
\hline
$\gamma^{\mathrm{G.f.a.}}_2$ & $\varepsilon_5$ $\ |\ $  $v_5$ & $0.25$ $\ |\ $  $0.25$ & $0.21$ $\ |\ $  $0.25$ & $0.25$ $\ |\ $  $0.25$ & $0.16$ $\ |\ $  $0.25$\\
$\gamma_2$ & $\varepsilon_5$ $\ |\ $  $v_5$ & $0.29$ $\ |\ $  $0.20$ & $-0.08$ $\ |\ $  $0.78$ & $0.09$ $\ |\ $  $0.15$ & $-0.33$ $\ |\ $  $0.33$\\
\end{tabular}
\end{ruledtabular}
\end{table*}

\end{document}